\documentclass{article}

\usepackage{arxiv}

\usepackage[cp1252]{inputenc}
\usepackage{textcomp}
\usepackage{amsmath}
\usepackage{amssymb}
\usepackage{graphicx}
\usepackage{array}
\usepackage[abs]{overpic}
\usepackage{color}
\usepackage{soul}
\usepackage{amsthm}
\usepackage{units}
\usepackage{pifont}
\usepackage{multirow}
\usepackage{stackengine}
\usepackage{hyperref}

\usepackage[FIGTOPCAP]{subfigure}

\usepackage{tikz}

\usetikzlibrary{angles,quotes}
\usepackage{gensymb}
\usepackage{pgfplots}
\usepgfplotslibrary{patchplots}
\pgfplotsset{compat=newest}

\newcommand*\Eval[3]{\left.#1\right\rvert_{#2}^{#3}}

\graphicspath{{figures/}}
\usepackage{booktabs}       
\usepackage{amsfonts}       
\usepackage{nicefrac}       
\usepackage{microtype}      
\usepackage{lipsum}
\usepackage{graphicx}
\graphicspath{{figures/}}

\usepackage[square,numbers]{natbib}
\usepackage{textcomp}
\usepackage{amsmath}
\usepackage{amssymb}
\usepackage{graphicx, subfigure}
\usepackage{array}
\usepackage[abs]{overpic}
\usepackage{tabularx} 
\usepackage{multirow}
\usepackage{hyperref}

\makeatletter

\usepackage{color}
\usepackage{soul}
\usepackage{amsthm}

\graphicspath{{figures/}}

\newcommand{\bs}{\boldsymbol}

\title{Physics-informed machine learning\\ in asymptotic homogenization of elliptic equations}

\author{
 Celal~Soyarslan\\
  Chair of Nonlinear Solid Mechanics \\ University of Twente, The Netherlands \\
  \texttt{c.soyarslan@utwente.nl} \\ 
   \And
 Marc~Pradas\\
  School of Mathematics and Statistics\\ The Open University, United Kingdom \\
  \texttt{marc.pradas@open.ac.uk} \\ 
}

\begin{document}
\maketitle

\begin{abstract}
We apply physics-informed neural networks (PINNs) to first-order two-scale periodic asymptotic homogenization of the property tensor in a generic elliptic equation. The problem of lack of differentiability of property tensors at the sharp phase interfaces is circumvented by making use of diffuse interface formulation. Periodic boundary conditions are incorporated strictly, through the introduction of an input-transfer layer (Fourier feature mapping), in which the sine and cosine of the inner product of position vectors and reciprocal lattice vectors are considered.
This, together with the absence of Dirichlet boundary conditions, results in a lossless boundary condition application. The only loss terms are then due to the differential equation itself, which removes the necessity of scaling the loss entries. In demonstrating the formulation's versatility based on the reciprocal lattice vectors, crystalline arrangements defined with Bravais lattices are used.  We also show that considering integer multiples of the reciprocal basis in the Fourier mapping leads to improved convergence of high-frequency functions.  We consider applications in one, two, and three dimensions. Periodic composites, composed of embeddings of monodisperse inclusions in the form of disks/spheres in the two-/three-dimensional matrix, are considered. For demonstration purposes, stochastic monodisperse disk arrangements are also considered.
\end{abstract}
\section{Introduction}
Machine learning (ML) has shown remarkable success in computer vision~\cite{Estevaetal2021}, medical diagnosis~\cite{RichensJ2020}, natural language processing~\cite{Leglazetal2021}, and financial applications~\cite{Dixonetal2020}. Following the pioneering work of Lagaris \emph{et al.,}~\cite{Lagaris1998}, the universal approximation property  of deep neural networks~\cite{Cybenko1989,HORNIK1989359,HORNIK1990551} has been used as an alternative method in scientific computation for solving partial differential equations (PDEs) in physics. In particular, physics-informed neural networks (PINNs) allow approximating the solution of PDEs using their strong form without the need for a weak form derivation and discretization, as is the case, e.g., for the finite elements method (FEM)~\cite{Lu_2021}. Unlike standard discretization techniques, the solution obtained from an artificial neural network (ANN) is differentiable with a closed analytic form that allows subsequent calculations~\cite{Lagaris1998}. That is, the function value and its derivatives can be computed exactly at any point of the problem domain with a trained network. In such a meshless approach, the solution is  approximated in a deep neural network whose hyper-parameters are found by minimizing the loss functions associated with the PDE. Hence, the problem of finding the solution to the PDE is  converted to an optimization problem.

The use of ANNs is nowadays extensive, from solving  integro-differential equations to fractional and stochastic differential equations, see, e.g.,~\cite{Lu_2021,ZHANG2019108850,Pang_2019}; and it is expected to become more prominent with the advent of accessible, scalable, and modular deep learning frameworks (such as, e.g., Tensorflow~\cite{Tensorflow2015-whitepaper} and PyTorch ~\cite{pytorch2019}), as well as  specialized libraries, e.g., DeepXDE~\cite{ Lu_2021} and SciANN~\cite{Raissi2019}. The increased efficiency of ANNs that results from reduced training time mediated by the improvement of parallel architectures is also exploited in applications in solid mechanics~\cite{HAGHIGHAT2021113741}, amongst others. For an extensive survey on the state of the art of  PINNs and additional libraries, see~\cite{ Cuomo2022}.
In comparison to  classical ML approaches, where the deep learning model is trained against a dataset with the aim of error minimization with respect to the dataset,  the loss function to be minimized in PINNs is composed of residual norms of the governing differential equation and associated initial, and boundary conditions (BCs) \cite{Lagaris1998,Lagaris2000,SIRIGNANO20181339}. Therefore, a key and nontrivial task in solving PDEs with ANNs is the application of appropriate BCs, the methods for which can be divided into being exact or approximate.

An extensive study on approximate penalty-based methods for the imposition of Dirichlet, Neumann, Robin, and periodic boundary conditions is given in~\cite{Chen2020review}. These methods penalize the loss term (residual) due to boundary conditions with a coefficient. However, this term, which is found through a trial and error process, influences the network's convergence and its accuracy~\cite{ Chen2020}. On the other hand, the exact imposition of BCs does not suffer from these weaknesses. For problems with  irregular boundaries, a radial basis function network can be used for the exact satisfaction of the boundary conditions~\cite{Lagaris2000}.

An essential class of BCs is periodic boundary conditions (PBCs), which are used, for example, in applications involving periodic structures in computational engineering~\cite{BARGMANN2018}, such as physical problems related to crystals; or periodic unit cell problems that are solved within asymptotic periodic homogenization~ \cite{Soyarslanetal2018,SOYARSLAN2019103098,Torquato2002, Fish2014},  amongst many others. When PBCs are applied in ANNs, an extensively used method is the penalty method that yields an approximate BC imposition. In \cite{Chen2020review}, periodicity is satisfied using penalty terms multiplied by prescribed parameters in the loss function. In this sense, the loss term represents the residual norms of the periodic conditions for the function and its first derivative. We should reiterate that penalty parameters are problematic, something generally known for penalty-based methods. For example, in shallow neural networks with one hidden layer, which use $\cos$ activation functions to create Fourier Neural Networks~\cite{NgomOana2020}, a key problem is the determination of the penalty coefficients and the requirement of knowing the period of the estimated function for the activation and loss functions. Other works on Fourier networks yielding Fourier series approximation properties are~\cite{ GallantWhite1988, Liu2013, Zhumekenovetal2019}. On the other hand, imposing exact C$^\infty$, or C$^k$ ($k>0$), PBCs can be achieved by using  an additional layer that transfers the input to either C$^\infty$ or C$^k$ periodic functions~\cite{DONG2021110242}. An alternative approach is to apply instead a transfer function to the solution that requires periodic conditions~\cite{Gokuzummca24020040}, which leads to  neural networks that satisfy the periodic boundary conditions of the microscopic problem. The work we present here is partly based on this approach.

In this work, we are interested in determining the effective (macroscopic) properties of composite material systems whose micro- (constituent-level) psychics is given by a generic elliptic PDE. It is worth noting that there are analytically derived bounds for this type of problem, e.g., (arithmetic) Voigt and the (harmonic) Reuss averages, as well as effective-medium theories such as the Maxwell, self-consistent, and differential effective-medium approximations. However,  these approximations use limited microstructural descriptors, such as the phase volume fraction and shape, and hence their validity is limited to dilute composite systems where the constituents possess low contrasting physical properties. Only full-field computational micro-macro transfer techniques provide the required accuracy for materials with random phase distributions creating phase interactions and high property contrast and to this end,
the finite element method has been conventionally used over the last years, see e.g.~\cite{LUKKASSEN1995519,Soyarslanetal2018,SOYARSLAN2019103098}.

We work on physics-informed deep learning with the exact imposition of PBCs for problems with arbitrary periodicity. To this end, we use the scientific ML and physics-informed library DeepXDE~\cite{Lu_2021}. All the problems studied here include cuboidal domains with orthogonal phases, and we impose PBCs by  adding a C$^\infty$ periodic layer to the neural network with $\cos$ and $\sin$ functions. Therefore, the training phase of the  neural network does not involve loss terms corresponding to the boundary conditions, and there is no need to consider any penalty factor.  This naturally leads to the deep neural network solution and its derivative being periodic. One advantage of our approach is that the emerging solution is valid outside the training region~\cite{ NgomOana2020}, which to our best knowledge,  has only been investigated partially for rectilinear domains in~\cite{DONG2021110242}. A detailed review of machine learning-mediated multiscale modeling and simulation is reported in \cite{Bishara2022}.

We consider first-order asymptotic homogenization problems that arise in a broad range of physical phenomena and are described in terms of elliptic PDEs of divergence type. Our numerical approach has two important elements that are particularly relevant to this type of problem:
\begin{enumerate}
    \item As we use reciprocal unit cell vectors,  our method can be applied to any form of crystal in 2D or 3D,  not necessarily rectilinear. This is crucial in homogenization problems as primitive unit cells of micro-structures can take various geometries. 
    \item Solutions of unit cell problems arising in homogenization are computed up to a constant. This is ideal for our PINNs environment as it allows us to use the loss term composed only of the PDE residual term without the need for any additional scaling.
\end{enumerate}

The rest of the paper is structured as follows. Section \ref{Sec:Theory} provides a summary of the theory behind first-order asymptotic homogenization and we outline the working principles of the PINNs. In Section \ref{Sec:Apps} we apply our approach to a selection of problems in 1D, 2D, and 3D. The results are compared for both local field distributions and the computed macroscopic property tensors. We show that using PINNs in asymptotic homogenization of property tensors provides highly accurate results with ample time gains, especially once complex phase geometries with rapid fluctuations are considered.
\section{Theory}
\label{Sec:Theory}
\subsection{Crystal Arrangements, Bravais Lattices, and Lattice-Periodic Functions}
\begin{figure}[htb!]
\centering
\subfigure[1D lattice]{
\begin{minipage}{1.0\textwidth}
\centering
\begin{tikzpicture}[scale=0.9]

\coordinate[] (T0) at (-1.5,0.);
\coordinate[] (T1) at (0.0,0.);
\coordinate[] (T2) at (1.5,0.);
\coordinate[] (T3) at (3.0,0.);
\coordinate[] (T4) at (4.5,0.);
\coordinate[] (T5) at (6.0,0.);
\coordinate[] (T6) at (7.5,0.);
\coordinate[] (T7) at (9.0,0.);


\draw[white,fill={rgb:red,221;green,109;blue,16},
opacity=0.75, line width=0.2mm] (0.75,0.075) -- (0.75,-0.075) -- (2.25,-0.075) -- (2.25,0.075) -- (0.75,0.075);

\draw[white,fill={rgb:red,1;green,103;blue,143}, opacity=0.75, line width=0.1mm] (4.5,0.075) -- (4.5,-0.075) -- (6.,-0.075) -- (6.,0.075) -- (4.5,0.075);

\draw[lightgray, fill=lightgray, line width=1mm , opacity=1.0] (T0) circle (2pt);
\draw[black, fill=black,line width=0.1mm , opacity=1.0] (T0) circle (0.2pt);
\draw[lightgray, fill=lightgray, line width=1mm , opacity=1.0] (T1) circle (2pt);
\draw[black, fill=black,line width=0.1mm , opacity=1.0] (T1) circle (0.2pt);
\draw[lightgray, fill=lightgray, line width=1mm , opacity=1.0] (T2) circle (2pt);
\draw[black, fill=black,line width=0.1mm , opacity=1.0] (T2) circle (0.2pt);
\draw[lightgray, fill=lightgray, line width=1mm , opacity=1.0] (T3) circle (2pt);
\draw[black, fill=black,line width=0.1mm , opacity=1.0] (T3) circle (0.2pt);
\draw[lightgray, fill=lightgray, line width=1mm , opacity=1.0] (T4) circle (2pt);
\draw[black, fill=black,line width=0.1mm , opacity=1.0] (T4) circle (0.2pt);
\draw[lightgray, fill=lightgray, line width=1mm , opacity=1.0] (T5) circle (2pt);
\draw[black, fill=black,line width=0.1mm , opacity=1.0] (T5) circle (0.2pt);
\draw[lightgray, fill=lightgray, line width=1mm , opacity=1.0] (T6) circle (2pt);
\draw[black, fill=black,line width=0.1mm , opacity=1.0] (T6) circle (0.2pt);
\draw[lightgray, fill=lightgray, line width=1mm , opacity=1.0] (T7) circle (2pt);
\draw[black, fill=black,line width=0.1mm , opacity=1.0] (T7) circle (0.2pt);
\draw [->] (1.5,0) -- (3.,0);

\draw[black, ->]  (1.5,0) -- node[ below] {  $\boldsymbol a$} (3.,0);

\node[black, text width=5cm, align=center, text opacity=1] at (1.5,0.5)
    {{$\mathcal{B}_1$}};
\node[black, text width=5cm, align=center, text opacity=1] at (5.25,0.5)
    {{$\mathcal{B}_2$}};

\draw[black, fill=black,line width=0.1mm , opacity=1.0] (1.5,0.) circle (1.5pt);

\end{tikzpicture} 
\end{minipage}
}\\
\subfigure[2D square lattice]{
\begin{minipage}{0.30\textwidth}
\centering
\begin{tikzpicture}[scale=0.9]
\coordinate[] (DA_b) at (2.5,1.25);
\coordinate[] (DB_b) at (3.75,1.25);
\coordinate[] (DC_b) at (2.5,2.5);
\coordinate[] (DD_b) at (3.75,2.5);
\coordinate[] (DA_r) at (0.625,0.625);
\coordinate[] (DB_r) at (1.875,0.625);
\coordinate[] (DC_r) at (0.625,1.875);
\coordinate[] (DD_r) at (1.875,1.875);
\coordinate[] (A_l) at (1.25,1.25);
\coordinate[] (B_l) at (1.25,2.5);
\coordinate[] (C_l) at (2.5,1.25);
\coordinate[] (A_lb) at (0.0,0.0);
\coordinate[] (B_lb) at (1.25,0.0);
\coordinate[] (C_lb) at (0.0,1.25);

\draw[black,dashed,fill={rgb:red,1;green,103;blue,143}, opacity=0.75, line width=0.1mm] (DA_b) -- (DB_b) -- (DD_b) -- (DC_b) -- (DA_b);
\draw[black,dashed,fill={rgb:red,221;green,109;blue,16}, opacity=0.75, line width=0.1mm] (DA_r) -- (DB_r) -- (DD_r) -- (DC_r) -- (DA_r);

\draw[lightgray, fill=lightgray, line width=1mm , opacity=1.0] (0,0) circle (2pt);
\draw[black, fill=black,line width=0.1mm , opacity=1.0] (0,0) circle (0.2pt);

\draw[lightgray, fill=lightgray, line width=1mm , opacity=1.0] (1.25,0) circle (2pt);
\draw[black, fill=black,line width=0.1mm , opacity=1.0] (1.25,0) circle (0.2pt);

\draw[lightgray, fill=lightgray, line width=1mm , opacity=1.0] (2.5,0) circle (2pt);
\draw[black, fill=black,line width=0.1mm , opacity=1.0] (2.5,0) circle (0.2pt);

\draw[lightgray, fill=lightgray, line width=1mm , opacity=1.0] (3.75,0) circle (2pt);
\draw[black, fill=black,line width=0.1mm , opacity=1.0] (3.75,0) circle (0.2pt);

\draw[lightgray, fill=lightgray, line width=1mm , opacity=1.0] (0.0,1.25) circle (2pt);
\draw[black, fill=black,line width=0.1mm , opacity=1.0] (0.0,1.25) circle (0.2pt);

\draw[lightgray, fill=lightgray, line width=1mm , opacity=1.0] (1.25,1.25) circle (2pt);
\draw[black, fill=black,line width=0.1mm , opacity=1.0] (1.25,1.25) circle (0.2pt);

\draw[lightgray, fill=lightgray, line width=1mm , opacity=1.0] (2.5,1.25) circle (2pt);
\draw[black, fill=black,line width=0.1mm , opacity=1.0] (2.5,1.25) circle (0.2pt);

\draw[lightgray, fill=lightgray, line width=1mm , opacity=1.0] (3.75,1.25) circle (2pt);
\draw[black, fill=black,line width=0.1mm , opacity=1.0] (3.75,1.25) circle (0.2pt);

\draw[lightgray, fill=lightgray, line width=1mm , opacity=1.0] (0.0,2.5) circle (2pt);
\draw[black, fill=black,line width=0.1mm , opacity=1.0] (0.0,2.5) circle (0.2pt);

\draw[lightgray, fill=lightgray, line width=1mm , opacity=1.0] (1.25,2.5) circle (2pt);
\draw[black, fill=black,line width=0.1mm , opacity=1.0] (1.25,2.5) circle (0.2pt);

\draw[lightgray, fill=lightgray, line width=1mm , opacity=1.0] (2.5,2.5) circle (2pt);
\draw[black, fill=black,line width=0.1mm , opacity=1.0] (2.5,2.5) circle (0.2pt);

\draw[lightgray, fill=lightgray, line width=1mm , opacity=1.0] (3.75,2.5) circle (2pt);
\draw[black, fill=black,line width=0.1mm , opacity=1.0] (3.75,2.5) circle (0.2pt);

\draw [->] (A_l) -- (B_l);
\draw [->] (A_l) -- (C_l);
\draw [->] (A_lb) -- (B_lb);
\draw [->] (A_lb) -- (C_lb);

\draw[black, ->]  (A_l) -- node[above left of=B_l, node distance=0.25in] {  $\boldsymbol a_2$} (B_l);
\draw[black,->]  (A_l) -- node[below right of=C_l, node distance=0.25in] { $\boldsymbol  a_1$} (C_l);

\coordinate (a) at (1,4);
\coordinate (b) at (3,5);
\coordinate (c) at (5,4);
\draw pic[draw,fill=black!30,angle radius=0.6cm,"{ $\varphi$}" shift={(0mm,0mm)}] {angle=B_lb--A_lb--C_lb};
\draw[black, ->]  (A_lb) -- node[below] { $\boldsymbol c_1$} (B_lb);
\draw[black,->]  (A_lb) -- node[left] { $\boldsymbol c_2$} (C_lb);

\node[white, text width=5cm, align=center, text opacity=1] at (1.25,0.9)
    {{$\mathcal{S}_1$}};
\node[white, text width=5cm, align=center, text opacity=1] at (3.125,1.875)
    {{$\mathcal{S}_2$}};

\draw[black, fill=black,line width=0.1mm , opacity=1.0] (A_l) circle (1.5pt);

\draw[black, fill=black,line width=0.1mm , opacity=1.0] (A_lb) circle (1.5pt);

\end{tikzpicture} 
\end{minipage}
}\hspace{0.1cm}
\subfigure[2D centered rectangular lattice]{
\begin{minipage}{0.60\textwidth}
\centering
\begin{tikzpicture}[scale=0.9]
\coordinate[] (A_lb) at (0.0, 0.0);
\coordinate[] (C_lb) at (0.0, 0.8660);
\coordinate[] (DA_r) at (1.5, 0.8660);
\coordinate[] (B_l)  at (1.5, 1.2990);
\coordinate[] (B_lb) at (3.0, 0.0);
\coordinate[] (DD_r) at (3.0, 0.4330);
\coordinate[] (A_l)  at (3.0, 0.8660);
\coordinate[] (DB_r) at (3.0, 1.2990);
\coordinate[] (DC_r) at (4.5, 0.8660);
\coordinate[] (C_l)  at (4.5, 1.2990);

\coordinate[] (DA_b) at (3.0, 0.0);
\coordinate[] (DB_b) at (4.5, 0.4330);
\coordinate[] (DC_b) at (6.0, 0.0);
\coordinate[] (DD_b) at (4.5,-0.4330);

\coordinate[] (nDA_r) at (2.18-1.5, 0.8660-0.4330);
\coordinate[] (nDB_r) at (2.32-1.5, 1.2990-0.4330);
\coordinate[] (nDC_r) at (3.68-1.5, 1.2990-0.4330);
\coordinate[] (nDD_r) at (3.82-1.5, 0.8660-0.4330);
\coordinate[] (nDE_r) at (3.68-1.5, 0.4330-0.4330);
\coordinate[] (nDF_r) at (2.32-1.5, 0.4330-0.4330);

\coordinate[] (nA_lb) at (0.0, -0.8660);
\coordinate[] (nB_lb) at (3.0, -0.8660);
\coordinate[] (nC_lb) at (0.0,  0.0);

\coordinate[] (A_l)  at (1.5, 0.8660-0.4330);
\coordinate[] (B_l)  at (0.0, 1.2990-0.4330);
\coordinate[] (C_l)  at (3.0, 1.2990-0.4330);

\coordinate[] (R_l) at (6.0,0.8660);
\coordinate[] (R_2) at (9.0,0.8660);
\coordinate[] (R_3) at (9.0,1.7320);
\coordinate[] (R_4) at (6.0,1.7320);

%
\draw[black,dashed,fill={rgb:red,133;green,203;blue,93}, opacity=0.75, line width=0.1mm] (R_l) -- (R_2) -- (R_3) -- (R_4) -- (R_l);
\draw[black,dashed,fill={rgb:red,1;green,103;blue,143}, opacity=0.75, line width=0.1mm] (DA_b) -- (DB_b) -- (DC_b) -- (DD_b) -- (DA_b);
\draw[black,dashed,fill={rgb:red,221;green,109;blue,16}, opacity=0.75, line width=0.1mm] (nDA_r) -- (nDB_r) -- (nDC_r) -- (nDD_r) -- (nDE_r) -- (nDF_r) -- (nDA_r);


\draw[lightgray, fill=lightgray, line width=1mm , opacity=1.0] (7.5,1.2990) circle (2pt);
\draw[black, fill=black,line width=0.1mm , opacity=1.0] (7.5,1.2990) circle (0.2pt);

\draw[lightgray, fill=lightgray, line width=1mm , opacity=1.0] (7.5,0.4330) circle (2pt);
\draw[black, fill=black,line width=0.1mm , opacity=1.0] (7.5,0.4330) circle (0.2pt);

\draw[lightgray, fill=lightgray, line width=1mm , opacity=1.0] (7.5,-0.4330) circle (2pt);
\draw[black, fill=black,line width=0.1mm , opacity=1.0] (7.5,-0.4330) circle (0.2pt);


\draw[lightgray, fill=lightgray, line width=1mm , opacity=1.0] (9.0,-0.8660) circle (2pt);
\draw[black, fill=black, line width=0.1mm , opacity=1.0] (9.0,-0.8660) circle (0.2pt);

\draw[lightgray, fill=lightgray, line width=1mm , opacity=1.0] (9.0,0) circle (2pt);
\draw[black, fill=black,line width=0.1mm , opacity=1.0] (9,0) circle (0.2pt);

\draw[lightgray, fill=lightgray, line width=1mm , opacity=1.0] (9.0,0.8660) circle (2pt);
\draw[black, fill=black,line width=0.1mm , opacity=1.0] (9.0,0.8660) circle (0.2pt);

\draw[lightgray, fill=lightgray, line width=1mm , opacity=1.0] (9.0,1.7320) circle (2pt);
\draw[black, fill=black,line width=0.1mm , opacity=1.0] (9.0,1.7320) circle (0.2pt);


\draw[lightgray, fill=lightgray, line width=1mm , opacity=1.0] (0.0,-0.8660) circle (2pt);
\draw[black, fill=black, line width=0.1mm , opacity=1.0] (0.0,-0.8660) circle (0.2pt);

\draw[lightgray, fill=lightgray, line width=1mm , opacity=1.0] (3.0,-0.8660) circle (2pt);
\draw[black, fill=black, line width=0.1mm , opacity=1.0] (3.0,-0.8660) circle (0.2pt);

\draw[lightgray, fill=lightgray, line width=1mm , opacity=1.0] (6.0,-0.8660) circle (2pt);
\draw[black, fill=black, line width=0.1mm , opacity=1.0] (6.0,-0.8660) circle (0.2pt);

\draw[lightgray, fill=lightgray, line width=1mm , opacity=1.0] (1.5,-0.4330) circle (2pt);
\draw[black, fill=black, line width=0.1mm , opacity=1.0] (1.5,-0.4330) circle (0.2pt);

\draw[lightgray, fill=lightgray, line width=1mm , opacity=1.0] (4.5,-0.4330) circle (2pt);
\draw[black, fill=black, line width=0.1mm , opacity=1.0] (4.5,-0.4330) circle (0.2pt);

\draw[lightgray, fill=lightgray, line width=1mm , opacity=1.0] (0.0,0) circle (2pt);
\draw[black, fill=black,line width=0.1mm , opacity=1.0] (0.0,0) circle (0.2pt);

\draw[lightgray, fill=lightgray, line width=1mm , opacity=1.0] (3.0,0) circle (2pt);
\draw[black, fill=black,line width=0.1mm , opacity=1.0] (3.0,0) circle (0.2pt);

\draw[lightgray, fill=lightgray, line width=1mm , opacity=1.0] (6.0,0) circle (2pt);
\draw[black, fill=black,line width=0.1mm , opacity=1.0] (6,0) circle (0.2pt);

\draw[lightgray, fill=lightgray, line width=1mm , opacity=1.0] (1.5,0.4330) circle (2pt);
\draw[black, fill=black,line width=0.1mm , opacity=1.0] (1.5,0.4330) circle (0.2pt);

\draw[lightgray, fill=lightgray, line width=1mm , opacity=1.0] (4.5,0.4330) circle (2pt);
\draw[black, fill=black,line width=0.1mm , opacity=1.0] (4.5,0.4330) circle (0.2pt);

\draw[lightgray, fill=lightgray, line width=1mm , opacity=1.0] (0.0,0.8660) circle (2pt);
\draw[black, fill=black,line width=0.1mm , opacity=1.0] (0.0,0.8660) circle (0.2pt);

\draw[lightgray, fill=lightgray, line width=1mm , opacity=1.0] (3.0,0.8660) circle (2pt);
\draw[black, fill=black,line width=0.1mm , opacity=1.0] (3.0,0.8660) circle (0.2pt);

\draw[lightgray, fill=lightgray, line width=1mm , opacity=1.0] (6.0,0.8660) circle (2pt);
\draw[black, fill=black,line width=0.1mm , opacity=1.0] (6.0,0.8660) circle (0.2pt);

\draw[lightgray, fill=lightgray, line width=1mm , opacity=1.0] (1.5,1.2990) circle (2pt);
\draw[black, fill=black,line width=0.1mm , opacity=1.0] (1.5,1.2990) circle (0.2pt);

\draw[lightgray, fill=lightgray, line width=1mm , opacity=1.0] (4.5,1.2990) circle (2pt);
\draw[black, fill=black,line width=0.1mm , opacity=1.0] (4.5,1.2990) circle (0.2pt);

\draw[lightgray, fill=lightgray, line width=1mm , opacity=1.0] (0.0,1.7320) circle (2pt);
\draw[black, fill=black,line width=0.1mm , opacity=1.0] (0.0,1.7320) circle (0.2pt);

\draw[lightgray, fill=lightgray, line width=1mm , opacity=1.0] (3.0,1.7320) circle (2pt);
\draw[black, fill=black,line width=0.1mm , opacity=1.0] (3.0,1.7320) circle (0.2pt);

\draw[lightgray, fill=lightgray, line width=1mm , opacity=1.0] (6.0,1.7320) circle (2pt);
\draw[black, fill=black,line width=0.1mm , opacity=1.0] (6.0,1.7320) circle (0.2pt);

\draw [->] (A_l) -- (B_l);
\draw [->] (A_l) -- (C_l);
\draw [->] (nA_lb) -- (nB_lb);
\draw [->] (nA_lb) -- (nC_lb);

\draw[black, ->]  (A_l) -- node[above left] {  $\boldsymbol a_2$} (B_l);
\draw[black,->]  (A_l) -- node[above right] { $\boldsymbol  a_1$} (C_l);

\coordinate (a) at (1,4);
\coordinate (b) at (3,5);
\coordinate (c) at (5,4);
\draw pic[draw,fill=black!30,angle radius=0.6cm,"{ $\varphi$}" shift={(0mm,0mm)}] {angle=nB_lb--nA_lb--nC_lb};
\draw[black,->]  (nA_lb) -- node[below right] {  $\boldsymbol c_1$} (nB_lb);
\draw[black,->]  (nA_lb) -- node[left] { $\boldsymbol  c_2$} (nC_lb);

\node[white, text width=5cm, align=center, text opacity=1] at (1.2,0.70-0.4330)  
    {{$\mathcal{R}^\mathrm{c}_1$}};
\node[white, text width=5cm, align=center, text opacity=1] at (4.5,0.0)
    {{$\mathcal{R}^\mathrm{c}_2$}};
\node[white, text width=5cm, align=center, text opacity=1] at (7.0,1.2990)
    {{$\mathcal{R}^\mathrm{c}_3$}};

\draw[black, fill=black,line width=0.1mm , opacity=1.0] (A_l) circle (1.5pt);

\draw[black, fill=black,line width=0.1mm , opacity=1.0] (nA_lb) circle (1.5pt);

\end{tikzpicture} 
\end{minipage}
}\\
\subfigure[3D simple cubic lattice]{
\begin{minipage}{1.0\textwidth}
\centering
\input{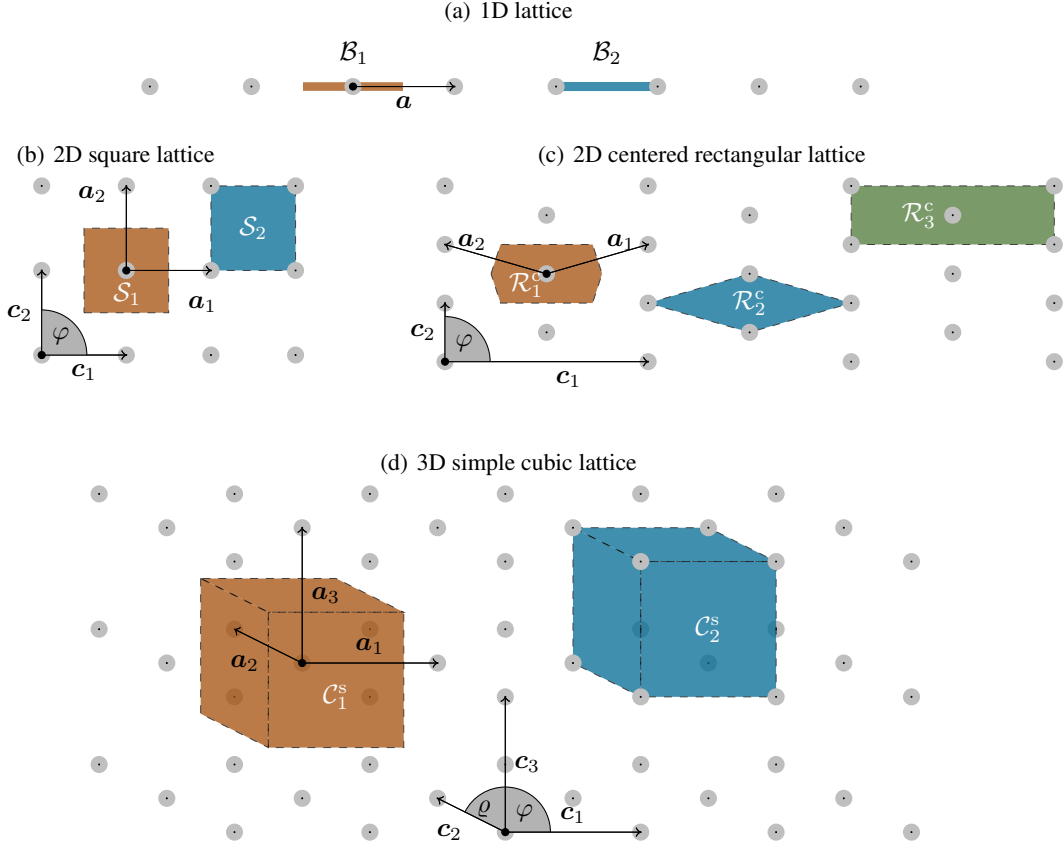}
\end{minipage}
}
\vspace{-10pt}
\caption{Selected Bravais lattices in 1D, 2D and 3D.  The choice of the primitive unit cell is not unique. $\mathcal{B}_i$, $\mathcal{S}_i$, $\mathcal{R}_i^\mathrm{c}$ and $\mathcal{C}_i^\mathrm{s}$ correspond to unit cells for one dimensional, two-dimensional square, centered rectangular and three-dimensional simple cubic lattices, respectively. Unit cells with $i=1,2$ constitute primitive unit cells, whereas for $i=1$ they are referred to as Wigner-Seitz cells.
$\mathcal{R}_3^\mathrm{c}$ is not a primitive unit cell as in the cell domain more than one lattice point is occupied. Instead, $\mathcal{R}_3^\mathrm{c}$ corresponds to a rectangular unit cell which has direct lattice vectors $\bs c_i$ for $i=1,2$.
For the 1D lattice, $a=|\bs a|$. For the 2D square lattice $\lvert \boldsymbol c_1 \rvert = \lvert \boldsymbol c_2 \rvert, \varphi = 90^{\circ}$. For the 2D centered rectangular lattice $\lvert \boldsymbol c_1 \rvert \neq \lvert \boldsymbol c_2 \rvert, \varphi = 90^{\circ}$.  Finally, for the 3D simple cubic lattice $|\boldsymbol{c}_1|=|\boldsymbol{c}_2|=|\boldsymbol{c}_3|,\,\varphi=\varrho=90^\circ$.}
 \label{fig:BravaisLattice1D2D3D}
\end{figure}

A Bravais lattice is a set of regularly spaced points, each of which has a position $\bs R$ as an integer combination of three translation vectors $\bs a_i$ for $i=1,2,3$, with $\bs R=m\,\bs a_1+n\,\bs a_2+o\,\bs a_3$, and $m,n,$ and $o$ are integers \cite{Kittel1953,Hofmann2008}. The translation vectors $\bs a_i$ are called direct lattice vectors or primitive translation vectors. The domain containing only one lattice point and the smallest volume $\bs a_1 \cdot \bs a_2\times \bs a_3$ that fills the entire space upon translation with integer combinations of $\bs a_i$ is the primitive unit cell. Fig.\ \ref{fig:BravaisLattice1D2D3D} demonstrates selected Bravais lattices in 1D, 2D, and 3D, which are used in subsequent application problems.

In one dimension, there exists only one Bravais lattice which is a regular grid of spacing  $a$ such that each point has a position $R=ma$,  where $m$ is an integer. In two dimensions, there are five Bravais lattices: oblique ($\mathcal{O}$), rectangular ($\mathcal{R}$), centered rectangular ($\mathcal{R}^\mathrm{c}$), hexagonal ($\mathcal{H}$) and square ($\mathcal{S}$). In three dimensions, there are in total 14 Bravais lattices. Exemplary lattices and corresponding unit cells in 1D, 2D, and 3D are  depicted in Fig.\ \ref{fig:BravaisLattice1D2D3D}. The choice of the primitive unit cell is not unique  unless it is a Wigner-Seitz cell created by Voronoi decomposition applied to the Bravais lattice. In Fig.~\ref{fig:BravaisLattice1D2D3D}, the unit cells $\mathcal{B}_1$, $\mathcal{S}_1$, $\mathcal{R}_1^\mathrm{c}$ and $\mathcal{C}_1^\mathrm{s}$ correspond to Wigner-Seitz cells whereas $\mathcal{B}_2$, $\mathcal{S}_2$, $\mathcal{R}_2^\mathrm{c}$ and $\mathcal{C}_2^\mathrm{s}$ constitute another arbitrary primitive unit cell. However, the unit cell $\mathcal{R}_3^\mathrm{c}$ for the centered rectangular lattice is not a primitive unit cell. Direct and reciprocal lattice vectors for the lattices given in Fig.~1 are given in Table\ \ref{T:bravais_lattices_1D2D3D}.

\begin{table}[htb!]
\label{T:bravais_lattices_1D2D3D}
\caption{Direct and reciprocal lattice vectors for each Bravais lattice given in Fig.\ \ref{fig:BravaisLattice1D2D3D}. For the demonstration of the direct lattice vector $\boldsymbol a$ and
the vector $\boldsymbol c$ see Fig.\ \ref{E:reciprocal}. For the 1D lattice, $a=|\bs a|=L$. For the 2D square lattice, $L=\rvert\boldsymbol c_1\rvert=\rvert\boldsymbol c_2\rvert$. For the 2D centered rectangular lattice, $L_1=\rvert\boldsymbol c_1\rvert$ and $L_2=\rvert\boldsymbol c_2\rvert$. Finally, for the 3D simple cubic lattice $L=\rvert\boldsymbol c_1\rvert=\rvert\boldsymbol c_2\rvert=\rvert\boldsymbol c_3\rvert$. Here the directions of the unit Cartesian basis $\bs e_i$ conform with $\bs c_i$ for $i=1,2,3$.}
\centering
\begin{tabular}{llll}
\hline
dimension & lattice type & direct lattice vectors  & reciprocal lattice vectors\\
\hline
1D & $\mathcal{B}$ & $a=L$   & $b=2\pi/L$\\
\hline
\multirow{2}{*}{2D} & \multirow{2}{*}{$\mathcal{S}$}
 & $\bs a_1=L\,\bs e_1$ &  $\bs b_1=2\pi/L\,\bs e_1$\\
 & & $\bs a_2=L\,\bs e_2$ &  $\bs b_2=2\pi/L\,\bs e_2$\\
\hline
\multirow{2}{*}{2D} & \multirow{2}{*}{$\mathcal{R}^\mathrm{c}$}
   & $\bs a_1=L_1/2\,\bs e_1+L_2/2 \,\bs e_2$
   & $\bs b_1=2\pi/L_1\,\bs e_1+2\pi/L_2 \,\bs e_2$\\
   & & $\bs a_2=-L_1/2\,\bs e_1+L_2/2 \,\bs e_2$
    & $\bs b_2=-2\pi/L_1\,\bs e_1+2\pi/L_2 \,\bs e_2$\\
\hline
\multirow{3}{*}{3D} & \multirow{3}{*}{$\mathcal{C}^\mathrm{s}$}
   & $\bs a_1=L\,\bs e_1$ &  $\bs b_1=2\pi/L\,\bs e_1$\\
   & & $\bs a_2=L\,\bs e_2$ &  $\bs b_2=2\pi/L\,\bs e_2$\\
   & & $\bs a_3=L\,\bs e_3$ &  $\bs b_3=2\pi/L\,\bs e_3$\\
\hline
\end{tabular}
\end{table}

The reciprocal Bravais lattice constitutes another set of regularly spaced points with  positions $\bs G=m'\,\bs b_1+n'\,\bs b_2+o'\,\bs b_3$ with $m',n'$ and $o'$ being integers. The noncoplanar vectors $\bs b_i$ for $i=1,2,3$ are called the reciprocal lattice vectors, and they can be derived from the lattice vectors
\begin{align}
\bs b_1=2\pi \dfrac{\bs a_2\times \bs a_3}{\bs a_1\cdot [\bs a_2\times\bs a_3]}\,,\quad
\bs b_2=2\pi \dfrac{\bs a_3 \times \bs a_1 }{\bs a_1\cdot [\bs a_2\times\bs a_3]}\quad\text{ and }\quad
\bs b_3=2\pi \dfrac{\bs a_1 \times \bs a_2 }{\bs a_1\cdot [\bs a_2\times\bs a_3]}\,,
\label{E:reciprocal}
\end{align}
such that $\bs a_i\cdot\bs b_j = 2\pi \delta_{ij}$.
Any lattice-periodic function $\bs \xi(\bs x)$ with $\bs \xi({\bs x})=\bs \xi({\bs x}+{\bs R})$ can be written as the following Fourier series, see, e.g., \cite{Kittel1953, Hofmann2008}
\begin{align}
\bs \xi(\bs x)=\sum_{\bs G} \breve{\bs \xi}_{\bs G}\, e^{i\bs G\cdot \bs x}\,,
\label{E:Fourier3D}
\end{align}
where $\bs x$ represents the vector position and $\breve{\bs \xi}_{\bs G}$ are generally complex Fourier coefficients.

\subsection{First-order Asymptotic Homogenization}
A generic form of elliptic PDEs is given as
\begin{align}
-\dfrac{\partial}{\partial x_i}\left(a_{ij}(\bs x)
\dfrac{\partial u}{\partial x_j}\right)=f(\bs x)\,,
\label{E:generic}
\end{align}
for $i,j=1,2,3$, where $\bs x$ denotes the three-dimensional (3D) position vector, $u$ is the quantity of interest, and $f$ is a source term. Equation\ \eqref{E:generic} arises in many different contexts~\cite{Torquato2002}. For example, in electrostatics, $u$ and $\bs a$ may refer to one of the electric potentials and to the dielectric constant, respectively; in magnetostatics, it is used to describe the magnetic (scalar) potential where $\bs a$ is  the magnetic permeability; and in continuum mechanics, it describes the pressure field of slow viscous fluid flow in porous materials where  $\bs a$ models the permeability of the medium.

We consider lattice-periodic two-phase material systems with linear physical properties that can be represented in terms of Fourier series, i.e.~$\bs a(\bs x)=\sum_{\bs G} \breve{\bs a}_{\bs G}\, e^{i\bs G\cdot \bs x}$, such that $\bs a(\bs x)=\bs a(\bs x+\bs R)$. For such systems, the material domain $\mathcal{V}$ corresponding to the primitive unit cell of the lattice encapsulates all material characteristics. Henceforth, let $\bs x$ denote the cell-scale (microscale) position that captures the fast variations of the field. It is related to the macroscale position $^\mathrm{M}\bs x$, which captures the slow variations of the field, as $\boldsymbol{x}=$ $^\textrm{M}\boldsymbol{x}/\epsilon$, where
$0<\epsilon \ll 1$ is  a small parameter representing the separation between scales.

Following formal steps of first-order two-scale asymptotic homogenization, see e.g.~\cite{Torquato2002,LUKKASSEN1995519}, the macroscopic property tensor $\bs{a}^\star$ is obtained through the following averaging operation
\begin{align}
a^\star_{im}=\dfrac{1}{|\mathcal{V}|}\,\int_{\mathcal{V}}a_{ik}\left[\delta_{km}+
\dfrac{\partial \chi^{m}(\bs{x})}{\partial x_k}\right]
\mathrm{d}V\,,
\label{E:propertyhomogenized}
\end{align}
for $i,m=1,2,3$, where $\chi^{m}(\bs{x})$ is a lattice-periodic corrector function that is the solution to the cell-problem
\begin{align}
-\dfrac{\partial}{\partial x_i}\left(a_{ij}(\bs x)
\dfrac{\partial \chi^m}{\partial x_j}\right)=
\dfrac{\partial a_{im}}{\partial x_i}\,.
\label{E:cell-problem}
\end{align}
Equation \eqref{E:cell-problem} corresponds to three problems indexed with $m=1,2,3$ each of which has the source term computed with the divergence of the $m^{\mathrm{th}}$ column vector of the property tensor $\bs a$. In the literature, two approaches are used to solve Eq.\ \eqref{E:cell-problem}, see, e.g., \cite{LUKKASSEN1995519}, where the unit cell is subjected to satisfy  Hill-Mandel boundary conditions, see e.g., \cite{Garboczi1998, ChatzigeorgiouJaviliSteinmann2014,JAVILI20134197, OstojaStarzewski2019}. The first approach solves the problem directly without any transformation.
This poses some challenges, particularly when  dealing with sharp phase interfaces at which $\partial a_{ij}/\partial x_i\to \infty$. Although this can be resolved through the incorporation of a diffuse interface formulation, the presence of random material microstructures with complex phase geometries  still  creates difficulties when implementing the problem into, e.g., commercial finite element packages. An alternative approach is to reformulate the problem such that it can be solved completely using Dirichlet boundary conditions, see, e.g., \cite{LUKKASSEN1995519,SOYARSLAN2023106188,SOYARSLAN2019103098}. To this end, the unit cell is supposed to be subjected to unitary gradients of the solution variable along with space directions. The total solution is then in the form of a superposition of a linearly varying field and a periodic fluctuation field, which corresponds to the corrector function.  In this work, we use the former method and solve Eq.\ \eqref{E:cell-problem} by PINNs using an appropriate regularization for the source term.

\subsection{Physics-Informed Neural Networks (PINNs)}
\label{Sec:PINNs}
Our theoretical approach for this part follows the works by Lu \emph{et al.}~\cite{Lu_2021} and Haghighat \emph{et al.}~\cite{HAGHIGHAT2021113741}.  The artificial neural networks used here are referred to as feed-forward neural networks (FNNs). In FNNs a recursive application of  linear and nonlinear transformations to the input 
proves to be sufficient in the solution of most PDEs \cite{Lu_2021}.

We consider an $L$-layer neural network denoted by $\mathcal N^{L}(\bs x; \bs \Theta ):\mathbb R^{\text{dim}_{\mathrm{in}}} \mapsto \mathbb R^{\text{dim}_{\mathrm{out}}}$, where $\text{dim}_{\mathrm{in}}$ and $\text{dim}_{\mathrm{out}}$ are the network input and output dimensions, respectively. The total number of hidden layers is $L-1$, and $\mathcal N^{L}(\bs x; \bs \Theta )$ represents the surrogate model that gives the approximate solution for the unknown $u$ of the differential equation.   Here, $\bs x$ denotes  the input vector, and $\bs \Theta:=\{\bs W, \bs b\}$ represent the set of trainable network parameters, where $\bs W$ and $\bs b$  denote the weight matrix and the  bias vector, respectively. The output vector is given by $\bs y$. Considering the input-output relations $\bs z^0 \leftarrow \bs x$ and $\bs y  \leftarrow \bs z^L$, the input, and the feed-forward network propagate the input vector through its layers via the following operation sequence
\begin{align}
\bs z^k&=\sigma(\bs W^k \bs z^{k-1}+\bs b^k)\,,\quad 1\leq k \leq L-1 \,,\\
\bs z^L&=\bs W^L \bs z^{L-1}+\bs b^L\,,
\label{E:nn}
\end{align}
where $\sigma$ is the nonlinear activation function, e.g., hyperbolic tangent ($\tanh$) or logistic sigmoid ($1/[1+\exp(-x)]$).  The outputs, the weight matrix, and the bias vector of each layer $k$ are, respectively, denoted by $\bs z^k \in \mathbb R^{q_k}$, $\bs W^k \in \mathbb R^{q_k\times q_{k-1}}$ and $\bs b^k \in \mathbb R^{q_k}$, where $q_k$ is the number of neurons.

Unlike conventional ANNs, PINNs aim at solving differential equations, which requires the computation of corresponding derivatives. The inputs of the network are problem coordinates or the independent variables representing the problem domain (space and time), whereas the output is the smooth field representing the solution of the PDE. In effect, the derivatives of network output with respect to network input are required, in which, the compositional structure of the network requires the chain rule. In  DeepXDE, this is achieved by using the method referred to as automatic differentiation with backpropagation~\cite{Lu_2021,Rumelhart1986}.

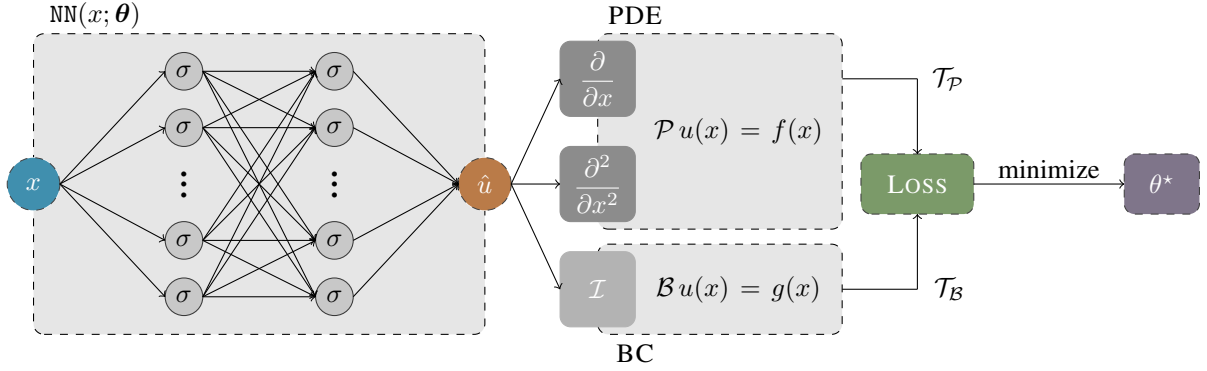
\begin{figure}[t!]
\hspace{-55pt}
\begin{tikzpicture}[scale=1.0]

\coordinate[] (xR) at (0.35,0);
\coordinate[] (xhiL) at (5.65,0);
\coordinate[] (xhiR) at (6.35,0);

\coordinate[] (o1L) at (7.0,1.4);
\coordinate[] (o1R) at (8.0,1.4);
\coordinate[] (o2L) at (7.0,0.0);
\coordinate[] (o2R) at (8.0,0.0);
\coordinate[] (o3L) at (7.0,-1.4);
\coordinate[] (o3R) at (8.0,-1.4);

\coordinate[] (BB1R) at (10.75,1.4);
\coordinate[] (BB2R) at (10.75,-1.4);

\coordinate[] (INT1) at (11.75,1.4);
\coordinate[] (INT2) at (11.75,-1.4);

\coordinate[] (LOSST) at (11.75, 0.4);

\node[text width=1.25cm, align=right, text opacity=1] at (11.75,1.4){$\mathcal{T}_\mathcal{P}$};
\node[text width=1.25cm, align=right, text opacity=1] at (11.75,-1.4){$\mathcal{T}_\mathcal{B}$};
\node[text width=5.0cm, align=center, text opacity=1] at (13.5,0.2){minimize};

\coordinate[] (LOSSB) at (11.75,-0.4);
\coordinate[] (LOSSR) at (12.5,0.);
\coordinate[] (THETAL) at (14.5,0);

\draw[black, -] (BB1R) -- (INT1);
\draw[black, ->] (INT1) -- (LOSST);
\draw[black, -] (BB2R) -- (INT2);
\draw[black, ->] (INT2) -- (LOSSB);
\draw[black, ->] (LOSSR) -- (THETAL);

\coordinate[] (n11L) at (1.75,-1.5);
\coordinate[] (n12L) at (1.75,-0.75);
\coordinate[] (n13L) at (1.75, 0.0);
\coordinate[] (n14L) at (1.75, 0.75);
\coordinate[] (n15L) at (1.75, 1.5);
\coordinate[] (n11R) at (2.25,-1.5);
\coordinate[] (n12R) at (2.25,-0.75);
\coordinate[] (n13R) at (2.25, 0.0);
\coordinate[] (n14R) at (2.25, 0.75);
\coordinate[] (n15R) at (2.25, 1.5);

\coordinate[] (n21L) at (3.75,-1.5);
\coordinate[] (n22L) at (3.75,-0.75);
\coordinate[] (n23L) at (3.75, 0.0);
\coordinate[] (n24L) at (3.75, 0.75);
\coordinate[] (n25L) at (3.75, 1.5);
\coordinate[] (n21R) at (4.25,-1.5);
\coordinate[] (n22R) at (4.25,-0.75);
\coordinate[] (n23R) at (4.25, 0.0);
\coordinate[] (n24R) at (4.25, 0.75);
\coordinate[] (n25R) at (4.25, 1.5);


\node[text width=2cm, align=center, text opacity=1] at (0.8,2.25){{$\texttt{NN}(x;\boldsymbol \theta)$}};
\draw[dashed, black, 
fill=gray!20, rounded corners](0.0,-2.0)--(6.0,-2.0)--(6.0,2)--(-0.0,2)--cycle;

\node[text width=5cm, align=center, text opacity=1] at (8.0,2.25){\textsc{PDE}};
\node[text width=5cm, align=center, text opacity=1] at (8.0,-2.25){\textsc{BC}};

\draw[dashed, black, 
fill=gray!20, rounded corners](7.5,-0.6)--(10.75,-0.6)--(10.75,2)--(7.5,2)--cycle;
\draw[dashed, black, 
fill=gray!20, rounded corners](7.5,-0.8)--(10.75,-0.8)--(10.75,-2)--(7.5,-2)--cycle;

\draw[dashed, black, 
fill={rgb:red,133;green,203;blue,93}, rounded corners, opacity=0.75](11.0,-0.4)--(12.5,-0.4)--(12.5,0.4)--(11,0.4)--cycle;
\node[white,text width=2cm, align=center, text opacity=1] at (11.75,0){\textsc{Loss}};
\draw[dashed, black, 
fill={rgb:red,165;green,137;blue,193}, rounded corners, opacity=0.75](14.5,-0.4)--(15.5,-0.4)--(15.5,0.4)--(14.5,0.4)--cycle;
\node[white, text width=2cm, align=center, text opacity=1] at (15.,0){{$\theta^\star$}};

\draw[white,fill=white, opacity=1.0, line width=0.1mm] (0,0) circle (0.35);
\draw[white,fill=white, opacity=1.0, line width=0.1mm] (6,0) circle (0.35);

\draw[gray!60, fill=gray!60, rounded corners](7.0,-1.9)--(8.0,-1.9)--(8.0,-0.9)--(7.0,-0.9)--cycle;
\draw[gray!90, fill=gray!90, rounded corners](7.0,-0.5)--(8.0,-0.5)--(8.0,0.5)--(7.0,0.5)--cycle;
\draw[gray!90, fill=gray!90, rounded corners](7.0,0.9)--(8.0,0.9)--(8.0,1.9)--(7.0,1.9)--cycle;

\node[white, text width=2cm, align=center, text opacity=1] at (7.5,1.4){{$\dfrac{\partial}{\partial x}$}};
\node[white, text width=2cm, align=center, text opacity=1] at (7.5,0){{$\dfrac{\partial^2}{\partial x^2}$}};
\node[white, text width=2cm, align=center, text opacity=1] at (7.5,-1.4){{$\mathcal{I}$}};

\node[text width=2.8cm, align=center, text opacity=1] at (9.35,-1.4){{$\mathcal{B}\,u(x)=g(x)$}};
\node[text width=2.8cm, align=center, text opacity=1] at (9.35,0.7){{$\mathcal{P}\,u(x)=f(x)$}};

\draw[black,dashed,fill={rgb:red,1;green,103;blue,143}, opacity=0.75, line width=0.1mm] (0,0) circle (0.35);
\node[white, text width=5cm, align=center, text opacity=1] at (0,0){{$x$}};

\draw[black, fill=lightgray, opacity=0.75] (2.0,-1.5) circle (0.25);
\node[text width=5cm, align=center, text opacity=1] at (2.0,-1.5){$\sigma$};
\draw[black, fill=lightgray, opacity=0.75] (2.0,-0.75) circle (0.25);
\node[text width=5cm, align=center, text opacity=1] at (2.0,-0.75){$\sigma$};
\node[text width=5cm, align=center, text opacity=1] 
at (2.0, 0.1){\LARGE $\vdots$};
\draw[black, fill=lightgray, opacity=0.75] (2.0, 0.75) circle (0.25);
\node[text width=5cm, align=center, text opacity=1] at (2.0,0.75){$\sigma$};
\draw[black, fill=lightgray, opacity=0.75] (2.0, 1.5) circle (0.25);
\node[text width=5cm, align=center, text opacity=1] at (2.0,1.5){$\sigma$};

\draw[black, fill=lightgray, opacity=0.75] (4.0,-1.5) circle (0.25);
\node[text width=5cm, align=center, text opacity=1] at (4.0,-1.5){$\sigma$};
\draw[black, fill=lightgray, opacity=0.75] (4.0,-0.75) circle (0.25);
\node[text width=5cm, align=center, text opacity=1] at (4.0,-0.75){$\sigma$};
\node[text width=5cm, align=center, text opacity=1] at (4.0, 0.1){\LARGE $\vdots$};
\draw[black, fill=lightgray, opacity=0.75] (4.0, 0.75) circle (0.25);
\node[text width=5cm, align=center, text opacity=1] at (4.0,0.75){$\sigma$};
\draw[black, fill=lightgray, opacity=0.75] (4.0, 1.5) circle (0.25);
\node[text width=5cm, align=center, text opacity=1] at (4.0,1.5){$\sigma$};

\draw[black,dashed,fill={rgb:red,221;green,109;blue,16}, opacity=0.75, line width=0.1mm] (6,0) circle (0.35);

\node[white, text width=5cm, align=center, text opacity=1] at (6,0.0){$\hat{u}$};

\draw[black, ->] (xR) -- (n11L);
\draw[black, ->] (xR) -- (n12L);
\draw[black, ->] (xR) -- (n14L);
\draw[black, ->] (xR) -- (n15L);

\draw[black, ->] (n11R) -- (n21L);
\draw[black, ->] (n11R) -- (n22L);
\draw[black, ->] (n11R) -- (n24L);
\draw[black, ->] (n11R) -- (n25L);

\draw[black, ->] (n12R) -- (n21L);
\draw[black, ->] (n12R) -- (n22L);
\draw[black, ->] (n12R) -- (n24L);
\draw[black, ->] (n12R) -- (n25L);

\draw[black, ->] (n14R) -- (n21L);
\draw[black, ->] (n14R) -- (n22L);
\draw[black, ->] (n14R) -- (n24L);
\draw[black, ->] (n14R) -- (n25L);

\draw[black, ->] (n15R) -- (n21L);
\draw[black, ->] (n15R) -- (n22L);
\draw[black, ->] (n15R) -- (n24L);
\draw[black, ->] (n15R) -- (n25L);

\draw[black, ->] (n21R) -- (xhiL);
\draw[black, ->] (n22R) -- (xhiL);
\draw[black, ->] (n24R) -- (xhiL);
\draw[black, ->] (n25R) -- (xhiL);

\draw[black, ->] (xhiR) -- (o1L);
\draw[black, ->] (xhiR) -- (o2L);
\draw[black, ->] (xhiR) -- (o3L);

\end{tikzpicture}
\vspace{-15.pt}
\caption{Shematic PINN with the conventional application of  BCs within the context of computational homogenization. $\mathcal{P}$ and $\mathcal{B}$  denote differential and boundary operators, respectively. $\mathcal{T}_\mathcal{P}$ and $\mathcal{T}_\mathcal{B}$ represent collocation points for the PDE and the boundary conditions, respectively. The image is adapted from \cite{Lu_2021}.}
 \label{fig:PINNsschematic}
\end{figure}

A schematic plot of a PINN for solving general PDEs is given in Fig.\ \ref{fig:PINNsschematic} for input $x$ and output $\hat{u}$, which approximates the actual solution $u$. Since the current problem is quasi-static, we do not consider time variations of the fields.  $\mathcal{P}$ and $\mathcal{B}$ are differential and boundary operators to be satisfied. For a given domain and boundary training sets, which are respectively denoted by $\mathcal{T}_\mathcal{P}$ and $\mathcal{T}_\mathcal{B}$, the neural network aims at minimizing the loss function $\mathcal{L}(\boldsymbol{\theta};\mathcal{T})$ by varying the network parameters $\boldsymbol{\theta}$. This process is called training. The converged parameter set giving the optimum output is denoted by $\boldsymbol{\theta}^\star$.  The loss function $\mathcal{L}(\boldsymbol{\theta};\mathcal{T})$ is defined in terms of the following weighted sum of the residuals of the PDE ($\mathcal{L}_\mathcal{P}(\boldsymbol{\theta};\mathcal{T}_\mathcal{P})$) and the boundary ($\mathcal{L}_\mathcal{B}(\boldsymbol{\theta};\mathcal{T}_\mathcal{B})$)
\begin{align}
\mathcal{L}(\boldsymbol{\theta};\mathcal{T})=\omega_\mathcal{P}\mathcal{L}_\mathcal{P}(\boldsymbol{\theta};\mathcal{T}_\mathcal{P})+\omega_\mathcal{B}\mathcal{L}_\mathcal{B}(\boldsymbol{\theta};\mathcal{T}_\mathcal{B})\,,
\label{E:cost_function}
\end{align}
where $\omega_\mathcal{P}$ and $\omega_\mathcal{B}$ are the loss weights associated with the PDE and boundary loss terms.
\subsection{PINNs with Fourier Features}
Fourier feature mappings were introduced in \cite{Rahimi2007} as a way to approximate shift-invariant kernels, and have been used in many machine learning applications ever since. The basic idea is to project  input points to the surface of a higher or lower dimensional hyperspace by using Fourier bases, thereby making the learning process of the neural network more efficient.

An example is the work of Ref.~\cite{Tanciketal2020}, where Fourier feature mappings are applied in computer vision and graphics applications. It is shown there that mapping the input coordinates with Fourier features enables ANNs to learn high-frequency functions that might not be accessible by standard networks. In particular, Fourier feature mappings can be used to substantially improve the learning efficiency by overcoming what is known as 'spectral bias', which is the bias of deep networks towards low-frequencies.

Spectral bias arises because of the rapid decay of the eigenvalues of the neural tangent kernel (NTK), which is the kernel that approximates the network's output. Such rapid decay in the eigenvalues leads, in turn, to a very slow convergence to the high-frequency components of the target function. Hence, as is shown in \cite{Tanciketal2020}, adding a Fourier feature mapping effectively transforms the NTK into a shift-invariant kernel while enabling it to tune its spectrum by choosing appropriate  frequencies in the Fourier mapping.

In our work, we use a similar approach in terms of Fourier features, with a mapping that projects the input data to a Fourier basis. As shown in the next section, one key element of our formulation is that the convergence of high-frequency functions is achieved by considering multiple integers of the reciprocal base vector, which is different from other applications where random multiples are used.
\subsection{PINNs in Periodic Homogenization}
In what follows, we show how the periodic boundary conditions are implemented \emph{exactly} without any need for an explicit imposition between periodically located boundary collocation points, eliminating the residual loss term relating  to periodic boundary conditions. To this end, we derive trigonometric basis representations of Eq.\ \eqref{E:Fourier3D} for one-, two-, and three-dimensions.
\subsubsection{One-Dimensional Bravais Lattice}
%
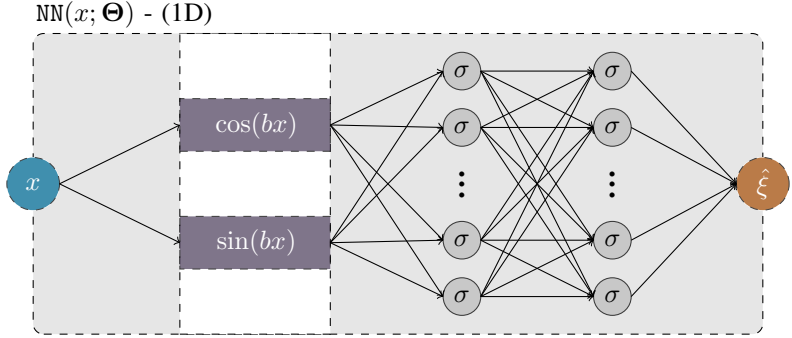
\begin{figure}[t!]
\begin{center}
\begin{tikzpicture}[scale=1.0]

\coordinate[] (xR) at (-3.35,0);
\coordinate[] (xhiL) at (5.65,0);
\coordinate[] (xhiR) at (6.35,0);

\coordinate[] (n01L) at (-1.75, 0.7833);
\coordinate[] (n01R) at (0.25,  0.7833);
\coordinate[] (n02L) at (-1.75,-0.7833);
\coordinate[] (n02R) at (0.25, -0.7833);

\coordinate[] (n11L) at (1.75,-1.5);
\coordinate[] (n12L) at (1.75,-0.75);
\coordinate[] (n13L) at (1.75, 0.0);
\coordinate[] (n14L) at (1.75, 0.75);
\coordinate[] (n15L) at (1.75, 1.5);
\coordinate[] (n11R) at (2.25,-1.5);
\coordinate[] (n12R) at (2.25,-0.75);
\coordinate[] (n13R) at (2.25, 0.0);
\coordinate[] (n14R) at (2.25, 0.75);
\coordinate[] (n15R) at (2.25, 1.5);

\coordinate[] (n21L) at (3.75,-1.5);
\coordinate[] (n22L) at (3.75,-0.75);
\coordinate[] (n23L) at (3.75, 0.0);
\coordinate[] (n24L) at (3.75, 0.75);
\coordinate[] (n25L) at (3.75, 1.5);
\coordinate[] (n21R) at (4.25,-1.5);
\coordinate[] (n22R) at (4.25,-0.75);
\coordinate[] (n23R) at (4.25, 0.0);
\coordinate[] (n24R) at (4.25, 0.75);
\coordinate[] (n25R) at (4.25, 1.5);


\node[text width=4cm, align=center, text opacity=1] at (-2.5,2.25){{$\texttt{NN}(x;\boldsymbol \Theta)$ - (1D)}};
\draw[dashed, black,
fill=gray!20, rounded corners](-3.70,-2.0)--(6.0,-2.0)--(6.0,2)--(-3.70,2)--cycle;

\draw[white,fill=white, opacity=1.0, line width=0.1mm] (-3.70,0) circle (0.35);
\draw[white,fill=white, opacity=1.0, line width=0.1mm] (6,0) circle (0.35);

\draw[dashed, black,
fill=white](-1.75,-2.0)--(0.25,-2.0)--(0.25,2)--(-1.75,2)--cycle;

\draw[dashed, black,
fill={rgb:red,165;green,137;blue,193},  opacity=0.75](-1.75,0.4333)--(0.25,0.4333)--(0.25,1.1333)--(-1.75,1.1333)--cycle;
\node[white,text width=2cm, align=center, text opacity=1] at (-0.75,0.7833){$\cos( b  x)$};
\draw[dashed, black,
fill={rgb:red,165;green,137;blue,193},  opacity=0.75](-1.75,-0.4333)--(0.25,-0.4333)--(0.25,-1.1333)--(-1.75,-1.1333)--cycle;
\node[white,text width=2cm, align=center, text opacity=1] at (-0.75,-0.7833){$\sin(b x)$};

\draw[black,dashed,fill={rgb:red,1;green,103;blue,143}, opacity=0.75, line width=0.1mm] (-3.70,0) circle (0.35);
\node[white, text width=5cm, align=center, text opacity=1] at (-3.70,0){{$x$}};

\draw[black, fill=lightgray, opacity=0.75] (2.0,-1.5) circle (0.25);
\node[text width=5cm, align=center, text opacity=1] at (2.0,-1.5){$\sigma$};
\draw[black, fill=lightgray, opacity=0.75] (2.0,-0.75) circle (0.25);
\node[text width=5cm, align=center, text opacity=1] at (2.0,-0.75){$\sigma$};
\node[text width=5cm, align=center, text opacity=1]
at (2.0, 0.1){\LARGE $\vdots$};
\draw[black, fill=lightgray, opacity=0.75] (2.0, 0.75) circle (0.25);
\node[text width=5cm, align=center, text opacity=1] at (2.0,0.75){$\sigma$};
\draw[black, fill=lightgray, opacity=0.75] (2.0, 1.5) circle (0.25);
\node[text width=5cm, align=center, text opacity=1] at (2.0,1.5){$\sigma$};

\draw[black, fill=lightgray, opacity=0.75] (4.0,-1.5) circle (0.25);
\node[text width=5cm, align=center, text opacity=1] at (4.0,-1.5){$\sigma$};
\draw[black, fill=lightgray, opacity=0.75] (4.0,-0.75) circle (0.25);
\node[text width=5cm, align=center, text opacity=1] at (4.0,-0.75){$\sigma$};
\node[text width=5cm, align=center, text opacity=1] at (4.0, 0.1){\LARGE $\vdots$};
\draw[black, fill=lightgray, opacity=0.75] (4.0, 0.75) circle (0.25);
\node[text width=5cm, align=center, text opacity=1] at (4.0,0.75){$\sigma$};
\draw[black, fill=lightgray, opacity=0.75] (4.0, 1.5) circle (0.25);
\node[text width=5cm, align=center, text opacity=1] at (4.0,1.5){$\sigma$};

\draw[black,dashed,fill={rgb:red,221;green,109;blue,16}, opacity=0.75, line width=0.1mm] (6,0) circle (0.35);

\node[white, text width=5cm, align=center, text opacity=1] at (6,0.0){$\hat{\xi}$};

\draw[black, ->] (xR) -- (n01L);
\draw[black, ->] (xR) -- (n02L);

\draw[black, ->] (n01R) -- (n11L);
\draw[black, ->] (n01R) -- (n12L);
\draw[black, ->] (n01R) -- (n14L);
\draw[black, ->] (n01R) -- (n15L);
\draw[black, ->] (n02R) -- (n11L);
\draw[black, ->] (n02R) -- (n12L);
\draw[black, ->] (n02R) -- (n14L);
\draw[black, ->] (n02R) -- (n15L);

\draw[black, ->] (n11R) -- (n21L);
\draw[black, ->] (n11R) -- (n22L);
\draw[black, ->] (n11R) -- (n24L);
\draw[black, ->] (n11R) -- (n25L);

\draw[black, ->] (n12R) -- (n21L);
\draw[black, ->] (n12R) -- (n22L);
\draw[black, ->] (n12R) -- (n24L);
\draw[black, ->] (n12R) -- (n25L);

\draw[black, ->] (n14R) -- (n21L);
\draw[black, ->] (n14R) -- (n22L);
\draw[black, ->] (n14R) -- (n24L);
\draw[black, ->] (n14R) -- (n25L);

\draw[black, ->] (n15R) -- (n21L);
\draw[black, ->] (n15R) -- (n22L);
\draw[black, ->] (n15R) -- (n24L);
\draw[black, ->] (n15R) -- (n25L);

\draw[black, ->] (n21R) -- (xhiL);
\draw[black, ->] (n22R) -- (xhiL);
\draw[black, ->] (n24R) -- (xhiL);
\draw[black, ->] (n25R) -- (xhiL);

\end{tikzpicture} 
\end{center}
\caption{Schematic representation of how to implement exact $C^\infty$ periodic boundary conditions  in one-dimensional problems through an added $C^\infty$ periodic input transfer layer.}
 \label{fig:NNmod1D}
\end{figure}
A reciprocal lattice, with lattice point positions $G=m'b$ can be identified where $m'$ is an integer and $b=2\pi/a$ such that $ab=2\pi$. For a scalar periodic function, say $\xi(x)$, one has $\xi(x)=\xi(x+R)$, where $n$ is an integer. In view of Eq.\ \eqref{E:Fourier3D}, $\xi(x)$ can be written as
\begin{align}
\xi(\bs x)=\sum_{G} \breve{\xi}_{G}\, e^{i G x}\,,
\label{E:Fourier1D}
\end{align}
where $\hat{\xi}_{G}$ are the complex Fourier coefficients. If $\xi(\bs x)$ is a real function, then the above series can be expressed in terms of trigonometric functions
\begin{align}
\xi(x)=\tfrac12 \alpha_0+\sum_{m=1}^{\infty} \left[\alpha_m\cos{(mbx)}+\beta_m\sin{(mbx)}\right]\,,
\label{E:FourierVtrig}
\end{align}
where $\alpha_m$ and  $\beta_m$ are the corresponding real Fourier coefficients. Using multiple-angle trigonometric identities, we can rewrite the above sum as:
\begin{align}
\xi(x)=&\tfrac12 \alpha_0+\alpha_1\cos(b x) +\beta_1\sin(bx) \nonumber \\
 &+ \sum_{m=2}^{\infty}\left[ \sum_{k=0}^{m} \overline{\alpha}_{mk}\cos^k(b x)\sin^{m-k}(b x)+
 \sum_{n=0}^{m} \overline{\beta}_{mn}\cos^n(b x)\sin^{m-n}(bx)\right]\,,
\label{E:FourierVtrig2}
\end{align}
where we have defined
\begin{align*}
\overline{\alpha}_{mk} = \begin{pmatrix}
n \\
k
\end{pmatrix} \alpha_m\cos\left[\frac{\pi}{2}(m-k)\right]\qquad\text{and}\qquad
\overline{\beta}_{mn} = \begin{pmatrix}
m \\
n
\end{pmatrix} \beta_n \sin\left[\frac{\pi}{2}(m-n)\right].
\end{align*}
Equation \eqref{E:FourierVtrig2} shows that $\xi(x)$ can be represented as
\begin{align}
    \xi({\bs x}) =\mathcal{F}_{\mathrm{1D}}\left[\cos(bx),\,\sin(bx)\right],
    \label{E:F1D}
\end{align}
where $\mathcal{F}_{\mathrm{1D}}$ is a nonlinear combination of $\cos(bx)$ and $\sin(bx)$.
Therefore, instead of approximating the nonlinear smooth function $\xi({\bs x})$ one can approximate $\mathcal{F}_{\mathrm{1D}}$ in an ANN by modifying the neural network architecture by adding the $C^\infty$  periodic basis functions as a transfer layer as demonstrated in Fig.\ \ref{fig:NNmod1D}. Here, the $\cos$ and $\sin$ functions in the added layer can be seen as activation functions prior to which no weights are applied. This provides a priori and exact satisfaction of the $C^\infty$ periodicity of the solution $\xi({\bs x})$. We note that a similar argument was given in \cite{LULUetal2021, Zhangetal2019}, who reported  that a periodic function with a given period can be decomposed into a weighted summation of the Fourier series basis function.

It is also worth remarking that without loss of generality, we can use integer multiples of reciprocal basis in Eq.~\eqref{E:F1D}, which still satisfy exact PBCs but also increase the learning process of high-frequency functions, as a consequence of Fourier feature mappings~\cite{Tanciketal2020}. Therefore, our approach provides a well-defined route to identify the number of required Fourier basis for improved convergence (see Section 3 for a detailed analysis of different applications). This is in contrast to the method used in \cite{DONG2021110242}, where they considered an arbitrary number of trigonometric functions in their representation, which, as we have shown above, is unnecessary. Indeed,  the  method proposed in~\cite{DONG2021110242} can always be reduced to our approach, and to test the validity of our method, some problems discussed in~\cite{DONG2021110242} had been solved in \ref{Apx:PBCs - tests}.
\subsubsection{Two- and Three-Dimensional Bravais Lattices}
\begin{figure}[t!]
\begin{center}
\begin{tikzpicture}[scale=1.0]

\coordinate[] (xR) at (-3.35,0);
\coordinate[] (xhiL) at (5.65,0);
\coordinate[] (xhiR) at (6.35,0);

\coordinate[] (n01L) at (-1.75, 1.41);
\coordinate[] (n01R) at (0.25, 1.41);
\coordinate[] (n02L) at (-1.75, 0.47);
\coordinate[] (n02R) at (0.25,  0.47);
\coordinate[] (n03L) at (-1.75, -0.47);
\coordinate[] (n03R) at (0.25, -0.47);
\coordinate[] (n04L) at (-1.75, -1.41);
\coordinate[] (n04R) at (0.25, -1.41);

\coordinate[] (n11L) at (1.75,-1.5);
\coordinate[] (n12L) at (1.75,-0.75);
\coordinate[] (n13L) at (1.75, 0.0);
\coordinate[] (n14L) at (1.75, 0.75);
\coordinate[] (n15L) at (1.75, 1.5);
\coordinate[] (n11R) at (2.25,-1.5);
\coordinate[] (n12R) at (2.25,-0.75);
\coordinate[] (n13R) at (2.25, 0.0);
\coordinate[] (n14R) at (2.25, 0.75);
\coordinate[] (n15R) at (2.25, 1.5);

\coordinate[] (n21L) at (3.75,-1.5);
\coordinate[] (n22L) at (3.75,-0.75);
\coordinate[] (n23L) at (3.75, 0.0);
\coordinate[] (n24L) at (3.75, 0.75);
\coordinate[] (n25L) at (3.75, 1.5);
\coordinate[] (n21R) at (4.25,-1.5);
\coordinate[] (n22R) at (4.25,-0.75);
\coordinate[] (n23R) at (4.25, 0.0);
\coordinate[] (n24R) at (4.25, 0.75);
\coordinate[] (n25R) at (4.25, 1.5);


\node[text width=4cm, align=center, text opacity=1] at (-2.5,2.25){{$\texttt{NN}(\boldsymbol x;\boldsymbol \Theta)$ - (2D)}};
\draw[dashed, black, 
fill=gray!20, rounded corners](-3.70,-2.0)--(6.0,-2.0)--(6.0,2)--(-3.70,2)--cycle;

\draw[white,fill=white, opacity=1.0, line width=0.1mm] (-3.70,0) circle (0.35);
\draw[white,fill=white, opacity=1.0, line width=0.1mm] (6,0) circle (0.35);

\draw[dashed, black, 
fill=white](-1.75,-2.0)--(0.25,-2.0)--(0.25,2)--(-1.75,2)--cycle;

\draw[dashed, black, 
fill={rgb:red,165;green,137;blue,193},  opacity=0.75](-1.75,1.06)--(0.25,1.06)--(0.25,1.76)--(-1.75,1.76)--cycle;
\node[white,text width=2cm, align=center, text opacity=1] at (-0.75,1.41){$\cos(
\boldsymbol b_1 \cdot \boldsymbol x)$};

\draw[dashed, black, 
fill={rgb:red,165;green,137;blue,193},  opacity=0.75](-1.75,0.12)--(0.25,0.12)--(0.25,0.82)--(-1.75,0.82)--cycle;
\node[white,text width=2cm, align=center, text opacity=1] at (-0.75,0.47){$\sin(
\boldsymbol b_1 \cdot \boldsymbol x)$};

\draw[dashed, black, 
fill={rgb:red,133;green,203;blue,93},  opacity=0.75](-1.75,-0.12)--(0.25,-0.12)--(0.25,-0.82)--(-1.75,-0.82)--cycle;
\node[white,text width=2cm, align=center, text opacity=1] at (-0.75,-0.47){$\cos(
\boldsymbol b_2 \cdot \boldsymbol x)$};

\draw[dashed, black, 
fill={rgb:red,133;green,203;blue,93},  opacity=0.75](-1.75,-1.06)--(0.25,-1.06)--(0.25,-1.76)--(-1.75,-1.76)--cycle;
\node[white,text width=2cm, align=center, text opacity=1] at (-0.75,-1.41){$\sin(
\boldsymbol b_2 \cdot \boldsymbol x)$};

\draw[black,dashed,fill={rgb:red,1;green,103;blue,143}, opacity=0.75, line width=0.1mm] (-3.70,0) circle (0.35);
\node[white, text width=5cm, align=center, text opacity=1] at (-3.70,0){{$\boldsymbol x$}};

\draw[black, fill=lightgray, opacity=0.75] (2.0,-1.5) circle (0.25);
\node[text width=5cm, align=center, text opacity=1] at (2.0,-1.5){$\sigma$};
\draw[black, fill=lightgray, opacity=0.75] (2.0,-0.75) circle (0.25);
\node[text width=5cm, align=center, text opacity=1] at (2.0,-0.75){$\sigma$};
\node[text width=5cm, align=center, text opacity=1] 
at (2.0, 0.1){\LARGE $\vdots$};
\draw[black, fill=lightgray, opacity=0.75] (2.0, 0.75) circle (0.25);
\node[text width=5cm, align=center, text opacity=1] at (2.0,0.75){$\sigma$};
\draw[black, fill=lightgray, opacity=0.75] (2.0, 1.5) circle (0.25);
\node[text width=5cm, align=center, text opacity=1] at (2.0,1.5){$\sigma$};

\draw[black, fill=lightgray, opacity=0.75] (4.0,-1.5) circle (0.25);
\node[text width=5cm, align=center, text opacity=1] at (4.0,-1.5){$\sigma$};
\draw[black, fill=lightgray, opacity=0.75] (4.0,-0.75) circle (0.25);
\node[text width=5cm, align=center, text opacity=1] at (4.0,-0.75){$\sigma$};
\node[text width=5cm, align=center, text opacity=1] at (4.0, 0.1){\LARGE $\vdots$};
\draw[black, fill=lightgray, opacity=0.75] (4.0, 0.75) circle (0.25);
\node[text width=5cm, align=center, text opacity=1] at (4.0,0.75){$\sigma$};
\draw[black, fill=lightgray, opacity=0.75] (4.0, 1.5) circle (0.25);
\node[text width=5cm, align=center, text opacity=1] at (4.0,1.5){$\sigma$};

\draw[black,dashed,fill={rgb:red,221;green,109;blue,16}, opacity=0.75, line width=0.1mm] (6,0) circle (0.35);

\node[white, text width=5cm, align=center, text opacity=1] at (6,0.0){$\hat{u}$};

\draw[black, ->] (xR) -- (n01L);
\draw[black, ->] (xR) -- (n02L);
\draw[black, ->] (xR) -- (n03L);
\draw[black, ->] (xR) -- (n04L);

\draw[black, ->] (n01R) -- (n11L);
\draw[black, ->] (n01R) -- (n12L);
\draw[black, ->] (n01R) -- (n14L);
\draw[black, ->] (n01R) -- (n15L);

\draw[black, ->] (n02R) -- (n11L);
\draw[black, ->] (n02R) -- (n12L);
\draw[black, ->] (n02R) -- (n14L);
\draw[black, ->] (n02R) -- (n15L);

\draw[black, ->] (n03R) -- (n11L);
\draw[black, ->] (n03R) -- (n12L);
\draw[black, ->] (n03R) -- (n14L);
\draw[black, ->] (n03R) -- (n15L);

\draw[black, ->] (n04R) -- (n11L);
\draw[black, ->] (n04R) -- (n12L);
\draw[black, ->] (n04R) -- (n14L);
\draw[black, ->] (n04R) -- (n15L);

\draw[black, ->] (n11R) -- (n21L);
\draw[black, ->] (n11R) -- (n22L);
\draw[black, ->] (n11R) -- (n24L);
\draw[black, ->] (n11R) -- (n25L);

\draw[black, ->] (n12R) -- (n21L);
\draw[black, ->] (n12R) -- (n22L);
\draw[black, ->] (n12R) -- (n24L);
\draw[black, ->] (n12R) -- (n25L);

\draw[black, ->] (n14R) -- (n21L);
\draw[black, ->] (n14R) -- (n22L);
\draw[black, ->] (n14R) -- (n24L);
\draw[black, ->] (n14R) -- (n25L);

\draw[black, ->] (n15R) -- (n21L);
\draw[black, ->] (n15R) -- (n22L);
\draw[black, ->] (n15R) -- (n24L);
\draw[black, ->] (n15R) -- (n25L);

\draw[black, ->] (n21R) -- (xhiL);
\draw[black, ->] (n22R) -- (xhiL);
\draw[black, ->] (n24R) -- (xhiL);
\draw[black, ->] (n25R) -- (xhiL);

\end{tikzpicture}
\end{center}
\caption{Schematic representation of how to implement exact $C^\infty$ periodic boundary conditions  in two-dimensional problems through an added $C^\infty$ periodic input transfer layer.}
 \label{fig:NNmod2D}
\end{figure}
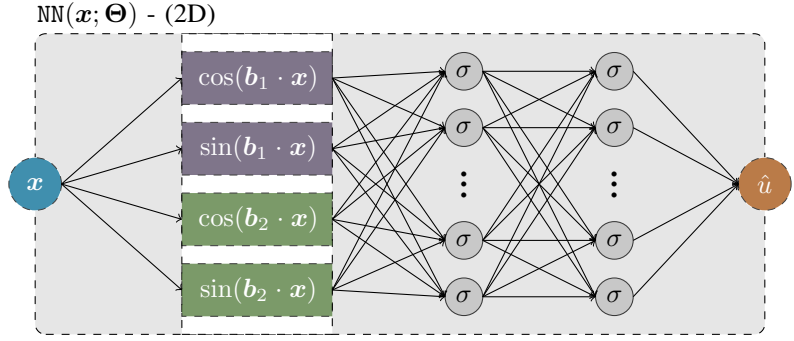

For a two-dimensional Bravais lattice, and considering that $\xi({\bs x})$ is a real periodic function which requires that $\overline{u_{\bs G}} = u_{-\bs G}$, Eq.\ \eqref{E:Fourier3D} can be represented with the following exponential Fourier series
\begin{align}
\xi({\bs x})& = \sum_{m,n=0}^{\infty}\cos\left(n{\bs b_1}\cdot {\bs x}\right)\left[\alpha_{m,n}\cos\left(n{\bs b_2}\cdot {\bs x} \right)+\beta_{m,n}\sin\left(n{\bs b_2}\cdot {\bs x}\right)\right] \nonumber \\
& + \sum_{m,n=0}^{\infty}\sin\left(m{\bs b_1}\cdot {\bs x}\right)\left[\gamma_{m,n}\cos\left(n {\bs b_2}\cdot{\bs x}\right)+\delta_{m,n}\sin\left(n {\bs b_2}\cdot{\bs x}\right)\right]\,,
\label{E:FourierVtrig2D}
\end{align}
where $\alpha_{m,n}$, $\beta_{m,n}$, $\gamma_{m,n}$, and $\delta_{m,n}$ are real constants. Using multiple-angle trigonometric identities, the above sum can be expressed in the form
\begin{align}
    \xi({\bs x}) =\mathcal{F}_{\mathrm{2D}}\left[
    \cos\left({\bs b_1}\cdot {\bs x}\right),\, \sin\left({\bs b_1}\cdot {\bs x}\right),\, \cos\left({\bs b_2}\cdot {\bs x}\right),\, \sin\left({\bs b_2}\cdot {\bs x}\right)\right],
    \label{E:F2D}
\end{align}
where $\mathcal{F}_{\mathrm{2D}}$ is a nonlinear function. As before, the problem of approximating the nonlinear smooth function $\xi({\bs x})$ is replaced by approximating $\mathcal{F}_{\mathrm{2D}}$ in an ANN by modifying the neural network architecture by adding the $C^\infty$  periodic basis functions as a transfer layer as demonstrated in Fig.\ \ref{fig:NNmod2D}.

It is important to remark that the pairs of trigonometric functions given in Eq.~\ref{E:F2D} provide a linearly independent irreducible basis for the solution, hence choosing these pairs of functions is sufficient to implement PBCs with neural networks in a periodic lattice of any shape. This is partly mentioned in~\cite{DONG2021110242} for the case of square lattices. However, we note that they do not justify why trigonometric functions can be used to implement PBCs, as we do here.

Extending from 2D to 3D is straightforward and suffice to say that a real periodic function  $\xi({\bs x})$ is represented by the corresponding nonlinear function $\mathcal{F}_{\mathrm{3D}}$ of the form
\begin{align}
    \xi({\bs x}) =\mathcal{F}_{\mathrm{3D}}\left[
    \cos\left({\bs b_1}\cdot {\bs x}\right),\, \sin\left({\bs b_1}\cdot {\bs x}\right),\,
    \cos\left({\bs b_2}\cdot {\bs x}\right),\, \sin\left({\bs b_2}\cdot {\bs x}\right),\,
    \cos\left({\bs b_3}\cdot {\bs x}\right),\, \sin\left({\bs b_3}\cdot {\bs x}\right)\right]\,.
    \label{E:F3D}
\end{align}
\subsubsection{Loss Contributions in Periodic Homogenization with PINNs}
In this work, we use PINNs to approximate the solution of Eq.\ \eqref{E:cell-problem} whose derivative is then used to compute the integral given in Eq.\ \eqref{E:propertyhomogenized}. We note that a solution up to a constant is sufficient to this end, eliminating the need to apply Dirichlet boundary conditions. With the a priori and exact satisfaction of PBCs by adding a $C^\infty$  periodic transfer layer, as demonstrated in the previous sections, the residual loss term corresponding  to the boundary conditions is zero, i.e.~$\mathcal{L}_\mathcal{B}(\boldsymbol{\theta};\mathcal{T}_\mathcal{B}) = 0$.

With the elimination of boundary terms, the conventional loss term is just that of the PDE residual $\text{min}\mathcal{L}(\boldsymbol{\theta};\mathcal{T})=\omega_\mathcal{P}\mathcal{L}_\mathcal{P}(\boldsymbol{\theta};\mathcal{T}_\mathcal{P})$. In the conventional application of the PINNs $\mathcal{L}_\mathcal{P}$ is the square norm of the PDE residuals
\begin{align}
\mathcal{L}_\mathcal{P}=\dfrac{1}{N}\sum_{k=1}^N \left(\left[
\dfrac{\partial}{\partial x_i}\left(a_{ij}(\bs x^k)
\dfrac{\partial \chi_\mathtt{NN}^m(\bs x^k; \bs \Theta)}{\partial x_j}\right)+
\dfrac{\partial a_{im}(\bs x^ k)}{\partial x_i}\right] \right)^2\,.
\label{E:nn_loss_conventional}
\end{align}
Alternatively, one may individually scale the loss contribution due to PDE at each material point with
\begin{align}
\mathcal{L}_\mathcal{P}=\dfrac{1}{N}\sum_{k=1}^N \left( \dfrac{1}{a(\bs x^ k)}\left[
\dfrac{\partial}{\partial x_i}\left(a_{ij}(\bs x^k)
\dfrac{\partial \chi_\mathtt{NN}^m(\bs x^k;\bs \Theta)}{\partial x_j}\right)+
\dfrac{\partial a_{im}(\bs x^ k)}{\partial x_i}\right] \right)^2\,,
\label{E:nn_loss_unnconventional}
\end{align}
assuming isotropic constituent properties with $\bs a(\bs x^ k)=a(\bs x^ k)\bs 1$. $\bs x^ k$ corresponds to the collocation point at which the PDE is computed.
\section{Applications}
\label{Sec:Apps}
In this section, we solve illustrative examples of regular and stochastic periodic material systems in 1D, 2D, and 3D. In the studied problems, the crystal structure is formed by attaching the basis containing a circular disk to every lattice point of the space lattice given in Fig.~\ref{fig:BravaisLattice1D2D3D}.
We consider a property contrast of $a_1/a_2=100$,  where $a_1$ corresponds to the inclusion property $a_\mathrm{i}$. This contrast is relatively large compared to the applications reported in the literature. In Ref. \cite{Gokuzummca24020040}, for example, in which PINNs in homogenization in electrostatics is investigated, a phase contrast of 10 was considered. The property of the matrix material is taken as $a_\mathrm{m}=a_2=1$. In \cite{Henkesetal2022}, a phase contrast below 10 was used for the elasticity moduli of the constituents while seeking out a physics-informed neural network solution for a continuum micromechanical problem.

\subsection{Numerical Details}
Unless otherwise stated, the neural network is chosen with a density of 50 and depth of 3, as these values were observed to suit the model to be trained well.  The learning rate is varied between 0.001 and 0.010. In the optimization step, the first 5000 epochs belong to adam optimizer, whereas the rest 25000 are to L-BFGS. Individually scaled and unscaled loss terms, see, Eqs.\ \eqref{E:nn_loss_unnconventional} and \eqref{E:nn_loss_conventional}, respectively, are considered.  Networks with and without the Fourier feature layer are also tested, observing that  without the Fourier feature layer, the application of periodic boundary conditions is only approximate. No additional scaling is applied to the corresponding loss terms.

In all of the problems of this study, the collocation points are generated using the Hammersley low-discrepancy sequence, which allows the training points to be more evenly distributed over the problem domain, leading to more accurate solutions. The number of collocation points should allow sampling from the smoothing region in the property field distribution. This region is the only one providing nonzero source terms, hence a nontrivial solution.

The model training with neural networks is accomplished on NVIDIA Tesla P100- and V100-GPUs. The training times vary depending on the batch size, the depth, and the density of the network.
\subsection{A One-dimensional Problem}
In 1D, Eq.\ \eqref{E:propertyhomogenized} becomes
\begin{align}
a^\star=\dfrac{1}{L}\,\int_{\mathcal{V}}a(x)\left[1+
\dfrac{\mathrm{d} \chi(x)}{\mathrm{d} x}\right]
\mathrm{d}x\, ,
\label{E:propertyhomogenized_1D}
\end{align}
where $a(x)$ is a scalar quantity and $x$ is the spatial coordinate. The one-dimensional composite unit cell domain amounts to  $\mathcal{V}:=\{x : -L/2 \leq x \leq L/2\}$, which is divided into two phases with corresponding domains $\mathcal{V}_1$ and $\mathcal{V}_2$, where $\mathcal{V}_1\bigcup \mathcal{V}_2=\mathcal{V}$.  Here, the $\mathcal{V}-$periodic corrector function $\chi(x)$, with $\chi(-L/2)=\chi(L/2)$, is the solution of the 1D cell-problem
\begin{align}
-\dfrac{\mathrm{d}}{\mathrm{d} x}\left(a(x)
\dfrac{\mathrm{d} \chi}{\mathrm{d} x}\right)=
\dfrac{\mathrm{d} a(x)}{\mathrm{d} x}\,.
\label{E:cell-problem_1D}
\end{align}
Integrating the above equation  twice and rearranging, we obtain
\begin{align}
a^\star=\left[\dfrac{1}{L}\int_{-L/2}^{L/2}\dfrac{1}{a(x)}\mathrm{d}x\right]^{-1},
\label{E:cell-problem_1D_rev3}
\end{align}
where we have used the periodicity of $\chi(x)$ with $\chi(-L/2)=\chi(L/2)$.

Letting its center correspond to the origin, we select $\mathcal{V}_1:=\{x : -r < x < r\}$ as the domain of phase 1. The remaining domain, or phase 2, represents the matrix material, and it is defined as  $\mathcal{V}_2:=\{x: -L/2 \leq x \leq -r\}\bigcup \{x : r \leq x \leq L/2\}$. In order to have a continuous derivative  at the interfaces located at $x=\pm r$, we regularize the $a(x)$ function with a smooth interface  as follows
\begin{align}
a(x)= a_1 +\dfrac{a_2-a_1}{2}\left[1-\tanh\left(\dfrac{x-r}{\xi}\right)\right]\,,
\label{E:regularization_1D_2}
\end{align}
where $a_1$ and $a_2$ are the properties of phase 1 and phase 2, respectively, $\xi$ is the numerical regularization (or smoothing) parameter. Putting this expression into Eq.~\eqref{E:cell-problem_1D_rev3} and evaluating the integral, we obtain
\begin{align}
a^\star=a_1 a_2 \dfrac{L}{\xi}
\Eval{\left[\log\left([[a_2-a_1]f(x)+a_1+a_2]^{[a_2-a_1]}\dfrac{[1-f(x)]^{a_1}}{[1+f(x)]^{a_2}}\right)\right]^{-1}}{0}{L/2}\,,
\label{E:analytical_solution_part_v1}
\end{align}
where we have defined $f(x)= \tanh\left((x-r)/\xi\right)$. Note that as $\xi\to 0$, $f(x)\to 1$ for $x>r$ and $f(x)\to-1$ for $x<r$. Therefore, as $\xi\to 0$ we obtain
\begin{align}
a^\star=\dfrac{a_1 a_2 L}{a_2[L-2r]+2a_1 r}\,,
\label{E:analytical_solution_part_v2}
\end{align}
which corresponds to  the conventional Reuss (harmonic) average for 1D composites in the sharp interface limit.
\subsubsection{PINN Implementation and Numerical Results}
As an illustrative example, we take $L=200\xi$ as the size of the  one-dimensional composite domain, with $\xi=1$, and consider inclusion volume fractions of $\phi_\mathrm{i}=\{0.10,0.20,\ldots,0.90\}$.

We can see from Eq.\ \eqref{E:regularization_1D_2} that the maximum absolute slope $|a'(x)|$ occurs at the inflection point $x=r$ with $|a'(r)|=|a_2-a_1|/2\xi$. Therefore, reducing $\xi$ or increasing the phase contrast to yield a higher property gap $|a_2-a_1|$ will have a similar influence on the property gradients and thus should be considered while selecting the number of collocation points. Here, the training batch size is selected as 512. However, in more general cases  with lower phase contrast, a lower number of collocation points also works.

The PINNs simulations will be compared to high-resolution finite element simulations  done with ABAQUS, which will be  used to generate the ground truth by discretizing the one-dimensional domain with 5000 elements.

\begin{figure*}[htb!]
\centering
\subfigure[with scaling]{
{{\includegraphics[width=0.45\textwidth,
trim=0 100 0 80, clip]{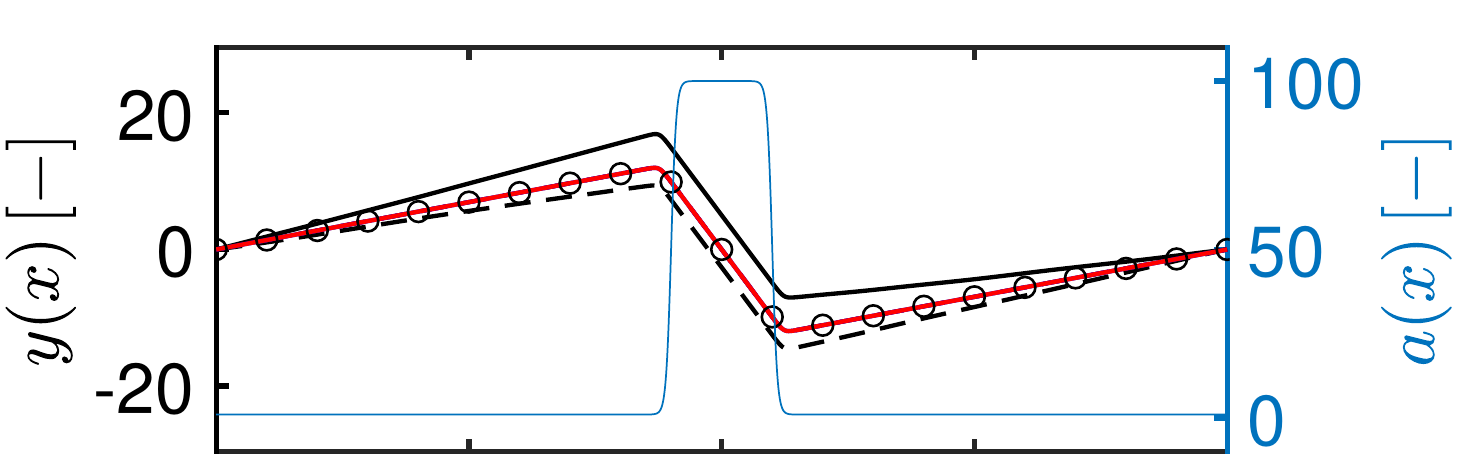}}}
}
\subfigure[without scaling]{
{{\includegraphics[width=0.45\textwidth,
trim=0 100 0  80, clip]{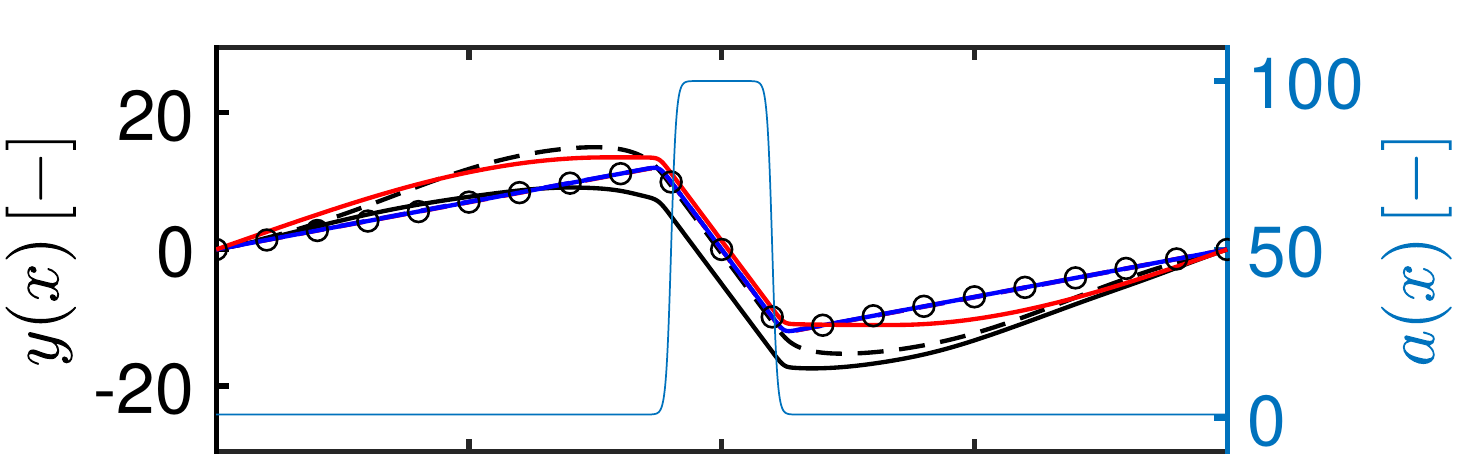}}}
}
{{\includegraphics[width=0.45\textwidth,
trim=0 100 0 80, clip]{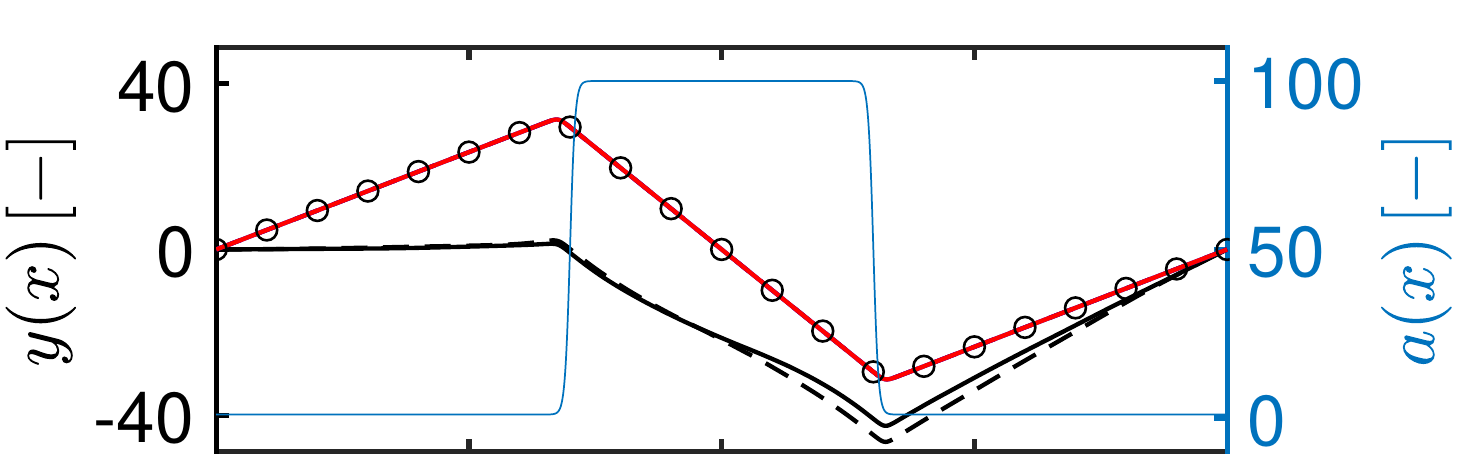}}}
{{\includegraphics[width=0.45\textwidth,
trim=0 100 0 80, clip]{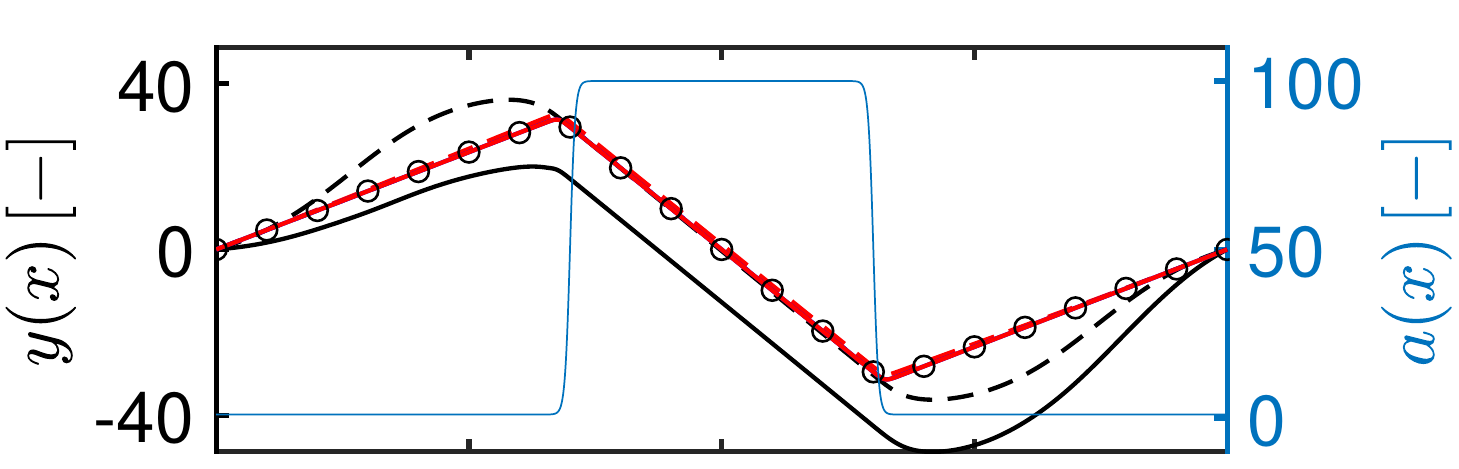}}}
{{\includegraphics[width=0.45\textwidth,
trim=0 100 0 80, clip]{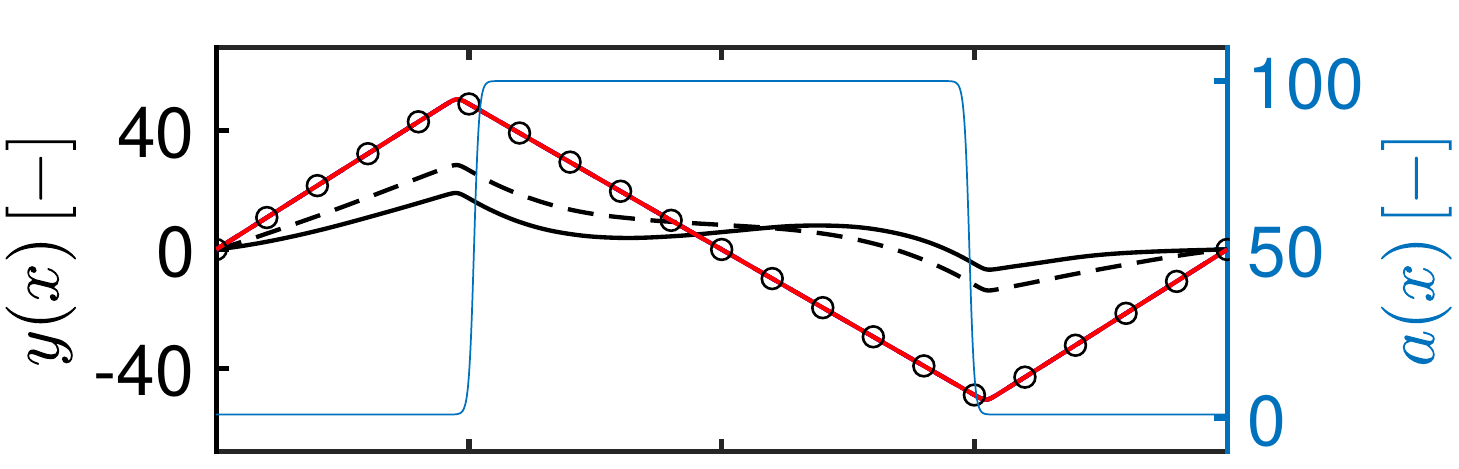}}}
{{\includegraphics[width=0.45\textwidth,
trim=0 100 0 80, clip]{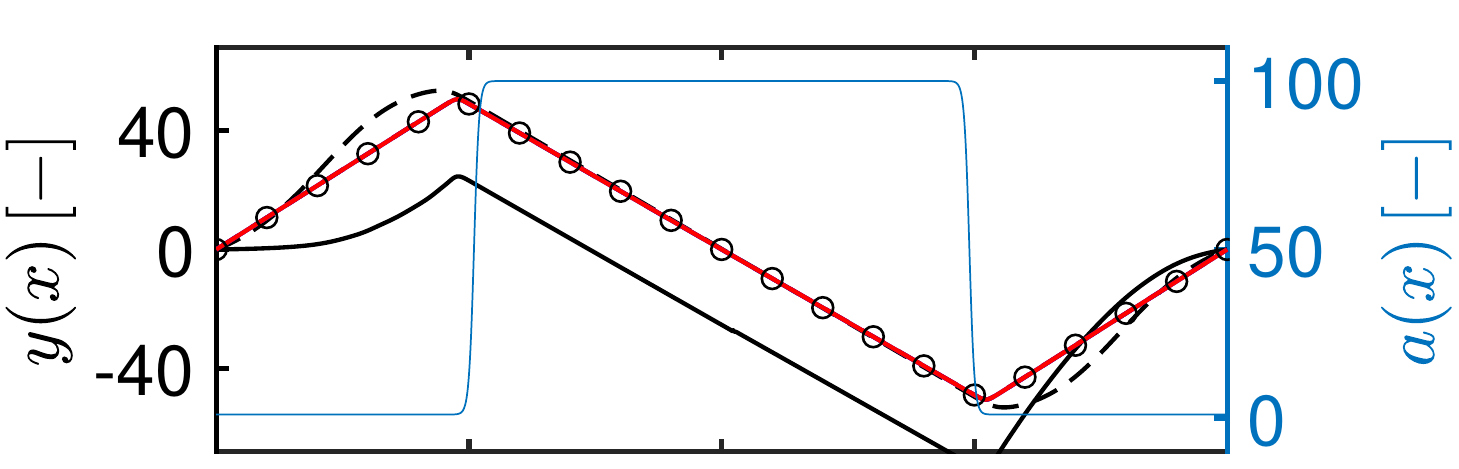}}}
{{\includegraphics[width=0.45\textwidth,
trim=0 100 0 80, clip]{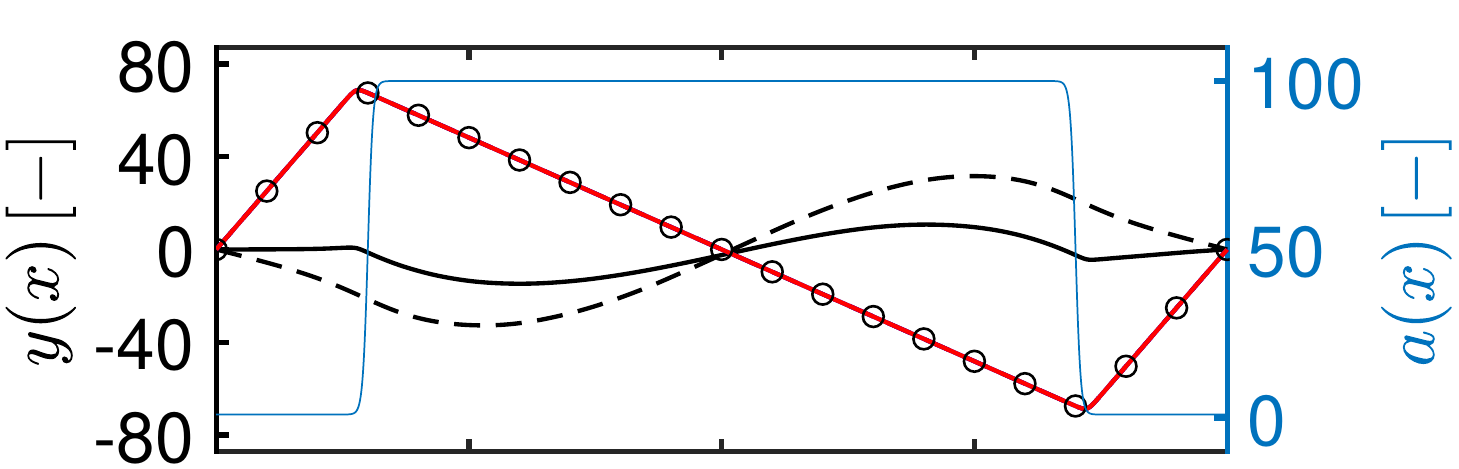}}}
{{\includegraphics[width=0.45\textwidth,
trim=0 100 0 80, clip]{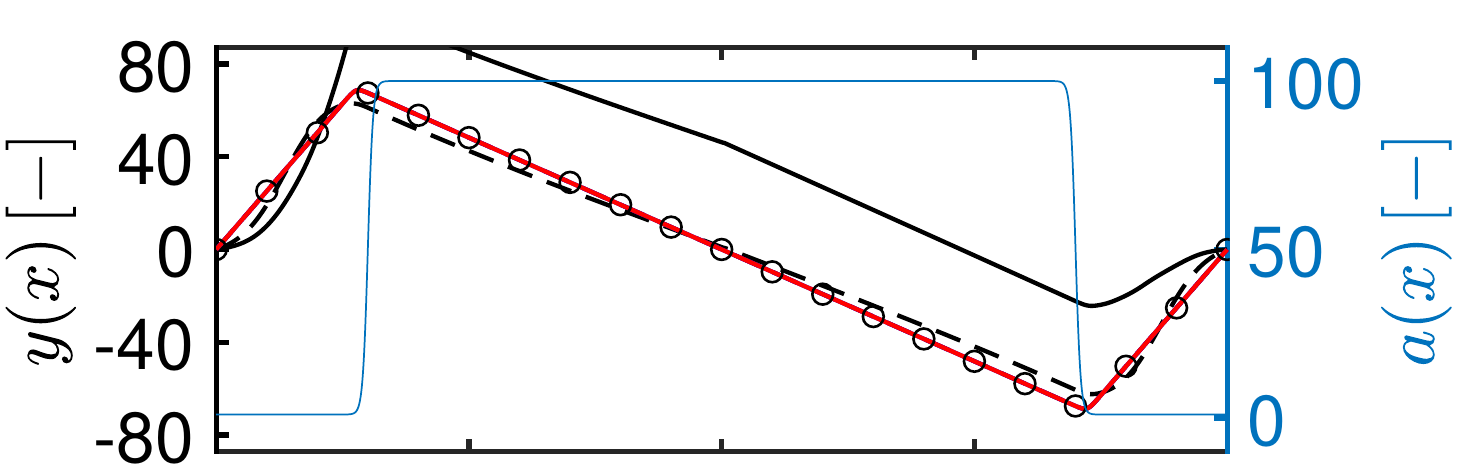}}}
{{\includegraphics[width=0.45\textwidth,
trim=0 40 0 80, clip]{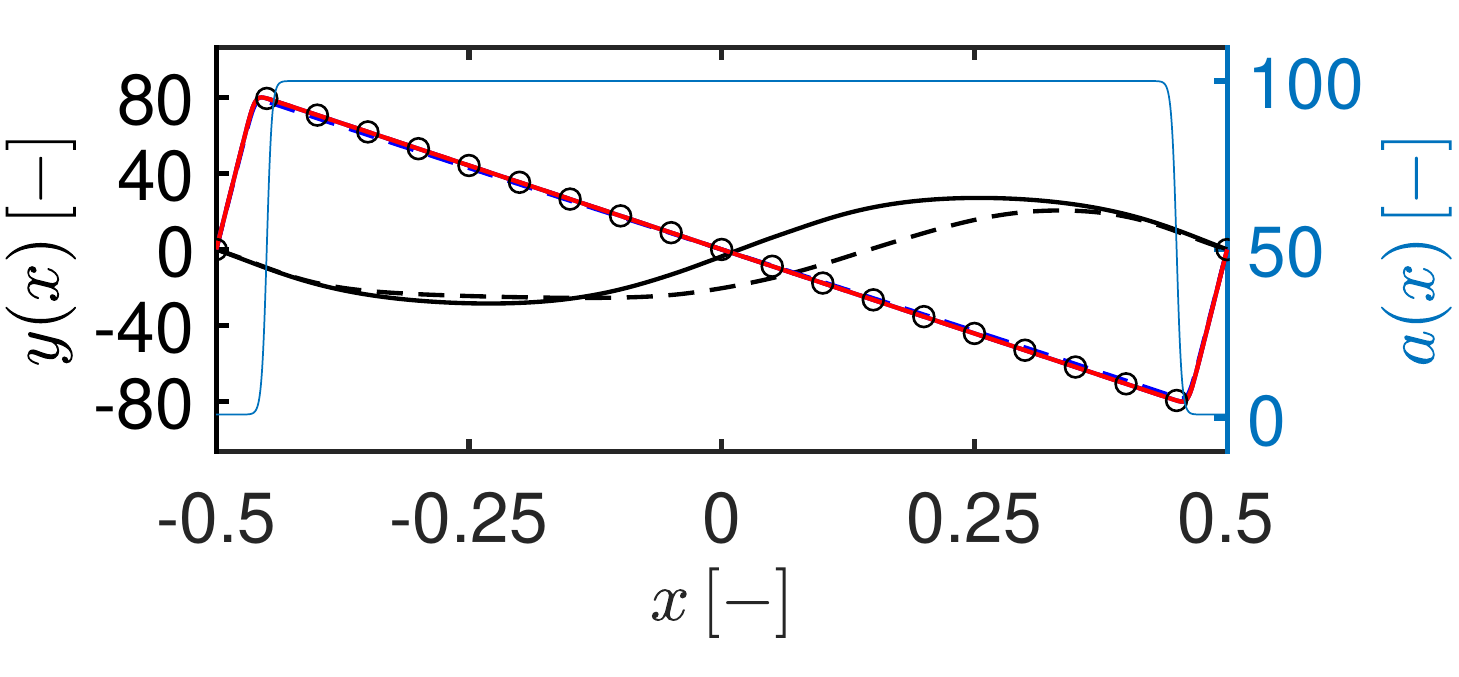}}}
{{\includegraphics[width=0.45\textwidth,
trim=0 40 0 80, clip]{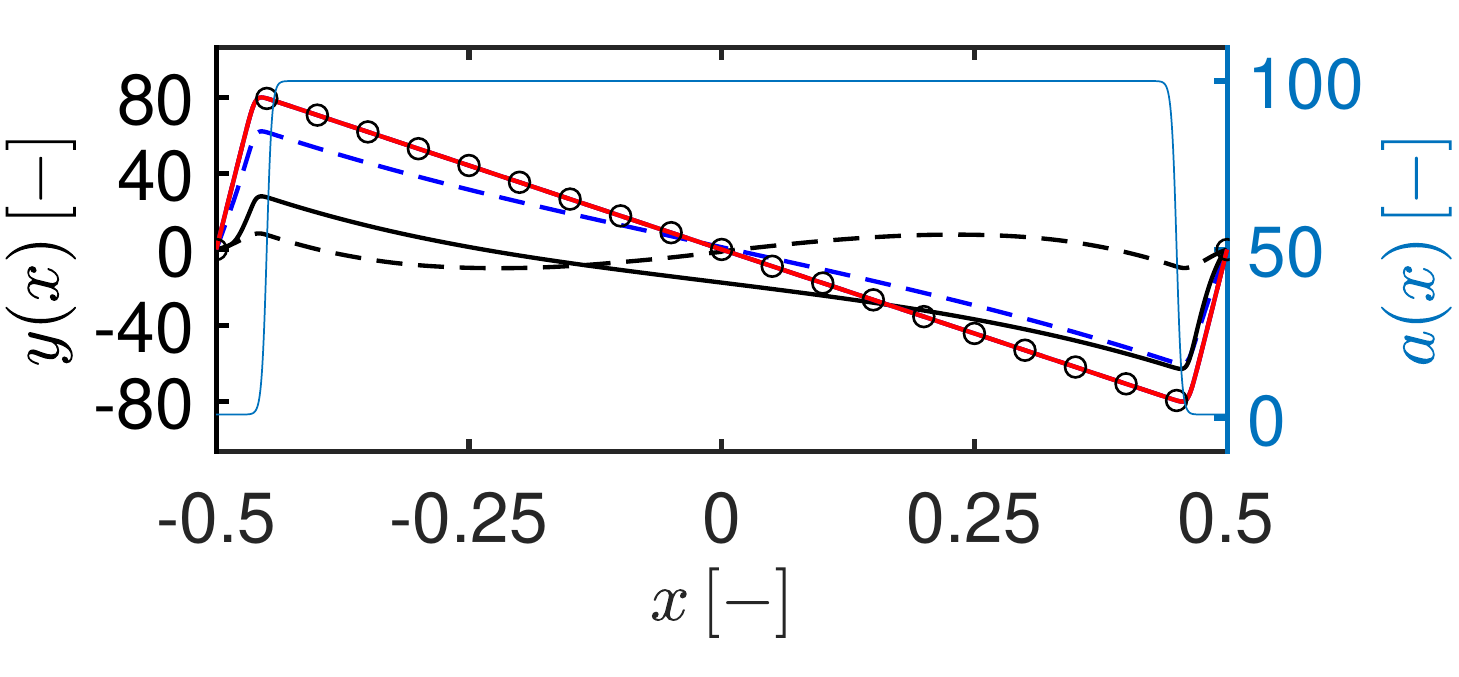}}}
\vspace{-5pt}
\caption{The ANN predictions for the solution field at the end of 30000 epochs and comparison to ABAQUS results (discrete circular markers). The light blue line represents the smoothed material property distribution. The left column belongs to the results with scaling and the right column without scaling. Dashed curves have a learning rate of 0.001 and continuous ones 0.01 for the adam optimizer. Black curves give results for no Fourier feature, whereas dark blue and red for low- and high-frequency Fourier features with a single and the first 10 integer multiples of the reciprocal base vector, respectively. }
\label{F:results_1D_y}
\end{figure*}

\begin{figure*}[htb!]
\centering
\subfigure[with scaling]{
{{\includegraphics[width=0.45\textwidth,
trim=0 100 0 80, clip]{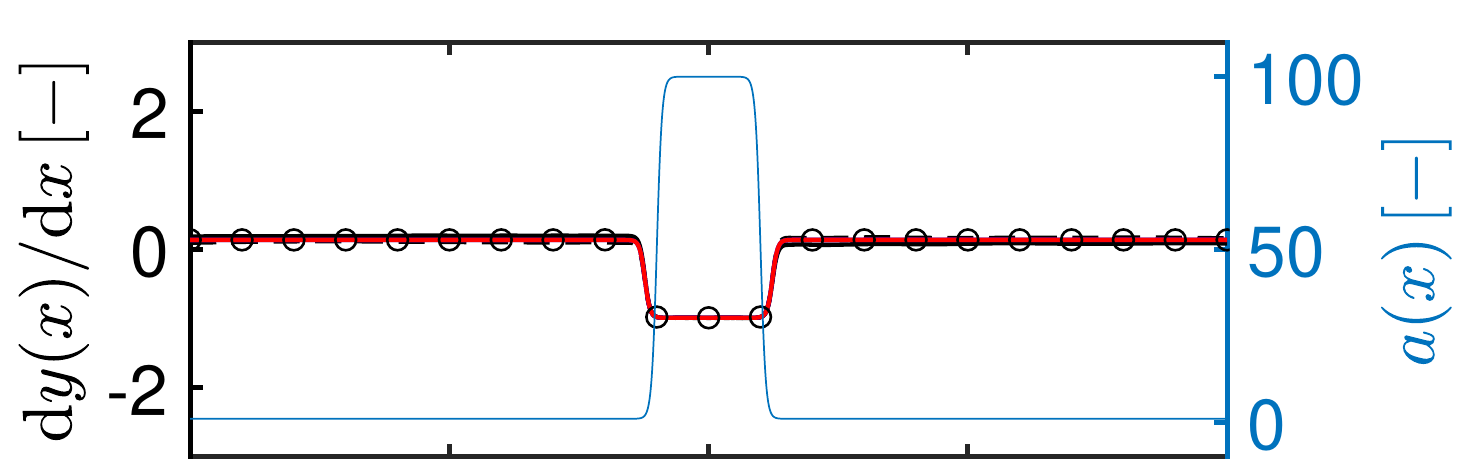}}}
}
\subfigure[without scaling]{
{{\includegraphics[width=0.45\textwidth,
trim=0 100 0 80, clip]{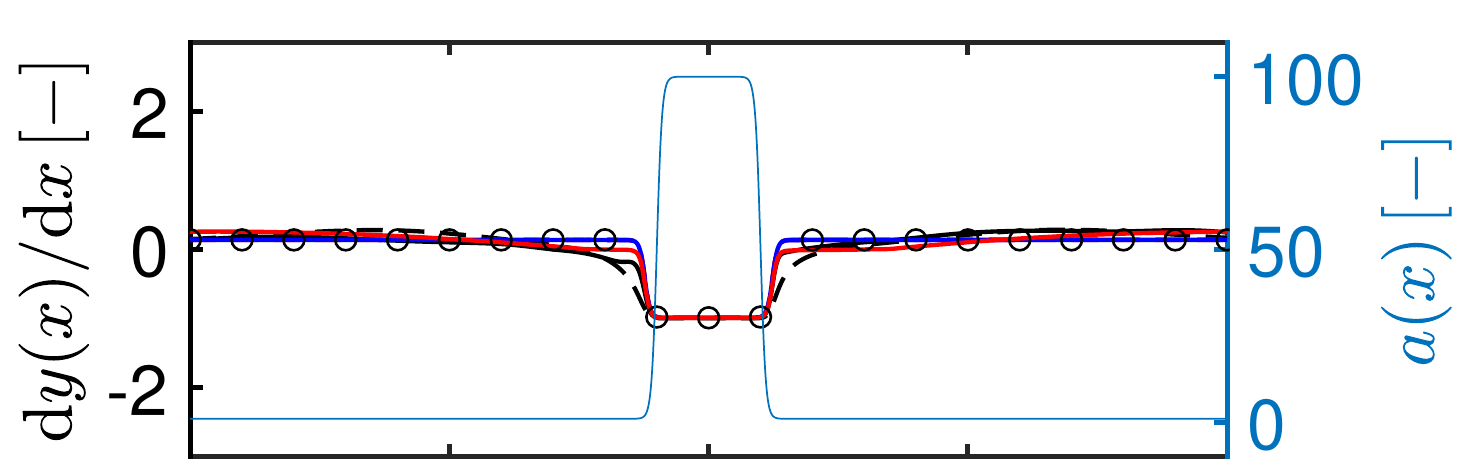}}}
}\\
\vspace{-5pt}
{{\includegraphics[width=0.45\textwidth,
trim=0 100 0 80, clip]{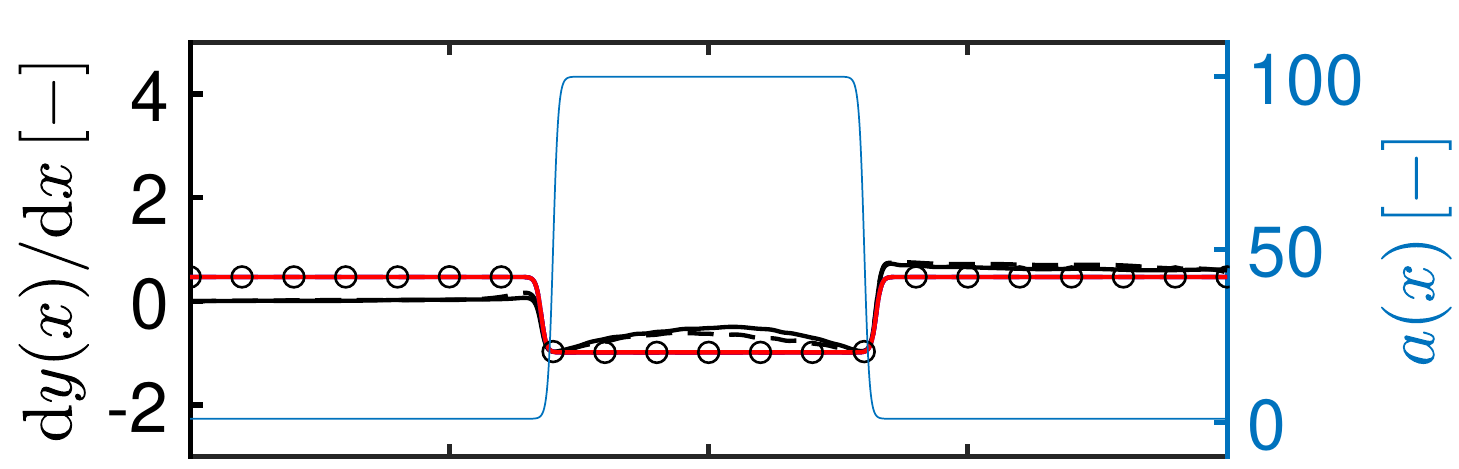}}}\,\,\,\,
{{\includegraphics[width=0.45\textwidth,
trim=0 100 0 80, clip]{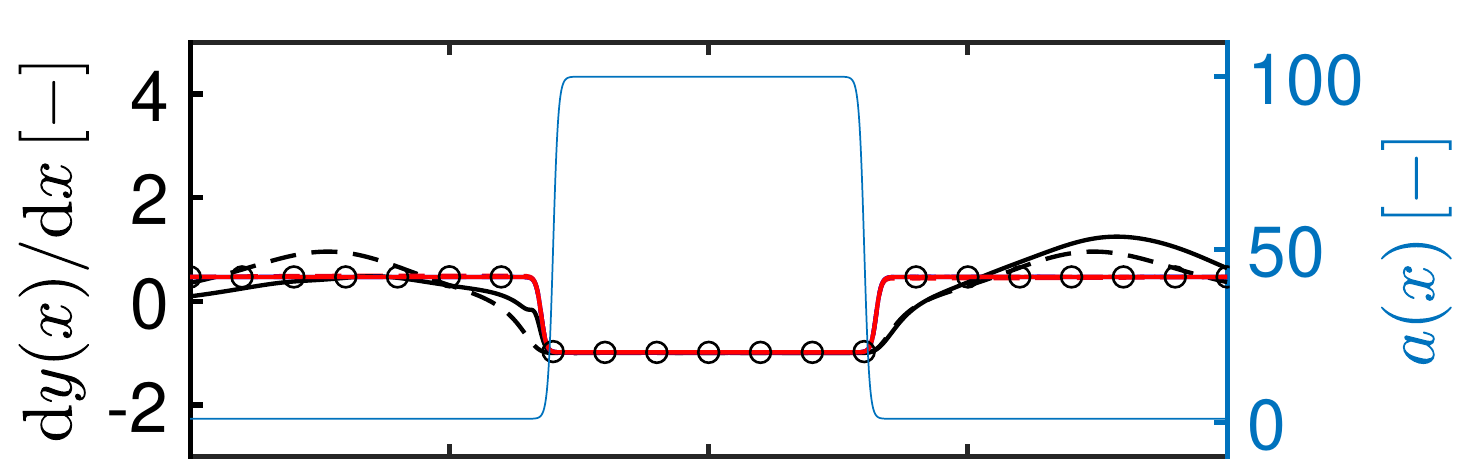}}}\\
{{\includegraphics[width=0.45\textwidth,
trim=0 100 0 80, clip]{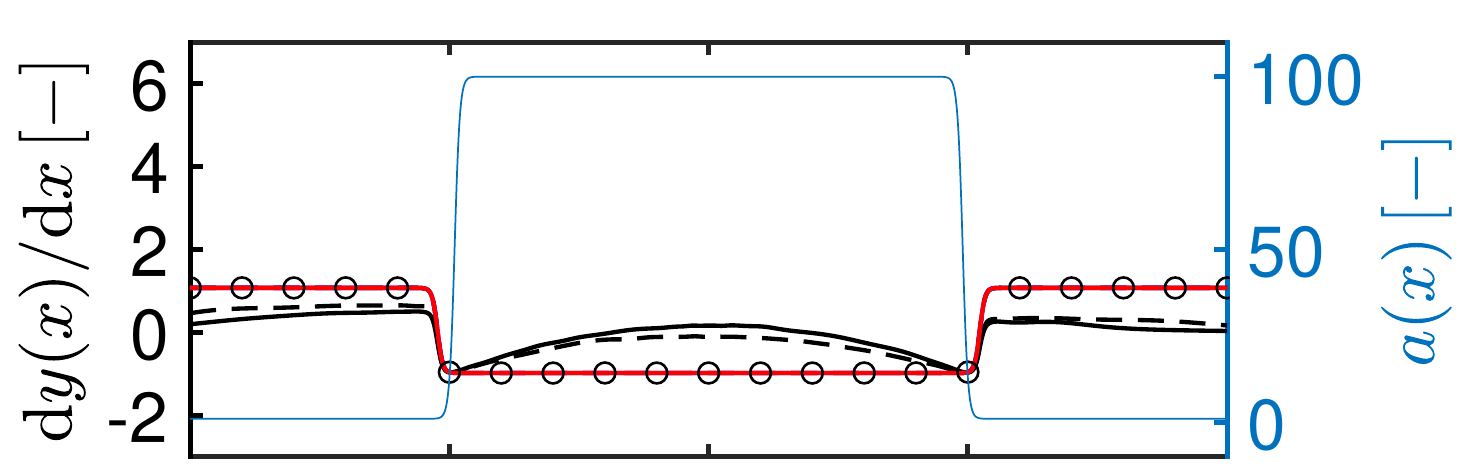}}}\,\,\,\,
{{\includegraphics[width=0.45\textwidth,
trim=0 100 0 80, clip]{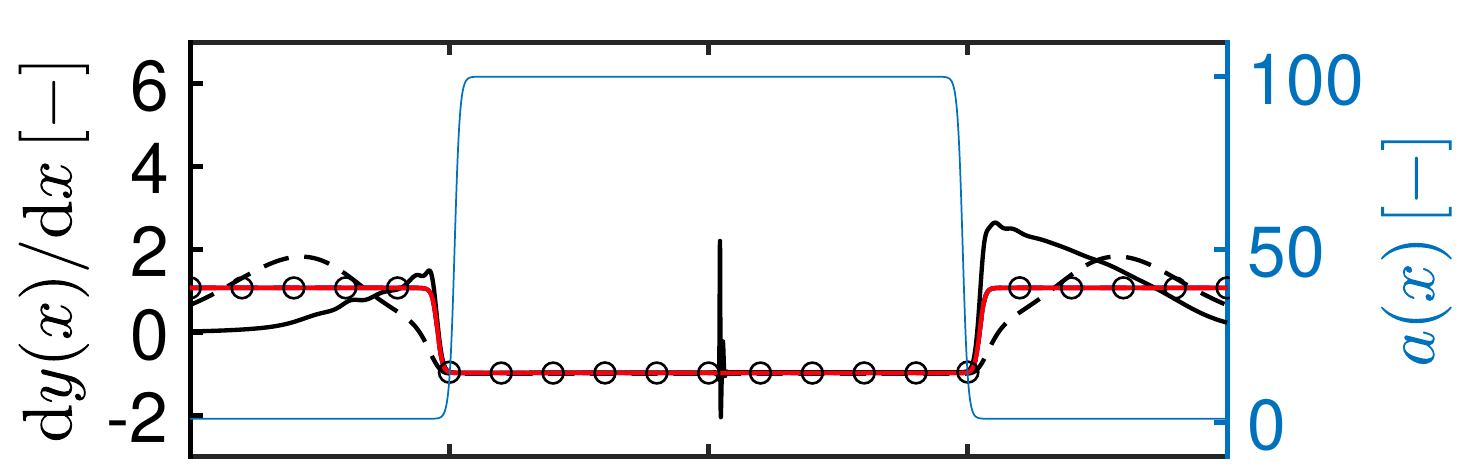}}}\\
{{\includegraphics[width=0.45\textwidth,
trim=0 100 0 80, clip]{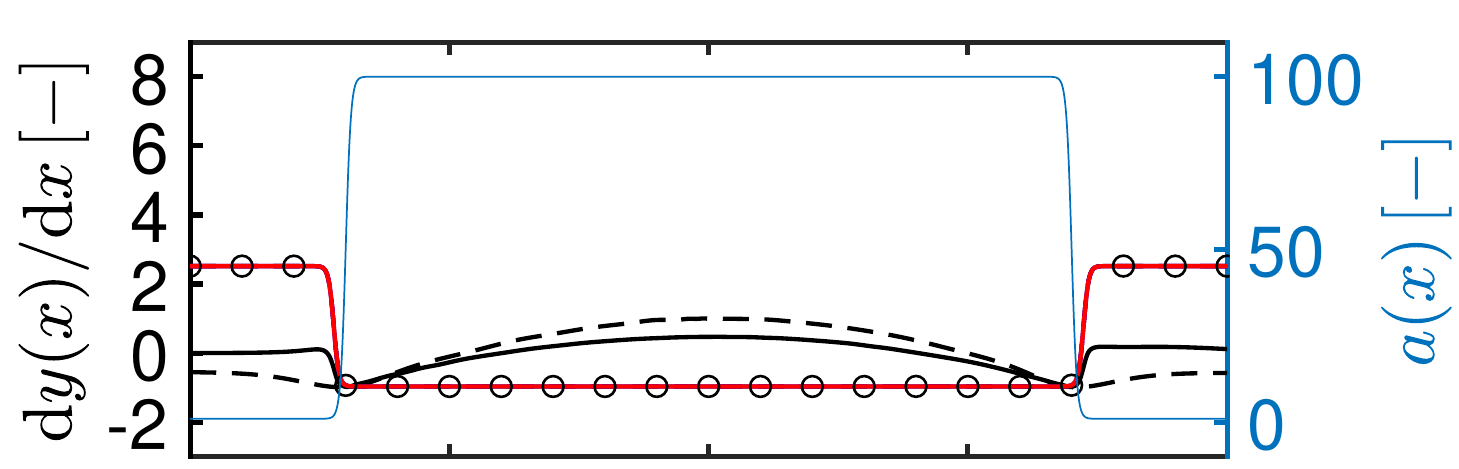}}}\,\,\,\,
{{\includegraphics[width=0.45\textwidth,
trim=0 100 0 80, clip]{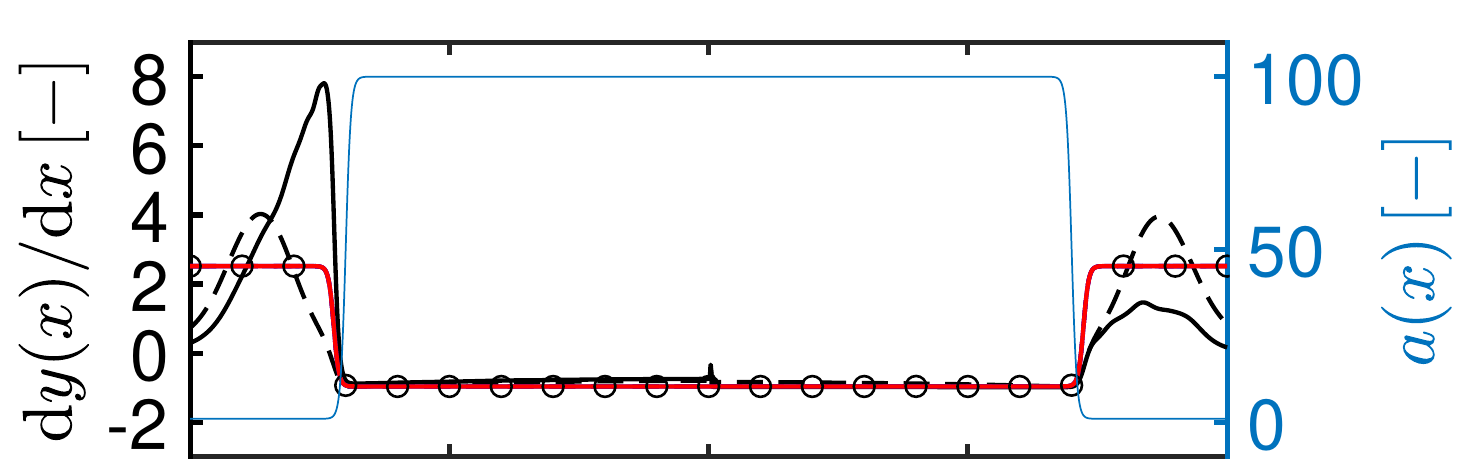}}}\\
{{\includegraphics[width=0.45\textwidth,
trim=0 40 0 80, clip]{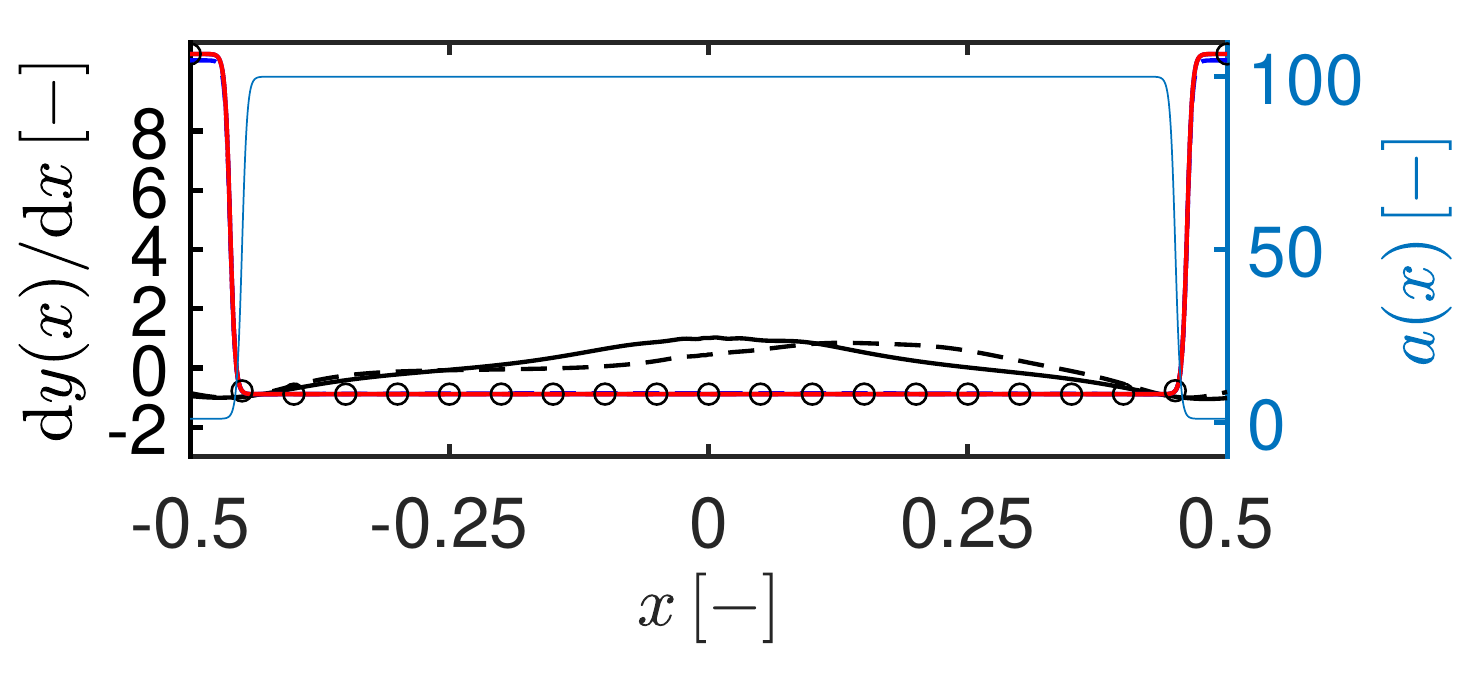}}}\,\,\,\,
{{\includegraphics[width=0.45\textwidth,
trim=0 40 0 80, clip]{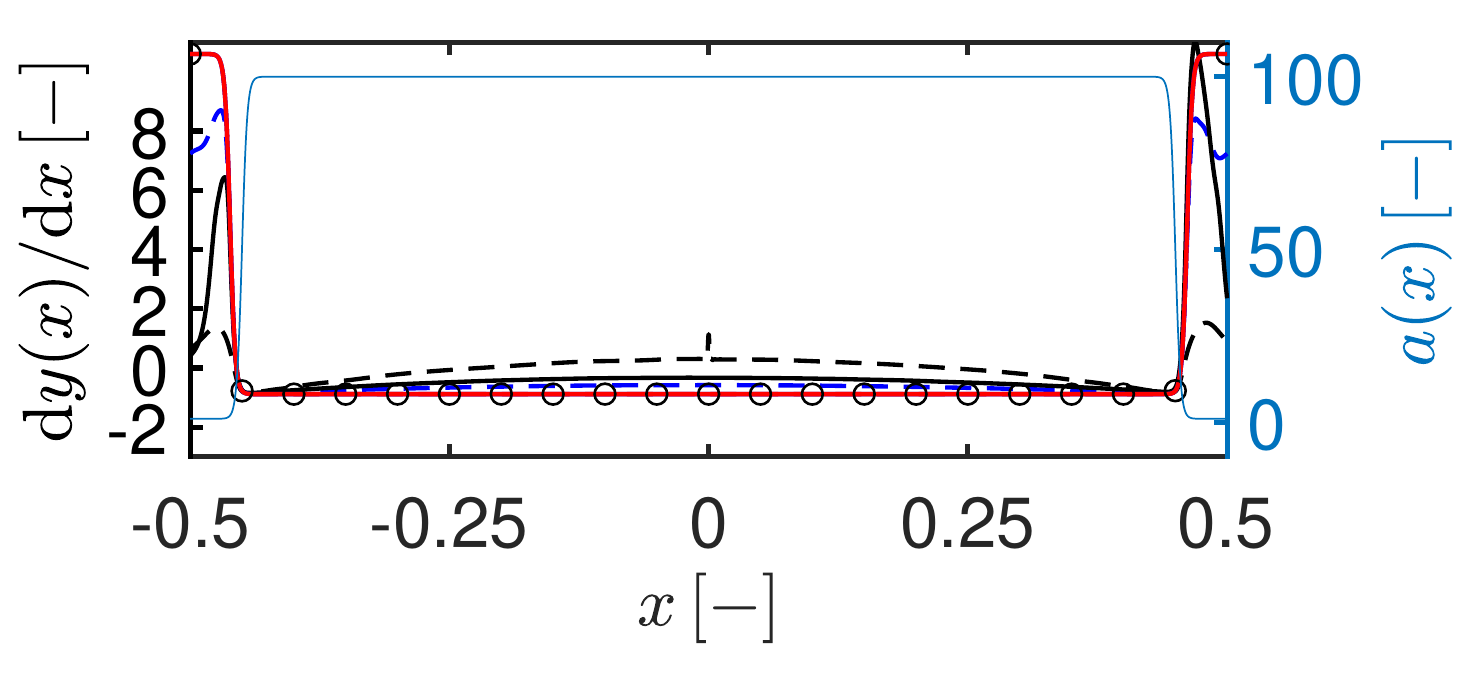}}}\\
\vspace{-5pt}
\caption{The ANN predictions for the solution field derivative at the end of 30000 epochs and comparison to ABAQUS results (discrete circular markers). The light blue line represents the smoothed material property distribution. The left column belongs to the results with scaling, and the right column is without scaling. Dashed curves have a learning rate of 0.001 and continuous ones 0.01 for the adam optimizer. Black curves give results for no Fourier feature, whereas dark blue and red for low- and high-frequency Fourier features with a single and the first 10 integer multiples of the reciprocal base vector, respectively.}
\label{F:results_1D_dydx}
\end{figure*}

The solution field and its spatial derivative are shown in Figs.\ \ref{F:results_1D_y} and \ref{F:results_1D_dydx}, respectively. The spatial derivative of the predicted solution is determined at selected collocation points using automatic differentiation in TensorFlow. Results from PINNs simulations are compared to finite element solutions (shown in  discrete circular markers). In both figures,  the left column corresponds to solutions for individual loss scaling with Eq.\ \eqref{E:nn_loss_unnconventional} and the right column without it. Our numerical tryouts show that the effect of this modification is marginal for the lower  phase contrasts.  However, we observe that the scaling increases accuracy for the case of low-volume fractions with  high contrast phase, as would be noticed by comparing top columns. Our results show that meaningful solutions for a wide range of phase volume fractions are only obtained when  Fourier features  are included.

\begin{figure*}[htb!]
\centering
\centering
\subfigure[with scaling]{
{{\includegraphics[width=0.45\textwidth,
trim=0 100 0 80, clip]{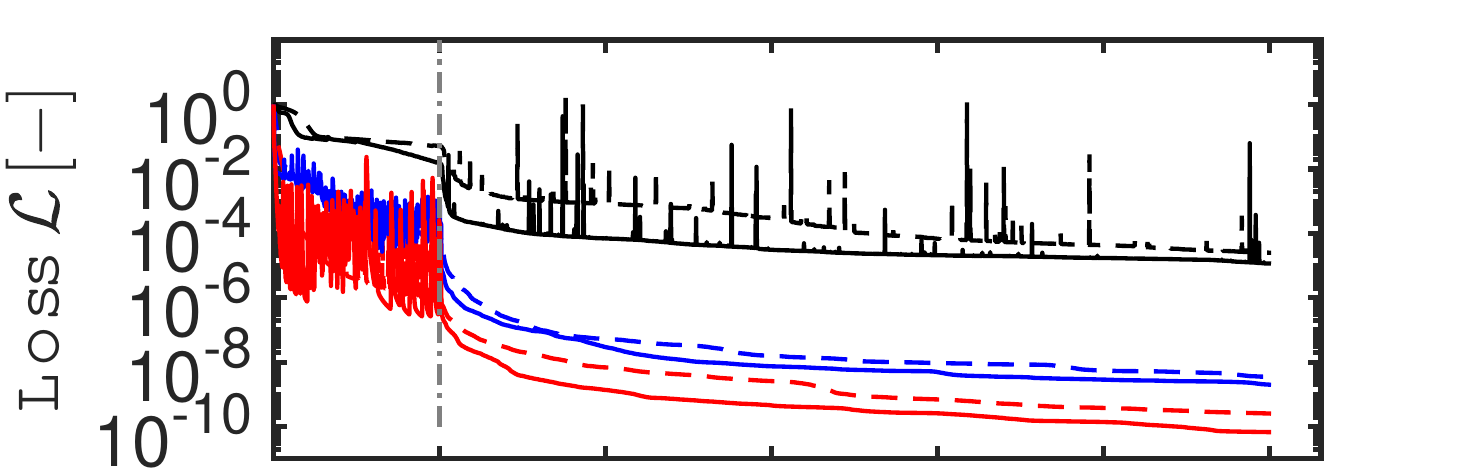}}}
}
\subfigure[without scaling]{
{{\includegraphics[width=0.45\textwidth,
trim=0 100 0 80, clip]{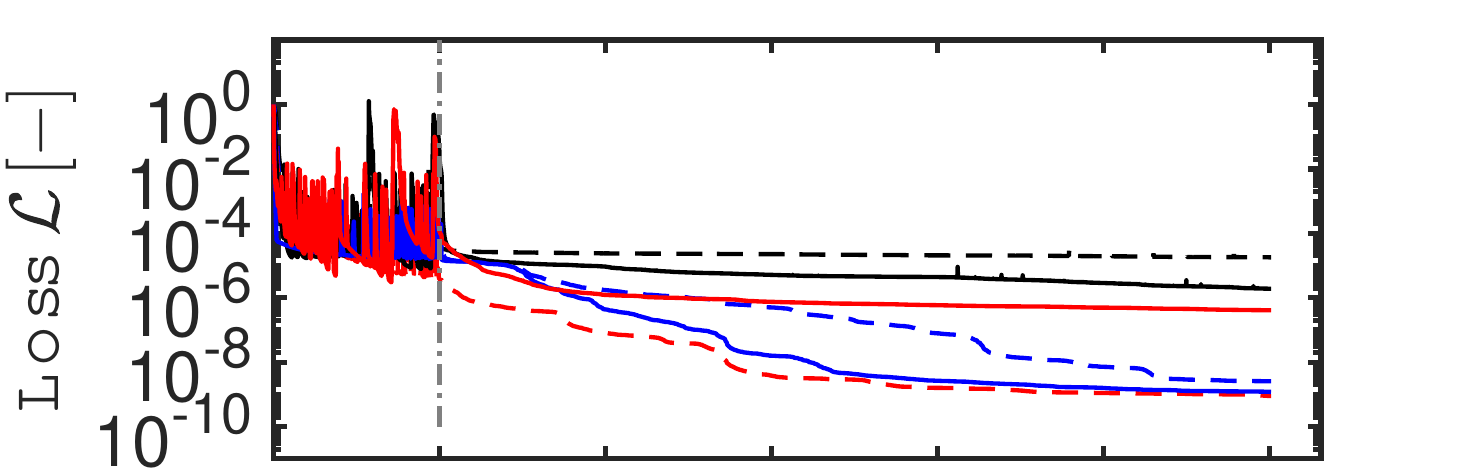}}}
}\\
\vspace{-5pt}
{{\includegraphics[width=0.45\textwidth,
trim=0 100 0 80, clip]{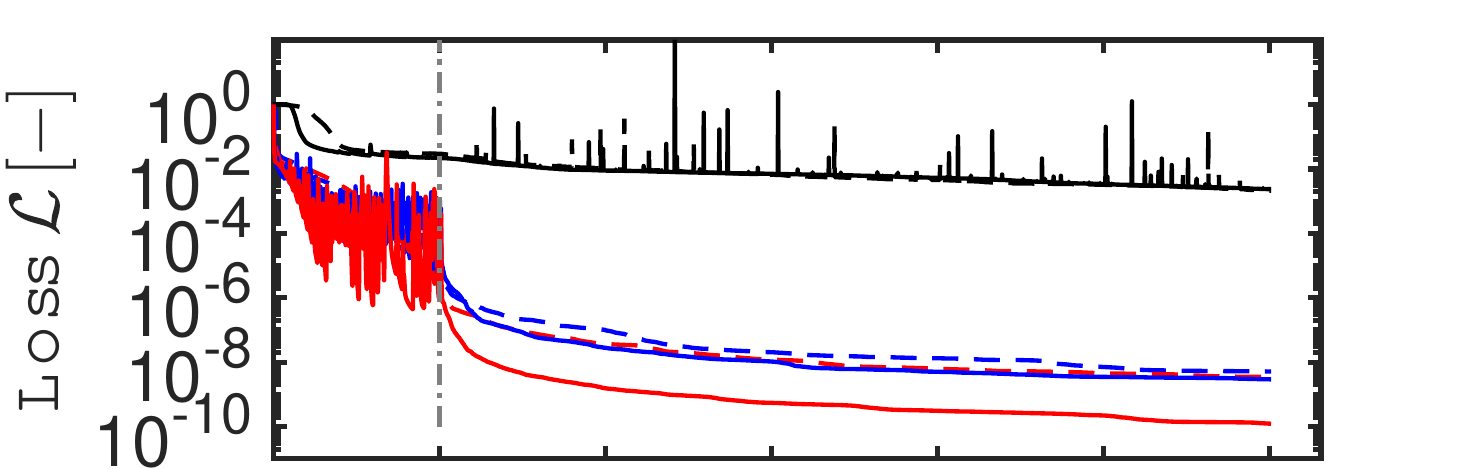}}}\,\,\,\,
{{\includegraphics[width=0.45\textwidth,
trim=0 100 0 80, clip]{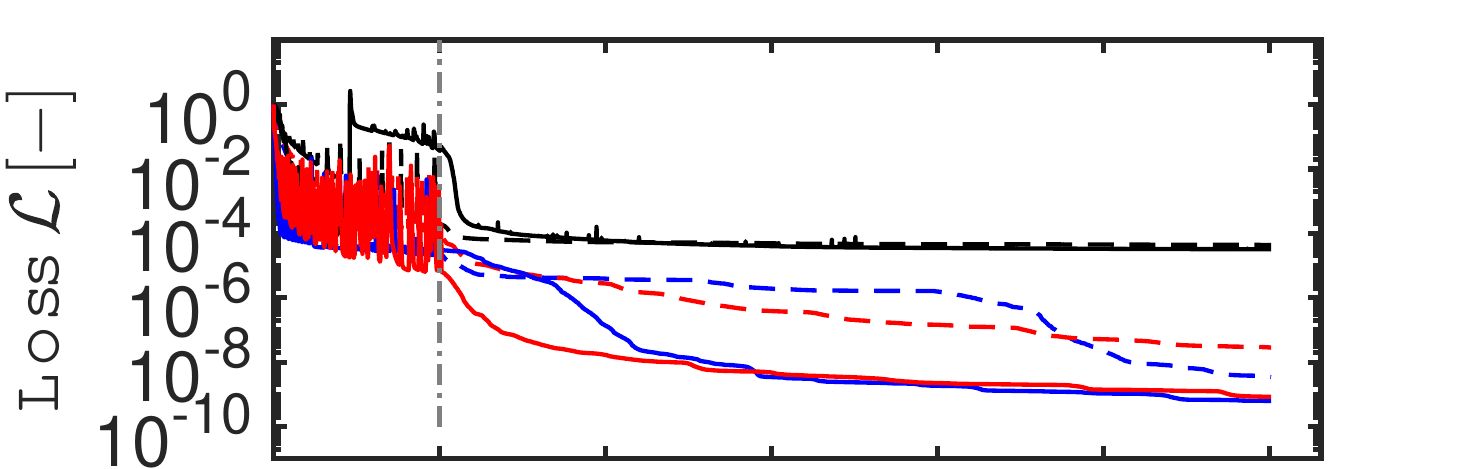}}}\\
{{\includegraphics[width=0.45\textwidth,
trim=0 100 0 80, clip]{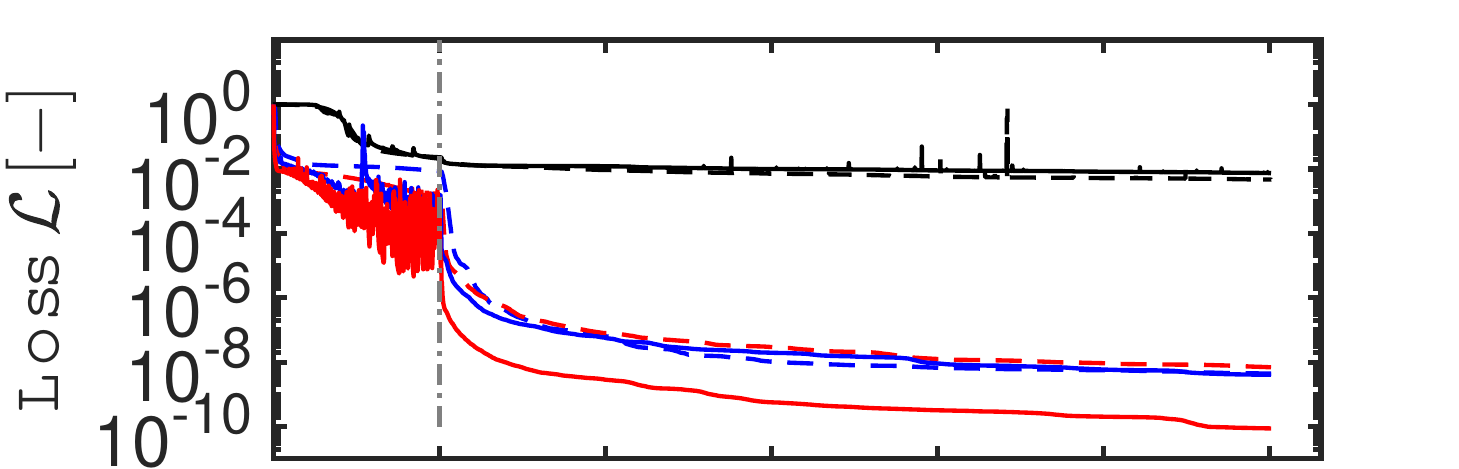}}}\,\,\,\,
{{\includegraphics[width=0.45\textwidth,
trim=0 100 0 80, clip]{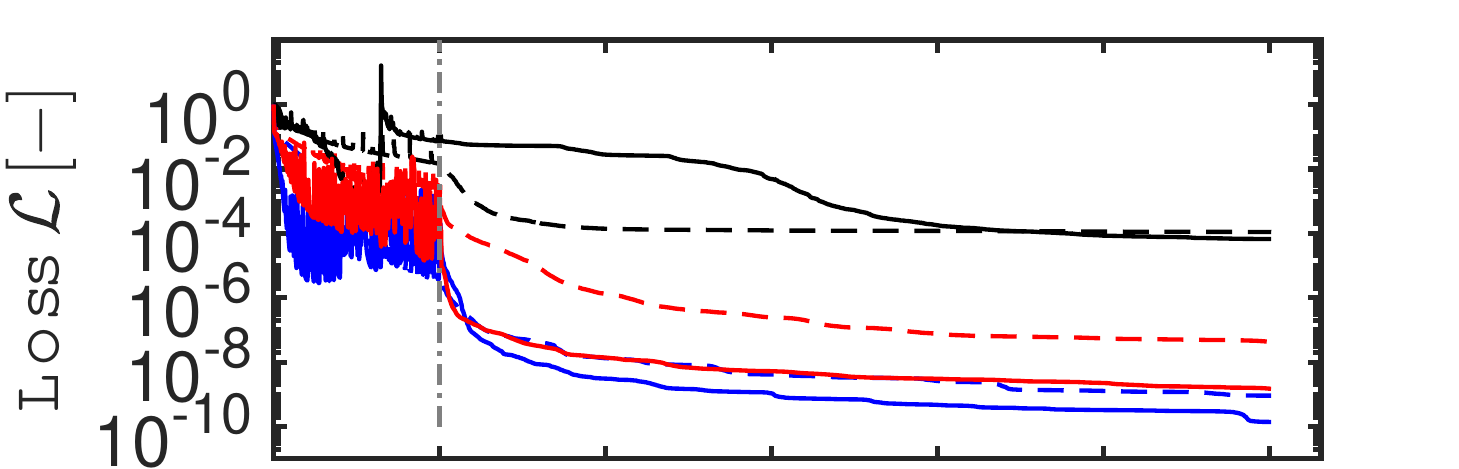}}}\\
{{\includegraphics[width=0.45\textwidth,
trim=0 100 0 80, clip]{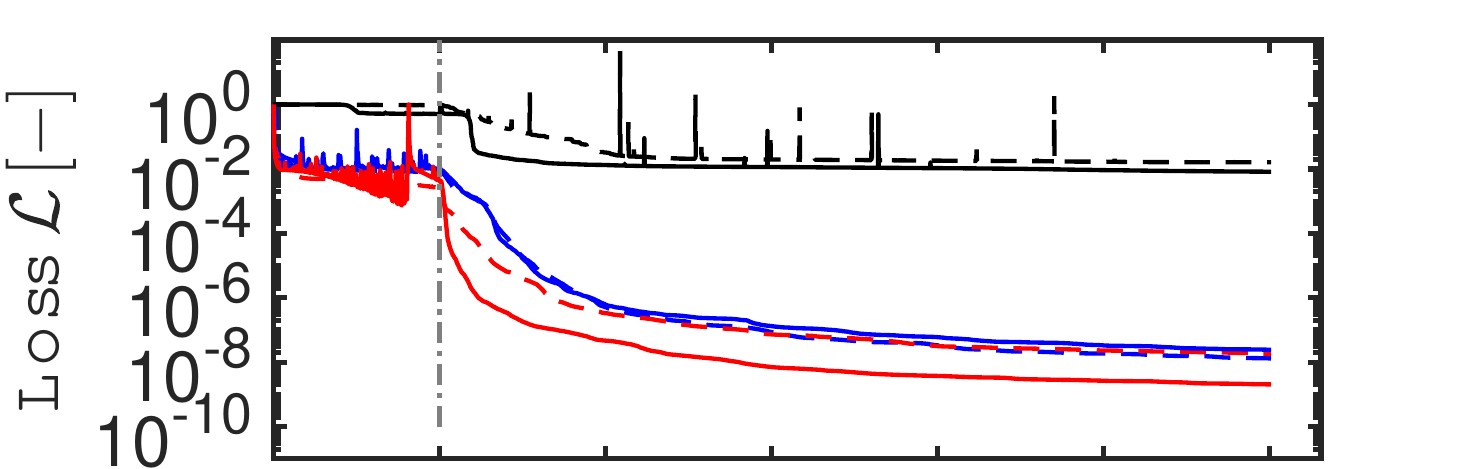}}}\,\,\,\,
{{\includegraphics[width=0.45\textwidth,
trim=0 100 0 80, clip]{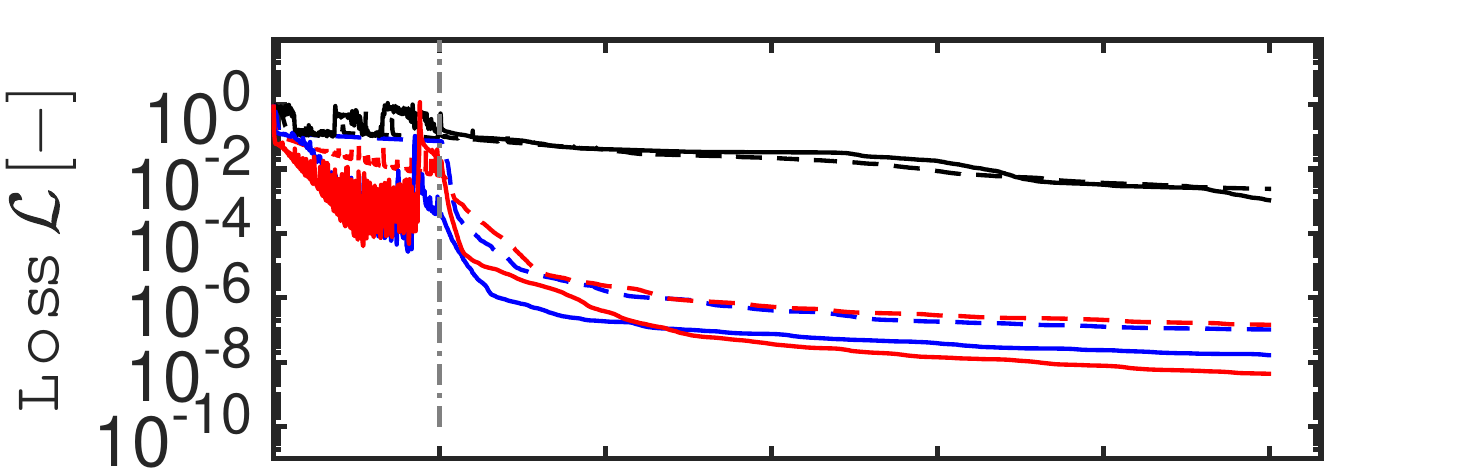}}}\\
{{\includegraphics[width=0.45\textwidth,
trim=0 40 0 80, clip]{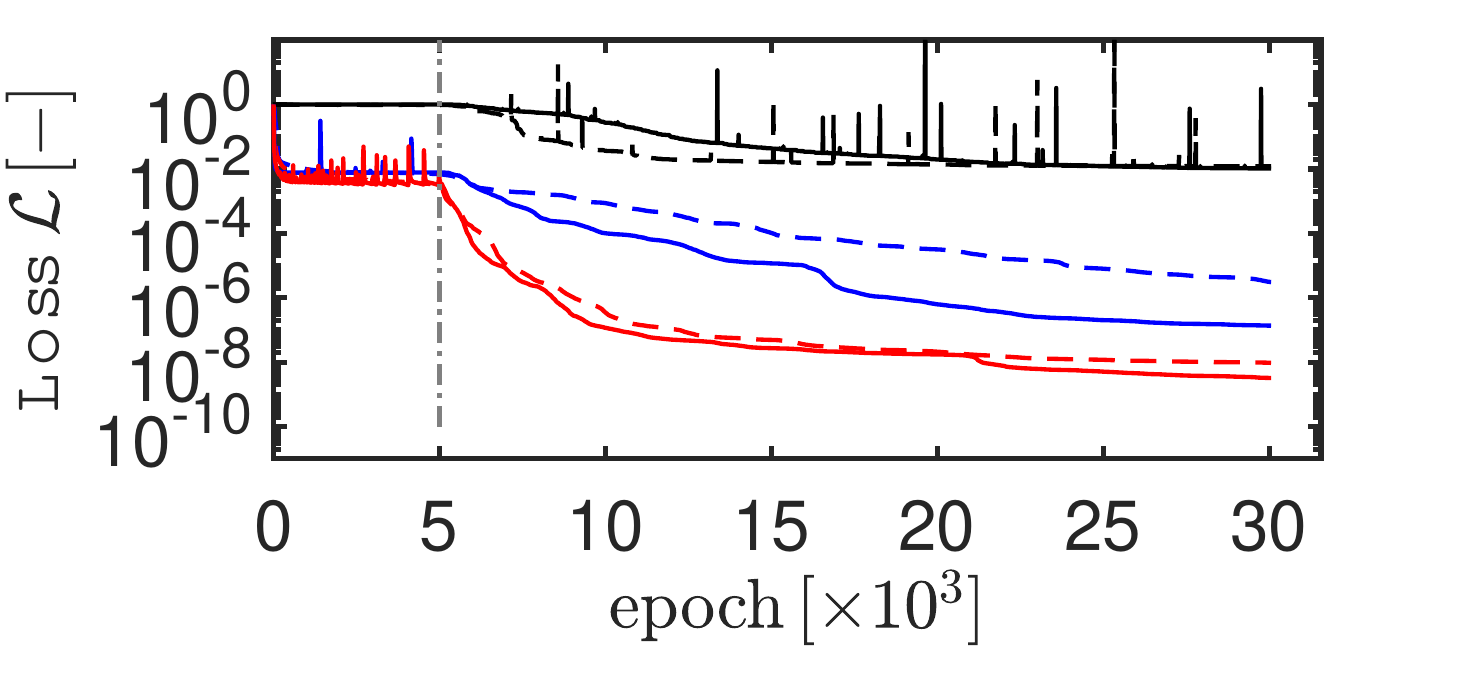}}}\,\,\,\,
{{\includegraphics[width=0.45\textwidth,
trim=0 40 0 80, clip]{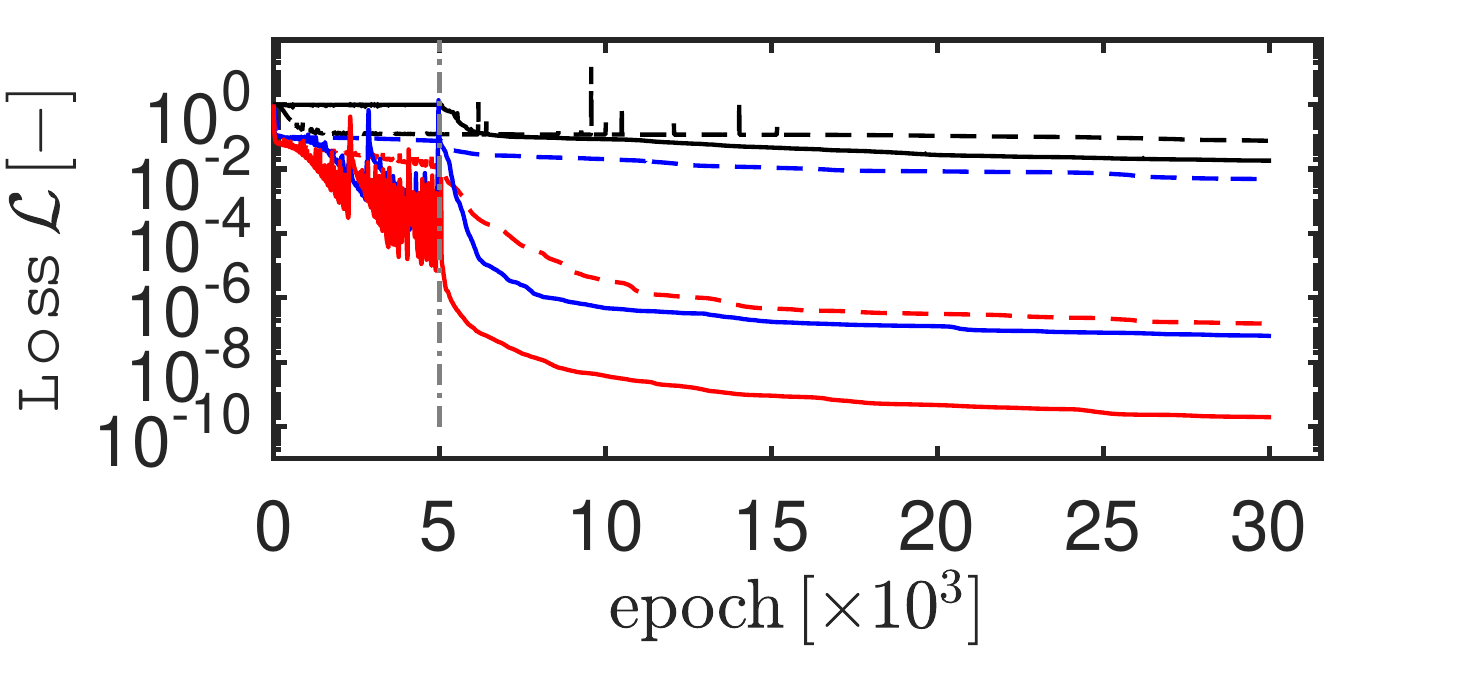}}}\\
\vspace{-5pt}
\caption{Training history of the loss function for the cell-problem PINN solution. The first 5000 epochs (marked with a vertical red dashed line) belong to adam optimizer, whereas the rest to L-BFGS. The left column belongs to the results with scaling and the right without scaling. Dashed results have a learning rate of 0.001 and continuous lines  of 0.01. Black curves give results for no Fourier feature, whereas dark blue and red for low- and high-frequency Fourier features with a single and the first 10 integer multiples of the reciprocal base vector, respectively.}
\label{F:results_1D_training_histories}
\end{figure*}

\begin{figure*}[htb!]
\centering
{{\includegraphics[width=0.18\textwidth,
trim=90 55 130 0, clip]{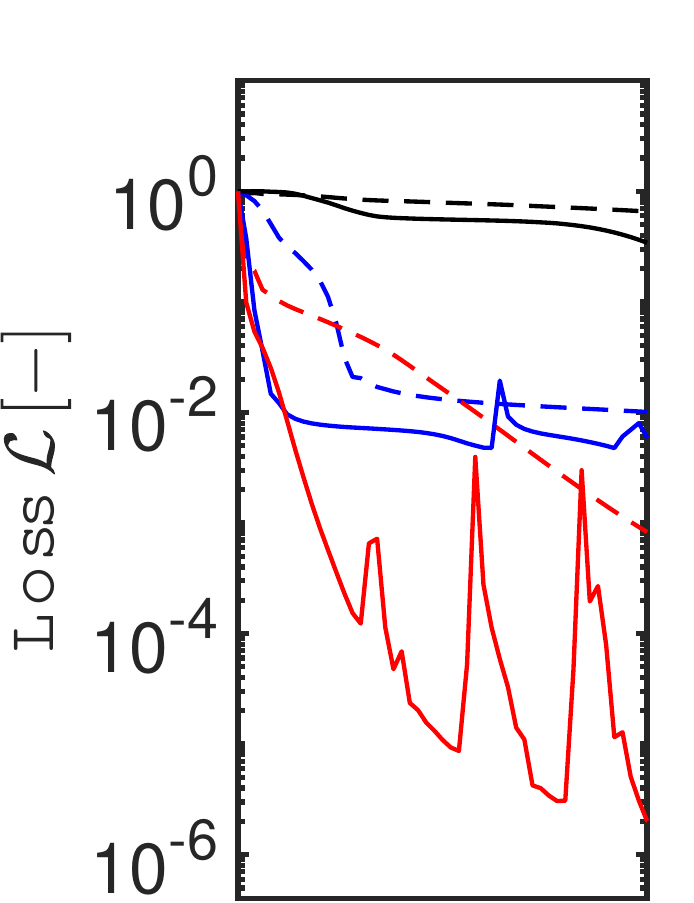}}}
{{\includegraphics[width=0.18\textwidth,
trim=90 55 130 0, clip]{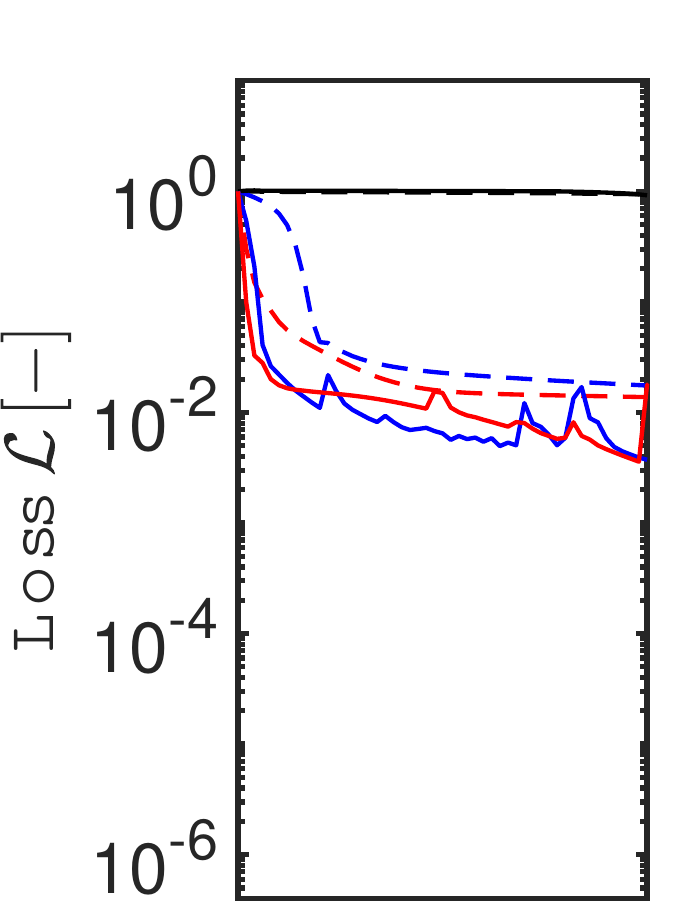}}}
{{\includegraphics[width=0.18\textwidth,
trim=90 55 130 0, clip]{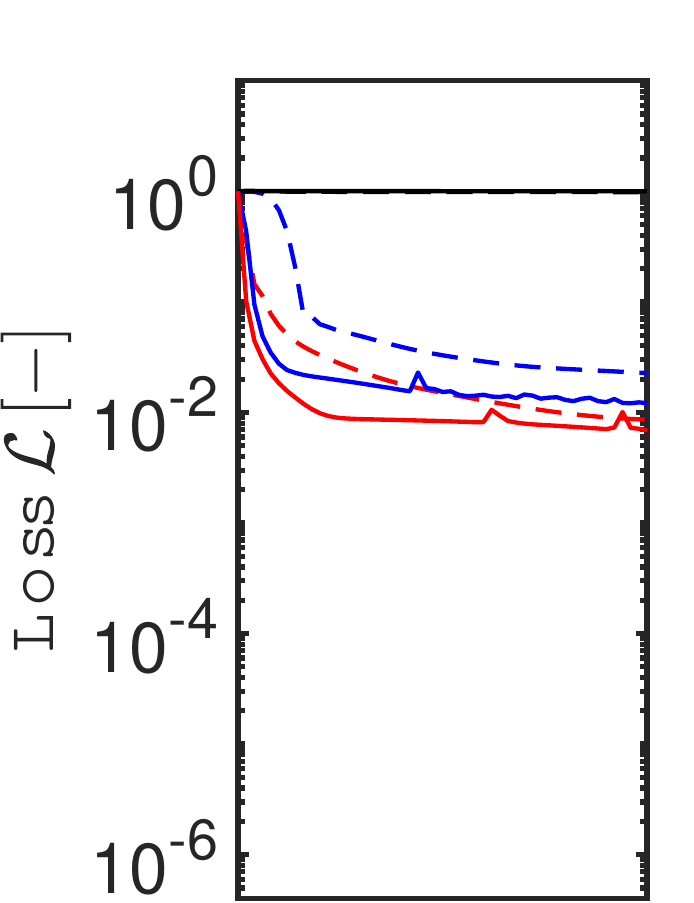}}}
{{\includegraphics[width=0.18\textwidth,
trim=90 55 130 0, clip]{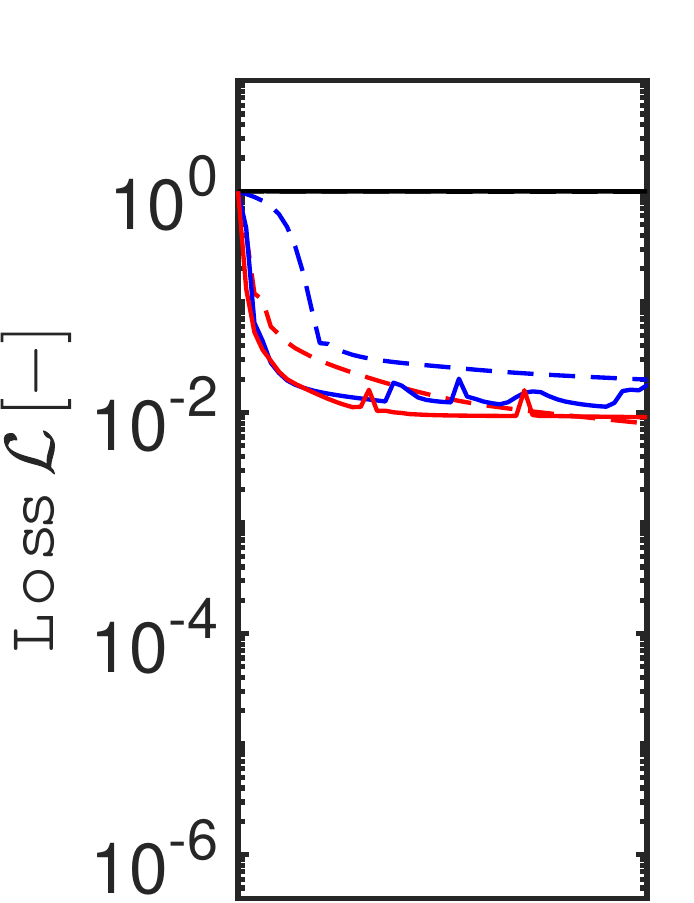}}}
{{\includegraphics[width=0.18\textwidth,
trim=90 55 130 0, clip]{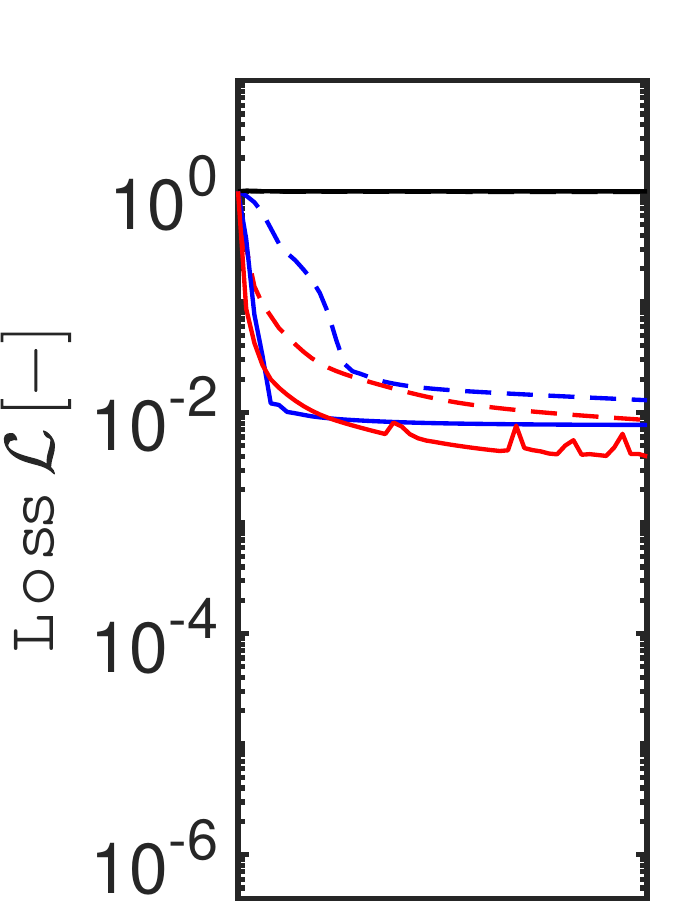}}}\\
\vspace{-1.pt}
{{\includegraphics[width=0.18\textwidth,
trim=90 0 130 0, clip]{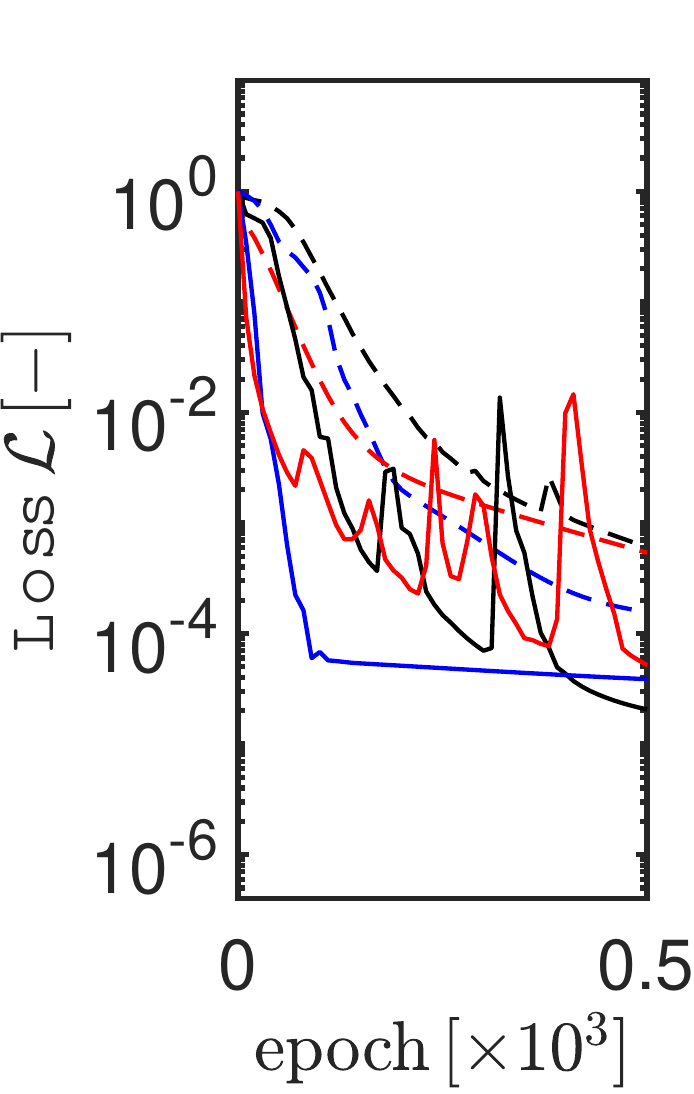}}}
{{\includegraphics[width=0.18\textwidth,
trim=90 0 130 0, clip]{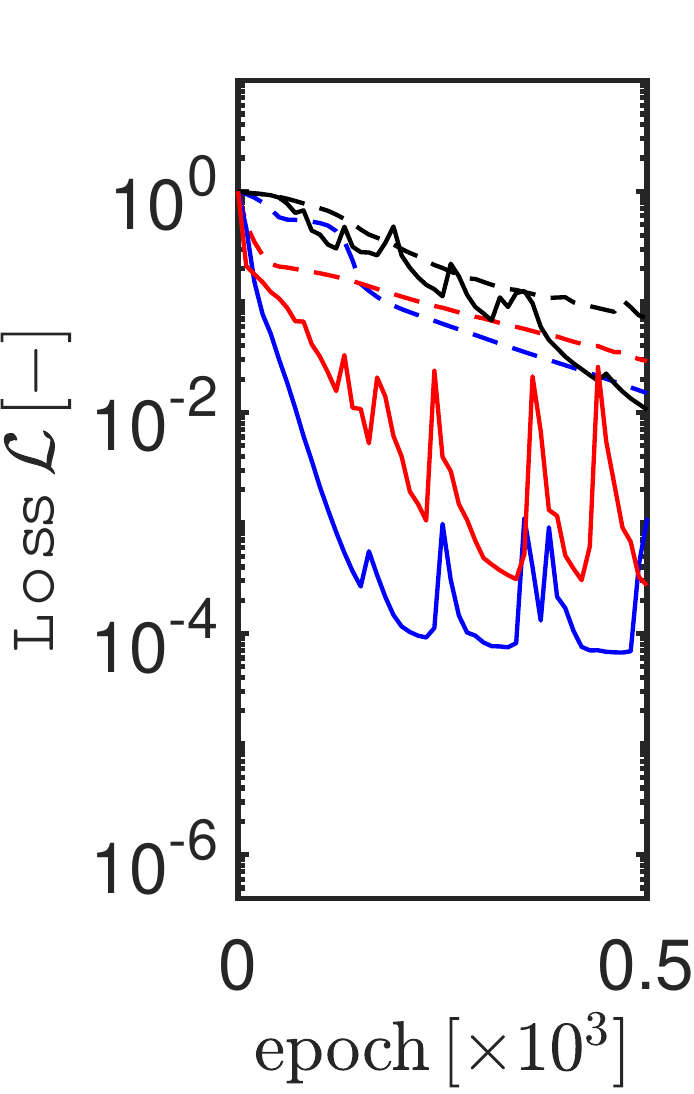}}}
{{\includegraphics[width=0.18\textwidth,
trim=90 0 130 0, clip]{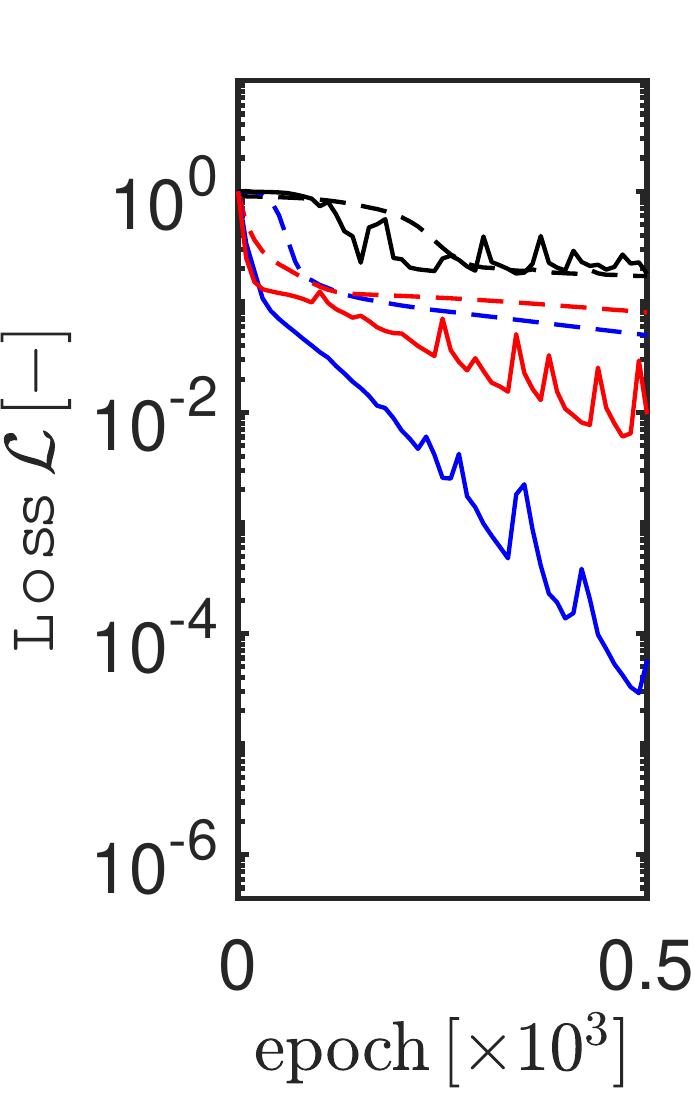}}}
{{\includegraphics[width=0.18\textwidth,
trim=90 0 130 0, clip]{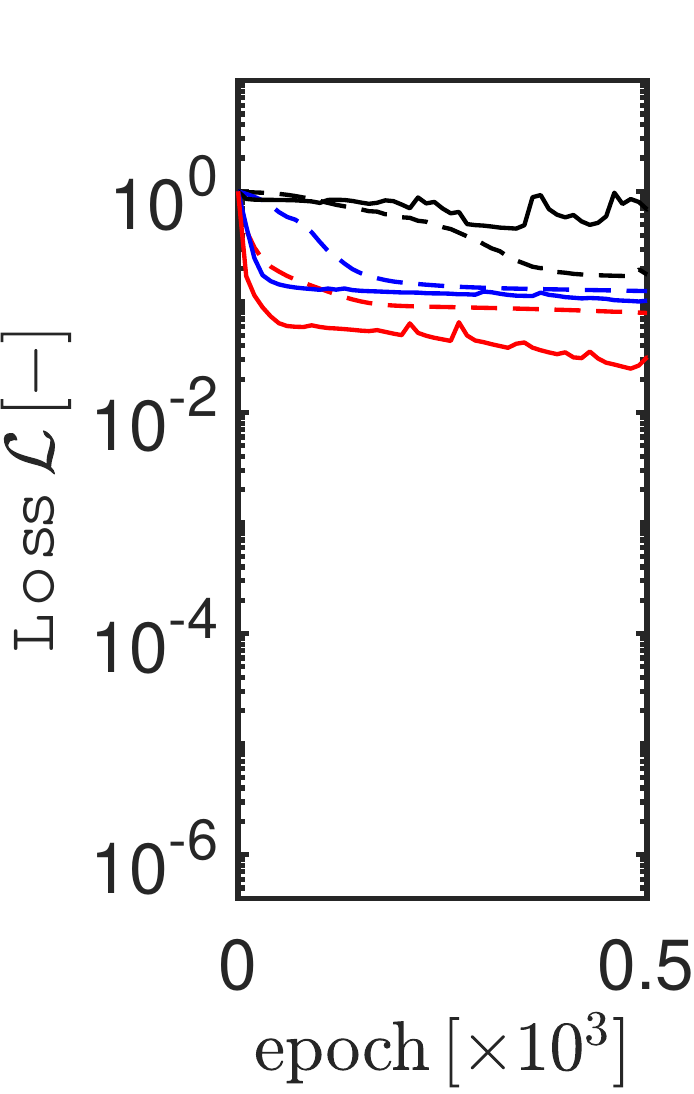}}}
{{\includegraphics[width=0.18\textwidth,
trim=90 0 130 0, clip]{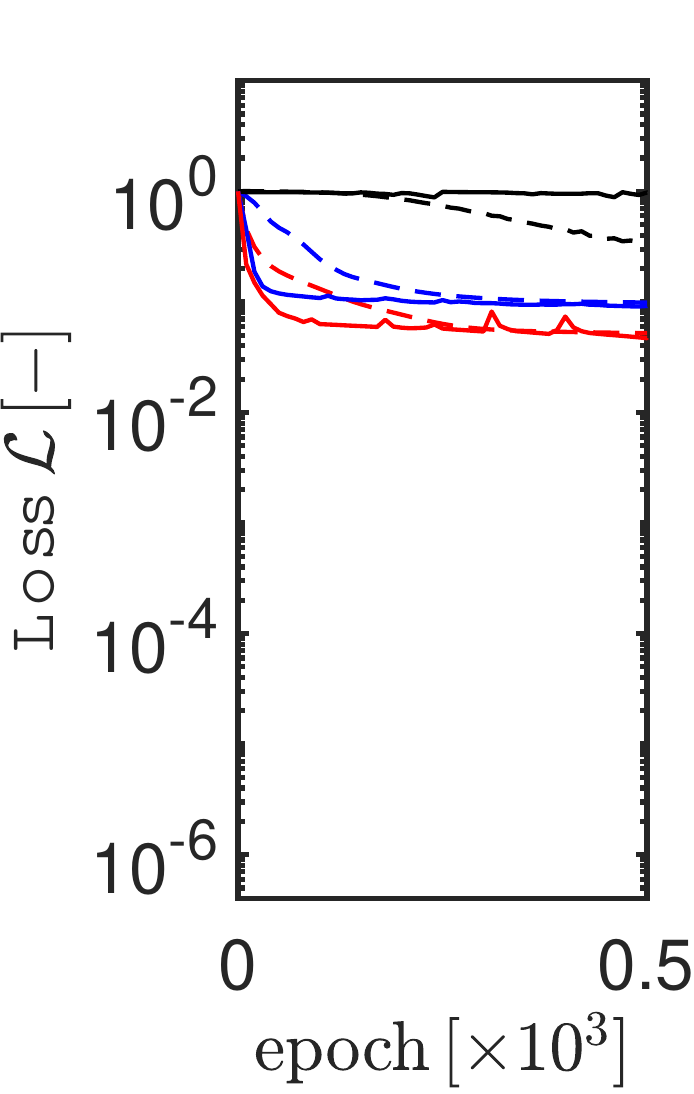}}}\\
\vspace{-5pt}
\caption{Training history of the loss function for the cell-problem PINN solution for the first 500 adam epochs. In the results, the top row belongs to the results with scaling and the bottom without scaling. Dashed results have a learning rate of 0.001 and continuous lines  of 0.01. Black curves give results for no Fourier feature, whereas dark blue and red for low- and high-frequency Fourier features with a single and the first 10 integer multiples of the reciprocal base vector, respectively.}
\label{F:results_1D_training_histories_first_500}
\end{figure*}

The training histories of the normalized loss function are given in Fig.\ \ref{F:results_1D_training_histories}. The loss for the approximate periodic boundary condition treatment does not provide an accurate prediction, no matter the learning rate. Thus, without Fourier features, the solution struggles to converge.
For the loss computations with the Fourier features, for which there is an exact imposition of the periodic boundary conditions, large fluctuations are observed in the stochastic gradient descent phase. On the other hand, in the L-BFGS optimization phase, the loss reduces monotonically. Figure \ref{F:results_1D_training_histories_first_500} demonstrates that although a lower learning rate generally shows better performance in terms of convergence, with a higher learning rate, convergence is reached at a higher speed.

\begin{figure*}[htb!]
\centering
\subfigure[with scaling]{
{{\includegraphics[height=0.22\textwidth,
trim=0 40 0 80, clip]{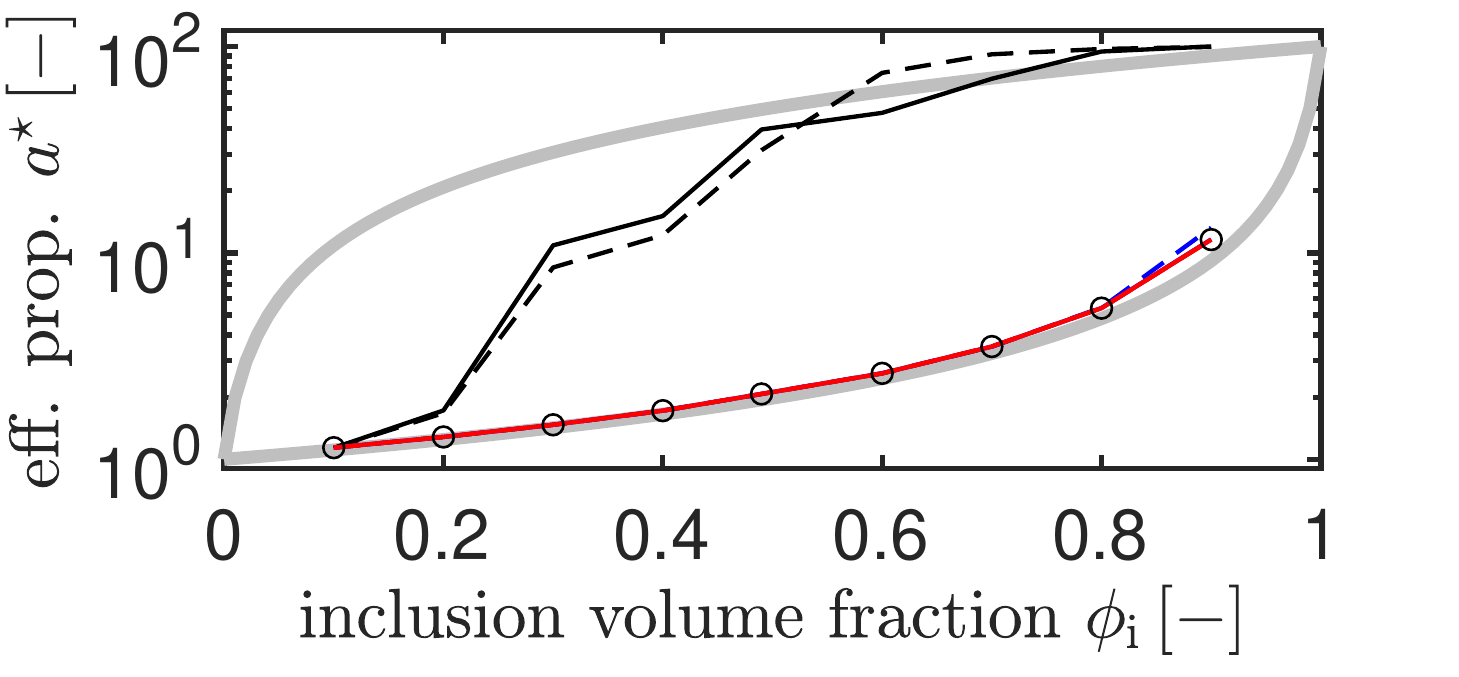}}}
}
\subfigure[without scaling]{
{{\includegraphics[height=0.22\textwidth,
trim=0 40 0 80, clip]{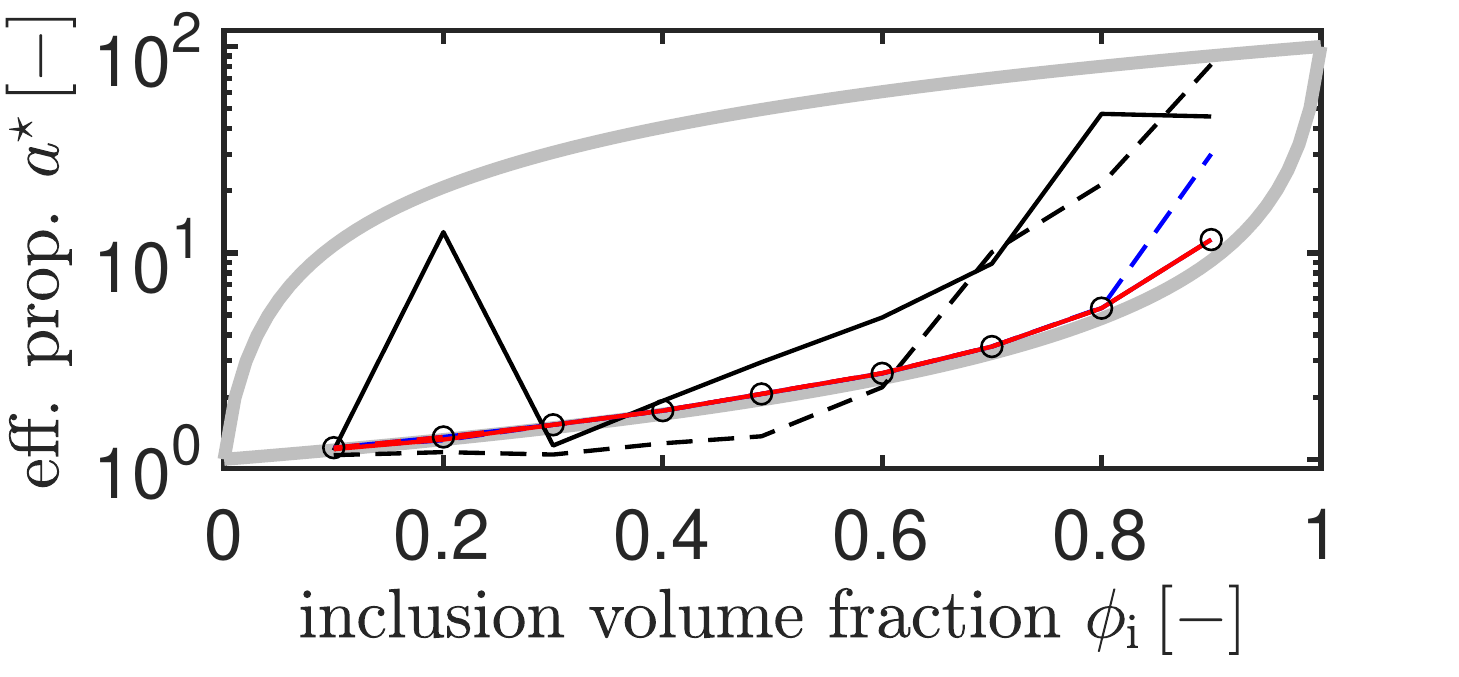}}}
}
\vspace{-5pt}
\caption{ANN solution of the effective properties $a^\star$ at the end of 30000 epochs and its comparison with Voigt (arithmetic average) and Reuss (geometric average) analytical bounds, and ABAQUS solution given in circle markers,  which is in agreement with the integral given in Eq.\ \eqref{E:analytical_solution_part_v1}.  The left column belongs to the results with scaling and the right without scaling. Dashed results have a learning rate of 0.001 and continuous lines  of 0.01. Black curves give results for no Fourier feature, whereas dark blue and red for low- and high-frequency Fourier features with a single and the first 10 integer multiples of the reciprocal base vector, respectively.}
\label{F:homogenization_results_1D}
\end{figure*}

The effective property predictions at the end of 30000 epochs as a function of inclusion volume fraction is given in Fig.\ \ref{F:homogenization_results_1D}.
In accordance with the solution field distributions and the loss diagrams, neural networks with a Fourier feature layer provide accurate predictions showing excellent agreement with the finite element solutions. The integrated, effective properties can be equivalently computed using Eq.\ \eqref{E:analytical_solution_part_v1}.
As anticipated, the computed effective properties are closer to the Reuss (harmonic) average. The reason for the discrepancy is the diffuse interface condition considered in the finite element and neural network solutions.

\subsection{Two-dimensional Problems}
We now consider the problem of finding the solution to Eq.\ \eqref{E:cell-problem} in a 2D system. At the microscale, we assume isotropy for each constituent such that $a_{im}(\boldsymbol x) = a(\boldsymbol x)\, \delta_{im}$. In effect, the source term $\partial a_{im}(\boldsymbol x)/\partial x_i$ satisfies the relation $\partial a_{im}(\boldsymbol x)/\partial x_i = \partial a(\boldsymbol x)/\partial x_m$. Finding the solution $\chi^{m}(\bs{x})$  for the  source term for $m$ allows computation of the $m^\mathrm{th}$ column vector of the effective property tensor $\boldsymbol a^\star$, see Eq.\ \eqref{E:propertyhomogenized}.

For 2D, the two load cases required are those with two source terms defined in terms of property field gradients $f^1=\partial a(\boldsymbol x)/\partial x_1$ and $f^2=\partial a(\boldsymbol x)/\partial x_2$. Thus, in view of Eq.\ \eqref{E:propertyhomogenized} with the chosen Cartesian basis, the computation of the components $a^\star_{11}$ and $a^\star_{21}$ requires the solution $\chi^{1}(\bs{x})$ for the source term  $f^1$ which gives
\begin{align}
a^\star_{11}=\dfrac{1}{|\mathcal{V}|}\,\int_{\mathcal{V}}a_{11}\left[1+
\dfrac{\partial \chi^{1}(\bs{x})}{\partial x_1}\right]
\mathrm{d}V\text{ and }
a^\star_{21}=\dfrac{1}{|\mathcal{V}|}\,\int_{\mathcal{V}}a_{22}\,
\dfrac{\partial \chi^{1}(\bs{x})}{\partial x_2}
\mathrm{d}V\,.
\label{E:propertyhomogenized_a11_and_a21_only}
\end{align}
The solution $\chi^{2}(\bs{x})$ obtained for the source term  $f^2$, on the other hand, allows computation of $a^\star_{22}$ and $a^\star_{12}=a^\star_{21}$, respectively, with
\begin{align}
a^\star_{22}=\dfrac{1}{|\mathcal{V}|}\,\int_{\mathcal{V}}a_{22}\left[1+
\dfrac{\partial \chi^{2}(\bs{x})}{\partial x_2}\right]
\mathrm{d}V\text{ and }\quad
a^\star_{12}=\dfrac{1}{|\mathcal{V}|}\,\int_{\mathcal{V}}a_{11}\,
\dfrac{\partial \chi^{2}(\bs{x})}{\partial x_1}
\mathrm{d}V\,.
\label{E:propertyhomogenized_a22_and_a12_only}
\end{align}
The matrix $a_\mathrm{m}$ and inclusion $a_\mathrm{i}$ properties are selected as $a_\mathrm{m}=1$ and $a_\mathrm{i}=2$, respectively. We consider the following regularized property distribution
\begin{align}
a(\boldsymbol x)= a_\mathrm{m} +\dfrac{a_\mathrm{i}-a_\mathrm{m}}{2}\left[1-\tanh\left(\dfrac{|\boldsymbol  x - \boldsymbol x_\mathrm{i}|-r}{\xi}\right)\right]\,.
\label{E:regularization_2D}
\end{align}
Here, $\boldsymbol x_\mathrm{i}$ denotes the position of the inclusion and $r$ its radius. As before, $\xi$ is the regularization parameter that controls the sharpness of the phase interface.

In the following, through the solution of the cell problems with the source terms $f^1$ and $f^2$ for the $\mathcal{V}-$periodic corrector functions $\chi^{1}(\bs{x})$ and $\chi^{2}(\bs{x})$, respectively, we compute $\bs{a}^\star$ for the selected microstructures. The gradients of $\chi^{1}(\bs{x})$ and $\chi^{2}(\bs{x})$, required in Eqs.\ \eqref{E:propertyhomogenized_a11_and_a21_only} and \eqref{E:propertyhomogenized_a22_and_a12_only}, are obtained from the differentiable ANN solution.
\subsubsection{Two-dimensional Square Lattice}
We start by considering a single primitive unit cell with a single central circular inclusion. Inclusion volume fractions of  $\phi_\mathrm{i}=\{0.1,0.2,0.3,0.4,0.5,0.6\}$ are considered. In view of Table\ \ref{T:bravais_lattices_1D2D3D} and Fig.\ \ref{fig:BravaisLattice1D2D3D}
we consider $L=\rvert\boldsymbol c_1\rvert=\rvert\boldsymbol c_2\rvert=200\xi$ with the regularization parameter $\xi=1$.

With Neumann's principle and the second-order tensor formalism for the property tensor, such  microstructures  possess isotropic macroscopic properties \cite{Neumann1885, Nye1985} leading to $\boldsymbol a^\star=a^\star_{11}\,\boldsymbol 1$ with $a^\star_{11}=a^\star_{22}$ and $a^\star_{12}=a^\star_{21}=0$. Thus, it suffices to find the solution to one of the source terms only. Here, we use $f^1=\partial a(\boldsymbol x)/\partial x_1$. Individually scaled partial differential equations are considered in the loss term. PBCs are exactly satisfied through low-
and high-frequency Fourier features with a single and the first 10 integer multiples of the reciprocal base vector. 25600 collocation points are used. For the adam optimizer, a learning rate of 0.010 is used.

\begin{figure*}[htb!]
\centering
{\frame{\includegraphics[height=0.14\textwidth,
trim=489 233 489 233, clip]{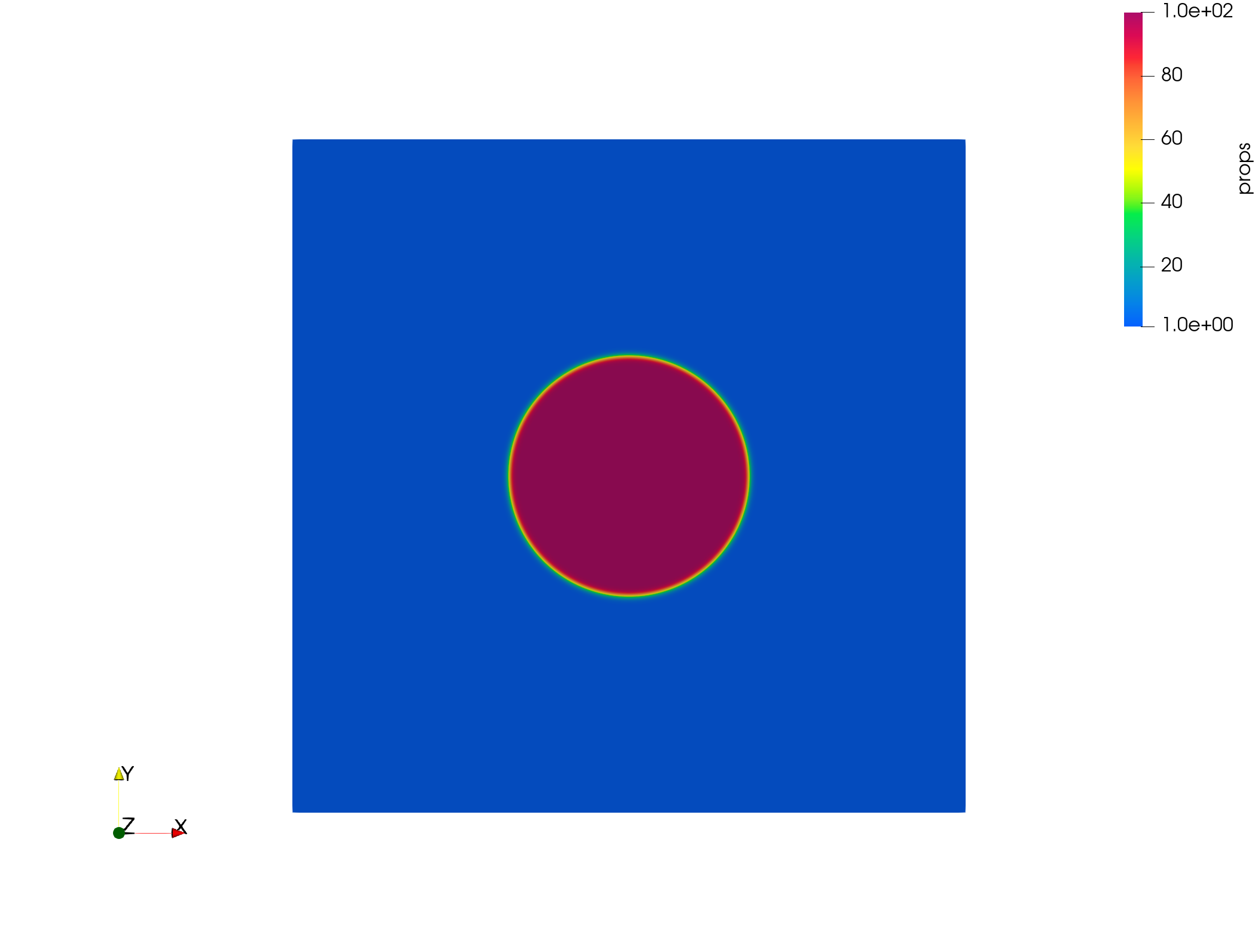}}}
{\frame{\includegraphics[height=0.14\textwidth,
trim=489 233 489 233, clip]{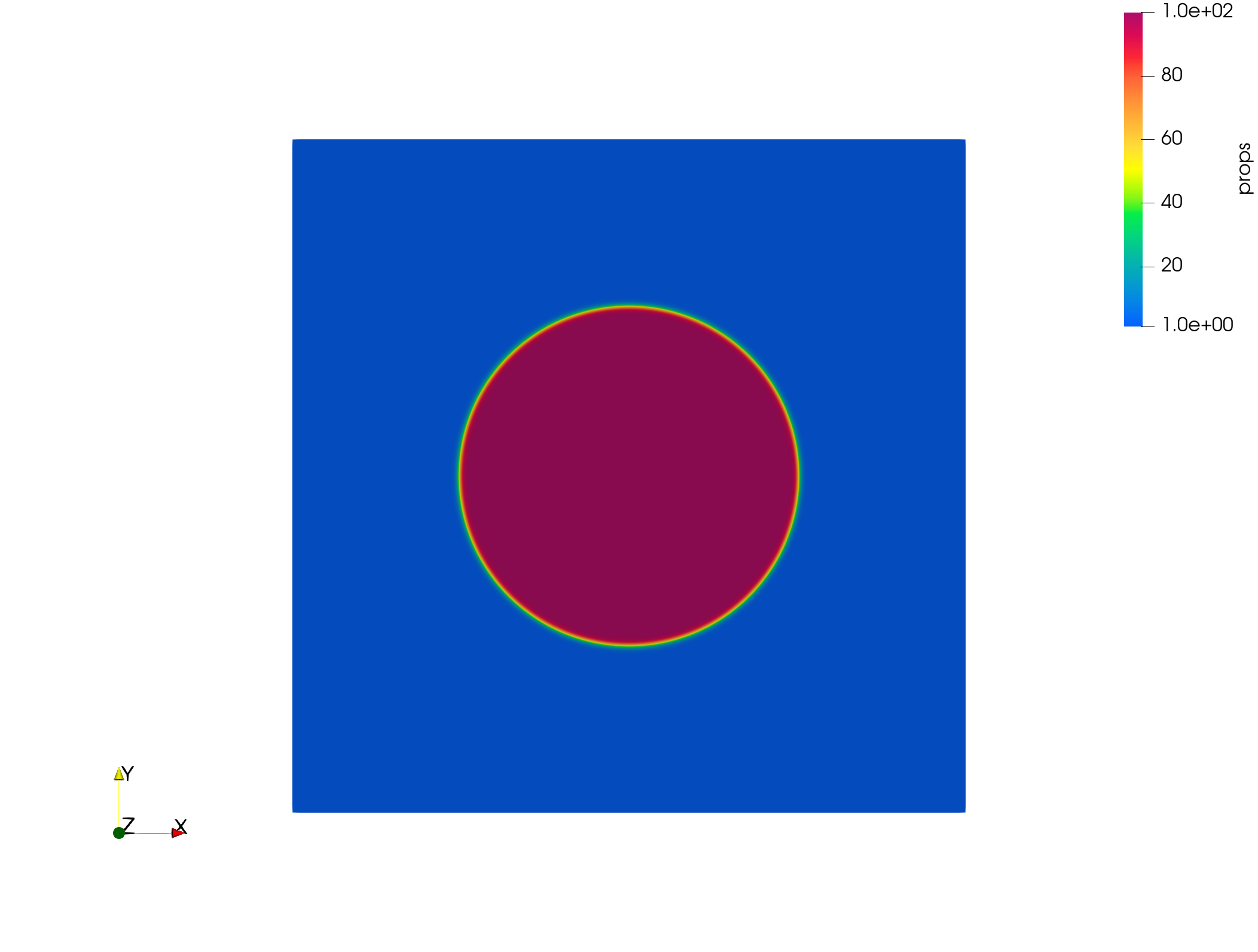}}}
{\frame{\includegraphics[height=0.14\textwidth,
trim=489 233 489 233, clip]{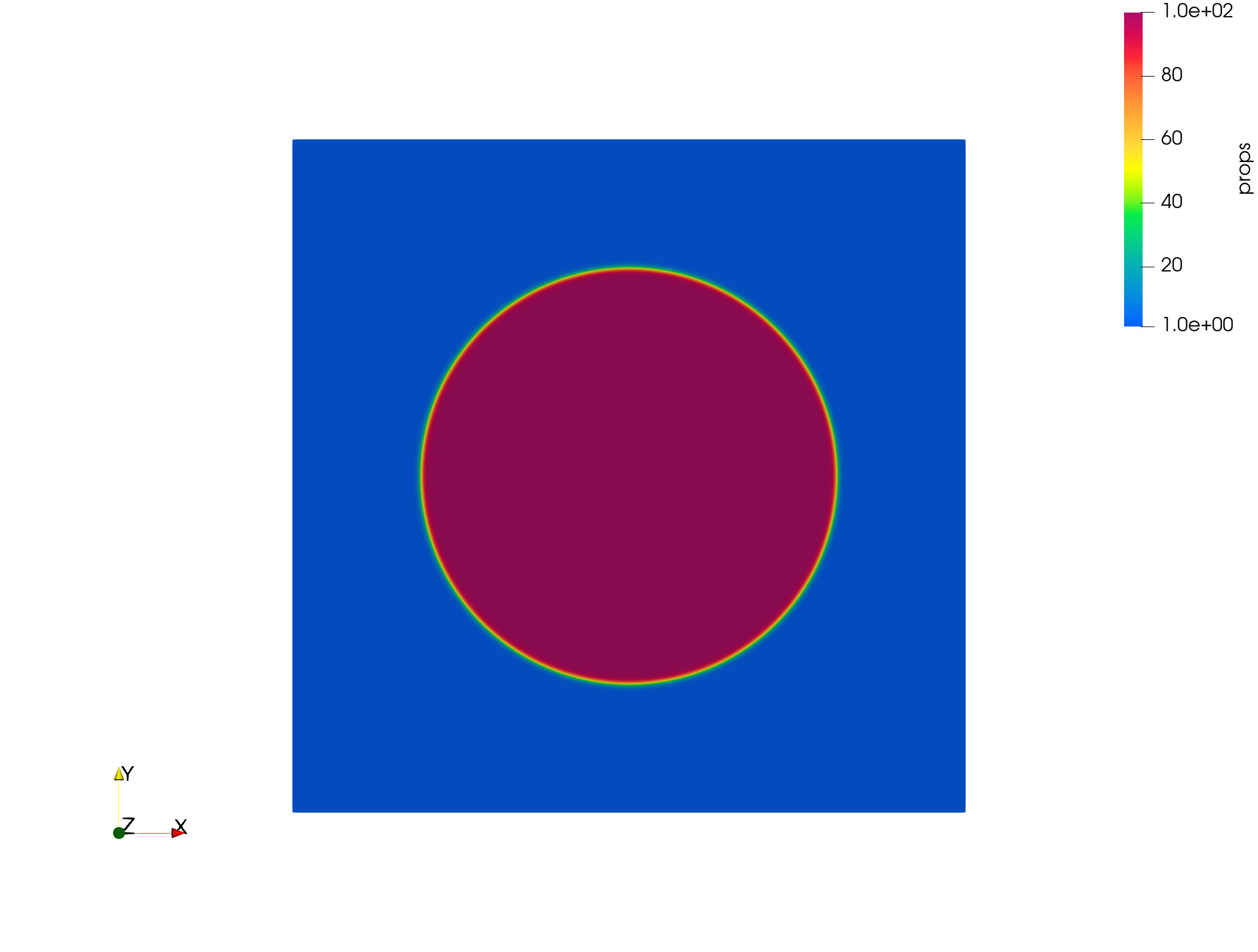}}}
{\frame{\includegraphics[height=0.14\textwidth,
trim=489 233 489 233, clip]{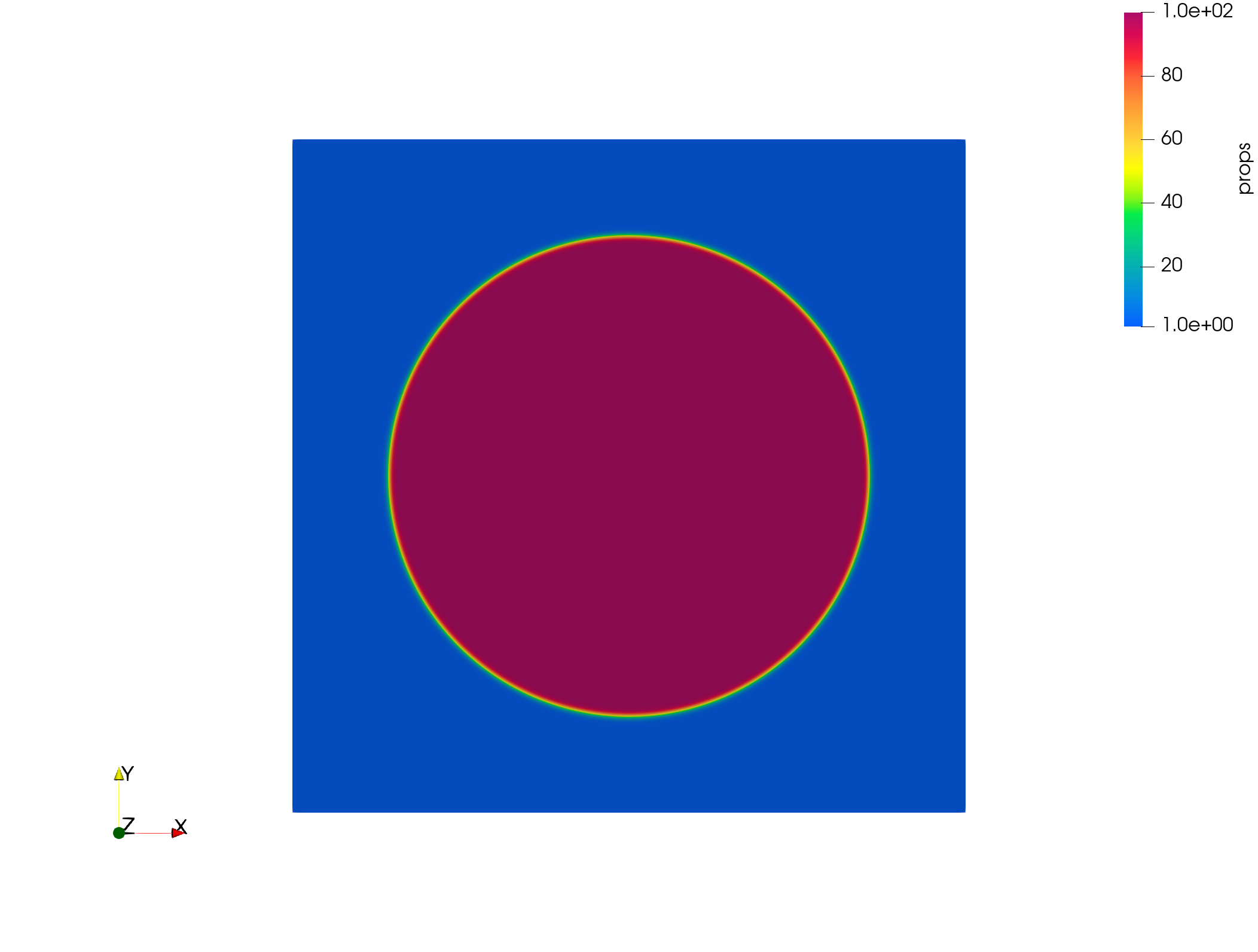}}}
{\frame{\includegraphics[height=0.14\textwidth,
trim=489 233 489 233, clip]{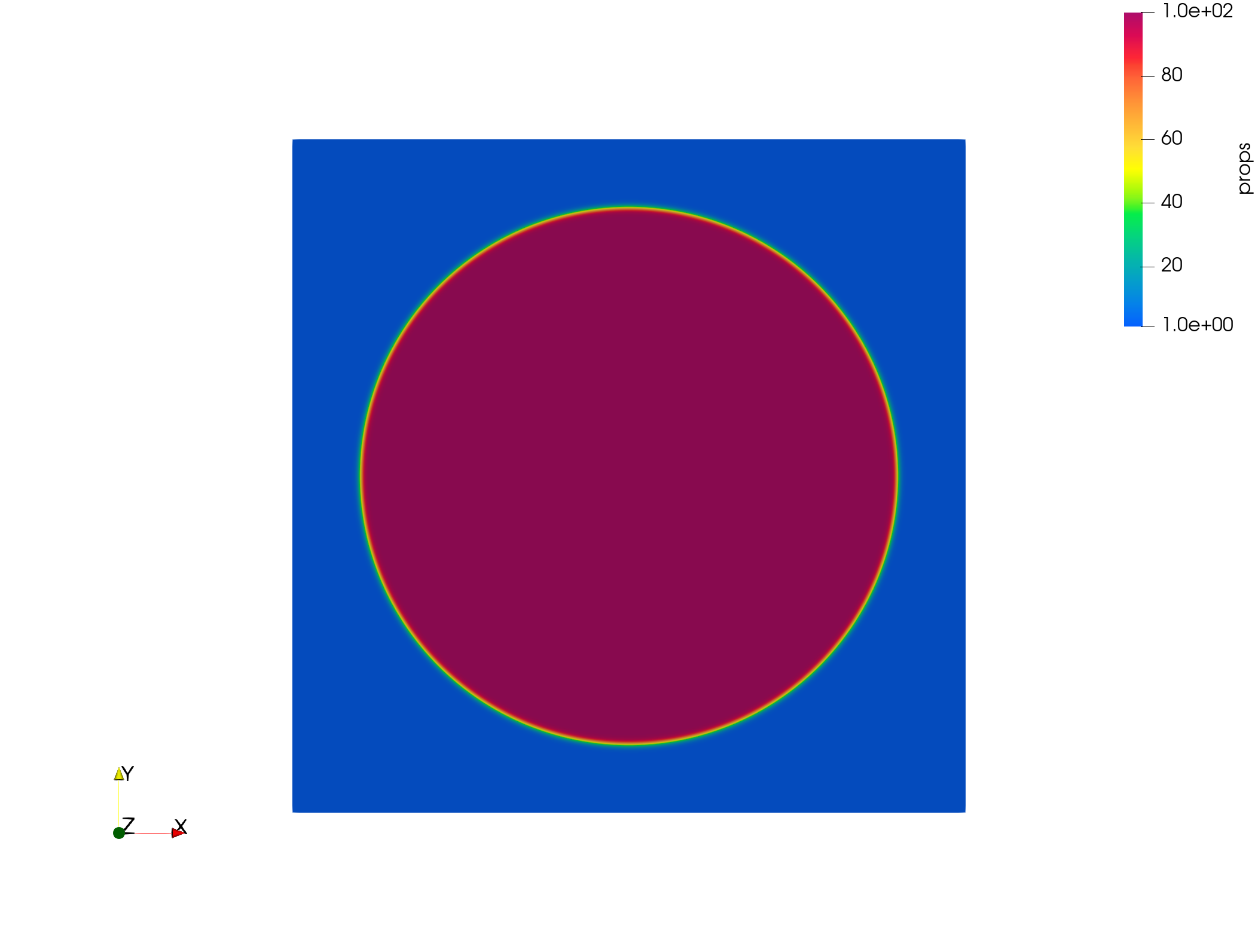}}}
{\frame{\includegraphics[height=0.14\textwidth,
trim=489 233 489 233, clip]{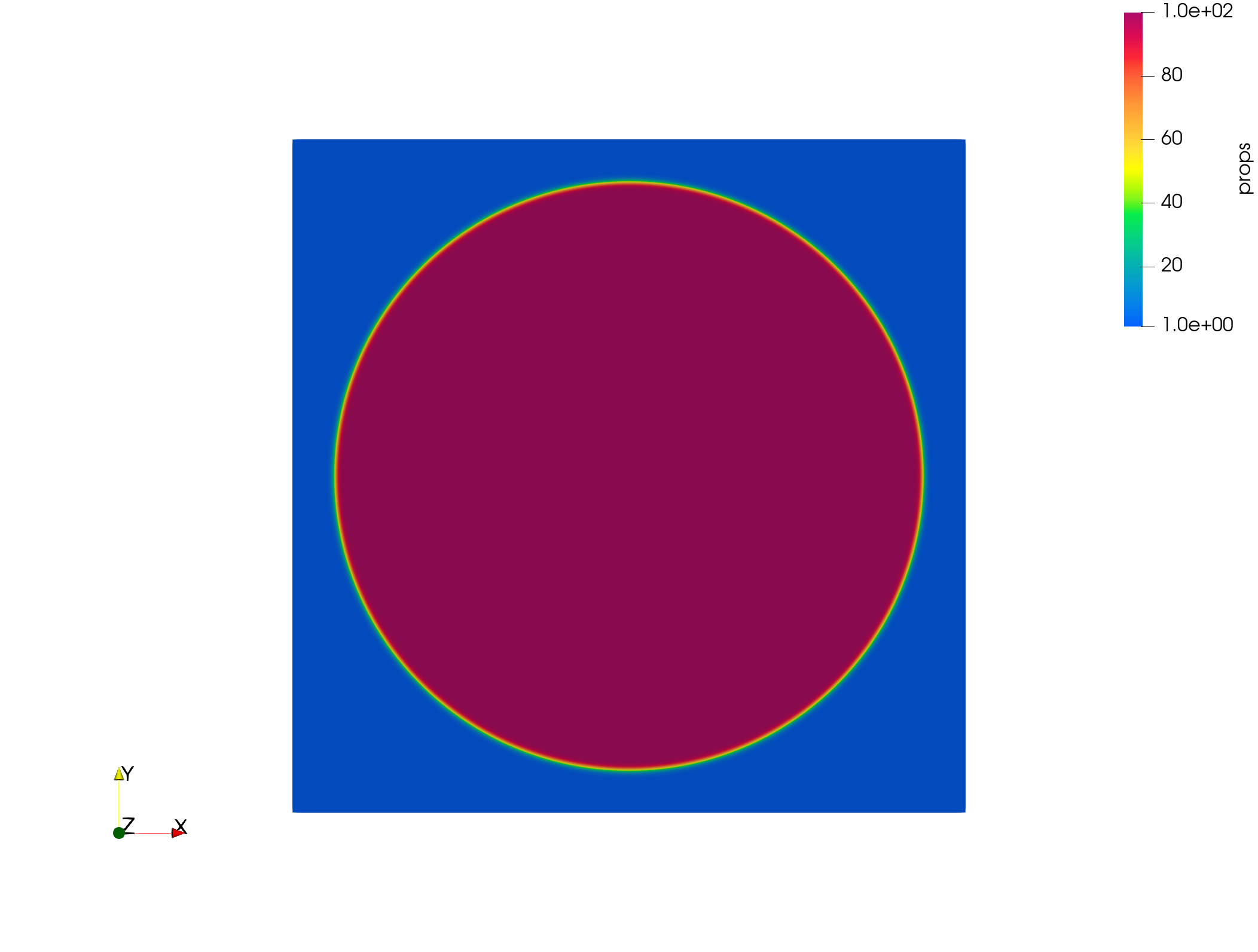}}}\\\vspace{0.1cm}
{\frame{\includegraphics[height=0.14\textwidth,
trim=489 233 489 233, clip]{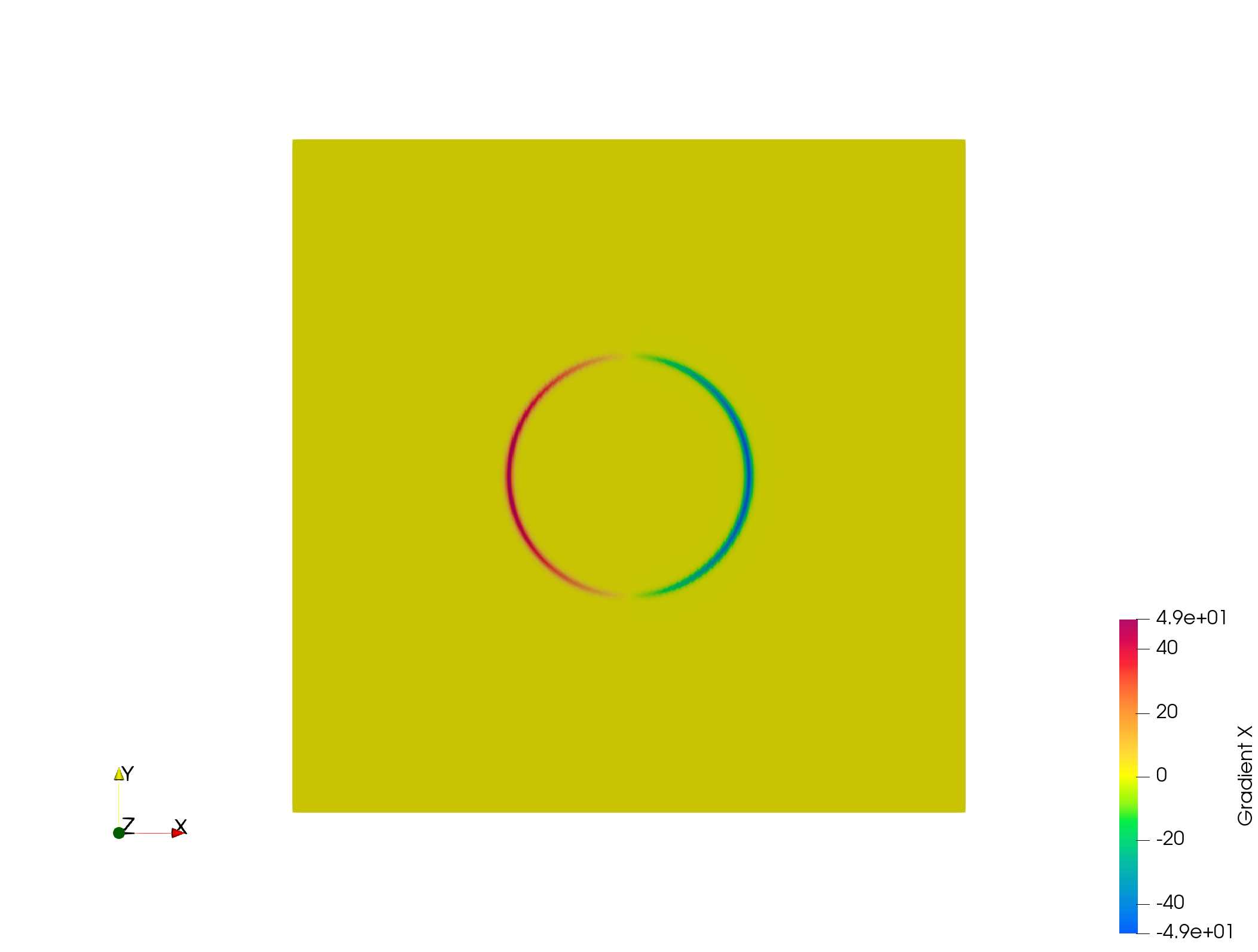}}}
{\frame{\includegraphics[height=0.14\textwidth,
trim=489 233 489 233, clip]{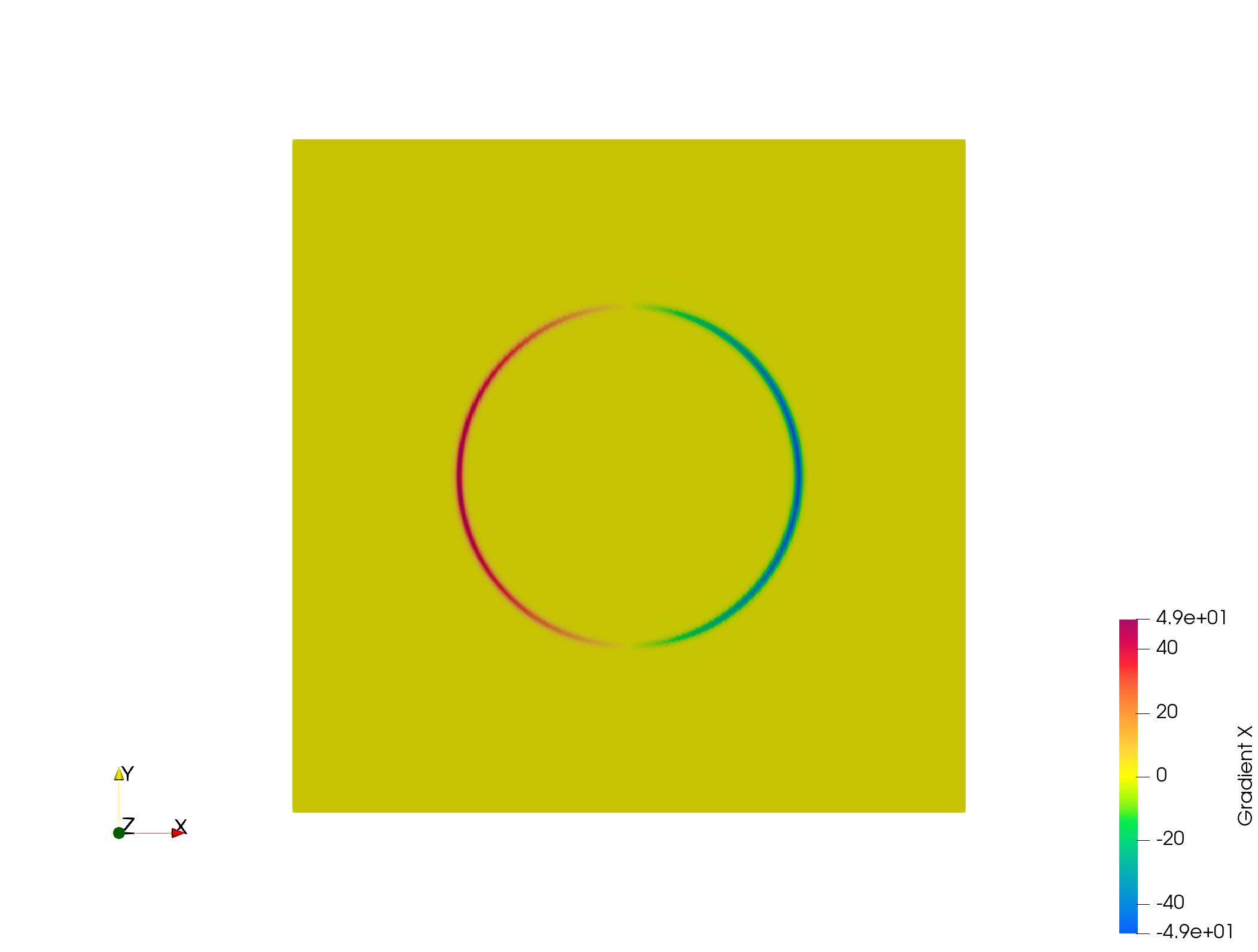}}}
{\frame{\includegraphics[height=0.14\textwidth,
trim=489 233 489 233, clip]{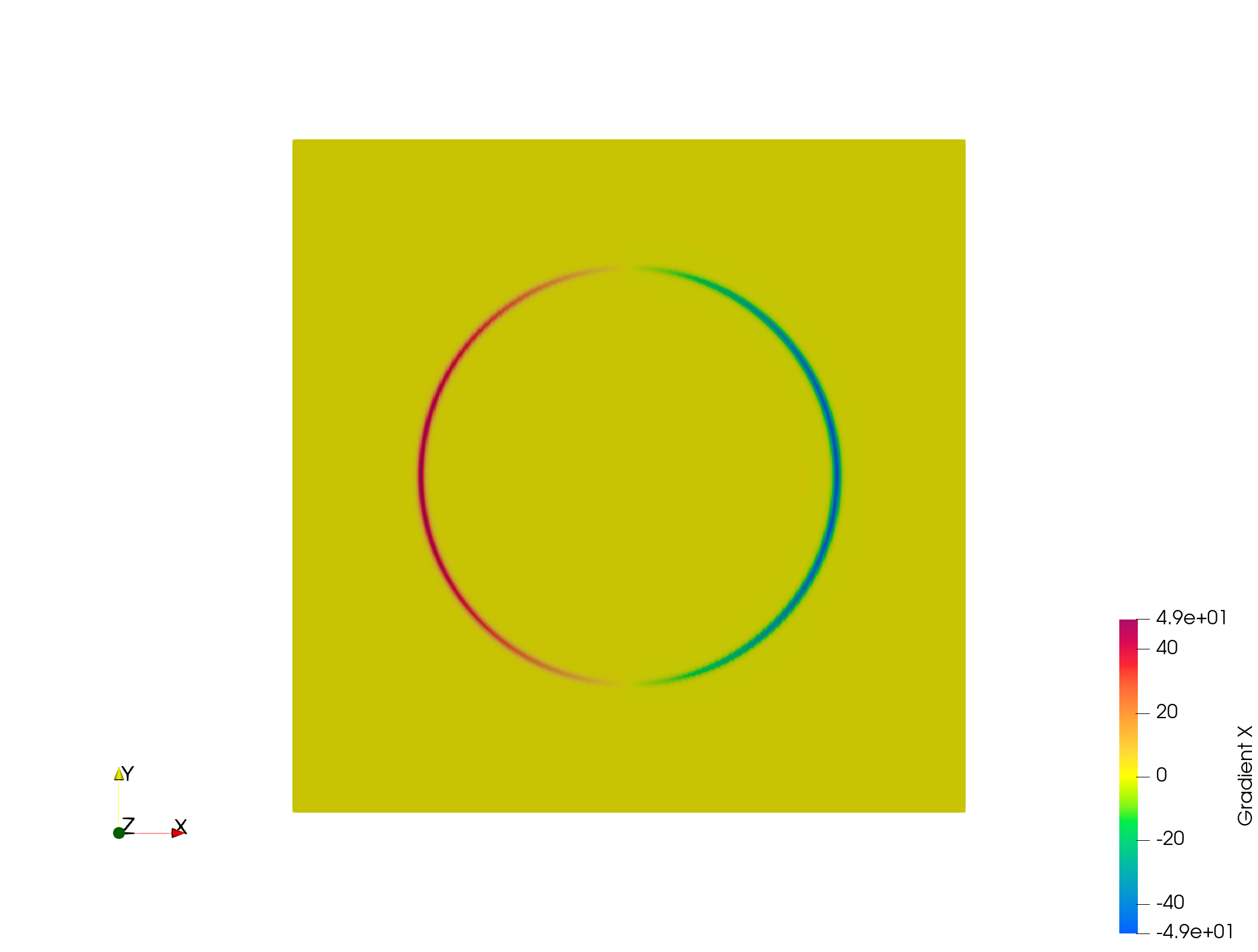}}}
{\frame{\includegraphics[height=0.14\textwidth,
trim=489 233 489 233, clip]{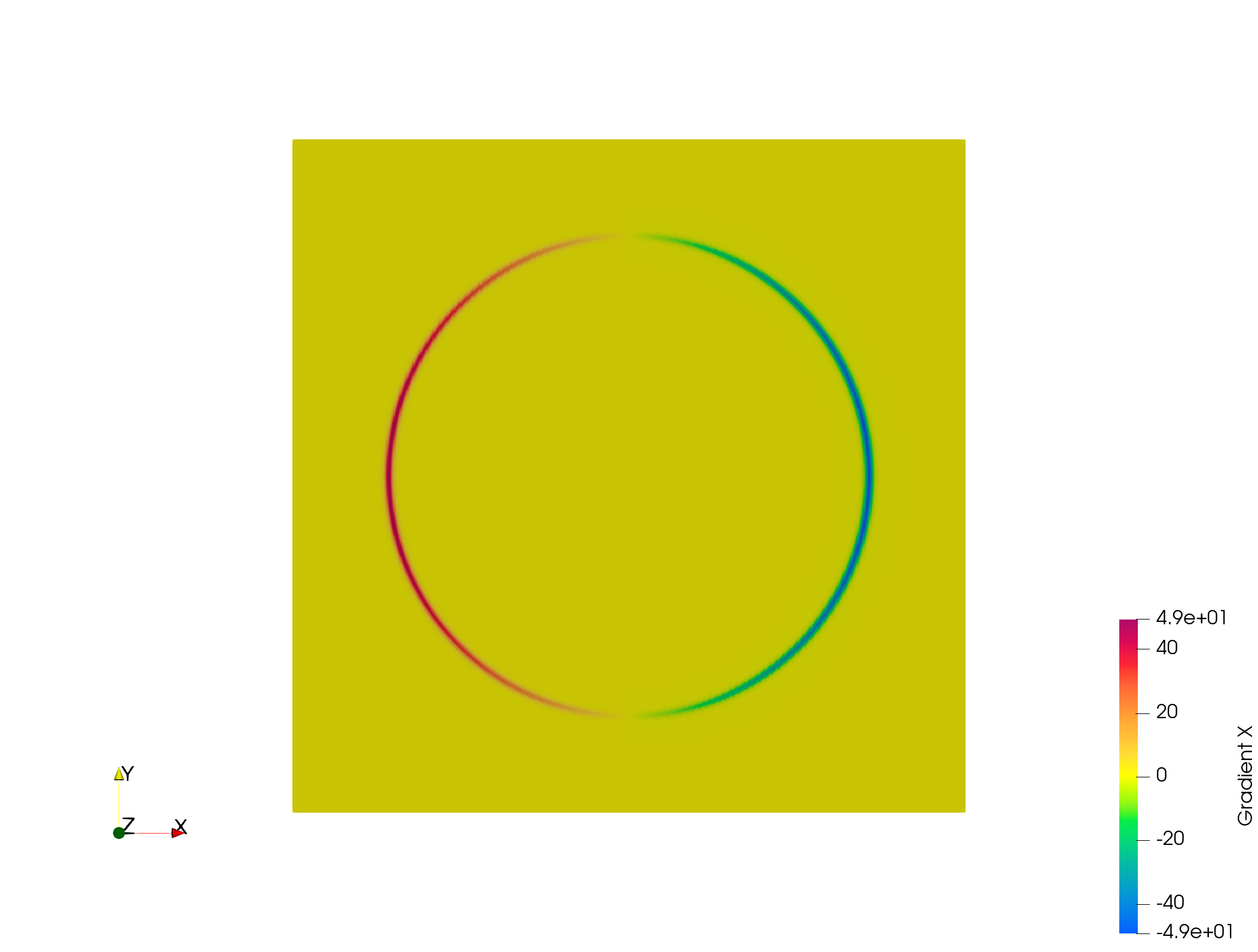}}}
{\frame{\includegraphics[height=0.14\textwidth,
trim=489 233 489 233, clip]{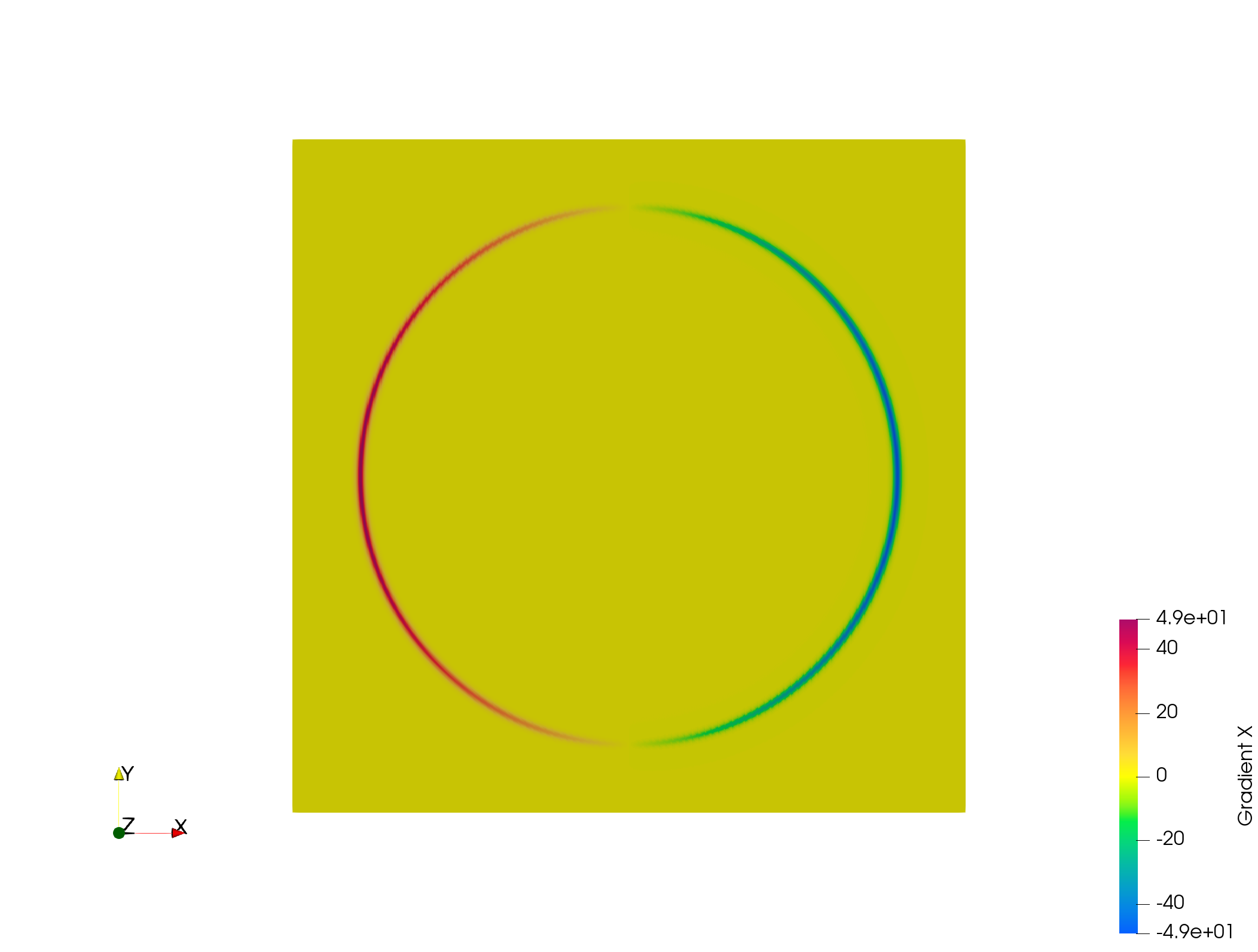}}}
{\frame{\includegraphics[height=0.14\textwidth,
trim=489 233 489 233, clip]{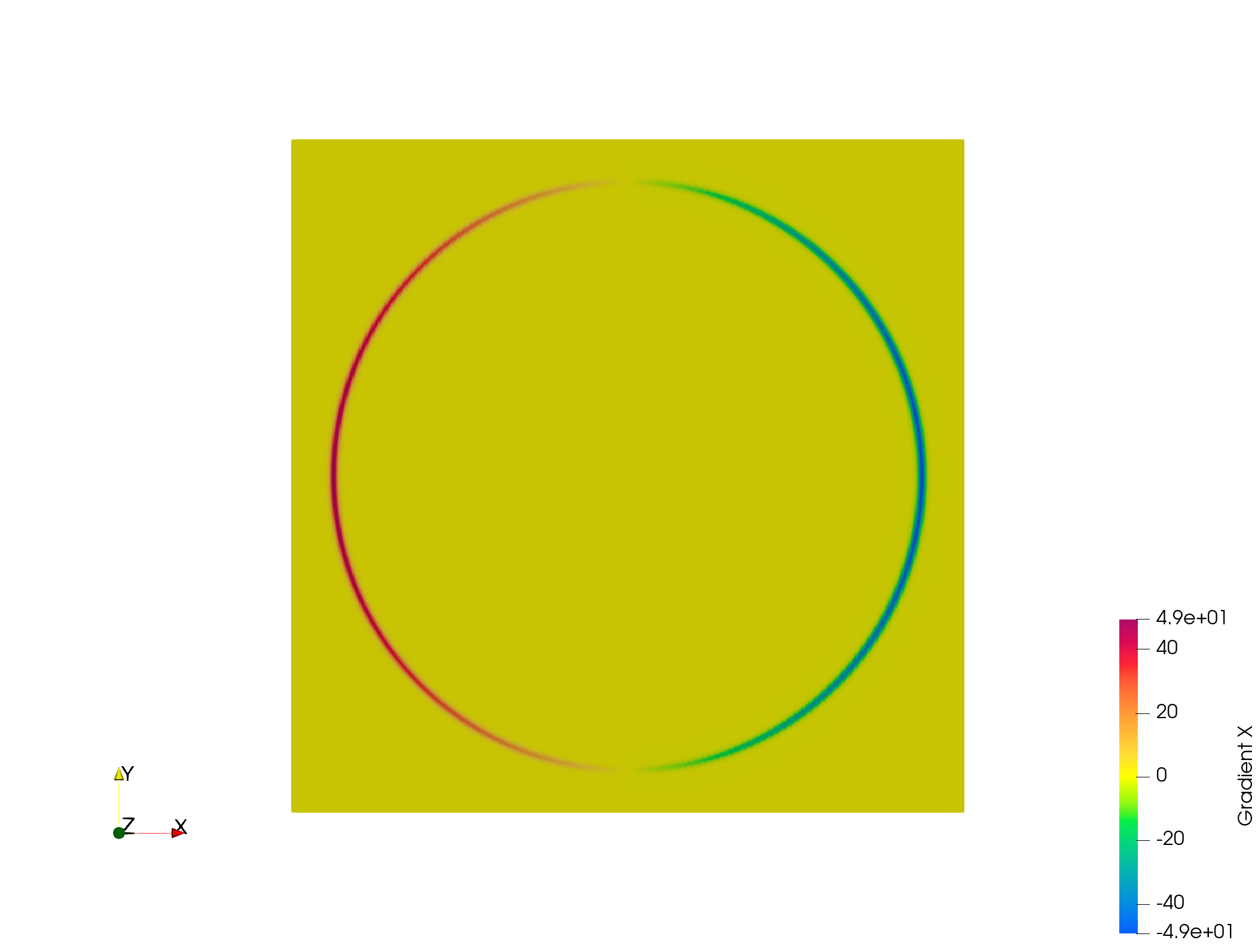}}}\\\vspace{0.1cm}
{\frame{\includegraphics[height=0.14\textwidth,
trim=489 233 489 233, clip]{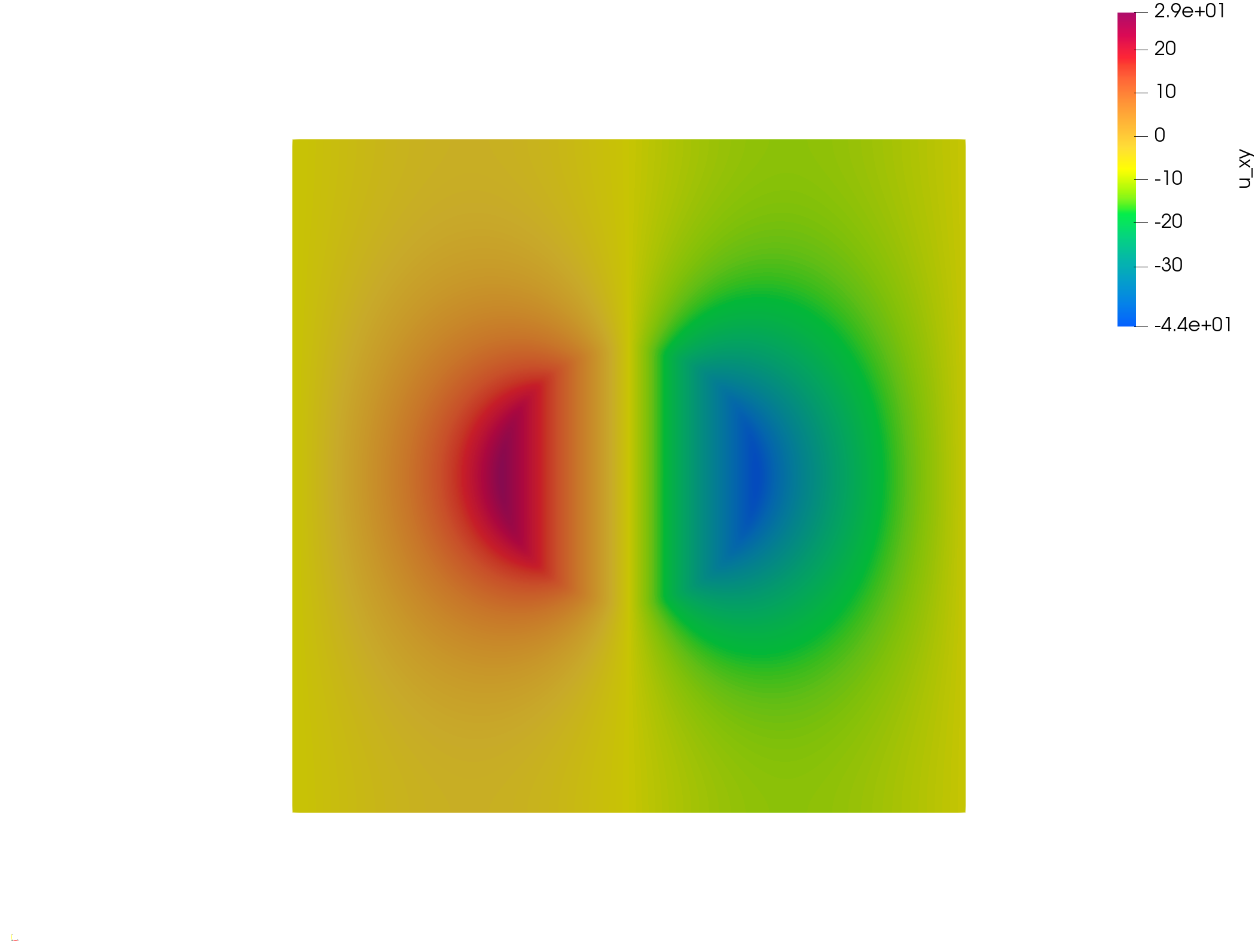}}}
{\frame{\includegraphics[height=0.14\textwidth,
trim=489 233 489 233, clip]{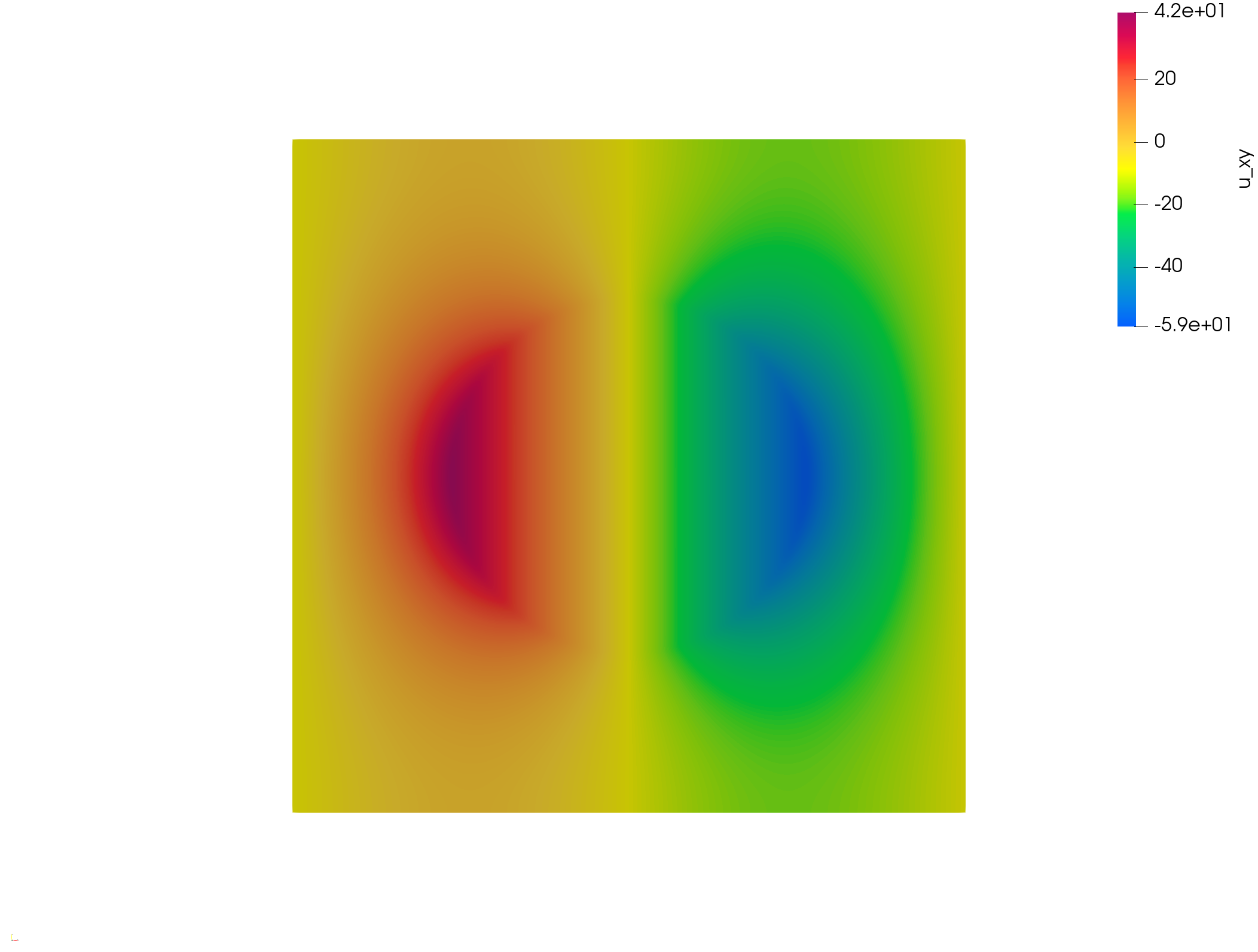}}}
{\frame{\includegraphics[height=0.14\textwidth,
trim=489 233 489 233, clip]{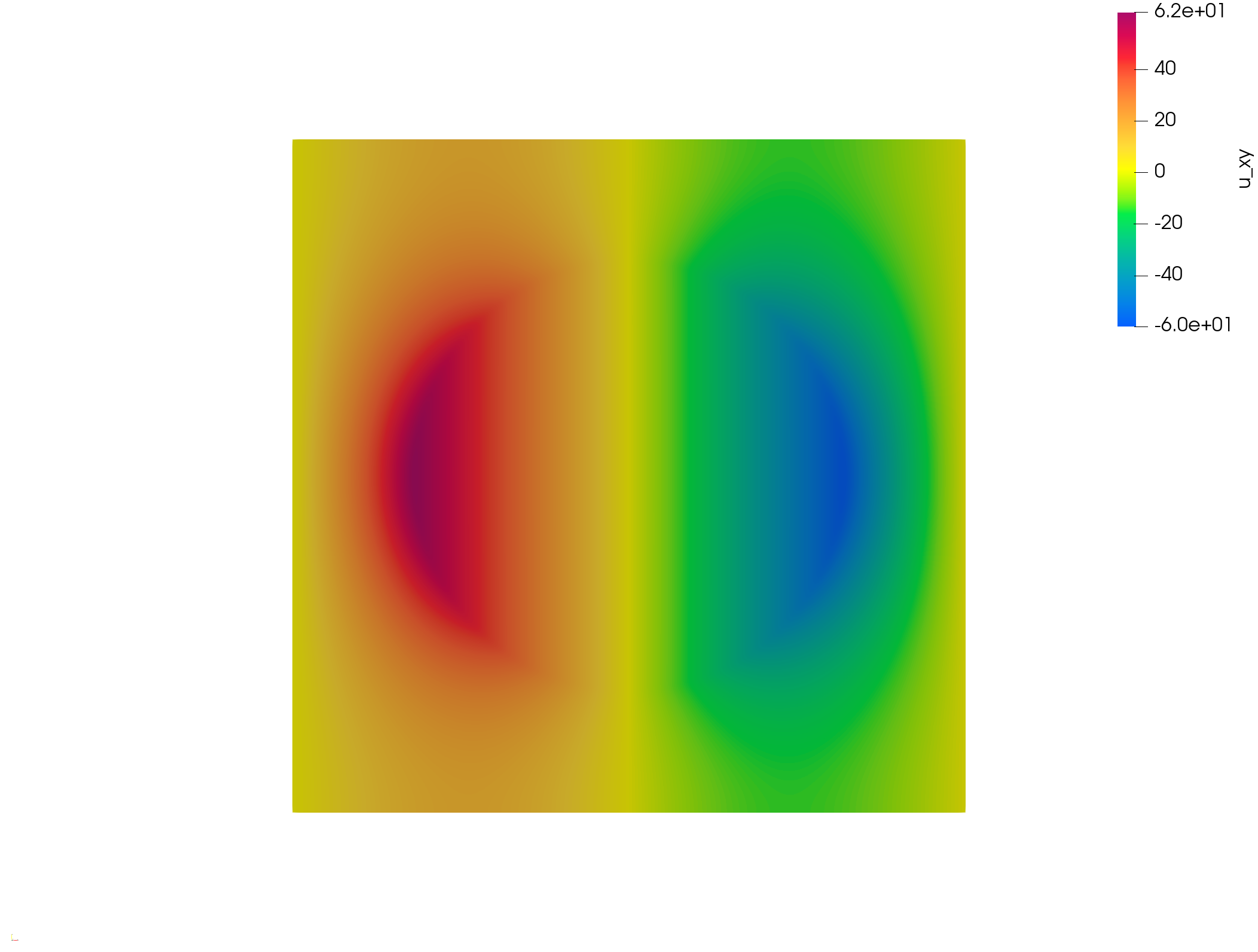}}}
{\frame{\includegraphics[height=0.14\textwidth,
trim=489 233 489 233, clip]{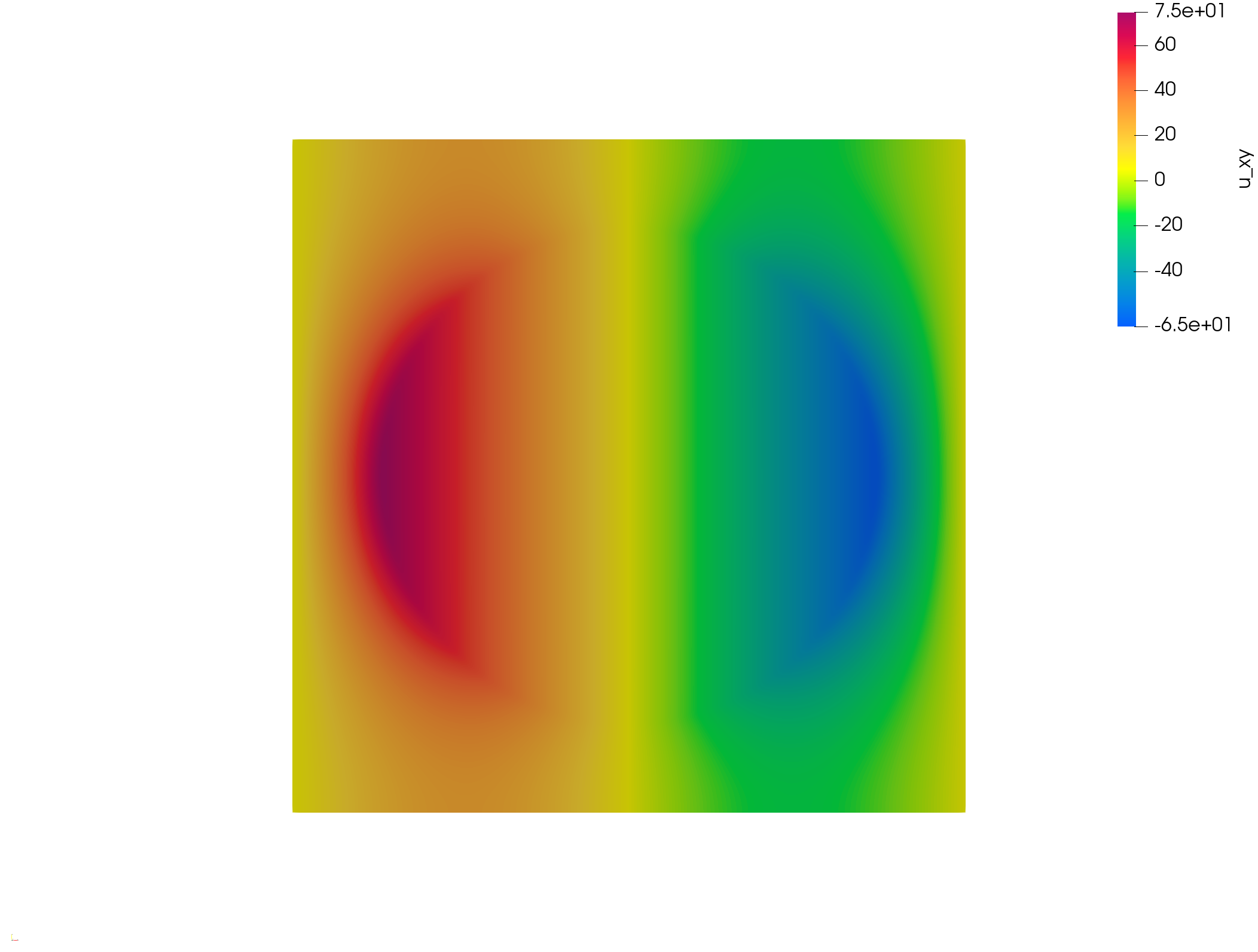}}}
{\frame{\includegraphics[height=0.14\textwidth,
trim=489 233 489 233, clip]{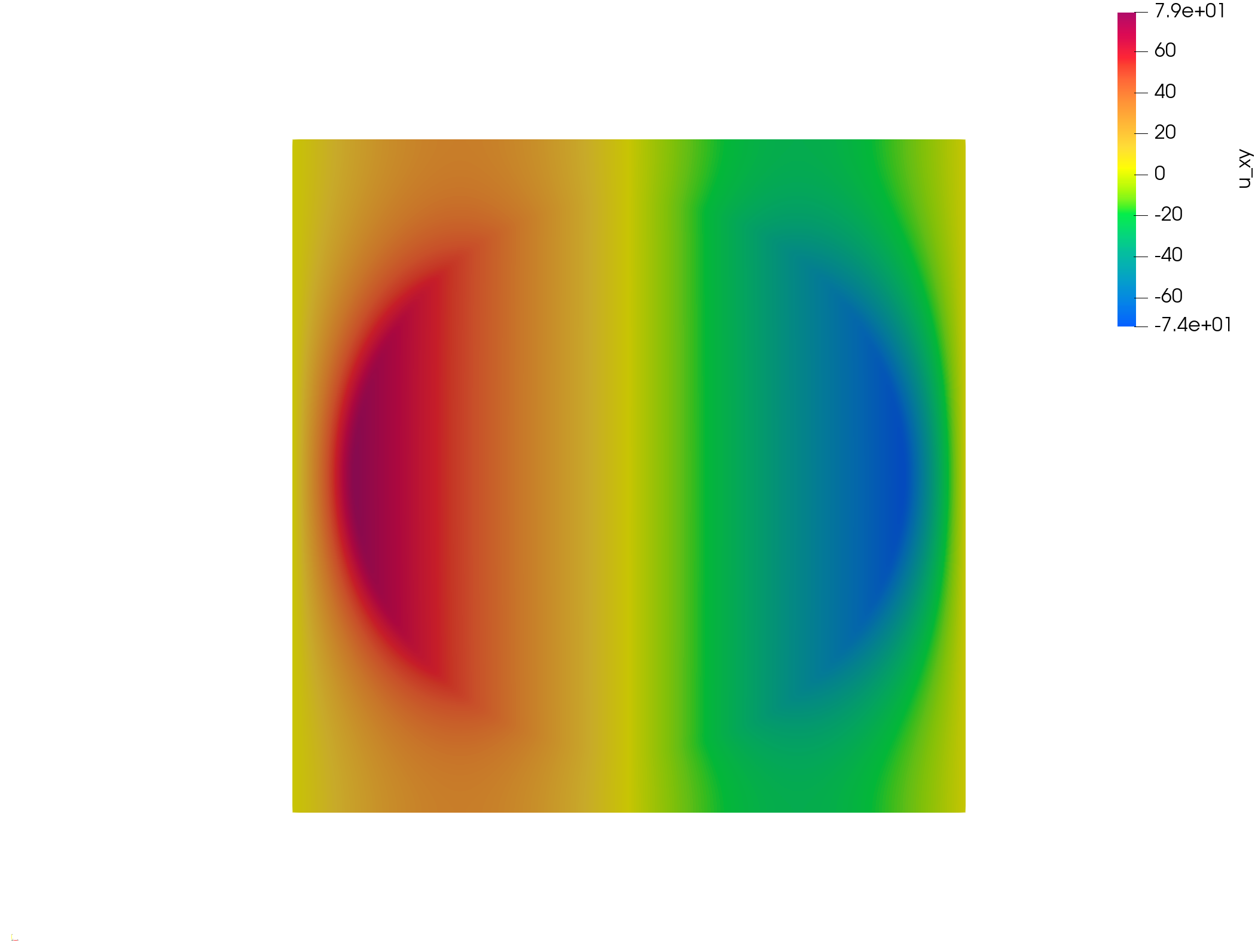}}}
{\frame{\includegraphics[height=0.14\textwidth,
trim=489 233 489 233, clip]{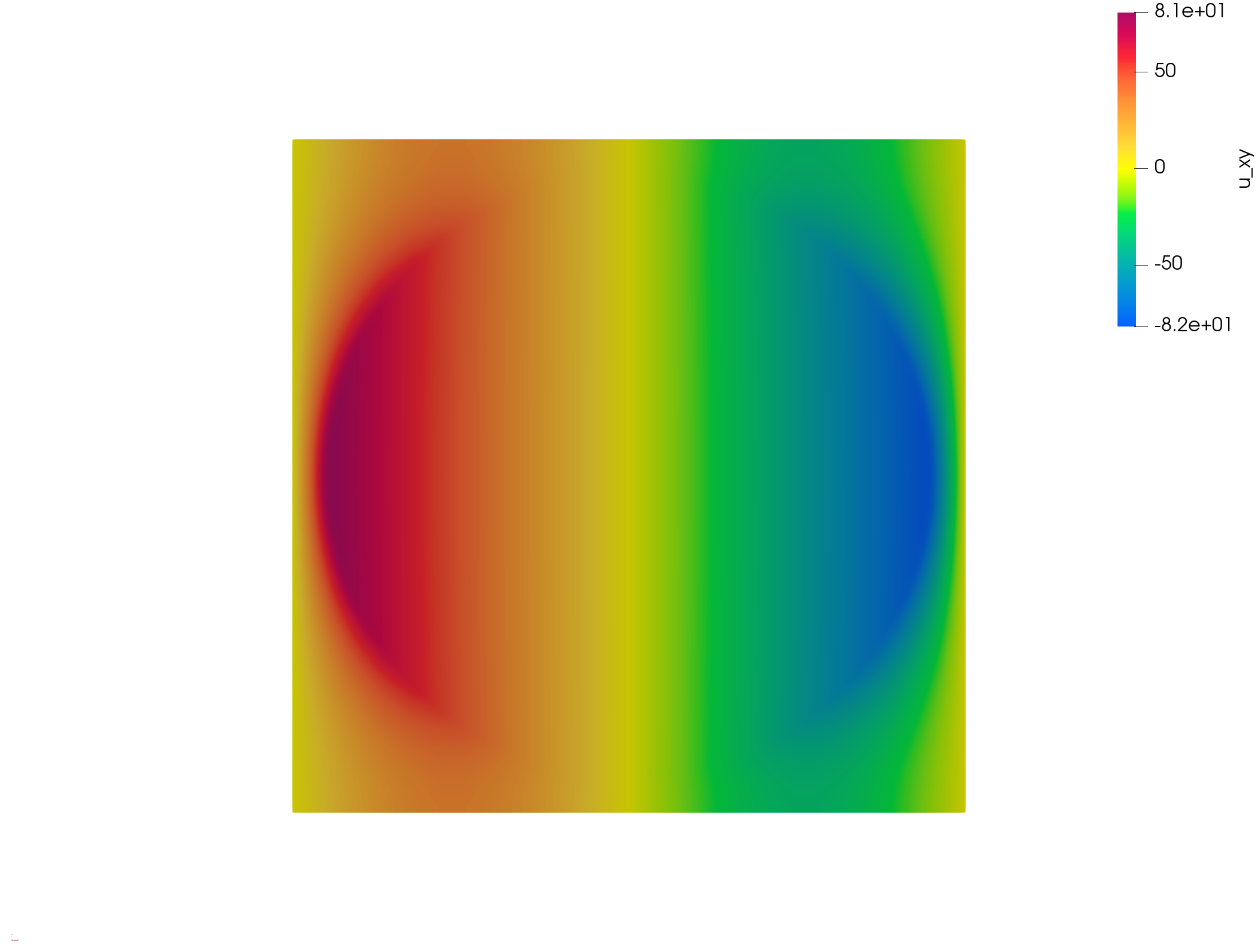}}}\\\vspace{0.1cm}
{\frame{\includegraphics[height=0.14\textwidth,
trim=489 233 489 233, clip]{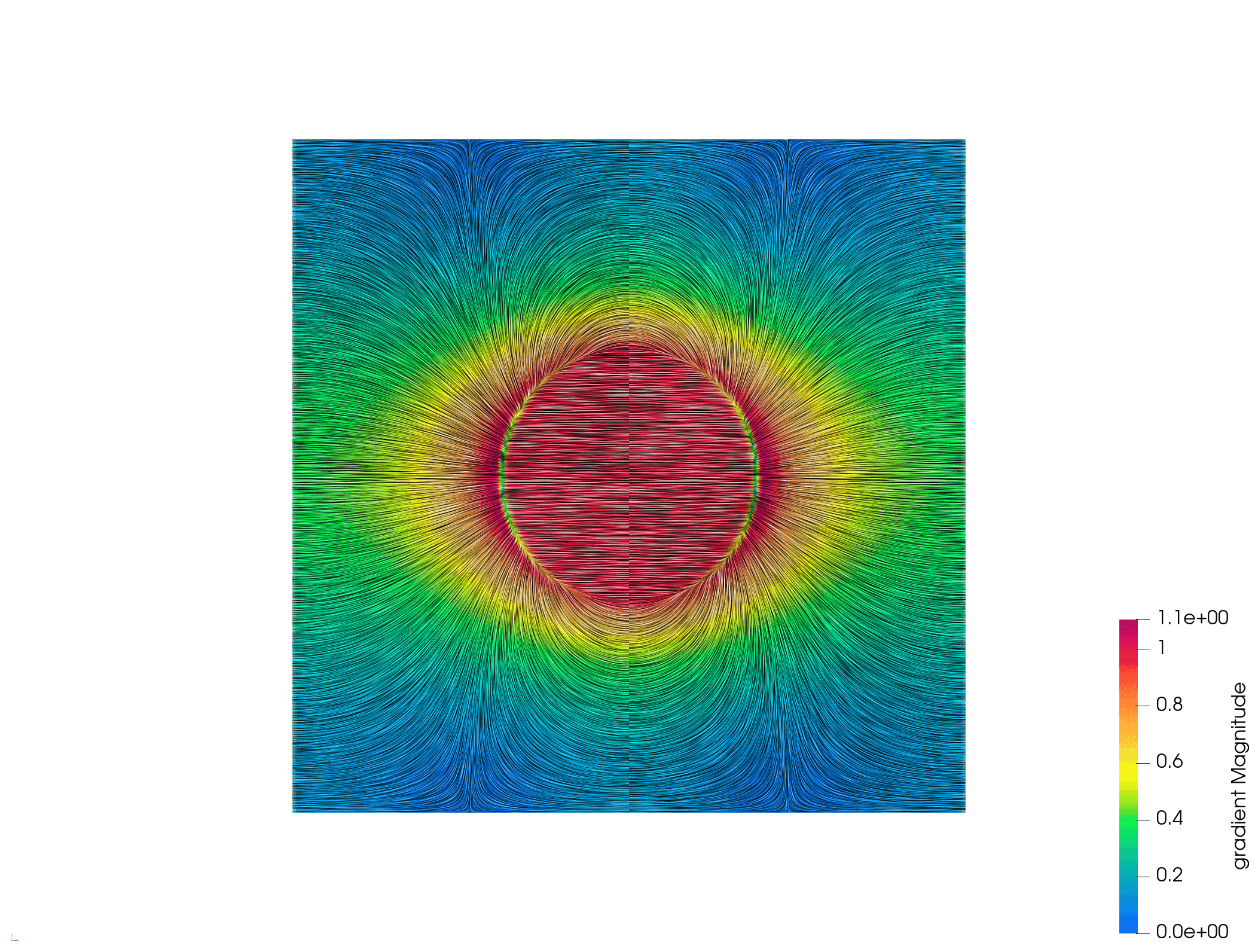}}}
{\frame{\includegraphics[height=0.14\textwidth,
trim=489 233 489 233, clip]{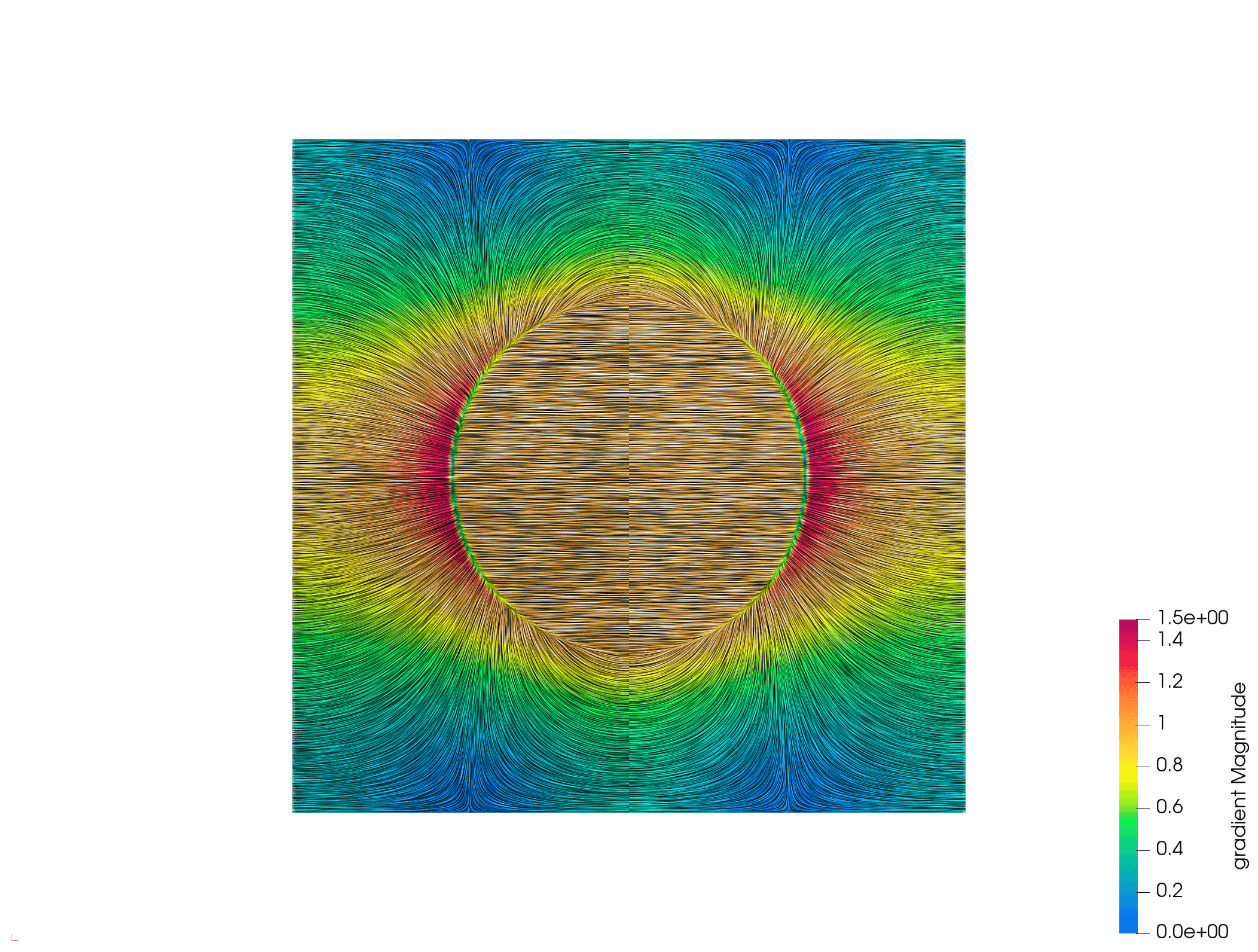}}}
{\frame{\includegraphics[height=0.14\textwidth,
trim=489 233 489 233, clip]{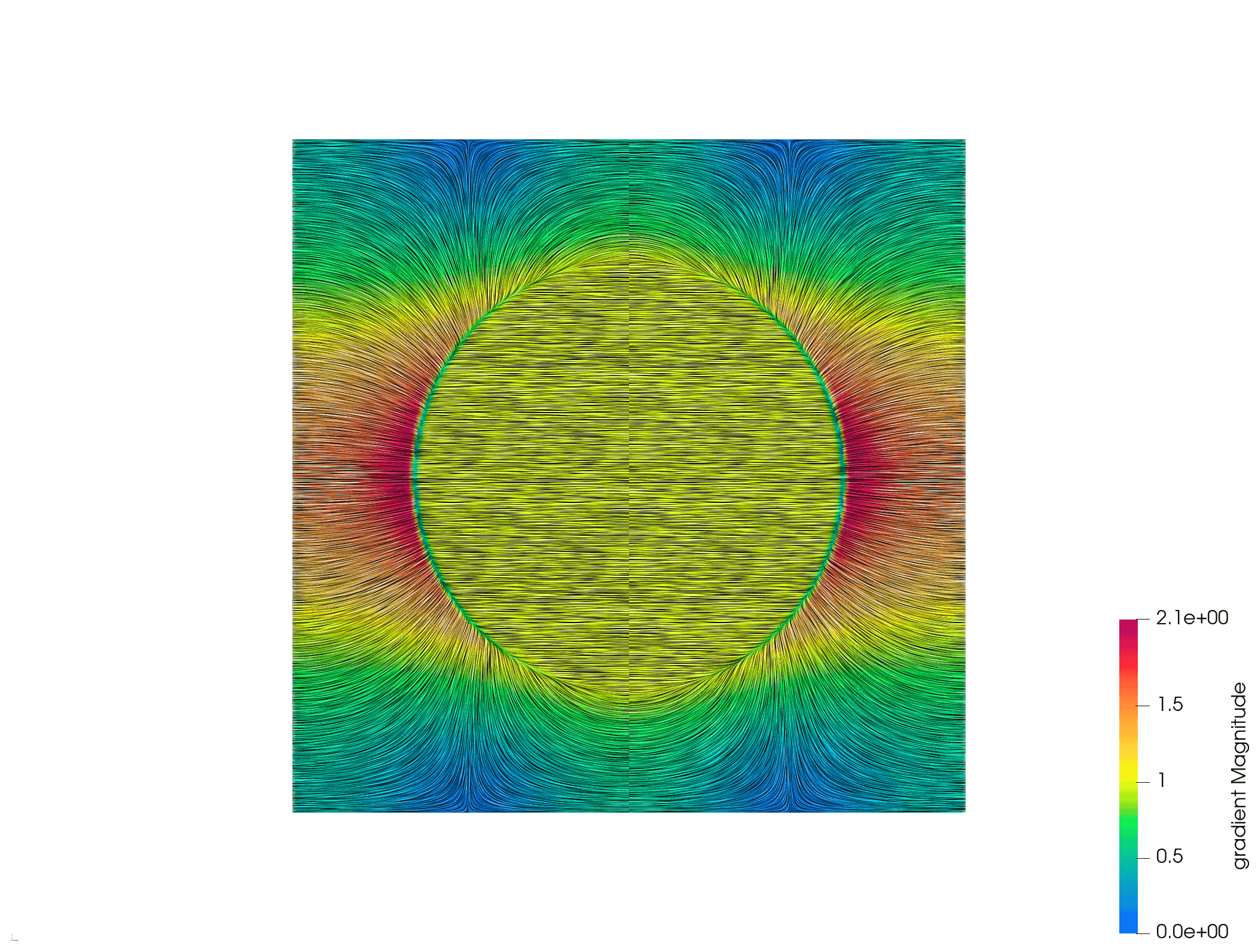}}}
{\frame{\includegraphics[height=0.14\textwidth,
trim=489 233 489 233, clip]{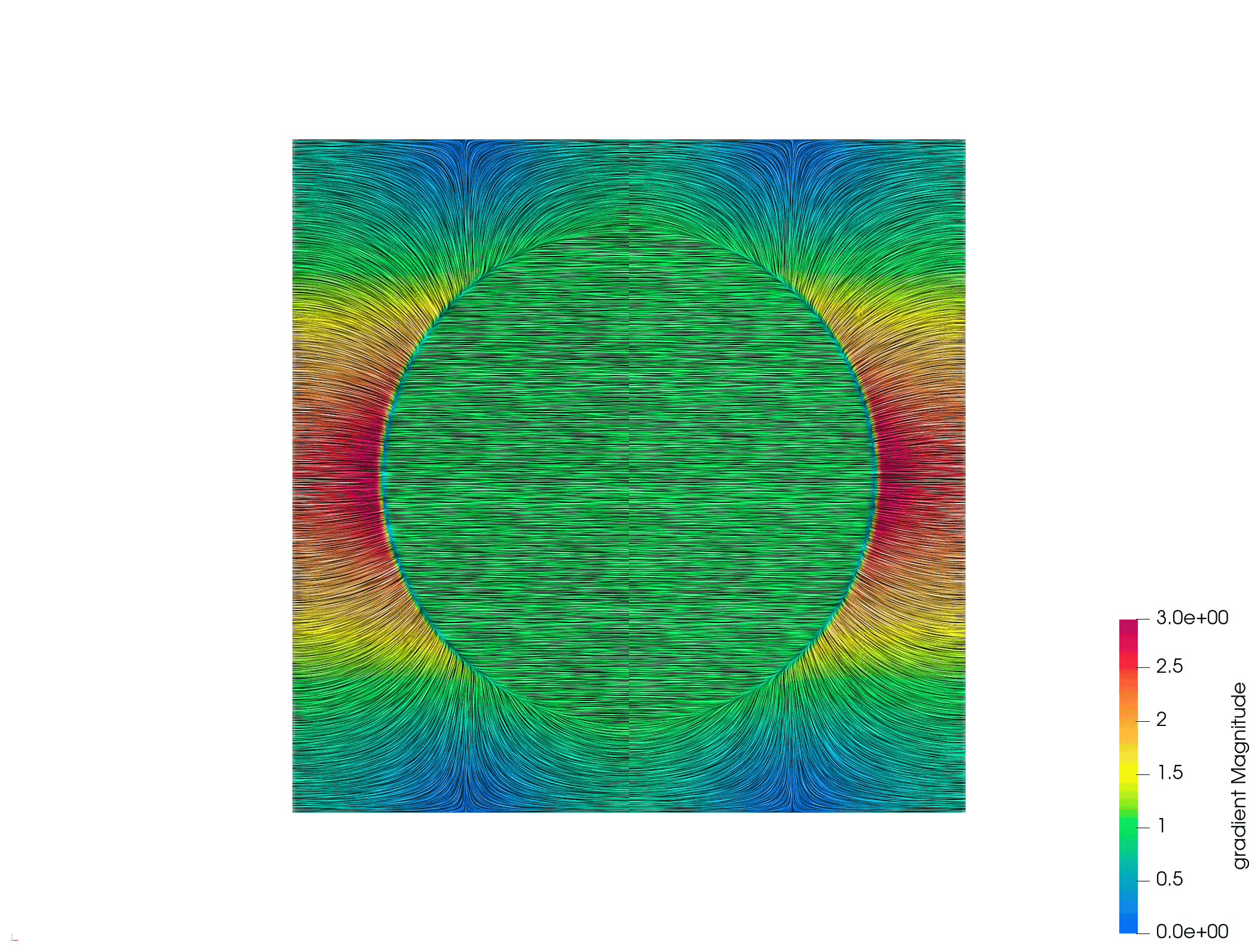}}}
{\frame{\includegraphics[height=0.14\textwidth,
trim=489 233 489 233, clip]{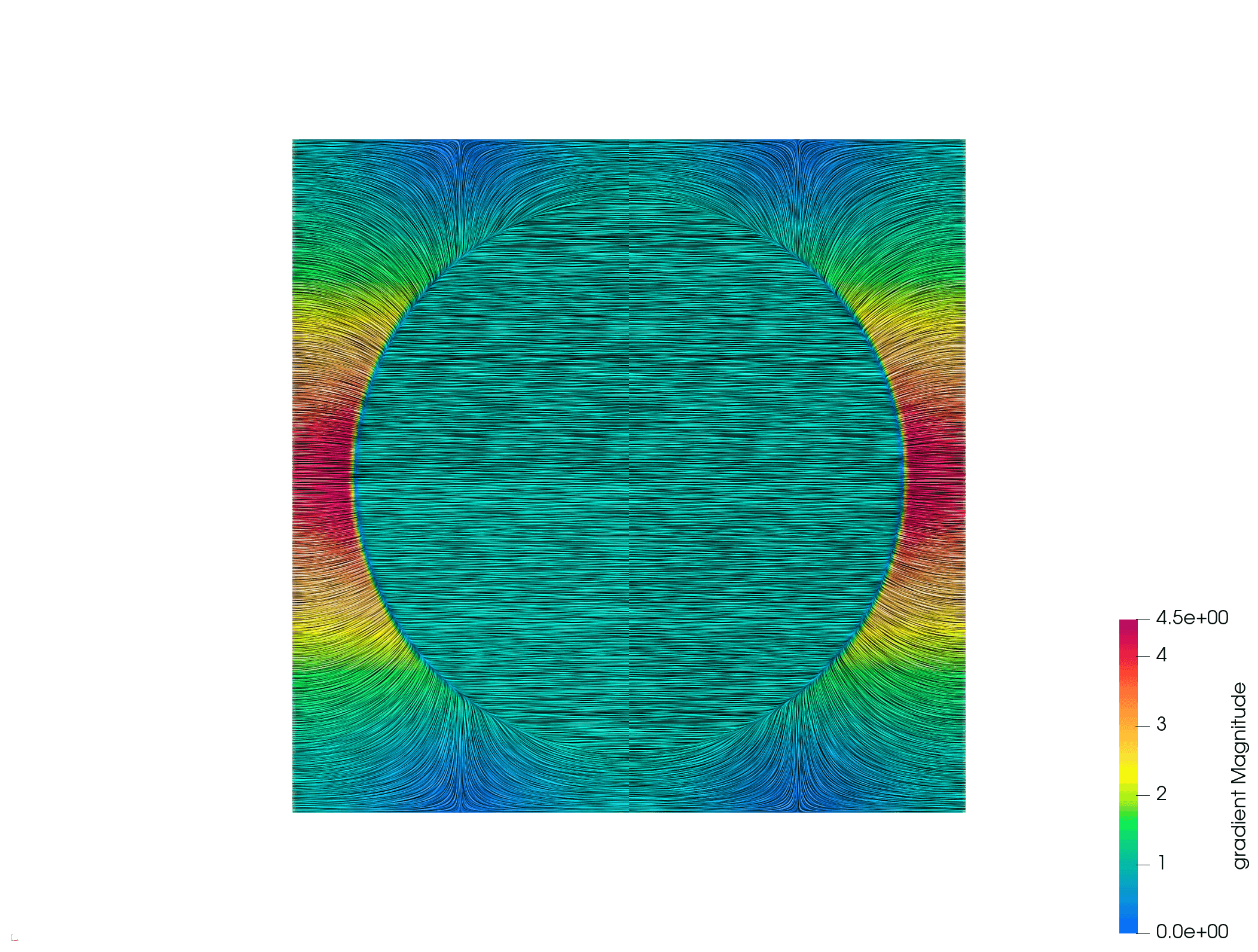}}}
{\frame{\includegraphics[height=0.14\textwidth,
trim=489 233 489 233, clip]{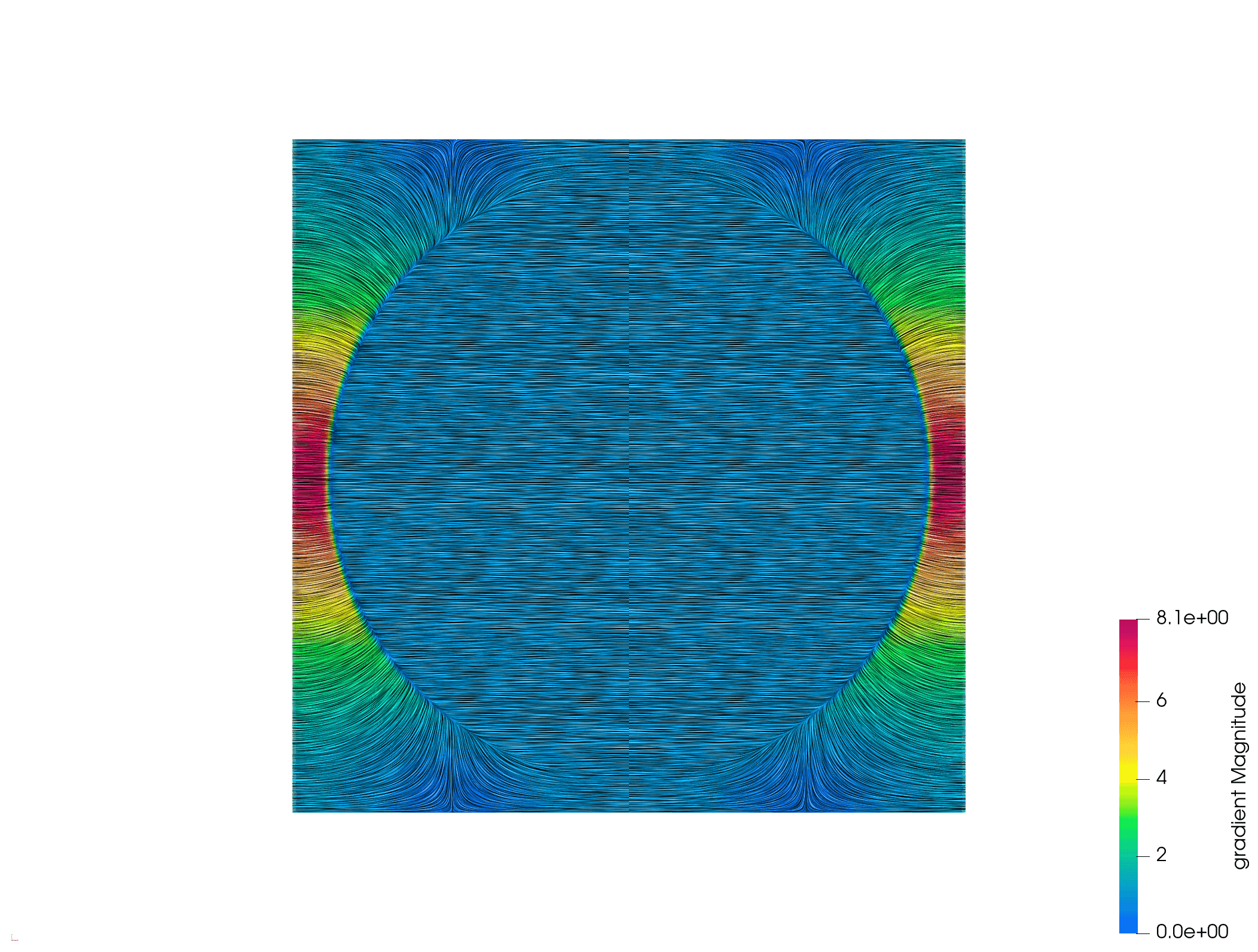}}}
min$\,\,$\frame{\includegraphics[width=0.30\textwidth, trim=0 15 0 15, clip]{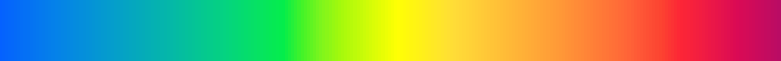}}$\,\,$max
\caption{\color{black} The contour plots for the input,
the property field $a(\boldsymbol x)$ with $\boldsymbol a(\boldsymbol x)=a(\boldsymbol x)\, \boldsymbol 1$ and  $\xi=1$ (row 1) and the source term $\partial a(\boldsymbol x)/\partial x_1$ (row 2), and the PINN solution,
$\chi(\boldsymbol x)$ (row 3) and $|\boldsymbol{\nabla}\chi|$ (row 4), fields of the cell-problem for periodic unit cell $\mathcal{S}_1$. Here high-frequency Fourier features with the first 10 integer multiples of the reciprocal base vector are considered. The unit cell constitutes a matrix with inclusions in the form of circular disks positioned at Bravais lattice points with corresponding inclusion volume fractions $\phi_\mathrm{i}=\{0.1,0.2,0.3,0.4,0.5,0.6\}$ demonstrated in ascending order from left to right.
The intervals [min, max] of the  contour plots are: for the first row $[0,100]$,
for the second row $[-49.5,49.5]$,
for the third row, from left to right,
$[0,73]$, $[0,101]$, $[0,122]$, $[0,140]$, $[0,154]$  and (d) $[0,164]$,
for the fourth row, from left to right,
$[0,1.1]$, $[0,1.5]$, $[0,2.1]$, $[0,3.0]$, $[0,4.5]$  and (d) $[0,8.1]$.}
\label{F:sq_volume_fractions_square_fields}
\end{figure*}

\begin{figure*}[htb!]
\centering
{\frame{\includegraphics[height=0.14\textwidth,
trim=489 233 489 233, clip]{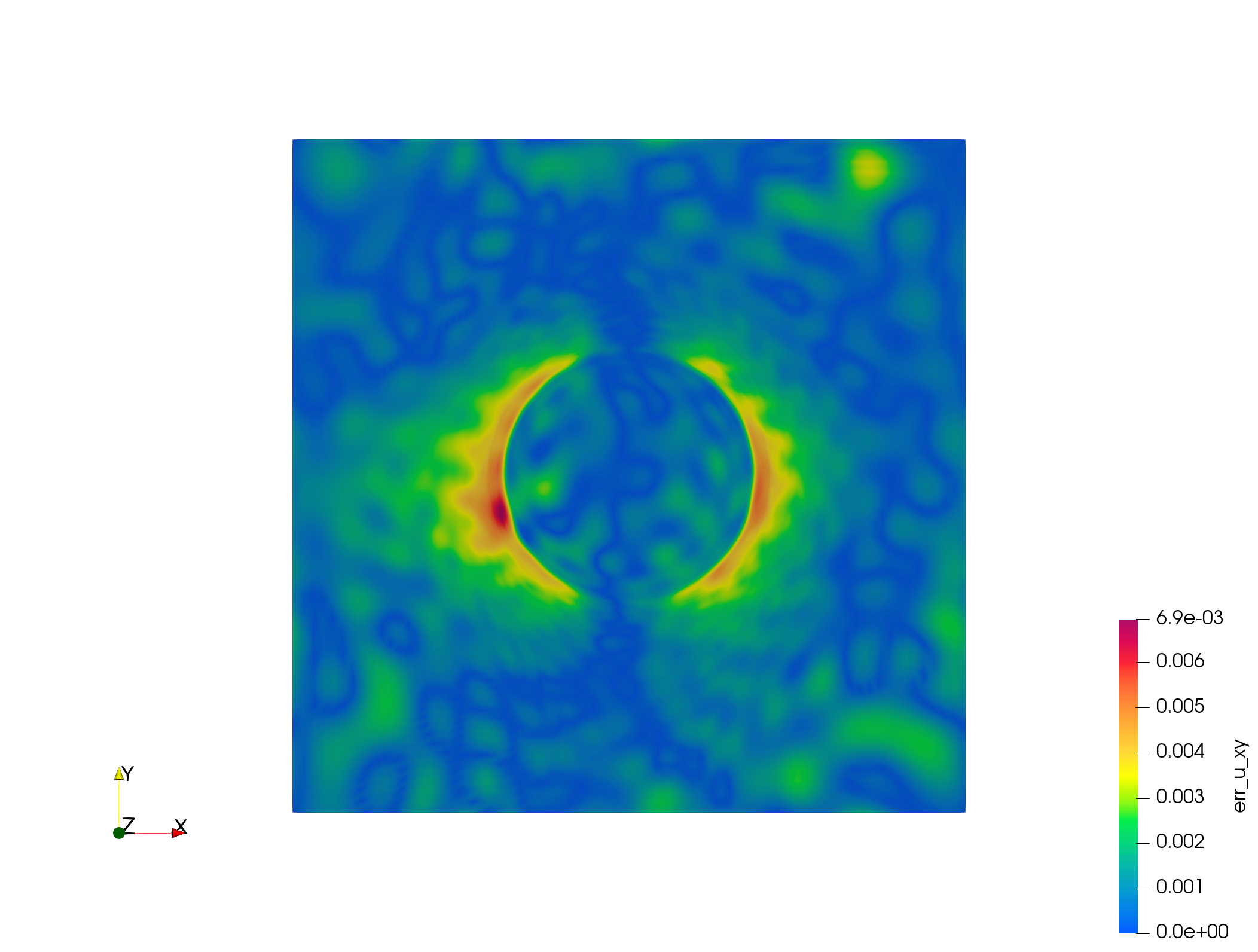}}}
{\frame{\includegraphics[height=0.14\textwidth,
trim=489 233 489 233, clip]{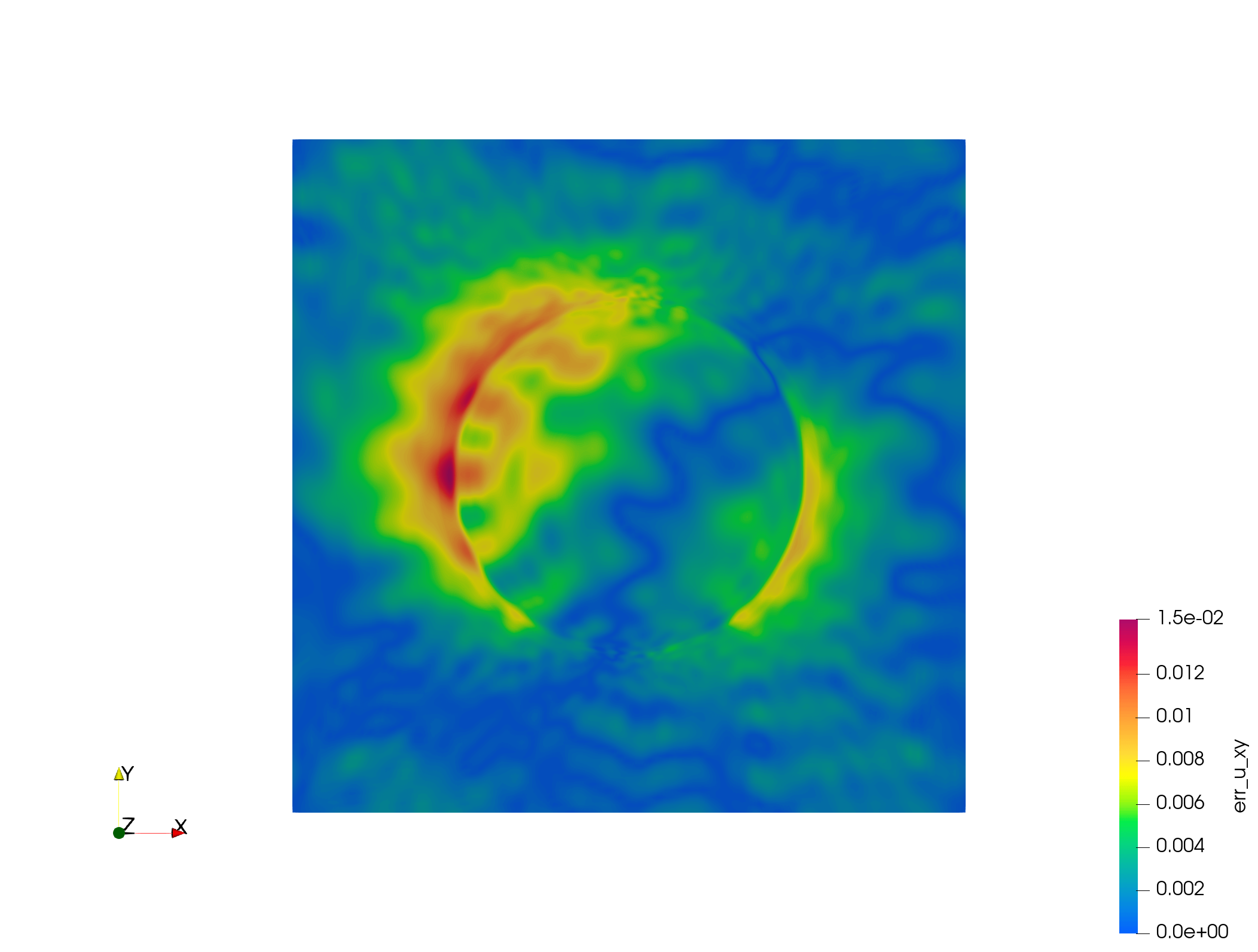}}}
{\frame{\includegraphics[height=0.14\textwidth,
trim=489 233 489 233, clip]{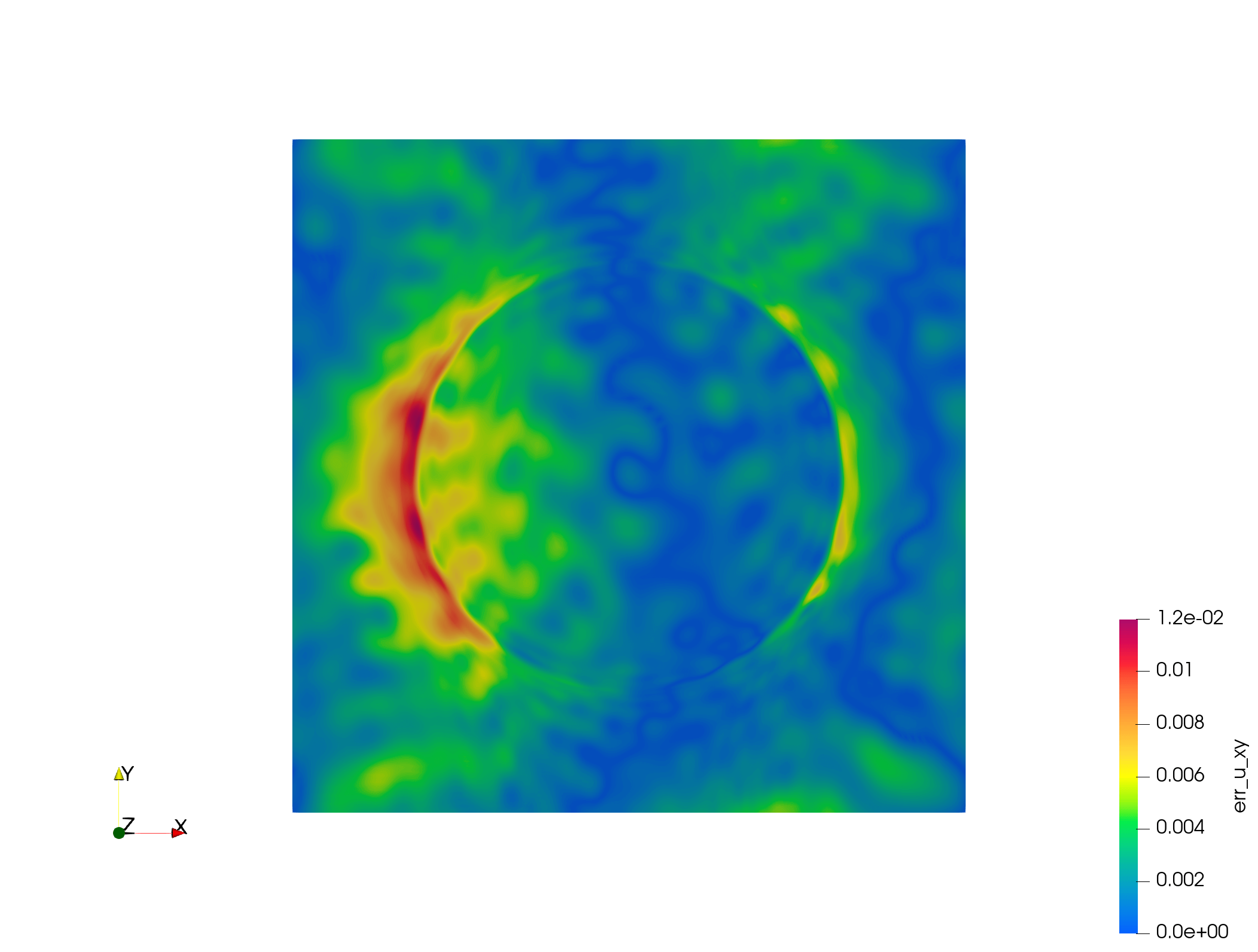}}}
{\frame{\includegraphics[height=0.14\textwidth,
trim=489 233 489 233, clip]{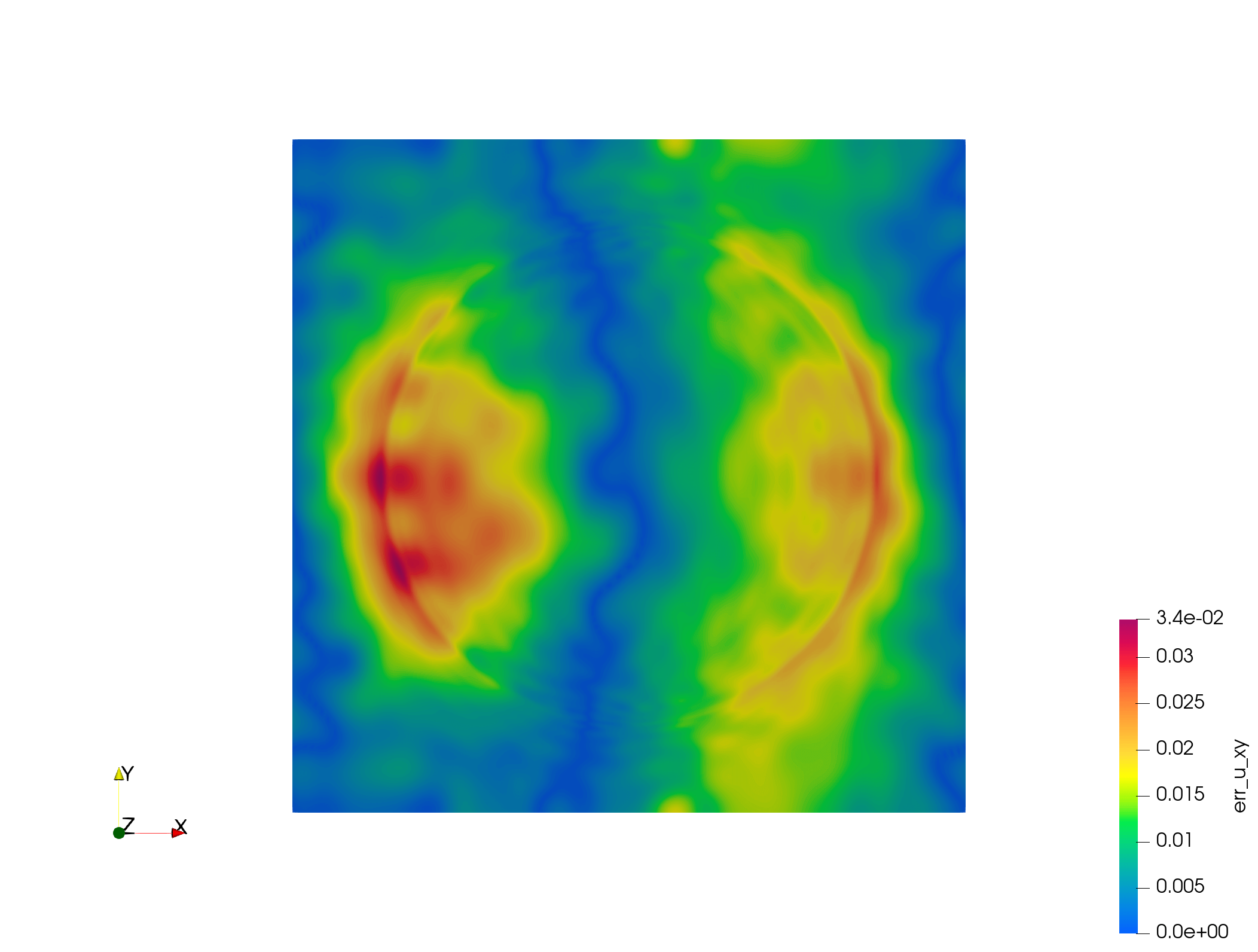}}}
{\frame{\includegraphics[height=0.14\textwidth,
trim=489 233 489 233, clip]{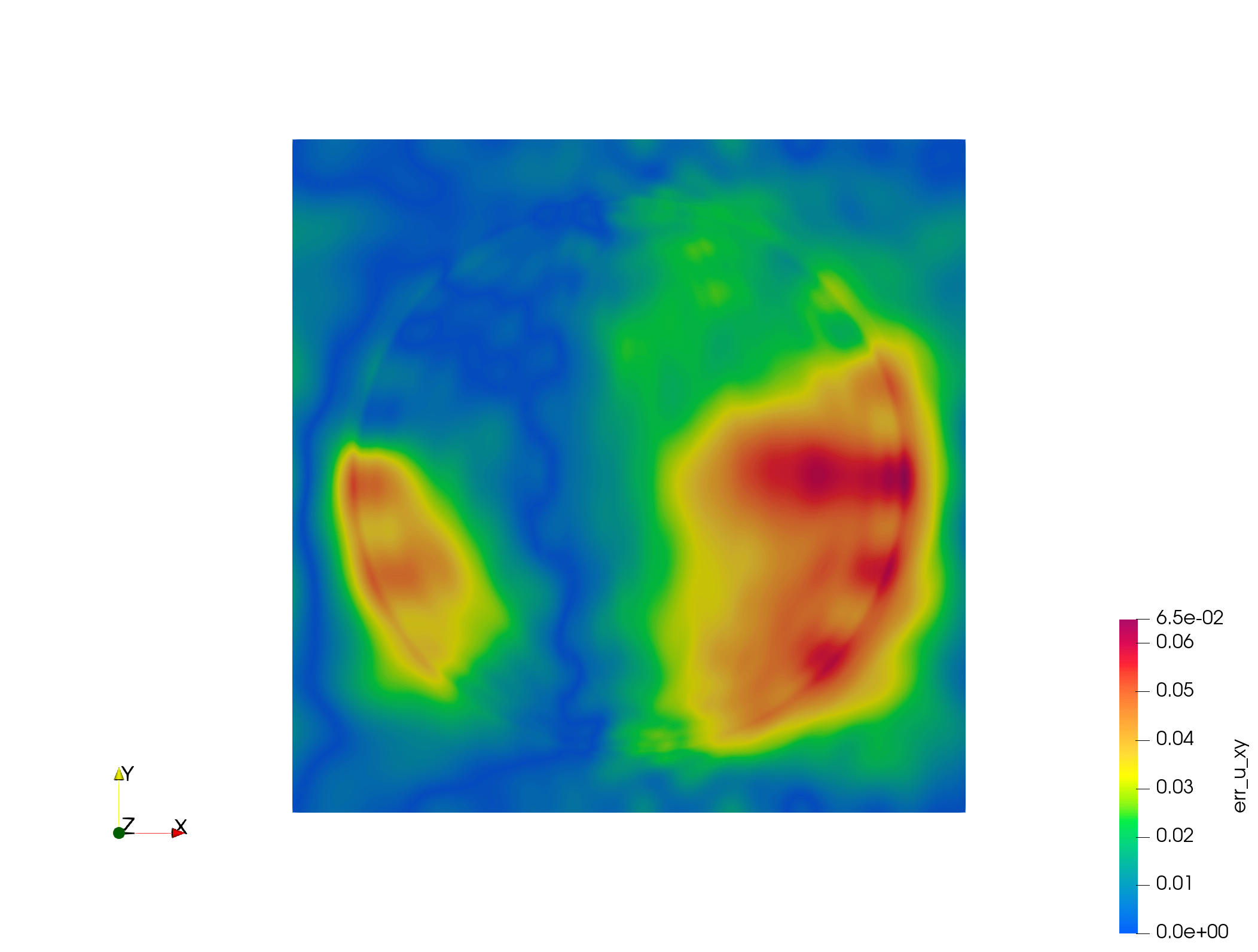}}}
{\frame{\includegraphics[height=0.14\textwidth,
trim=489 233 489 233, clip]{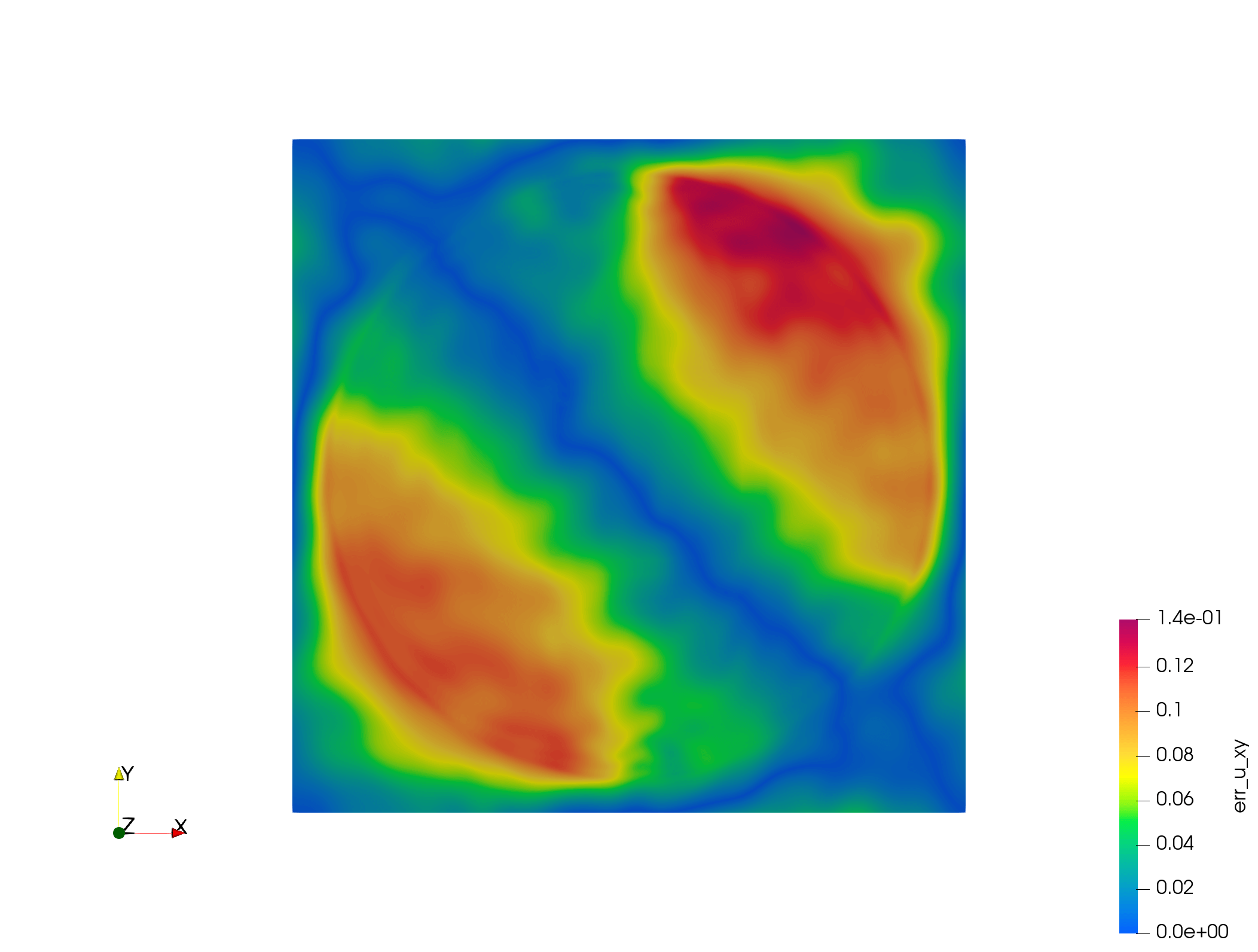}}}\\\vspace{0.1cm}
{\frame{\includegraphics[height=0.14\textwidth,
trim=489 233 489 233, clip]{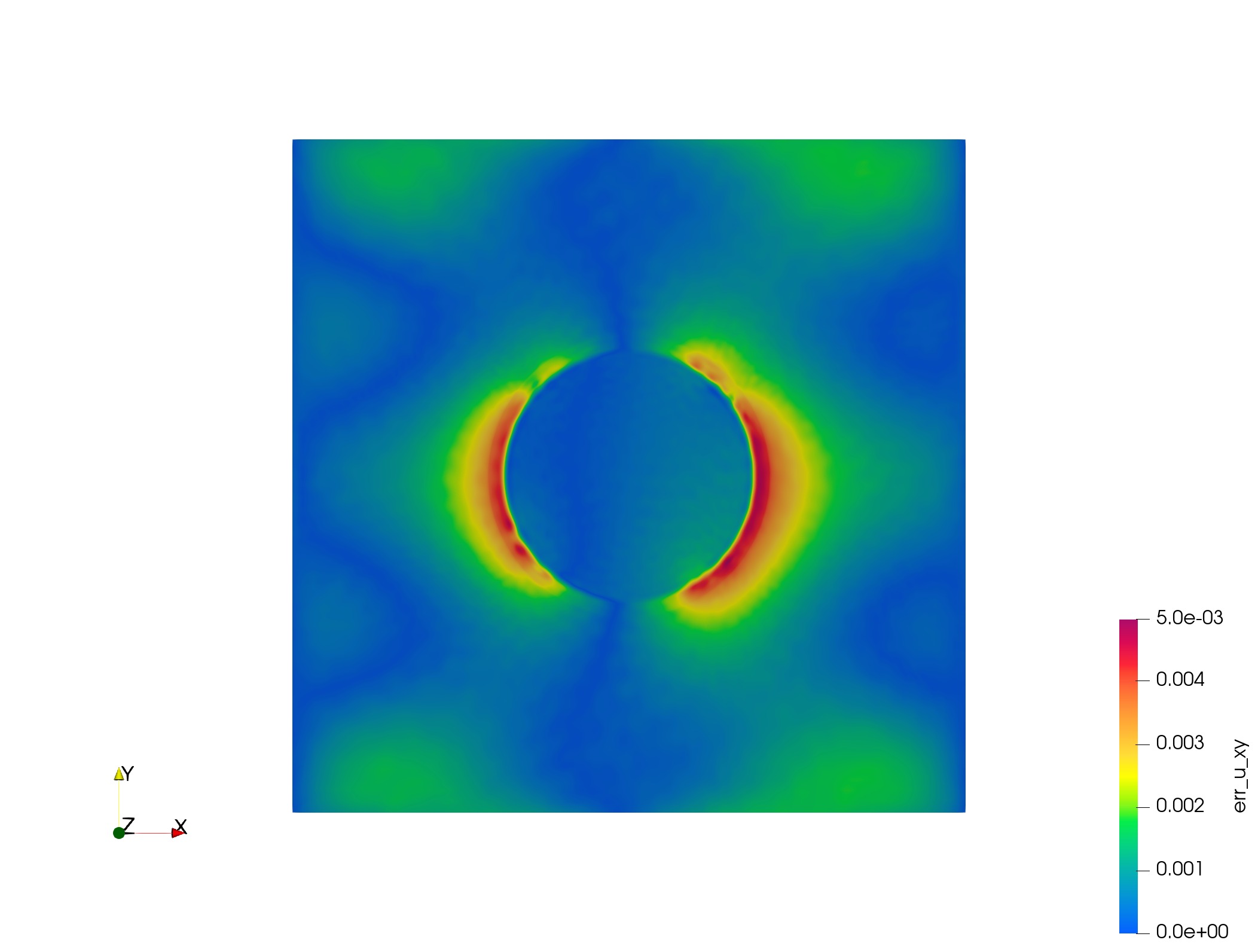}}}
{\frame{\includegraphics[height=0.14\textwidth,
trim=489 233 489 233, clip]{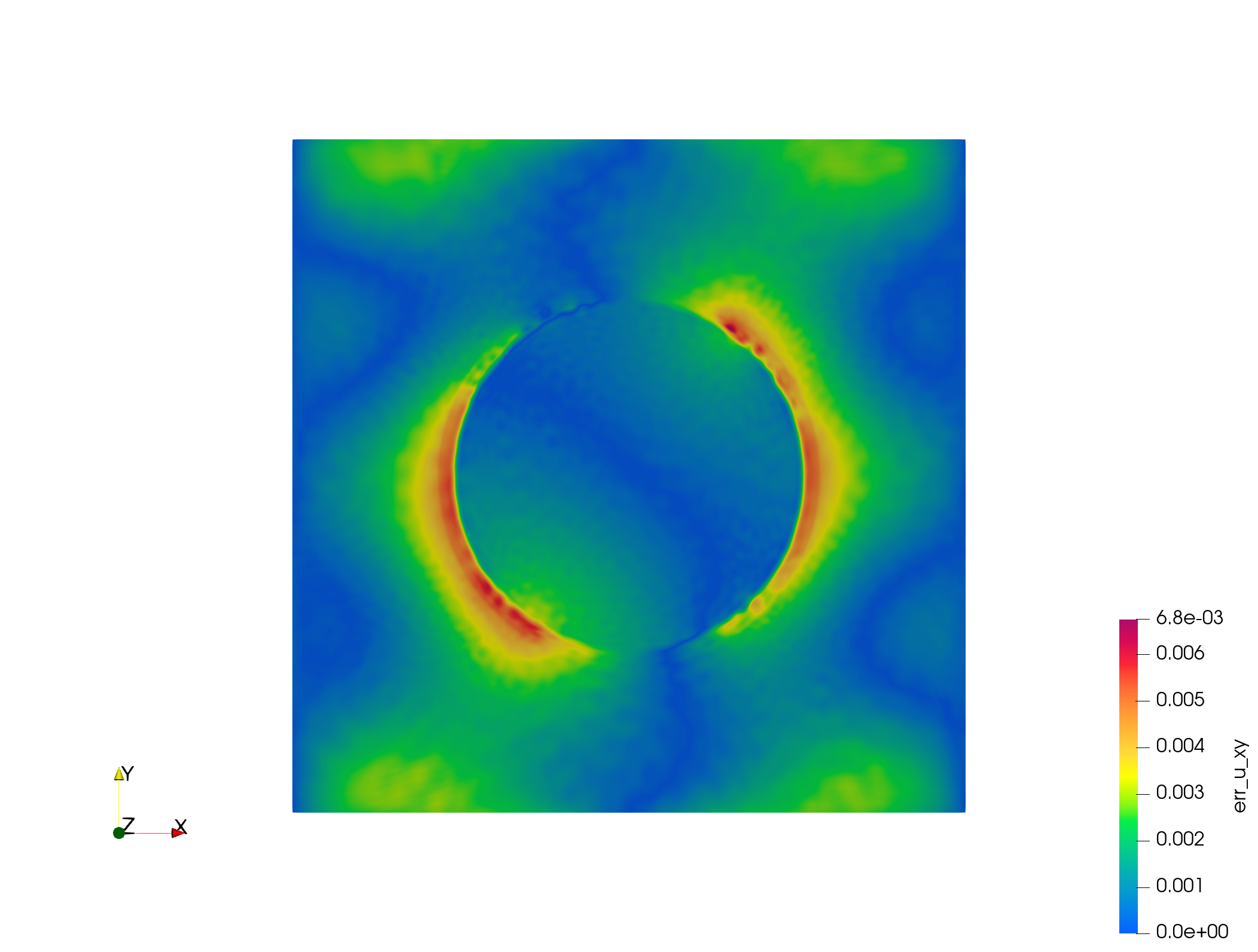}}}
{\frame{\includegraphics[height=0.14\textwidth,
trim=489 233 489 233, clip]{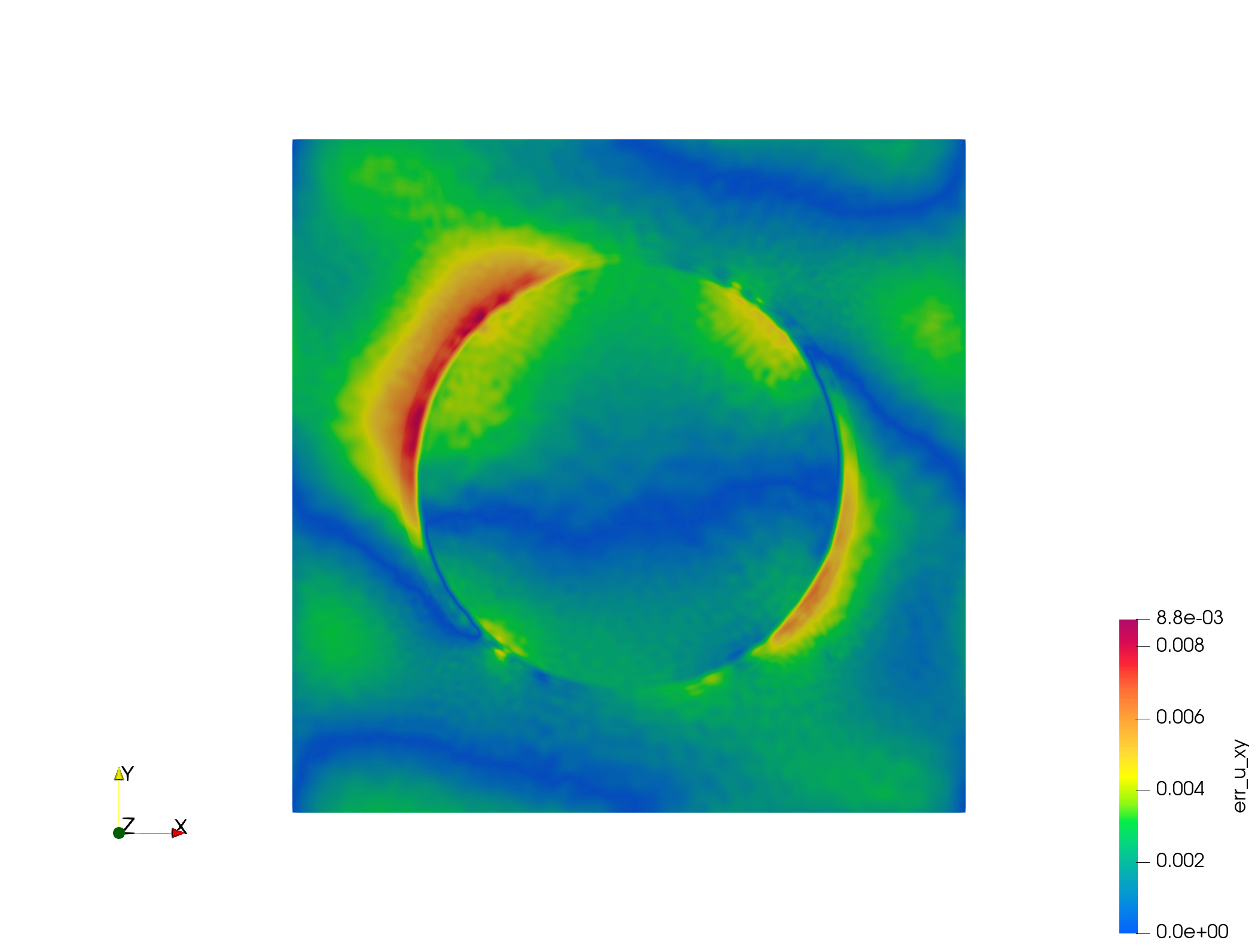}}}
{\frame{\includegraphics[height=0.14\textwidth,
trim=489 233 489 233, clip]{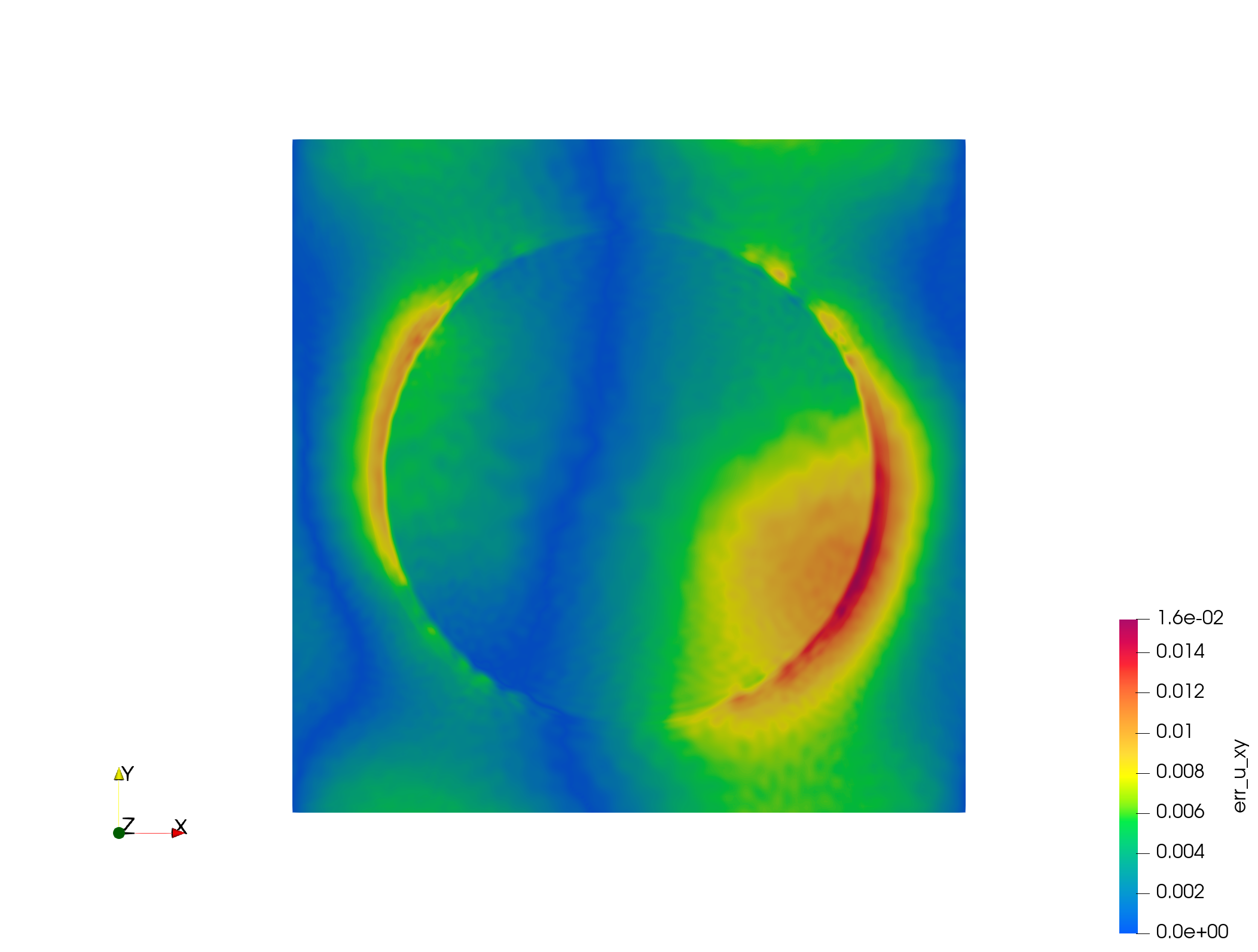}}}
{\frame{\includegraphics[height=0.14\textwidth,
trim=489 233 489 233, clip]{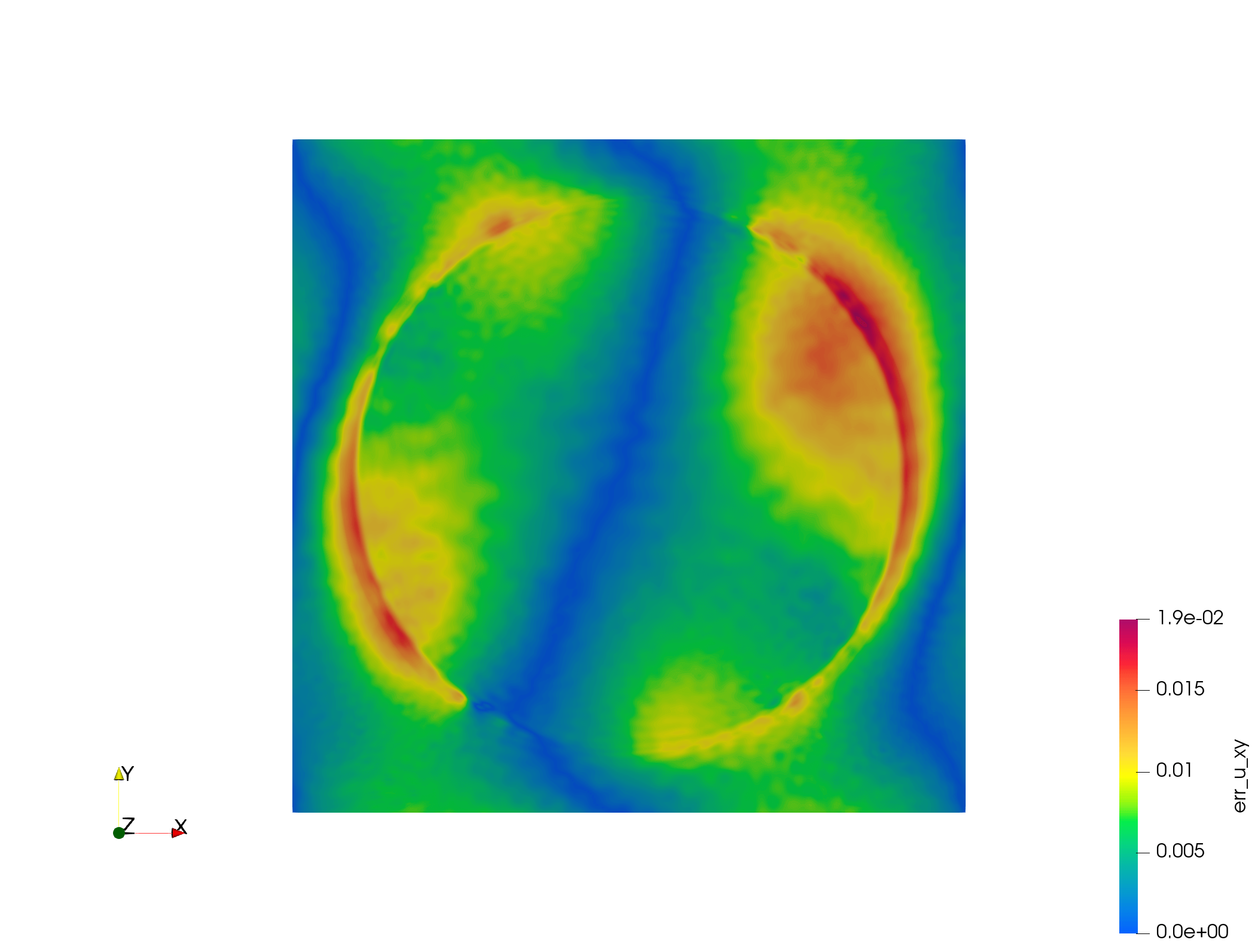}}}
{\frame{\includegraphics[height=0.14\textwidth,
trim=489 233 489 233, clip]{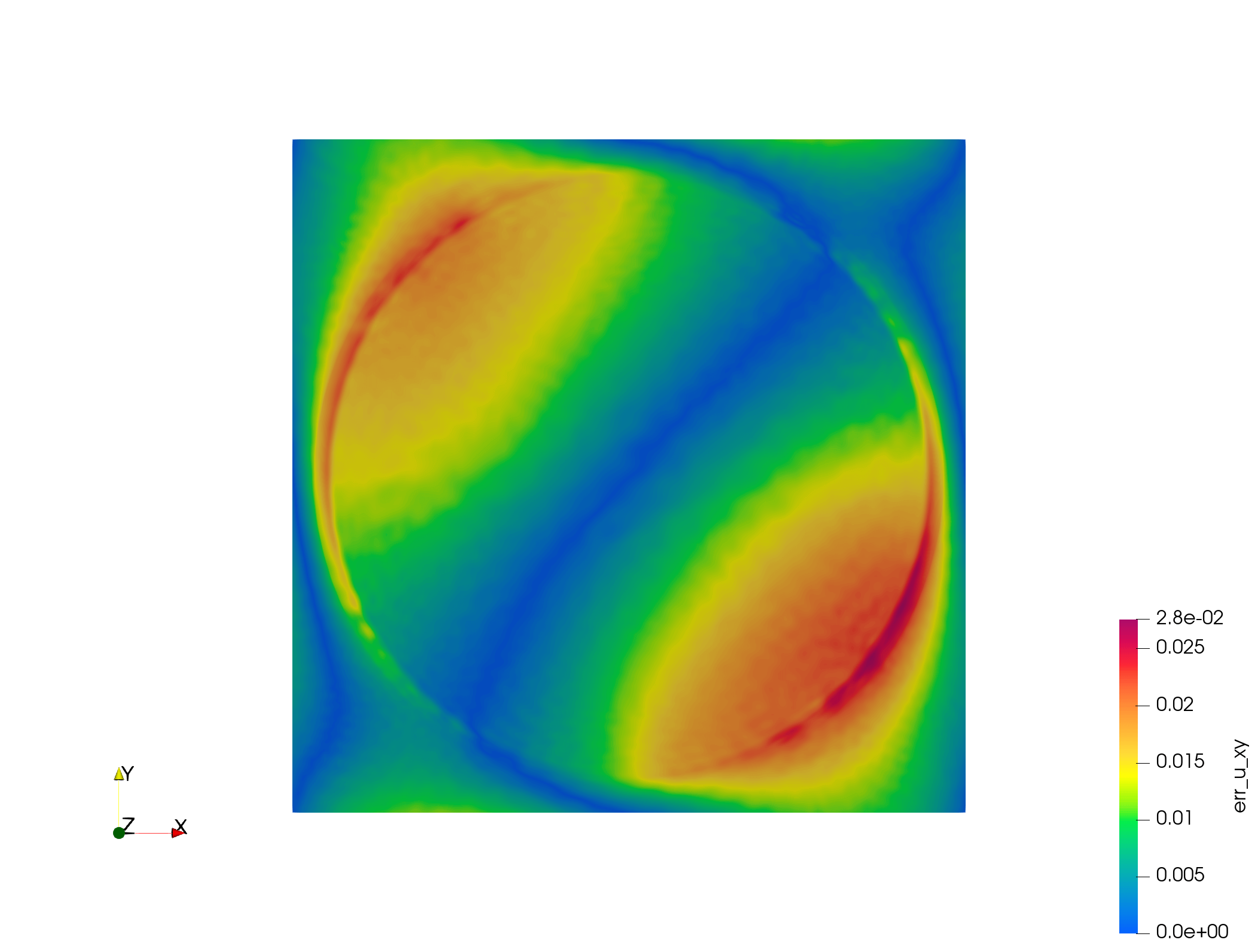}}}\\
min$\,\,$\frame{\includegraphics[width=0.30\textwidth, trim=0 15 0 15, clip]{legend_2}}$\,\,$max
\caption{The contour plots for the absolute error in the ANN prediction for
$\chi(\boldsymbol x)$ in comparison to the finite element solution. The top and the bottom rows give  low- and high-frequency Fourier features with a single and the first 10 integer multiples of the reciprocal base vector, respectively.}
\label{F:sq_volume_fractions_abq_errors}
\end{figure*}

\begin{figure*}[htb!]
\centering
{{\includegraphics[width=0.45\textwidth,
trim=0 100 0 80, clip]{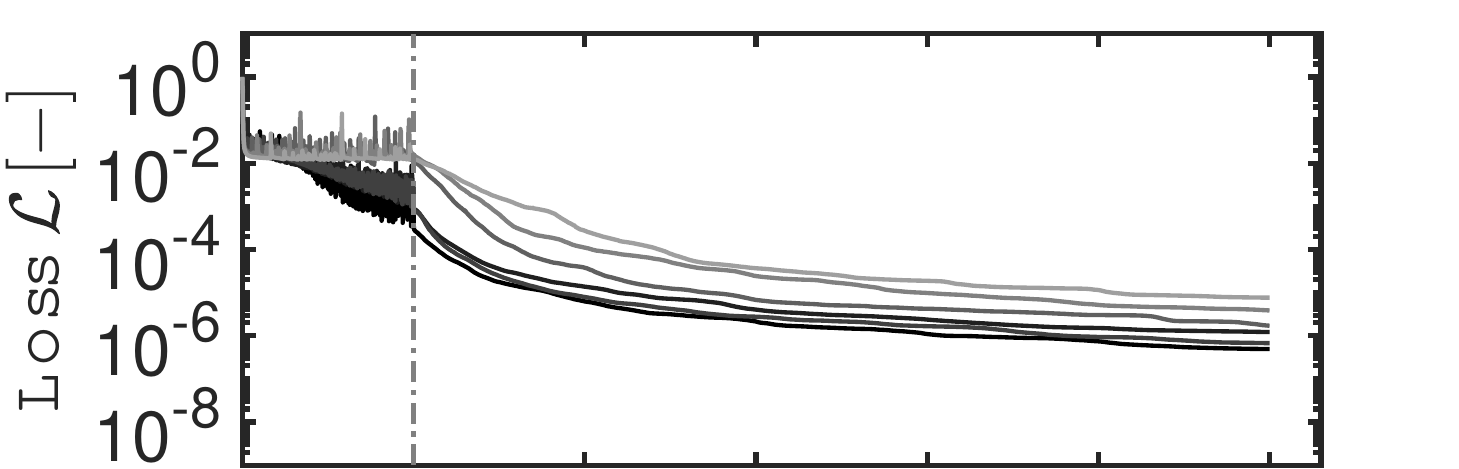}}}
{{\includegraphics[width=0.45\textwidth,
trim=0 100 0 80, clip]{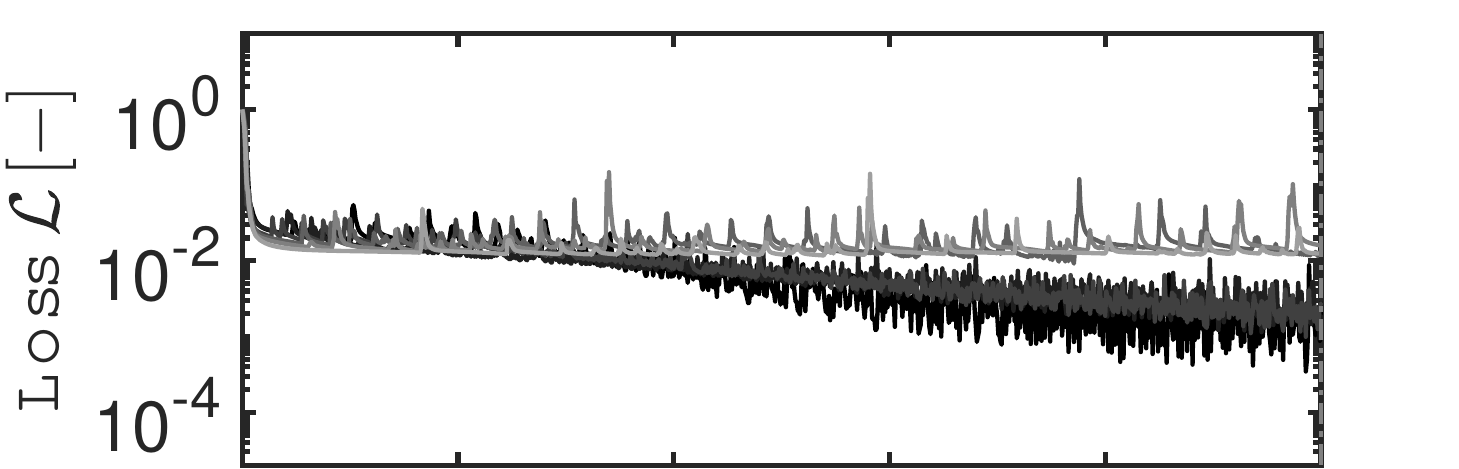}}}\\
{{\includegraphics[width=0.45\textwidth,
trim=0 40 0 80, clip]{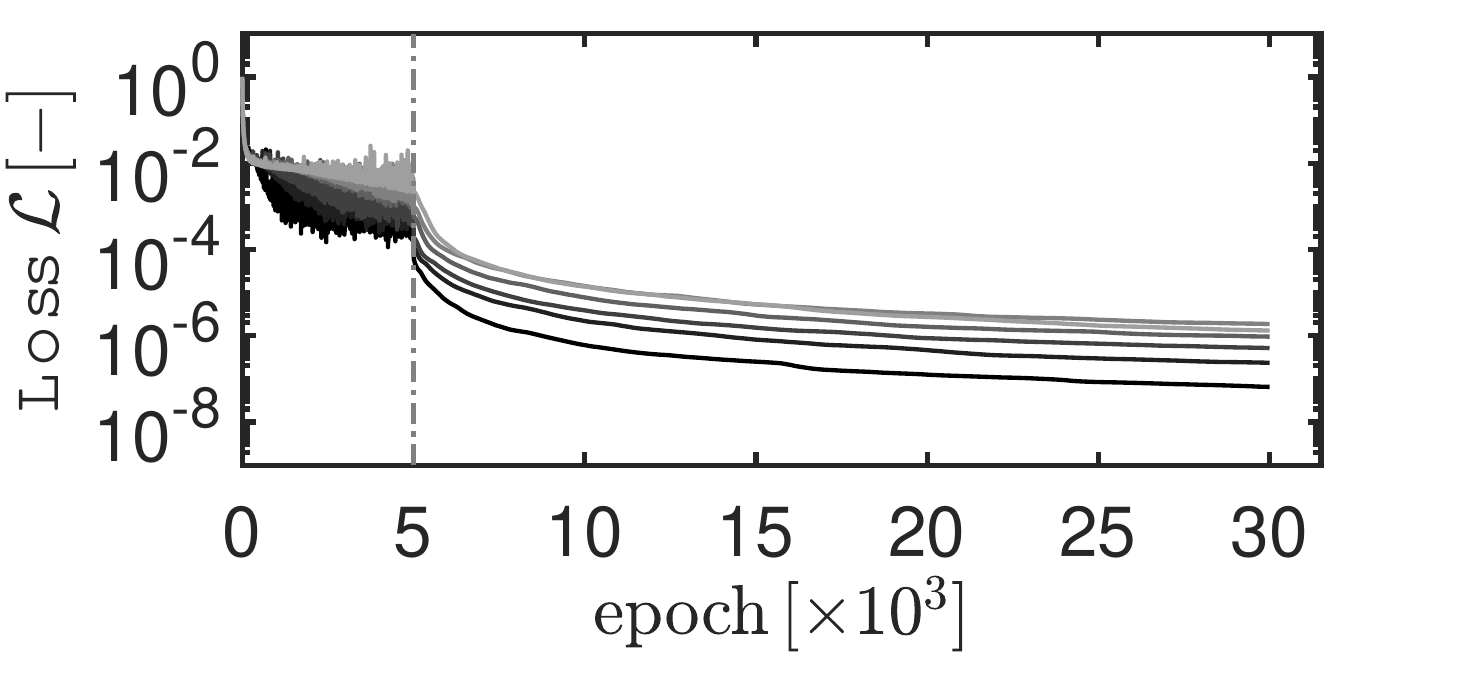}}}
{{\includegraphics[width=0.45\textwidth,
trim=0 40 0 80, clip]{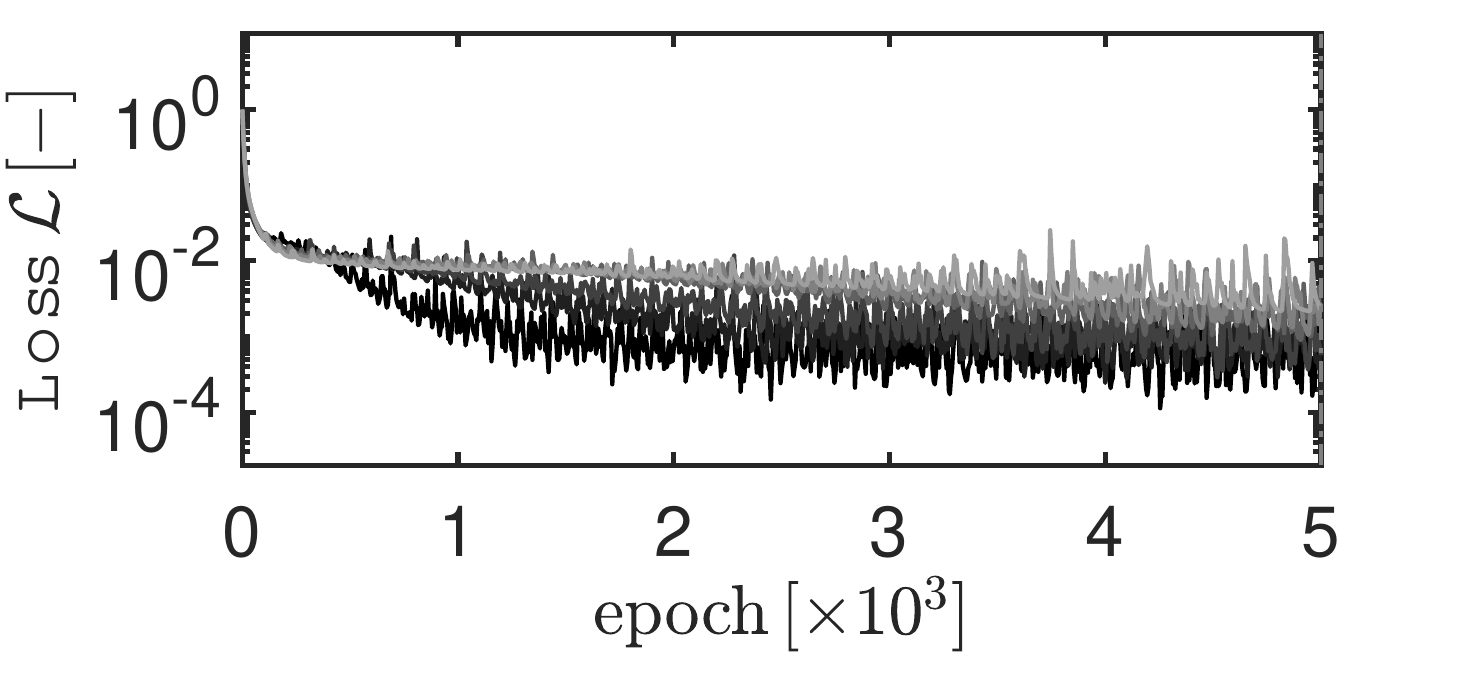}}}\\
\vspace{-10pt}
\caption{Training history of the normalized loss function for the ANN solutions. The first 5000 epochs (marked with a vertical dashed line) belong to adam optimizer, whereas the rest to L-BFGS. From dark to light colors, the volume fraction changes from 0.10 to 0.60. The top and the bottom rows give  low- and high-frequency Fourier features with a single and the first 10 integer multiples of the reciprocal base vector, respectively.}
\label{F:sq_results_training_histories}
\end{figure*}

Fig.\ \ref{F:sq_volume_fractions_square_fields} depicts the contour plots for the property field $a(\boldsymbol x)$, the source term $\partial a(\boldsymbol x)/\partial x_1$, and the ANN solution fields $\chi(\boldsymbol x)$  and $|\boldsymbol{\nabla}\chi|$ at the end of 30000 epochs for high-frequency Fourier features. The results are compared to high-resolution finite element simulations conducted with ABAQUS; see Fig.\ \ref{F:sq_volume_fractions_abq_errors}. The finite element model includes 260000 linear quadrilateral elements and 260401 nodes. A local element refinement is applied at the diffuse interface zone. As in the case of 1D, the finite element solutions are assumed to correspond to the ground truth. It is observed that
high-frequency Fourier features with the first 10 integer multiples of the reciprocal base vector lead to a better agreement with the ground truth. The training histories for the normalized loss function given in Fig.\ \ref{F:sq_results_training_histories} support this observation.

\begin{figure*}[htb!]
\centering
\subfigure[low-frequency]{
{{\includegraphics[height=0.22\textwidth,
trim=0 40 0 80, clip]{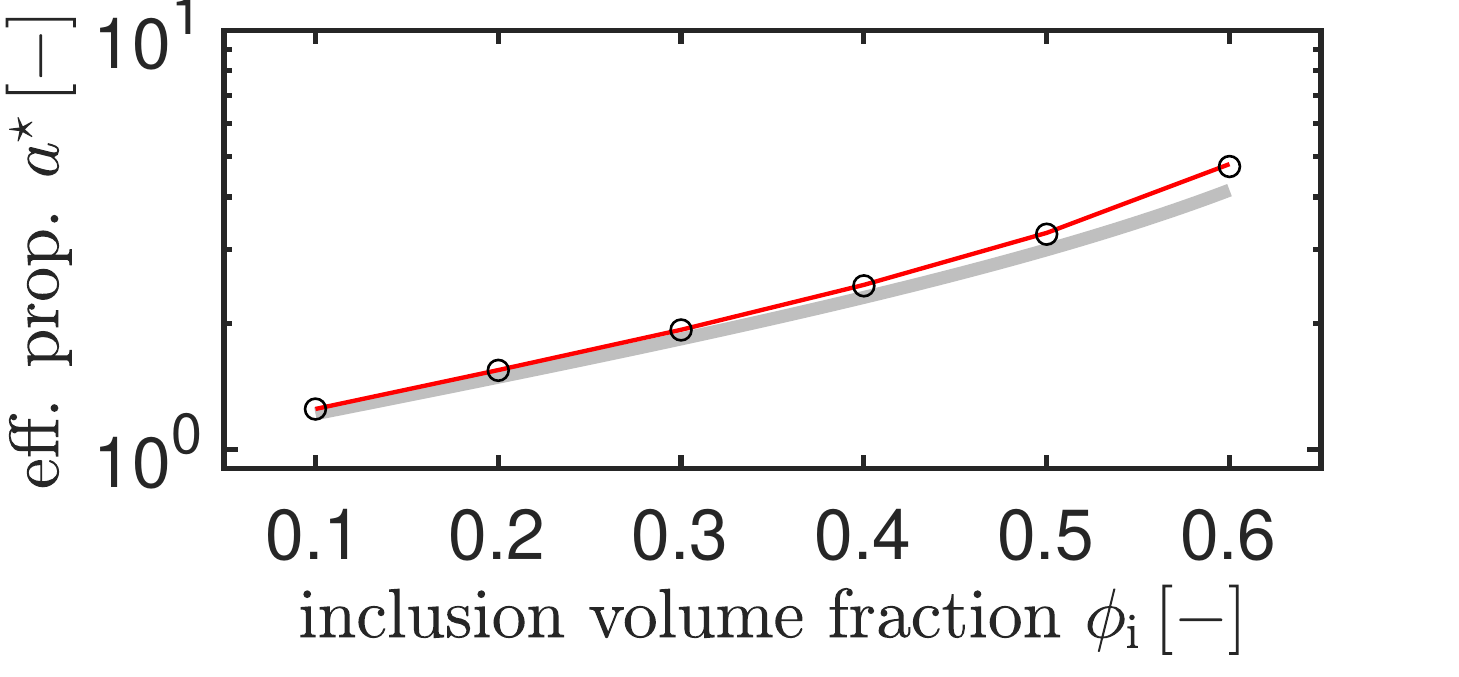}}}}
\subfigure[high-frequency]{
{{\includegraphics[height=0.22\textwidth,
trim=0 40 0 80, clip]{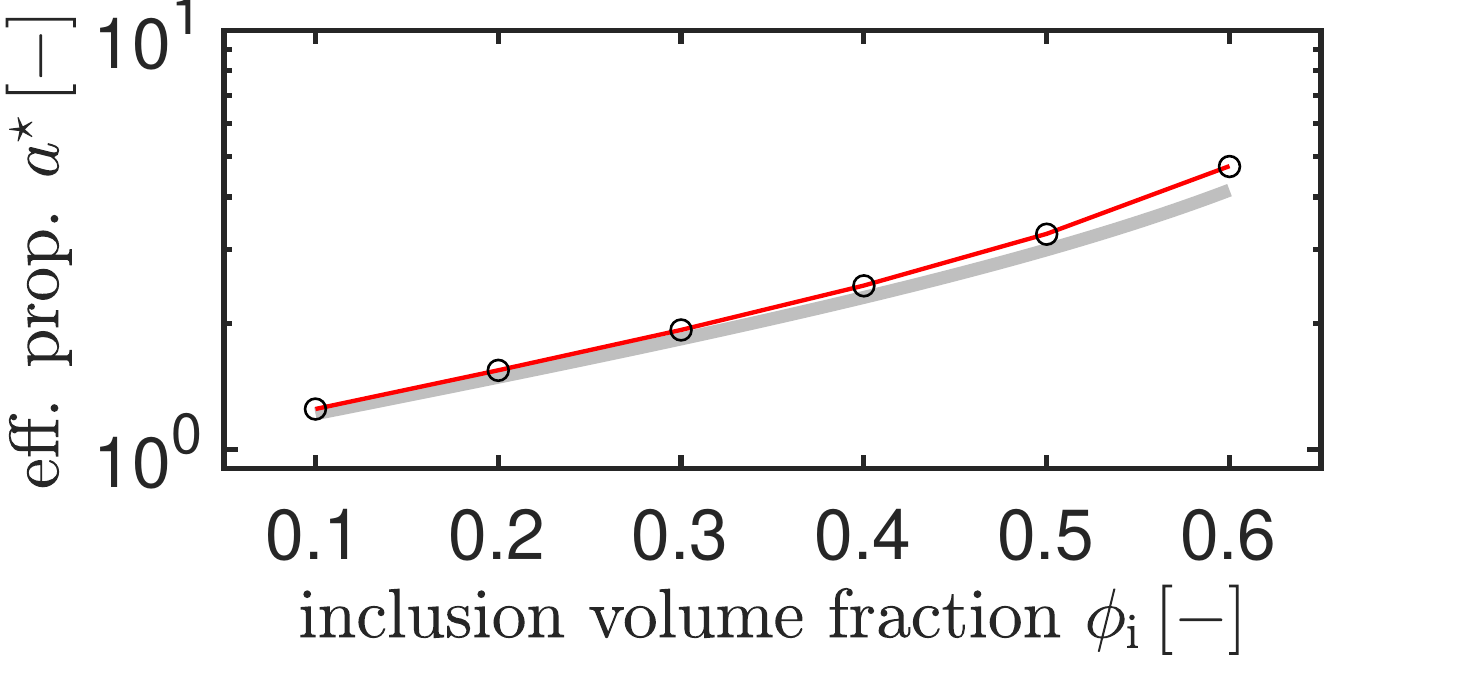}}}}
\vspace{-10pt}
\caption{ANN prediction for the effective properties $a^\star$ and comparison with ABAQUS solution as well as the truncated series expansion solution of Godin\ \cite{Godin2013} given in \cite{Ren2018} for sharp interface conditions; see \ref{section:2D_analytics}. The left and the right rows give low- and high-frequency Fourier features with a single and the first 10 integer multiples of the reciprocal base vector, respectively.}
\label{F:eff_perm_square}
\end{figure*}

Fig.\ \ref{F:eff_perm_square} shows the ANN predictions for the effective property $a^\star_{11}$ and its comparison to the ABAQUS solution as well as the analytical truncated series expansion solution of Godin\ \cite{Godin2013} cited in \cite{Ren2018}; see \ref{section:2D_analytics}. The ANN predictions agree excellently with the finite element simulation results for both low- and high-frequency Fourier feature solutions. The discrepancy between the analytical solutions is, as before, the sharp interface assumption considered in the latter.

\begin{figure*}[htb!]
\centering
\subfigure[$a(\boldsymbol x)$]{
{{\includegraphics[height=0.22\textwidth,
trim=489 233 489 233, clip]{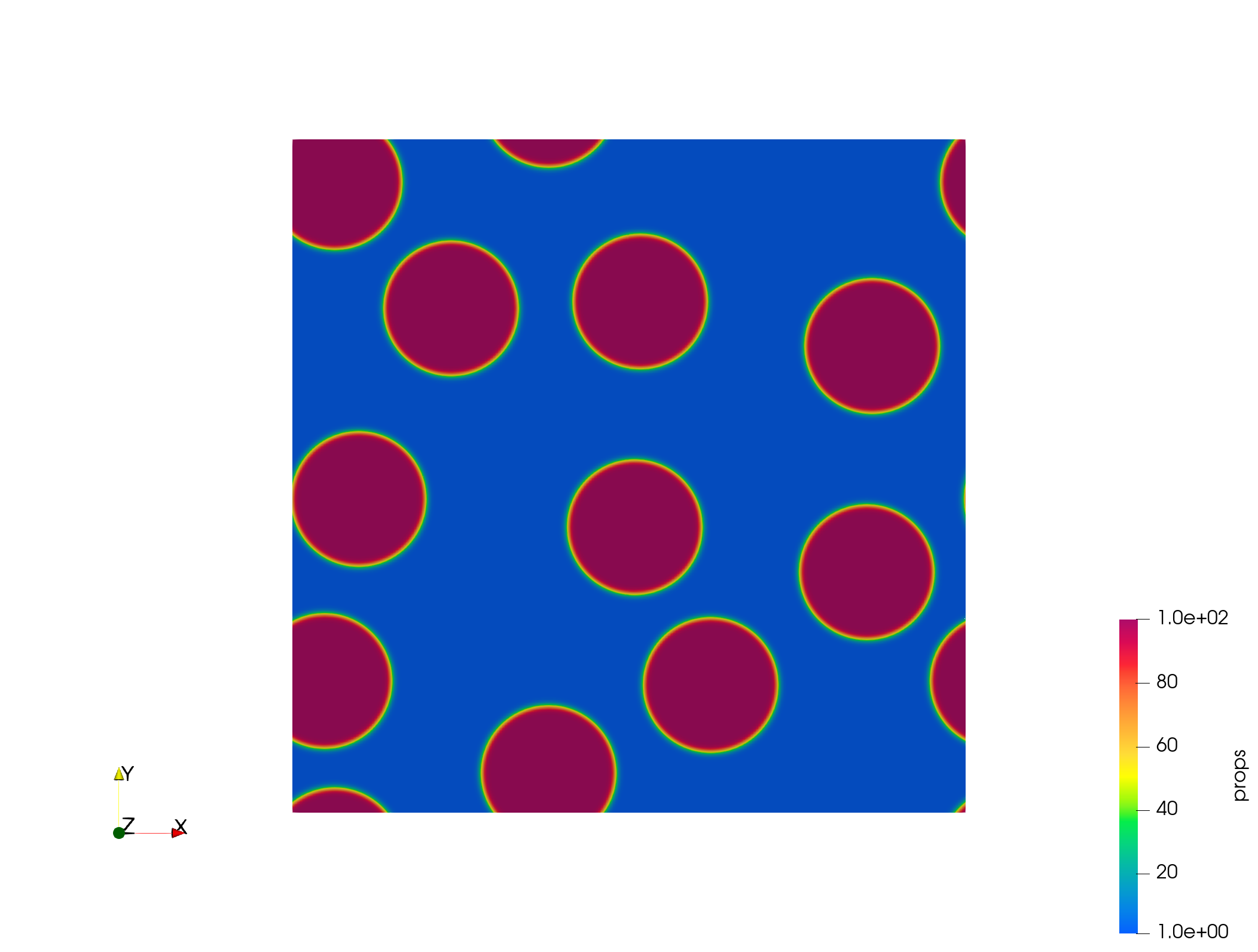}}}
}\hspace{0.05cm}
\subfigure[$f^1=\partial a(\boldsymbol x)/\partial x_1$]{
{{\includegraphics[height=0.22\textwidth,
trim=489 233 489 233, clip]{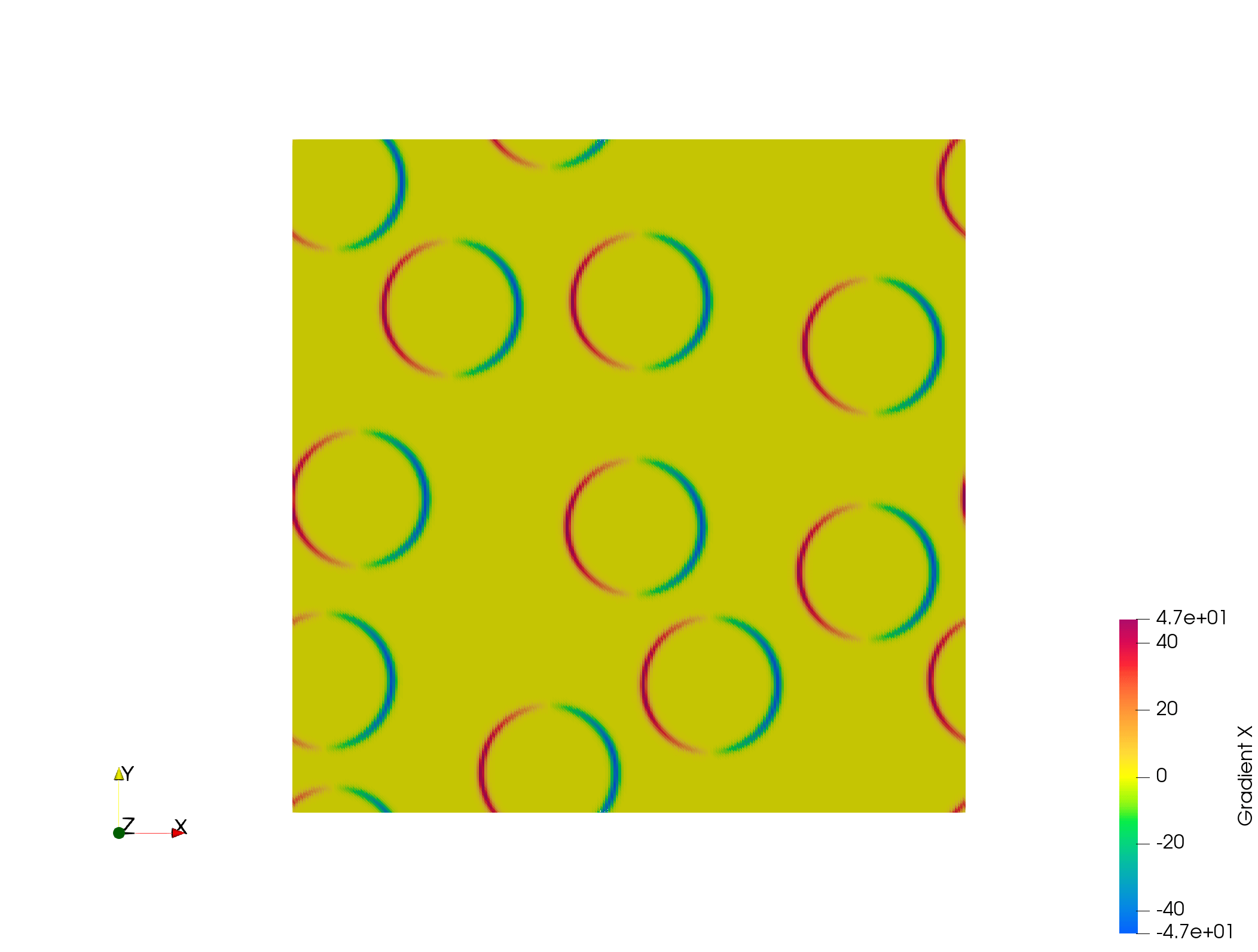}}}
}\hspace{0.05cm}
\subfigure[$f^2=\partial a(\boldsymbol x)/\partial x_2$]{
{{\includegraphics[height=0.22\textwidth,
trim=489 233 489 233, clip]{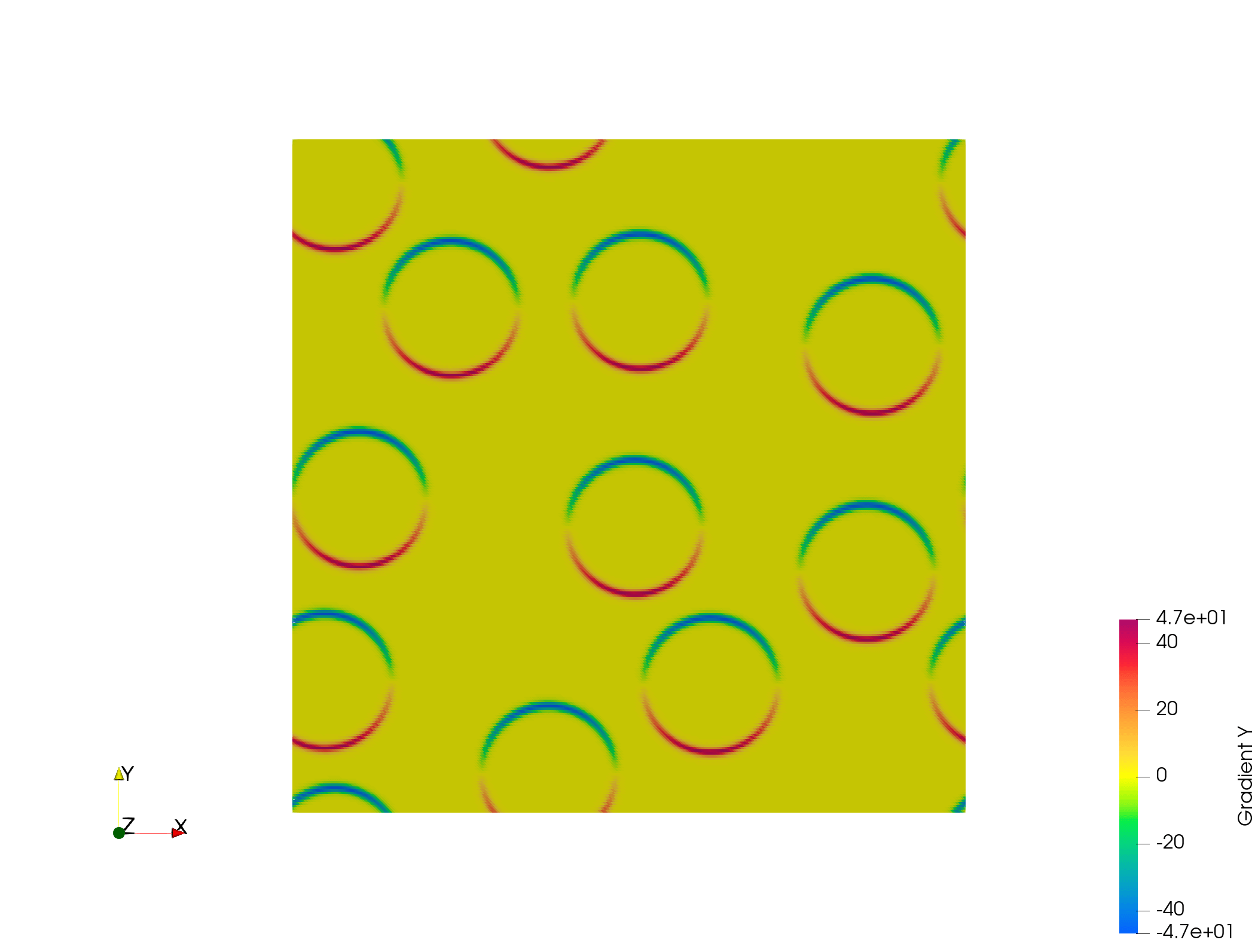}}}
}\\
min$\,\,$\frame{\includegraphics[width=0.30\textwidth, trim=0 15 0 15, clip]{legend_2}}$\,\,$max
\caption{The contour plots for the property (a) and source terms as property derivatives along $x-$ and $y-$directions for the square periodic unit cell with random disk distribution. The intervals [min, max] of the  contour plots are (a) $[1,100]$, (b) and (c) $[-49.5,49.5]$.}
\label{F:structure_2D_random}
\end{figure*}

We continue our studies on the 2D square lattice with periodic and stochastic disk distributions resulting in an inclusion fraction slightly above $\phi_i=0.30$. The phase distribution and the source terms are depicted in see Fig.\ \ref{F:structure_2D_random}. Since a random structure with a relatively small volume element is considered, an anisotropic property is anticipated.  Thus, solutions to both source terms are considered, yielding the complete property tensor components. 65536 collocation points with 50$\times$3 and 100$\times$6 ANN architectures are used in training. In both cases, high-frequency Fourier features with the first 10 integer multiples of the reciprocal base vector are considered.

\begin{figure*}[htb!]
\centering
\subfigure[ $\chi(\boldsymbol x)$ for $f^1$]{
{{\includegraphics[height=0.22\textwidth,
trim=489 233 489 233, clip]{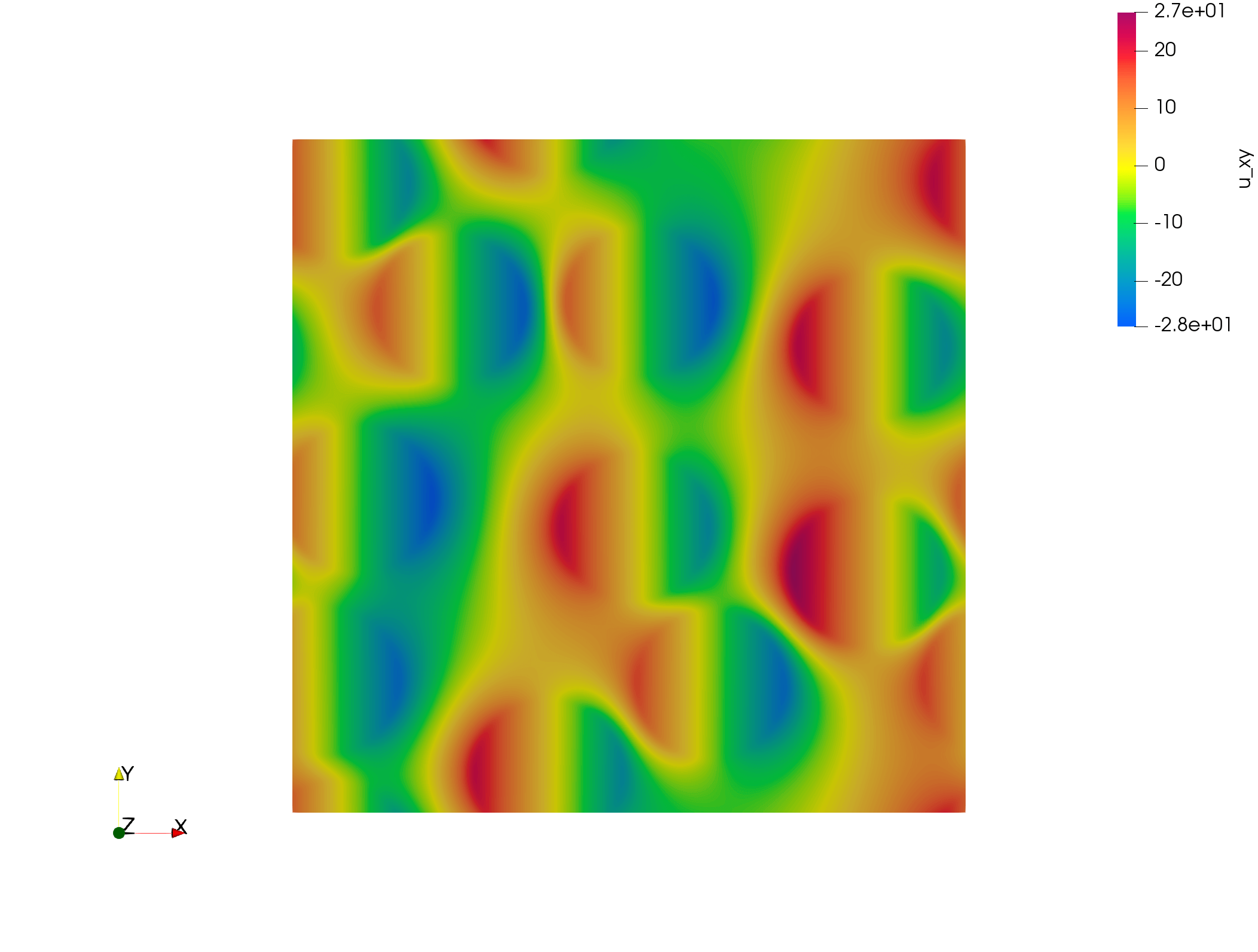}}}
}\hspace{0.05cm}
\subfigure[$|\boldsymbol{\nabla}\chi(\boldsymbol x)|$ for $f^1$]{
{{\includegraphics[height=0.22\textwidth,
trim=489 233 489 233, clip]{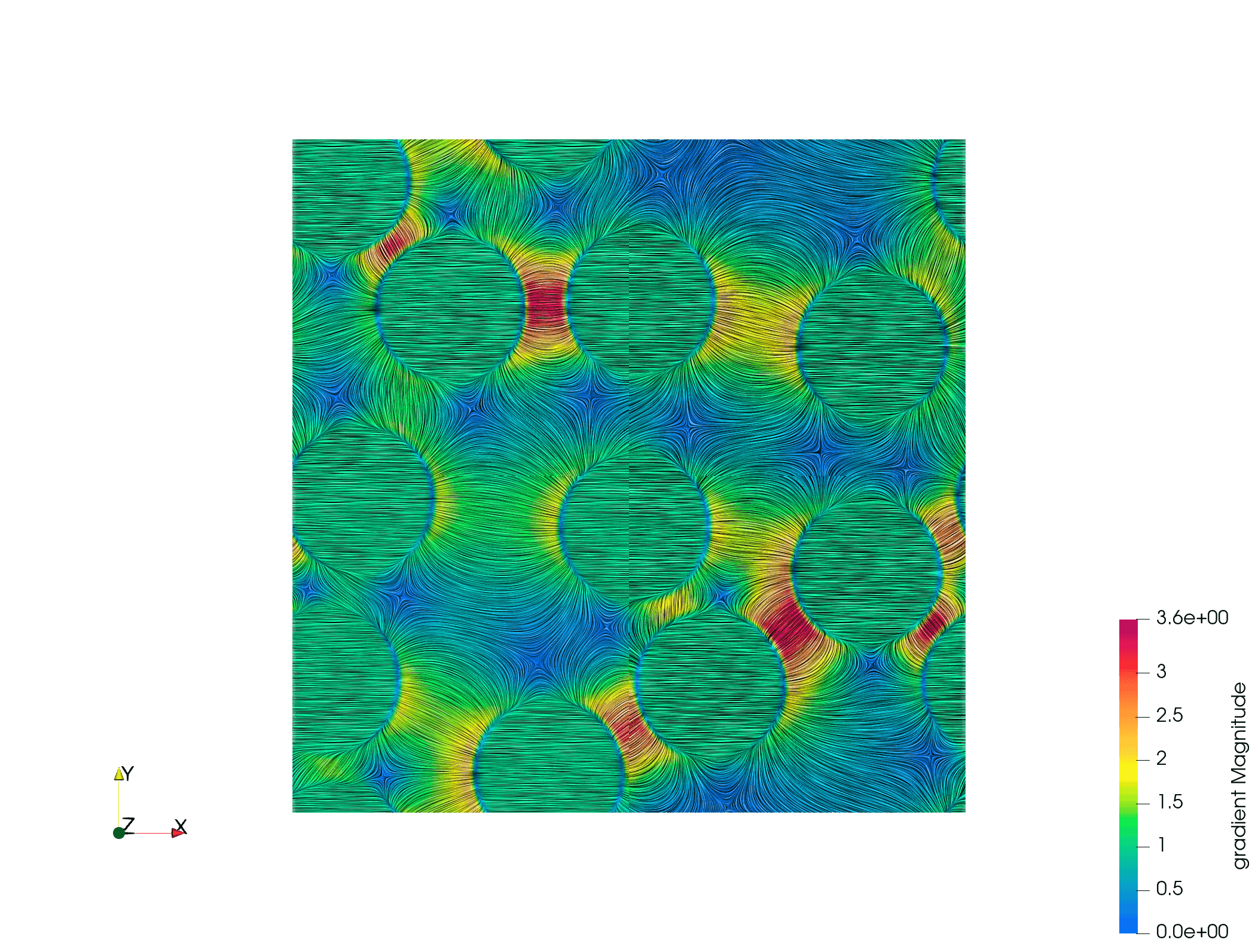}}}
}\hspace{0.25cm}
\subfigure[ $\chi(\boldsymbol x)$ for $f^2$]{
{{\includegraphics[height=0.22\textwidth,
trim=489 233 489 233, clip]{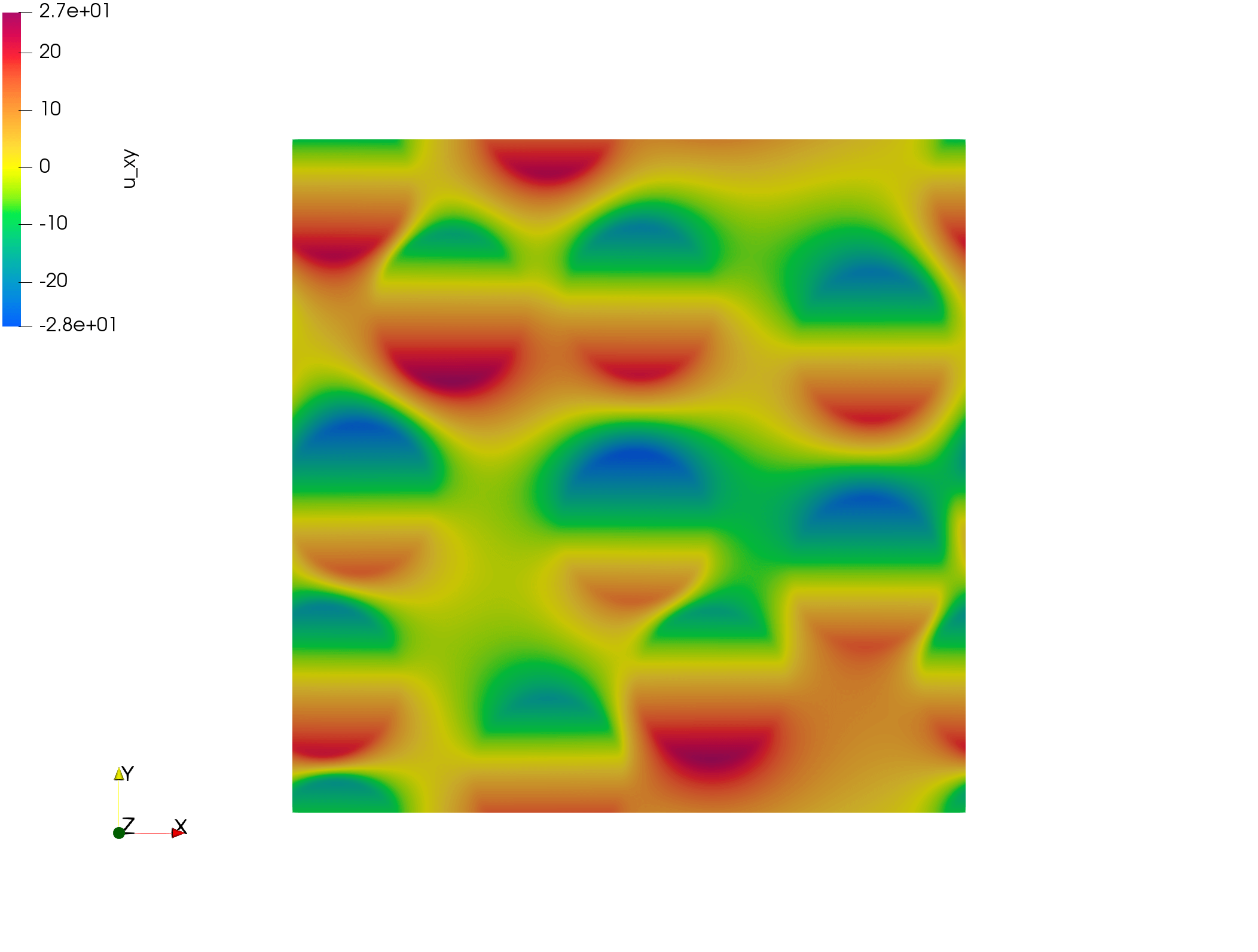}}}
}\hspace{0.05cm}
\subfigure[$|\boldsymbol{\nabla}\chi(\boldsymbol x)|$ for $f^2$]{
{{\includegraphics[height=0.22\textwidth,
trim=489 233 489 233, clip]{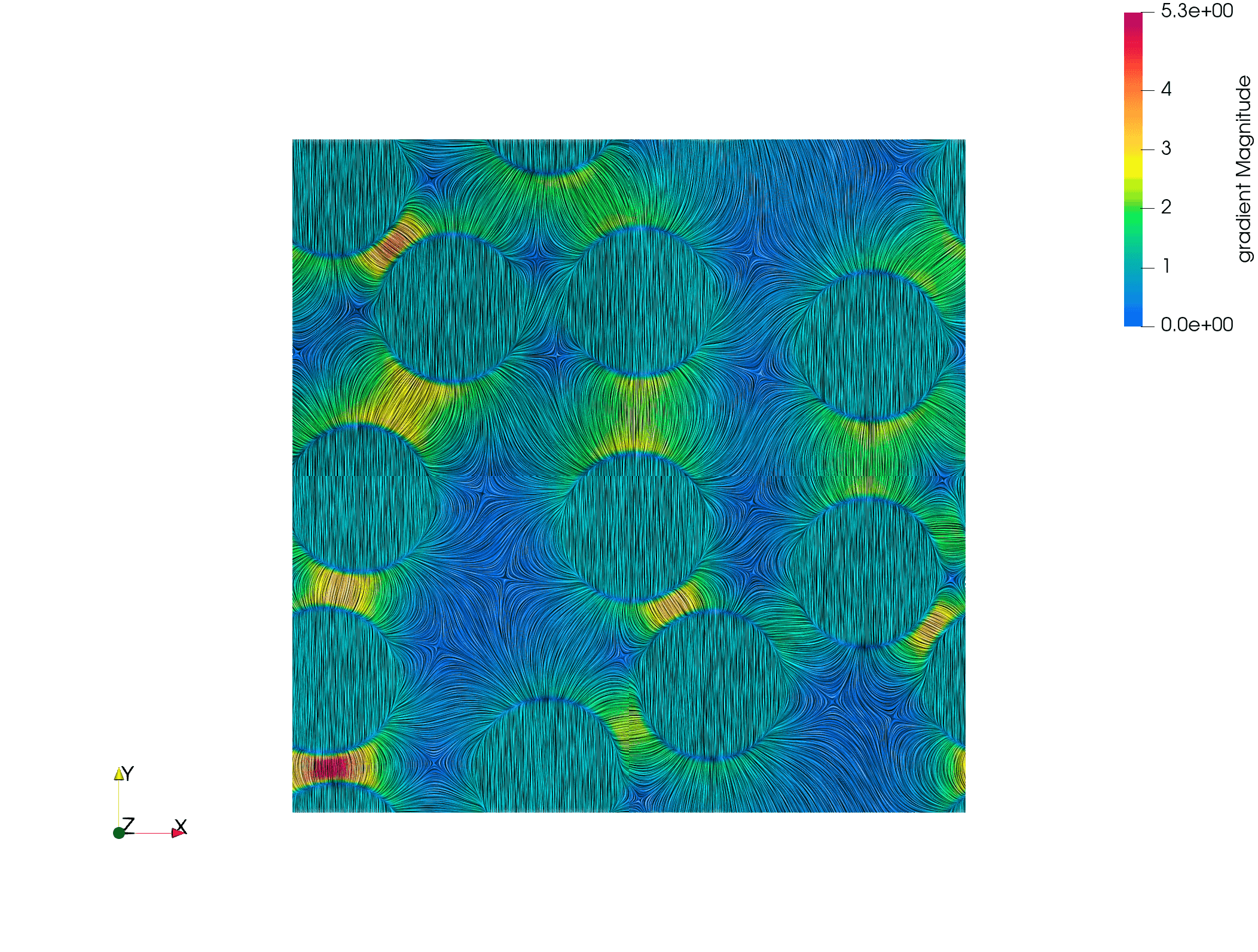}}}
}\\
min$\,\,$\frame{\includegraphics[width=0.30\textwidth, trim=0 15 0 15, clip]{legend_2}}$\,\,$max
\caption{The contour plots for the ANN solution $\chi(\boldsymbol x)$ for (a) and (c)
and the flux norm  $|\boldsymbol{\nabla}\chi(\boldsymbol x)|$ for (b) and (d) fields of the cell-problem for the square periodic unit cell. The intervals [min, max] of the  contour plots are (a) $[0,55]$, (b) $[0,3.6]$ (c) $[0,55]$,  and (d) $[0,5.3]$. Results are obtained using a 100$\times$6 architecture with high-frequency Fourier features with the first 10 integer multiples of the reciprocal base vector.}
\label{F:results_2D_random}
\end{figure*}

\begin{figure*}[htb!]
\centering
{{\includegraphics[width=0.45\textwidth,
trim=0 100 0 80, clip]{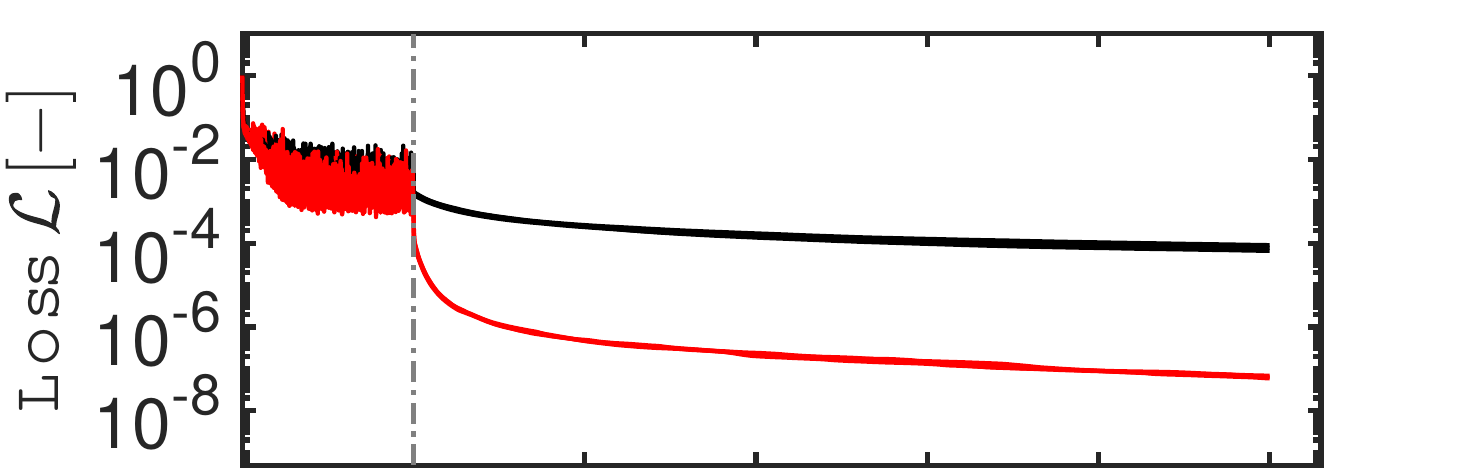}}}
{{\includegraphics[width=0.45\textwidth,
trim=0 100 0 80, clip]{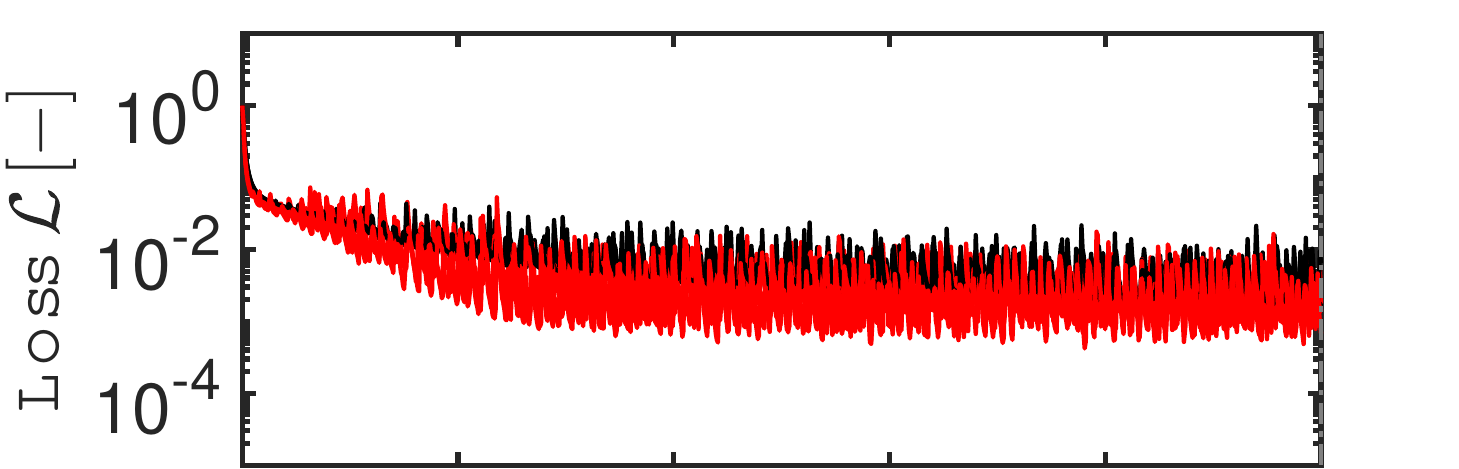}}}\\
{{\includegraphics[width=0.45\textwidth,
trim=0 40 0 80, clip]{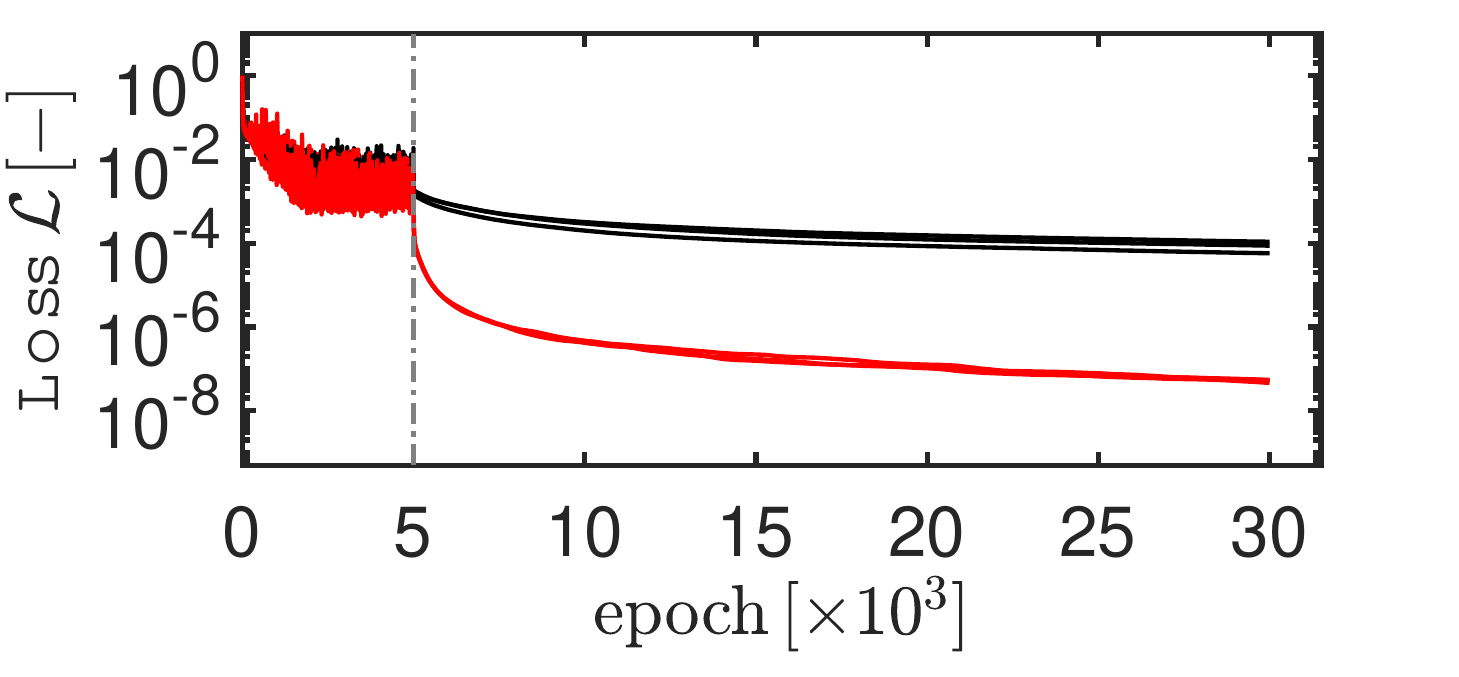}}}
{{\includegraphics[width=0.45\textwidth,
trim=0 40 0 80, clip]{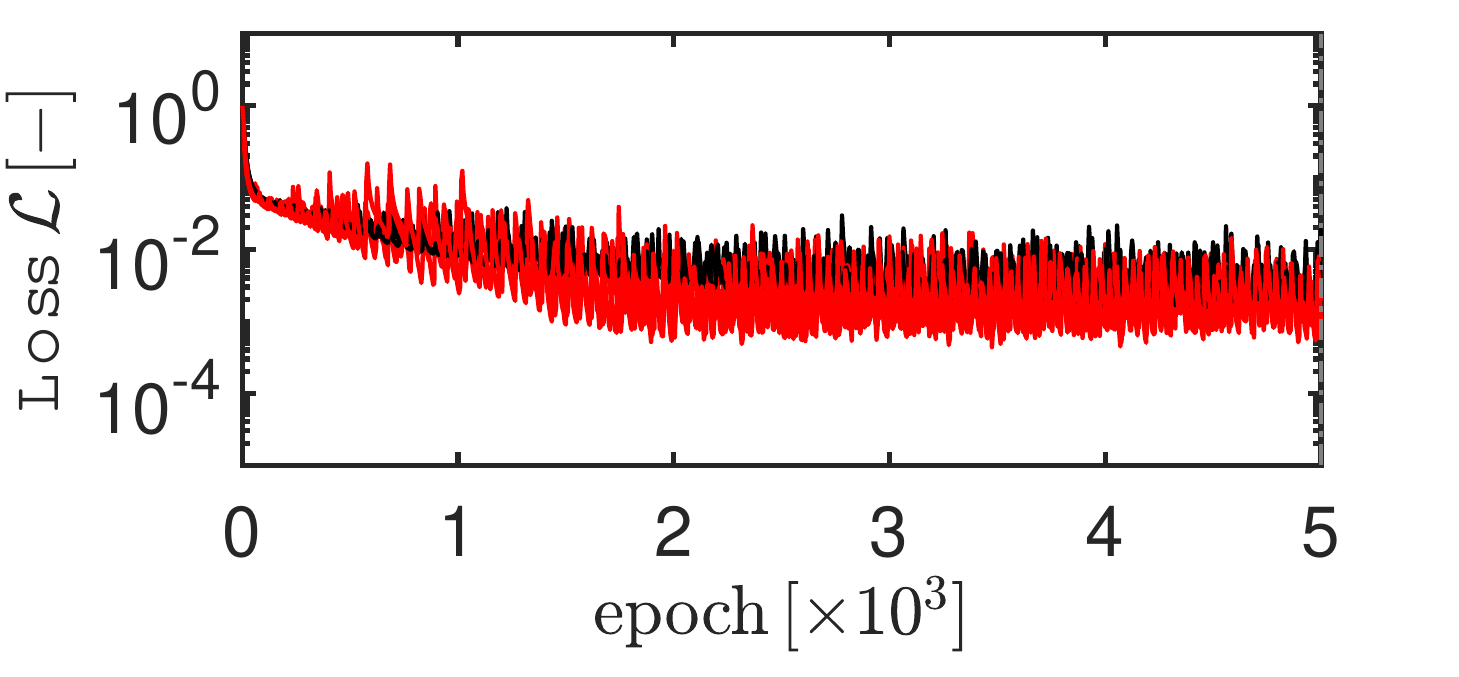}}}\\
\vspace{-10pt}
\caption{Training history of the normalized loss function for the ANN solutions with 50$\times$3 (black) and 100$\times$6 (red) architectures for the source terms $f^1$ and $f^2$ for three random trials each. The first 5000 epochs (marked with a vertical dashed line) belong to adam optimizer, whereas the rest to L-BFGS. High-frequency Fourier features with the first 10 integer multiples of the reciprocal base vector are used.}
\label{F:results_sq_random_training_histories}
\end{figure*}

Considering the stochastic nature of the ANN training process and the absence of ground truth, we reverted to repeated solutions. For each architecture, the ANN is trained three times. The computed effective property tensor components with a chosen basis that conforms to square dimensions are given in Table\ \ref{T:sq_rand_effectiveprops}1 in Appendix \ref{S:tabulated_results}. As also demonstrated by the training scaled loss history plots given in Fig.\ \ref{F:results_sq_random_training_histories}, 100$\times$6 ANN architecture gives results with higher accuracy in precision with the following representations in terms of mean and standard deviation  in four decimal points  for each component with $a_{11}=2.2969\pm0.0018$,
$a_{12}=-0.0261\pm0.0002$,
$a_{21}=-0.0296\pm0.0010$,
$a_{22}=2.4012\pm0.0022$.
Corresponding  solution fields for the applied source terms are given in Fig.\ \ref{F:results_2D_random}. In light of our computations, the following representation in two decimal points constitutes an accurate representation of the symmetric effective property tensor in matrix form,  revealing a small anisotropy.
\begin{equation}
\left[ \bs{a}^{\star}\right]
=\left[
\begin{array}{rr}
2.30  & -0.03 \\
-0.03  & 2.40 \\
\end{array}%
\right]\,.
\end{equation}

\subsubsection{Two-dimensional Centered Rectangular Lattice}
In this part, our applications focus on the centered rectangular lattice with $L_1/L_2=3$ where  $L_1=\rvert\boldsymbol c_1$ and $L_2=\rvert\boldsymbol c_2$. We consider the inclusion volume fraction of $\phi_\mathrm{i}=\{0.3\}$. The area of the primitive unit cell is arranged to be equal to the square unit cell considered in the previous example with $L_1\, L_2=L^2$. This provides identical radii of the circular inclusion for both cases for the same inclusion volume fraction. We investigate hexagonal, rhombic, and rectangular unit cells, which are referred to as $\mathcal{R}^\mathrm{c}_1$, $\mathcal{R}^\mathrm{c}_2$ and $\mathcal{R}^\mathrm{c}_3$, respectively; see Fig.\   \ref{fig:BravaisLattice1D2D3D}.
 The direct and the reciprocal unit cell vectors given for $\mathcal{R}^\mathrm{c}_1$ and $\mathcal{R}^\mathrm{c}_2$ are not orthogonal; see Table\ \ref{T:bravais_lattices_1D2D3D}. For the rectangular unit cell $\mathcal{R}^\mathrm{c}_3$, however, they are; with $\bs a_i=\bs c_i$ and $\bs b_i=2\pi\bs c_i/|\bs c_i|^2$ for $i=1,2$. In the ANN training, 65536 collocation points are used for $\mathcal{R}^\mathrm{c}_1$ and $\mathcal{R}^\mathrm{c}_2$ whereas for $\mathcal{R}^\mathrm{c}_3$ which has double the area, the number of collocation points are doubled.
Learning rates of 0.001 and 0.010 with 50$\times$3 and 100$\times$6 ANN architectures. Like before, considering the stochastic nature of the ANN training process and the absence of ground truth, we reverted to repeated solutions. For each architecture, the ANN is trained three times. For $\mathcal{R}^\mathrm{c}_1$ and $\mathcal{R}^\mathrm{c}_2$, higher accuracy was recorded with a learning rate of 0.001 with 100$\times$6 ANN architecture possessing low-frequency Fourier features.  For $\mathcal{R}^\mathrm{c}_3$, on the other hand,  a learning rate of 0.010 with 50$\times$3 ANN architecture possessing high-frequency Fourier features with the first 10 integer multiples of the reciprocal base vector performed best once the consequential training loss is concerned, see Fig.\  \ref{F:cr_results_training_histories}.

Fig.\ \ref{F:sq_volume_fractions_square_fields} depicts the contour plots for the property field $a(\boldsymbol x)$, the source terms $\partial a(\boldsymbol x)/\partial x_1$ and $\partial a(\boldsymbol x)/\partial x_2$, and the corresponding ANN solution fields $\chi(\boldsymbol x)$  and $|\boldsymbol{\nabla}\chi|$ at the end of 30000 epochs for the best-performing hyperparameter set for each case. The utilization of Fourier features in conjunction with reciprocal unit cells allows for the satisfaction of periodicity in non-rectangular problem domains. As observed from the contour plots, the choice of the selected unit cell does not change the solution.

The computed effective property tensor components are given in Table\ \ref{T:cr_results}2 in Appendix \ref{S:tabulated_results} for $\mathcal{R}^\mathrm{c}_1$,  $\mathcal{R}^\mathrm{c}_2$ and $\mathcal{R}^\mathrm{c}_3$. As demonstrated in Fig.\ \ref{F:results_sq_random_training_histories}, the training losses for all of the unit cells are  sufficiently small at the end of 30000 epochs.  Nevertheless, learning trends differ due to the difference in the selected ANN hyperparameters. An orthotropic effective property matrix, when using a chosen basis that conforms to rectangular dimensions, is expected, with non-zero diagonal terms and zero off-diagonal terms, resulting in $a^\star_{11} \neq a^\star_{22}$ and $a^\star_{12}=a^\star_{21}=0$. For $\mathcal{R}^\mathrm{c}_1$, a conversion of the tabulated values to mean and standard deviation in four decimal points for each component results in $a_{11}=1.7402\pm0.0042$,
$a_{12}=0.0018\pm0.0021$,
$a_{21}=0.0042\pm0.0016$,
$a_{22}=2.2820\pm0.0023$.
For $\mathcal{R}^\mathrm{c}_2$, a conversion of the tabulated values to mean and standard deviation for each component results in $a_{11}=1.7423\pm0.0025$,
$a_{12}=-0.0004\pm0.0012$,
$a_{21}=0.0010\pm0.0013$,
$a_{22}=2.2846\pm0.0026$.
Finally, for $\mathcal{R}^\mathrm{c}_3$, this gives $a_{11}=1.7378\pm0.0008$,
$a_{12}=0.0000\pm0.0000$,
$a_{21}=0.0004\pm0.0014$,
$a_{22}=2.2946\pm0.0023$.
These results show a good agreement between the solutions devising different unit cells. In light of our computations, the following representation in two decimal points emerges as a good approximation for the symmetric effective property tensor in matrix form, revealing anisotropy,
\begin{equation}
\left[ \bs{a}^{\star}\right]
=\left[
\begin{array}{ccc}
1.74  &  0.00 \\
0.00  & 2.29 \\
\end{array}%
\right]\,.
\end{equation}

\begin{figure*}[htb!]
\centering
{{\includegraphics[height=0.10\textwidth,
trim=350 375 350 375, clip]{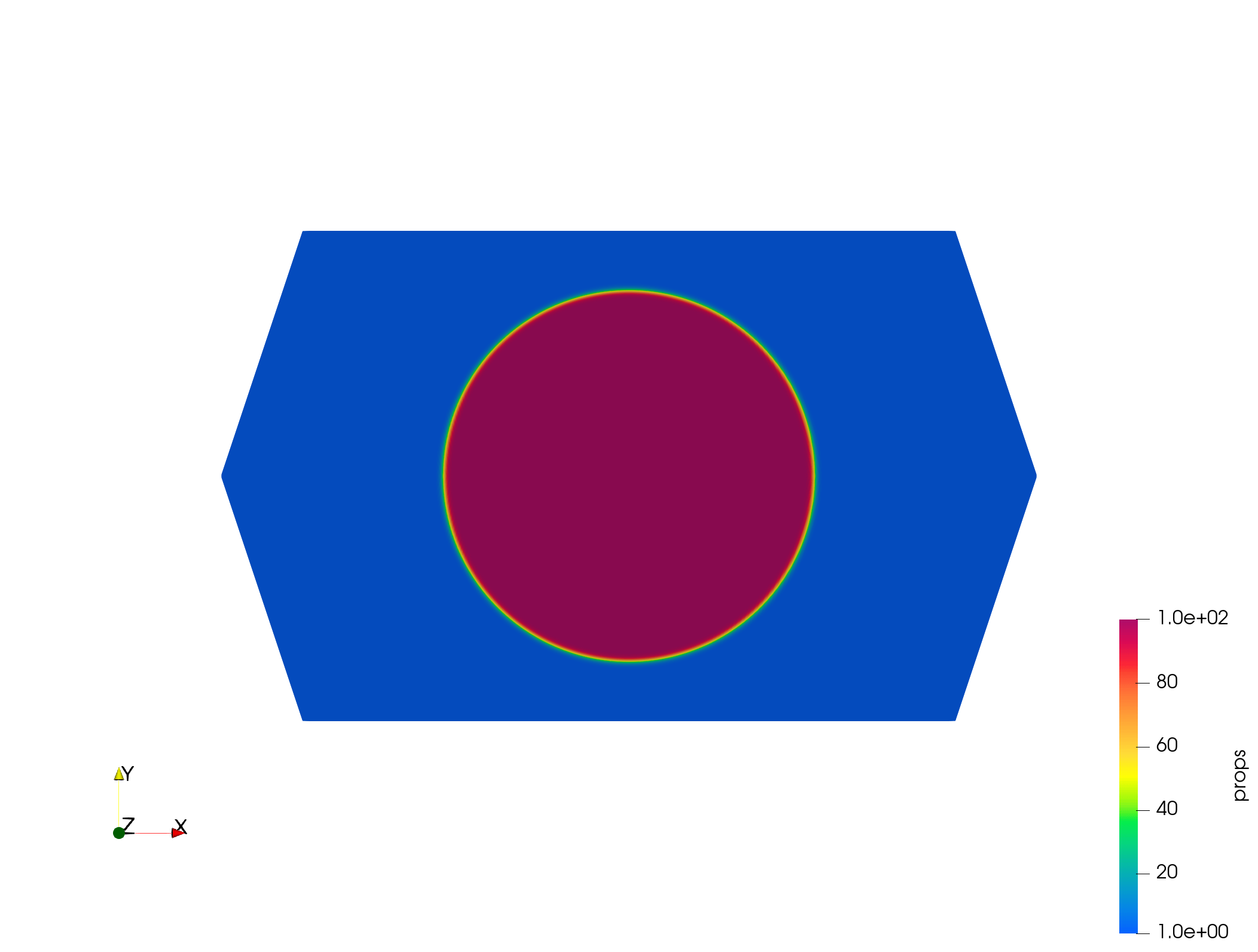}}}\hspace{0.5cm}
{{\includegraphics[height=0.10\textwidth,
trim=300 540 300 540, clip]{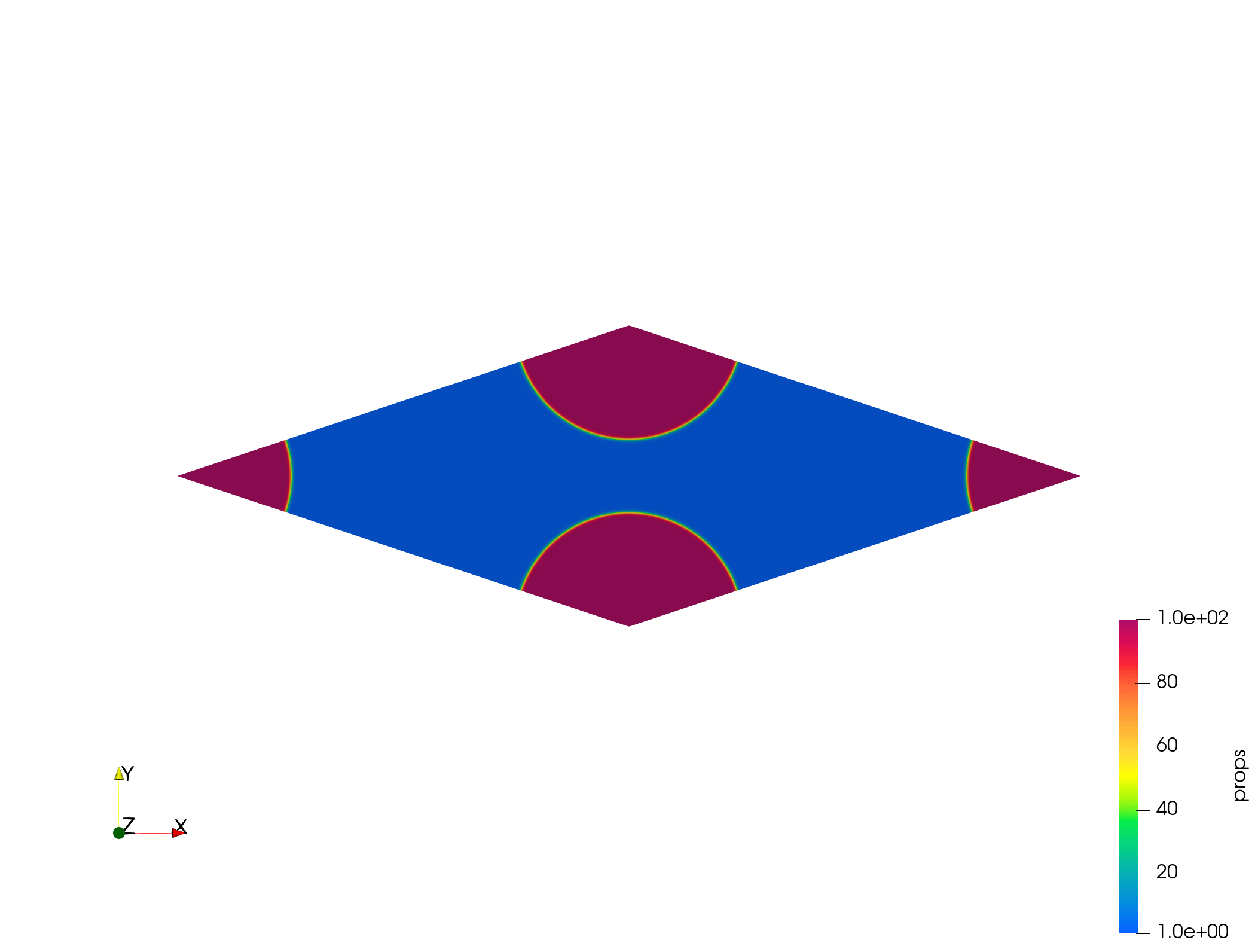}}}\hspace{0.5cm}
{{\includegraphics[height=0.10\textwidth,
trim=300 540 300 540, clip]{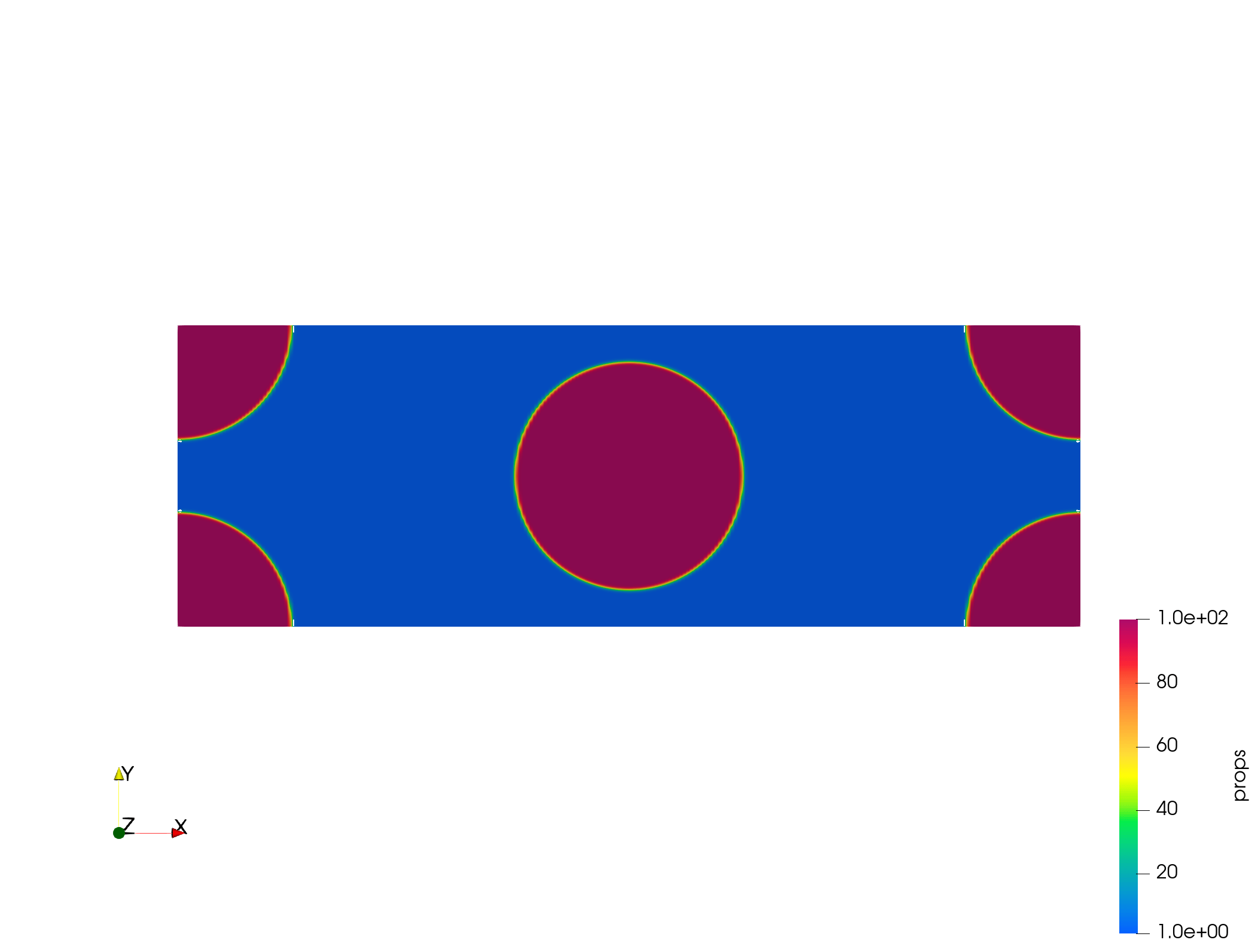}}}\\\vspace{0.75cm}
{{\includegraphics[height=0.10\textwidth,
trim=350 375 350 375, clip]{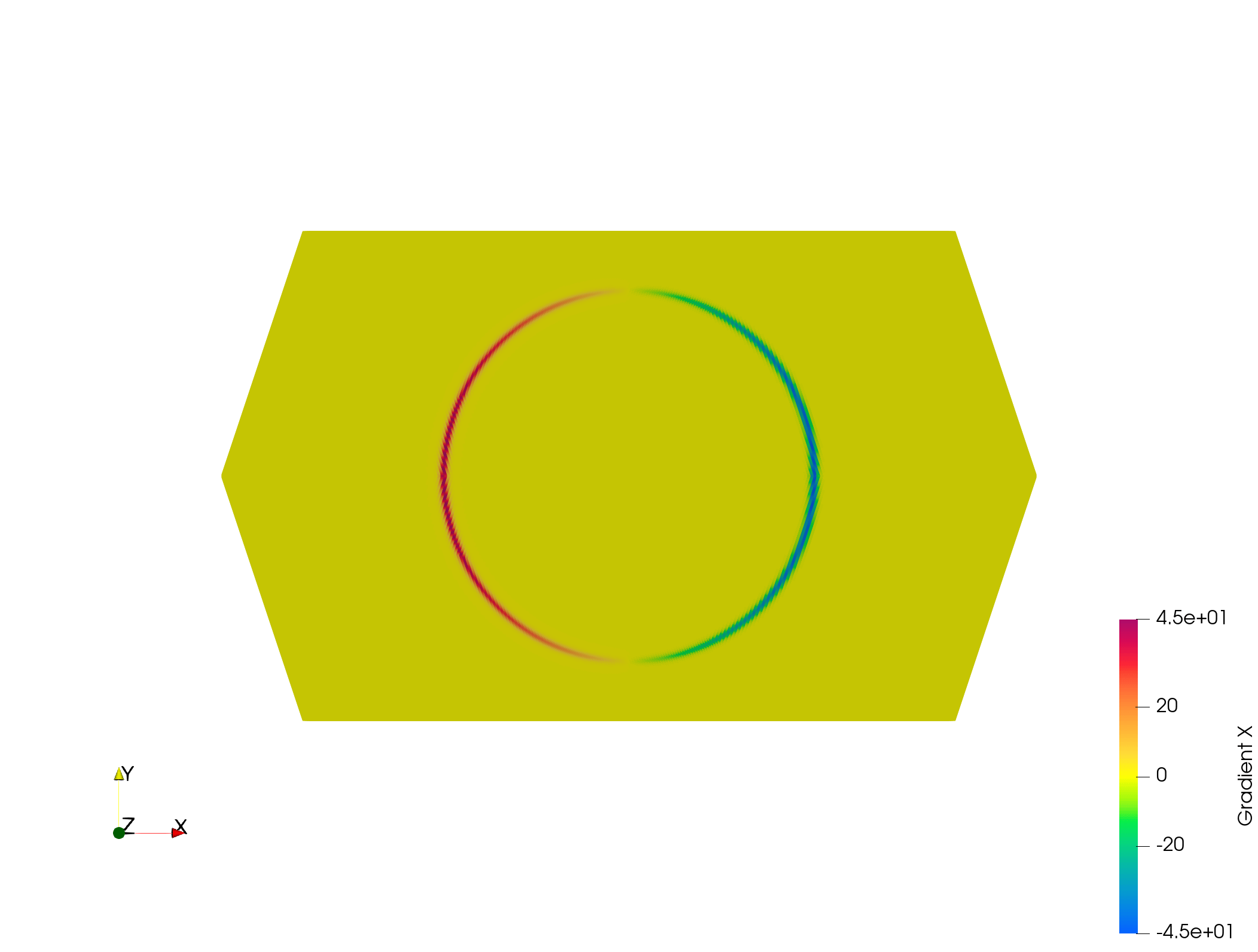}}}\hspace{0.5cm}
{{\includegraphics[height=0.10\textwidth,
trim=300 540 300 540, clip]{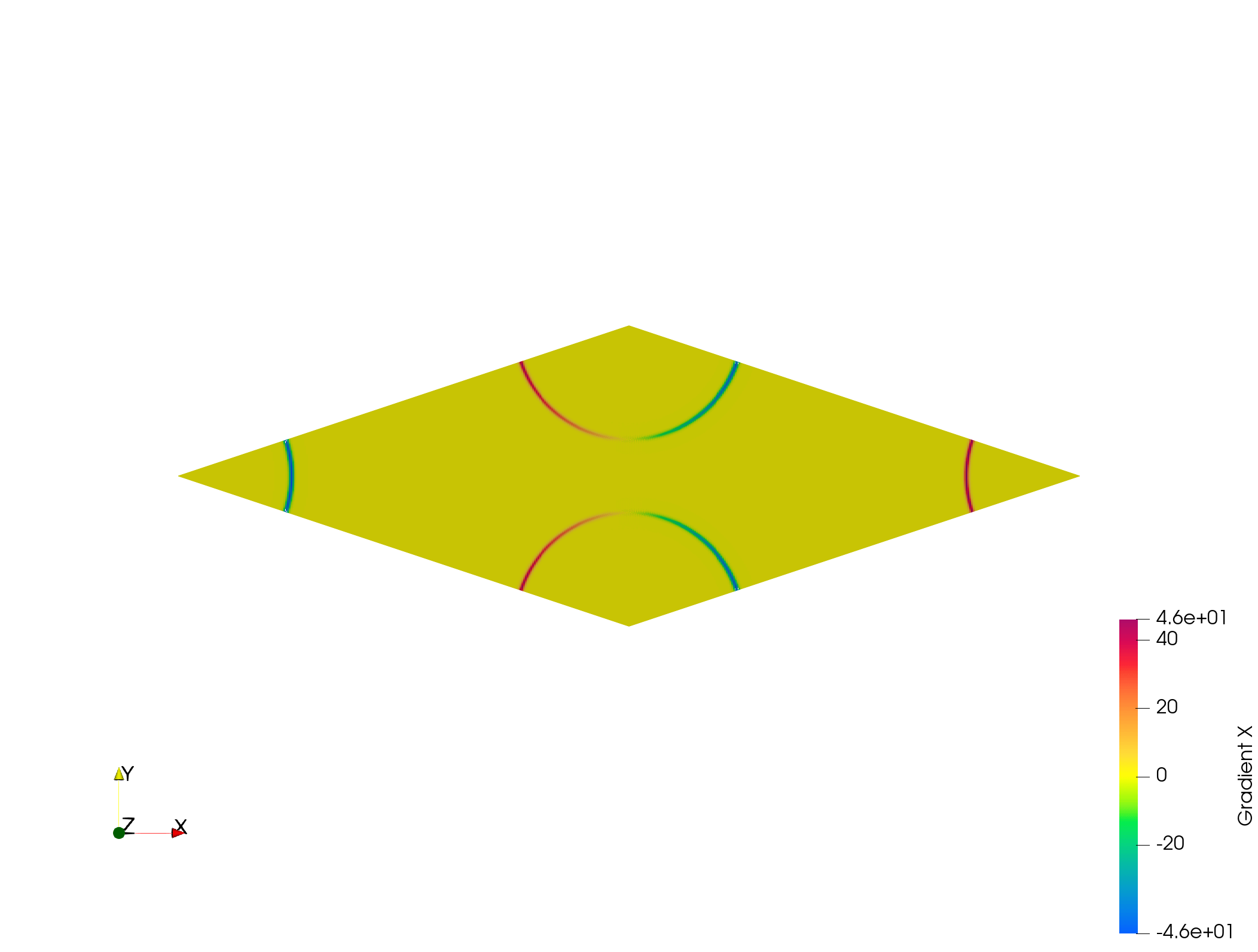}}}\hspace{0.5cm}
{{\includegraphics[height=0.10\textwidth,
trim=300 540 300 540, clip]{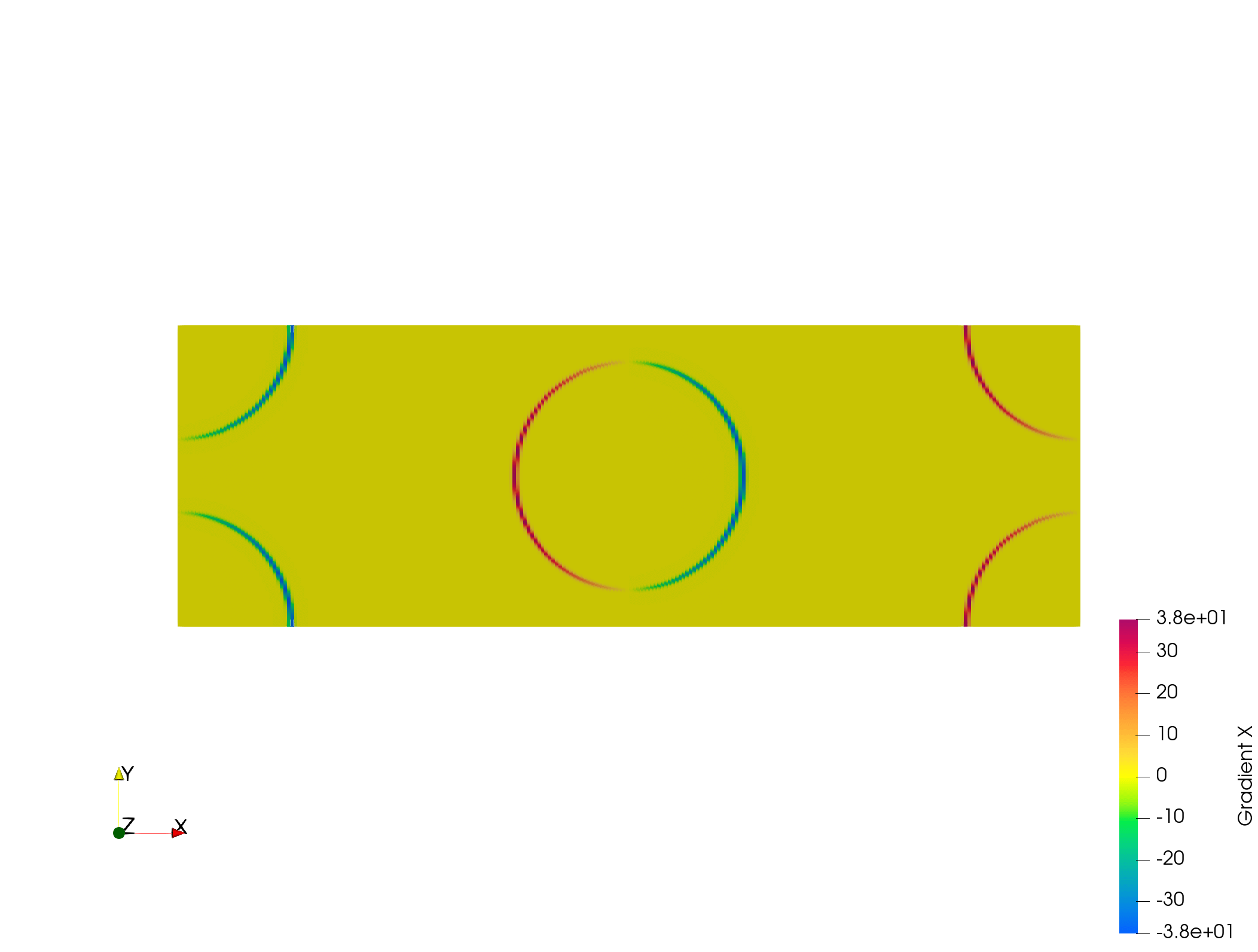}}}\\\vspace{0.1cm}
{{\includegraphics[height=0.10\textwidth,
trim=350 375 350 375, clip]{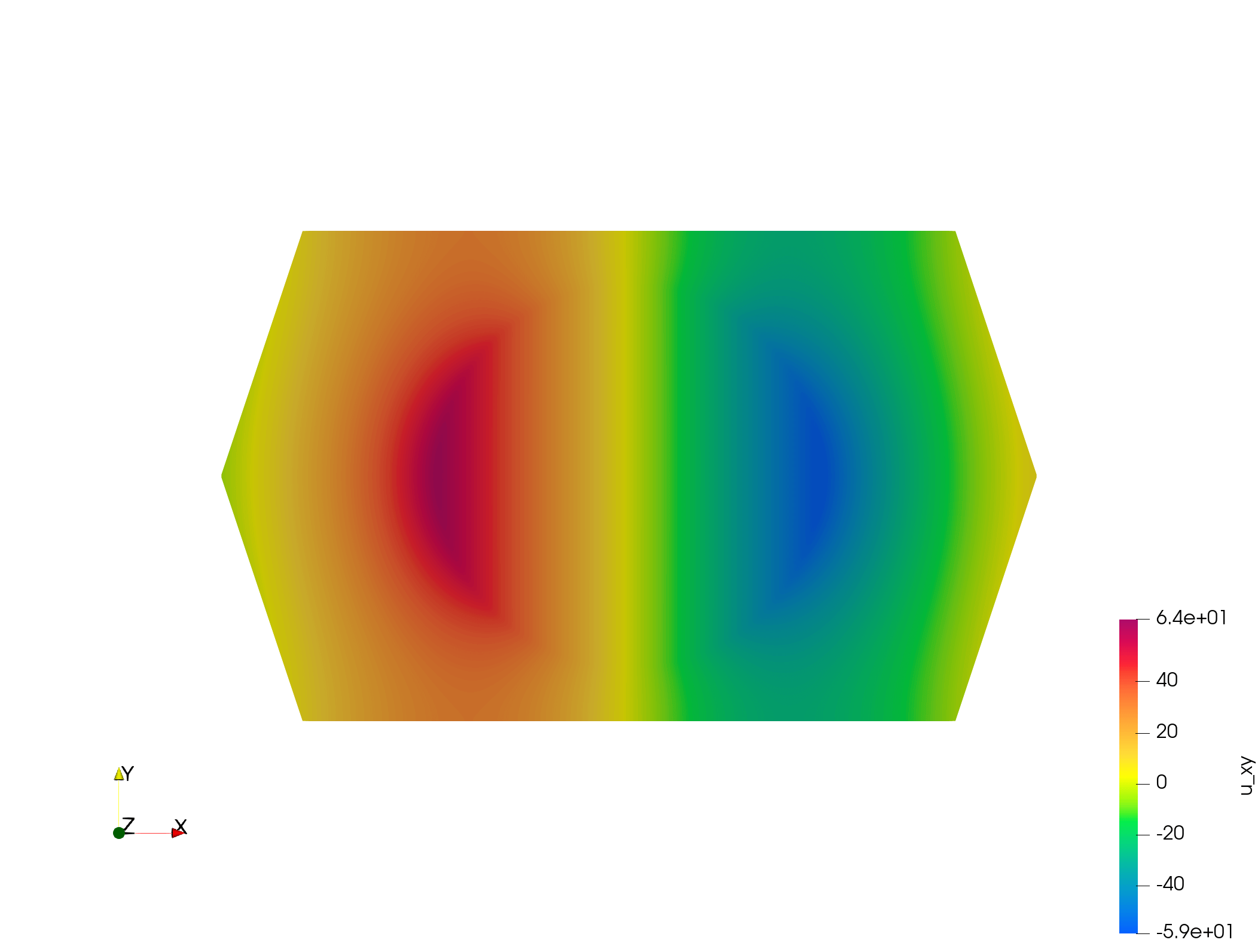}}}\hspace{0.5cm}
{{\includegraphics[height=0.10\textwidth,
trim=300 540 300 540, clip]{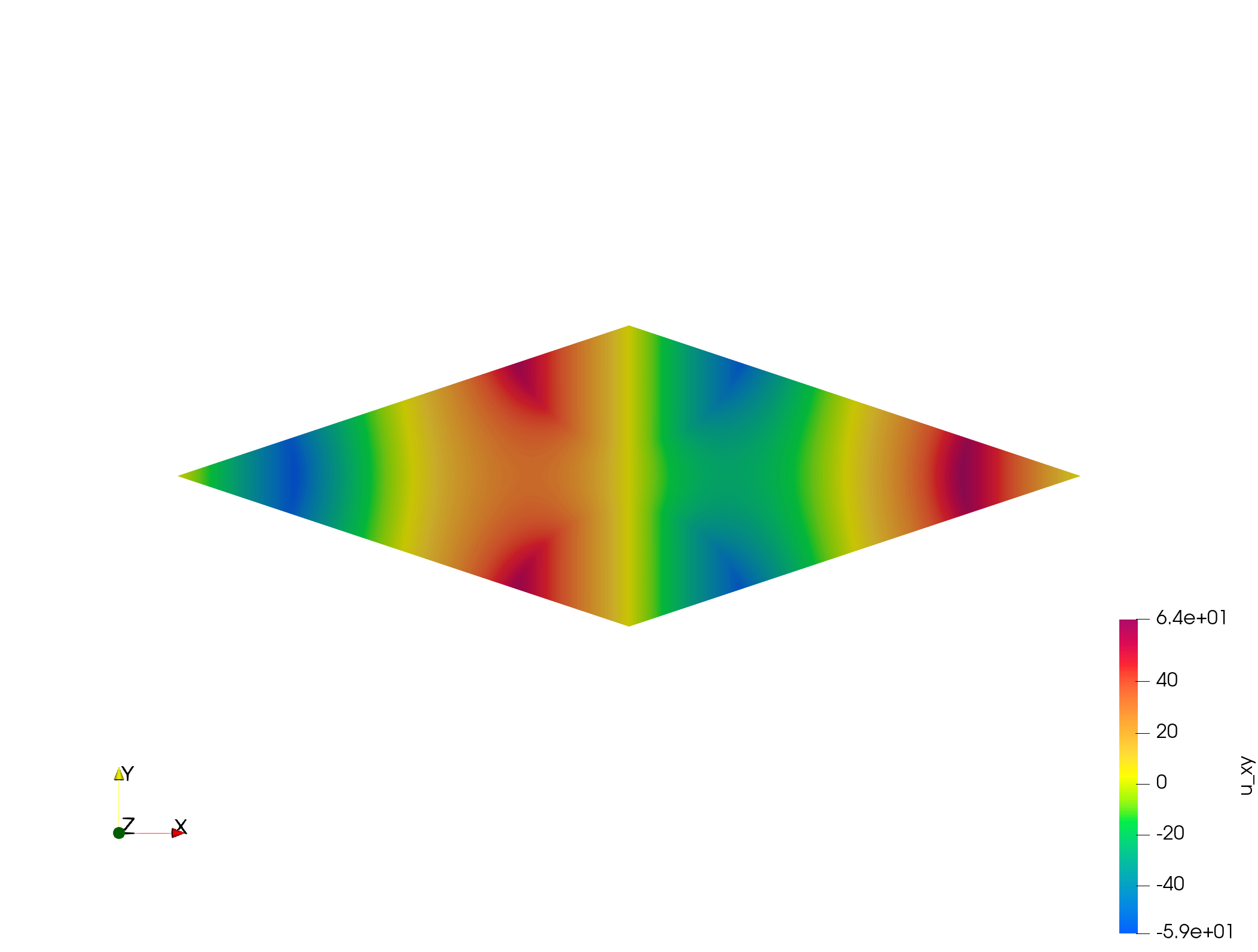}}}\hspace{0.5cm}
{{\includegraphics[height=0.10\textwidth,
trim=300 540 300 540, clip]{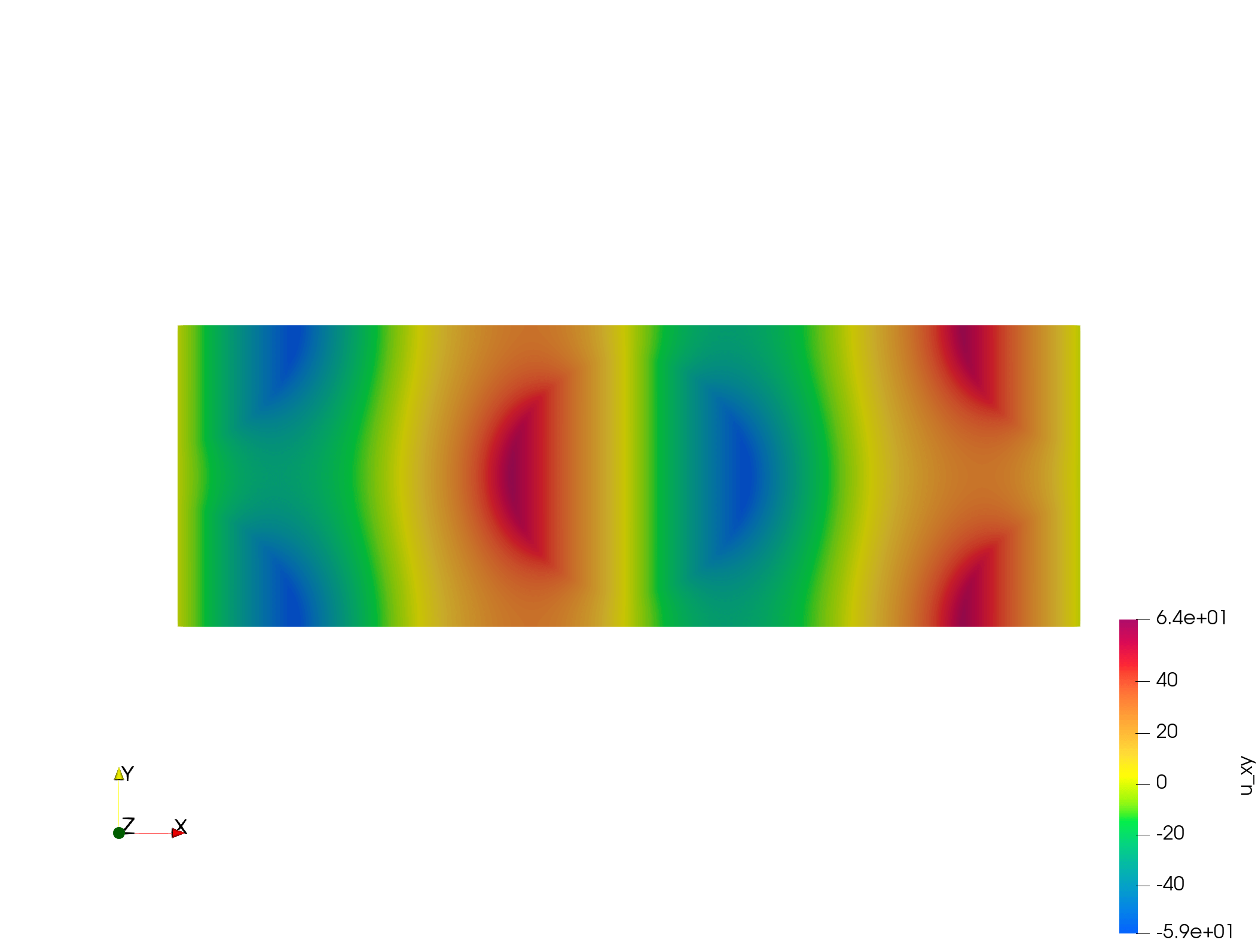}}}\\\vspace{0.1cm}
{{\includegraphics[height=0.10\textwidth,
trim=350 375 350 375, clip]{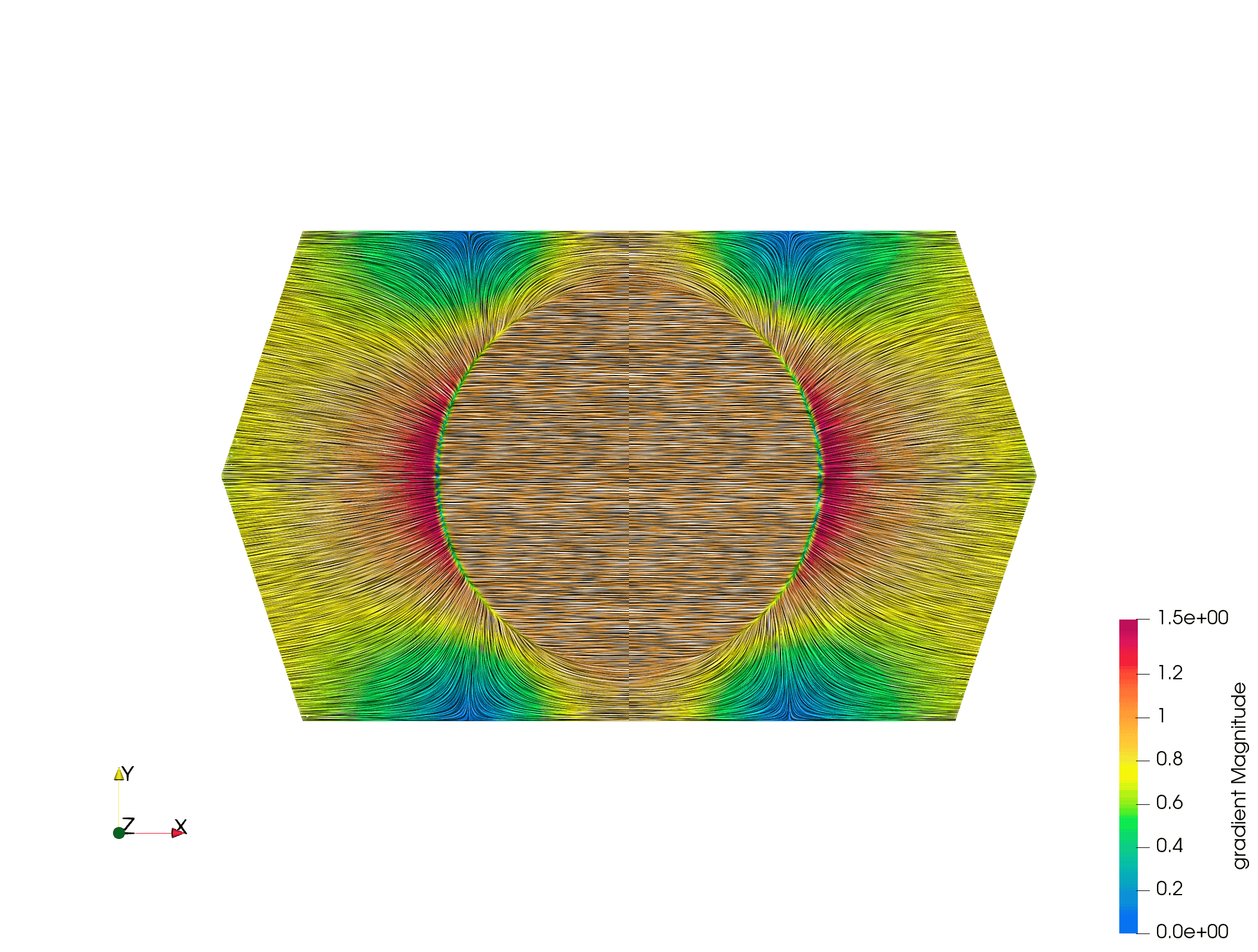}}}\hspace{0.5cm}
{{\includegraphics[height=0.10\textwidth,
trim=300 540 300 540, clip]{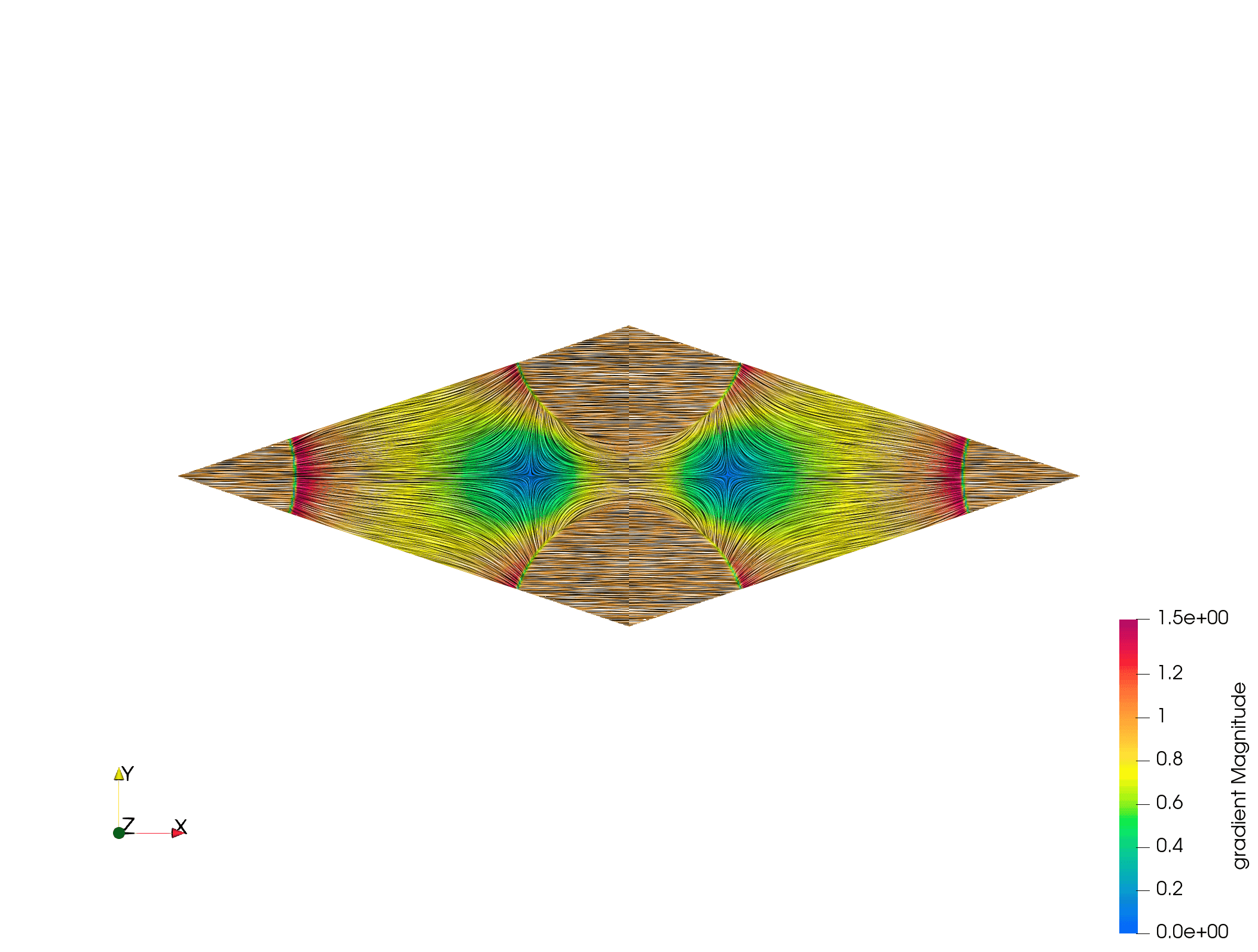}}}\hspace{0.5cm}
{{\includegraphics[height=0.10\textwidth,
trim=300 540 300 540, clip]{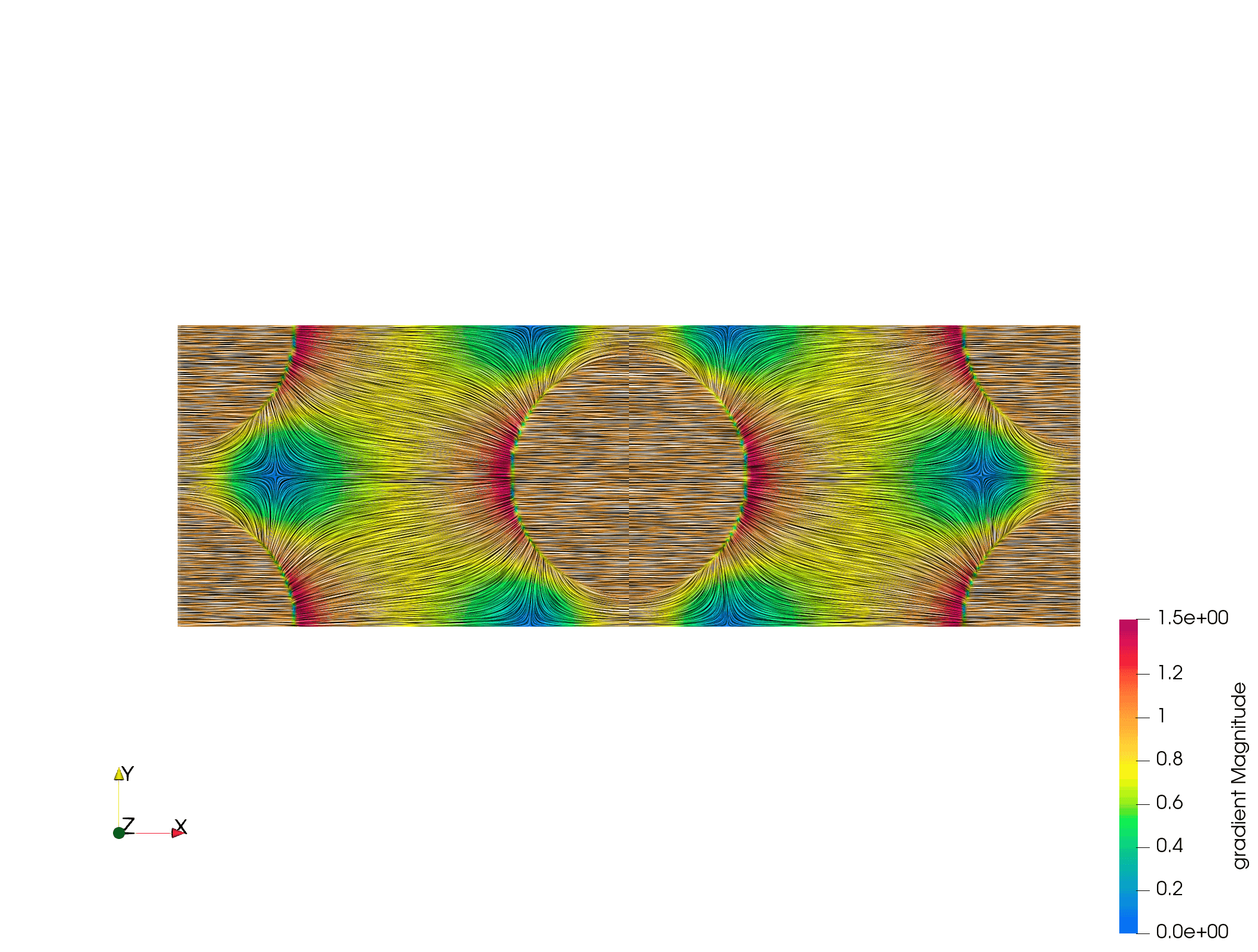}}}\\\vspace{0.75cm}
{{\includegraphics[height=0.10\textwidth,
trim=350 375 350 375, clip]{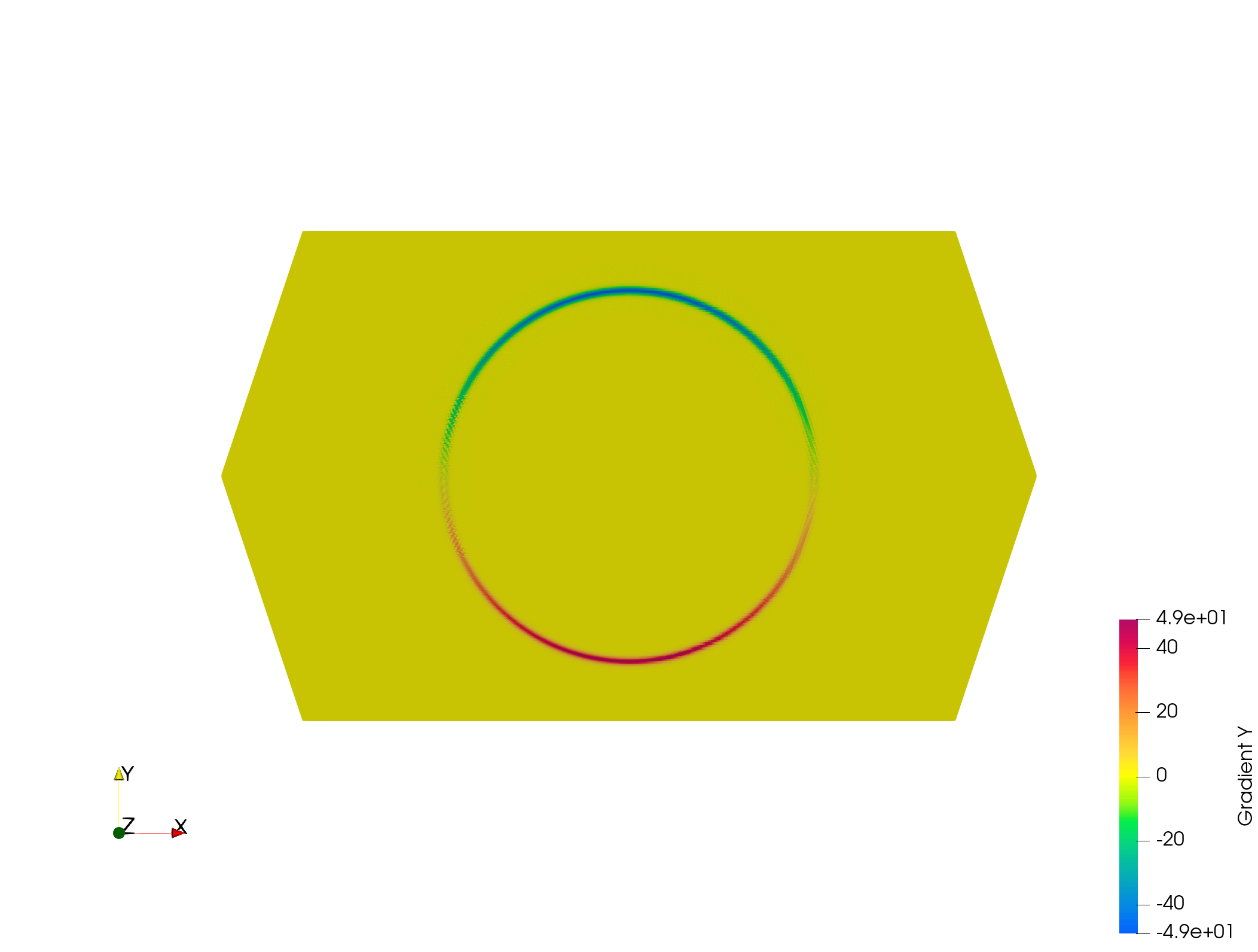}}}\hspace{0.5cm}
{{\includegraphics[height=0.10\textwidth,
trim=300 540 300 540, clip]{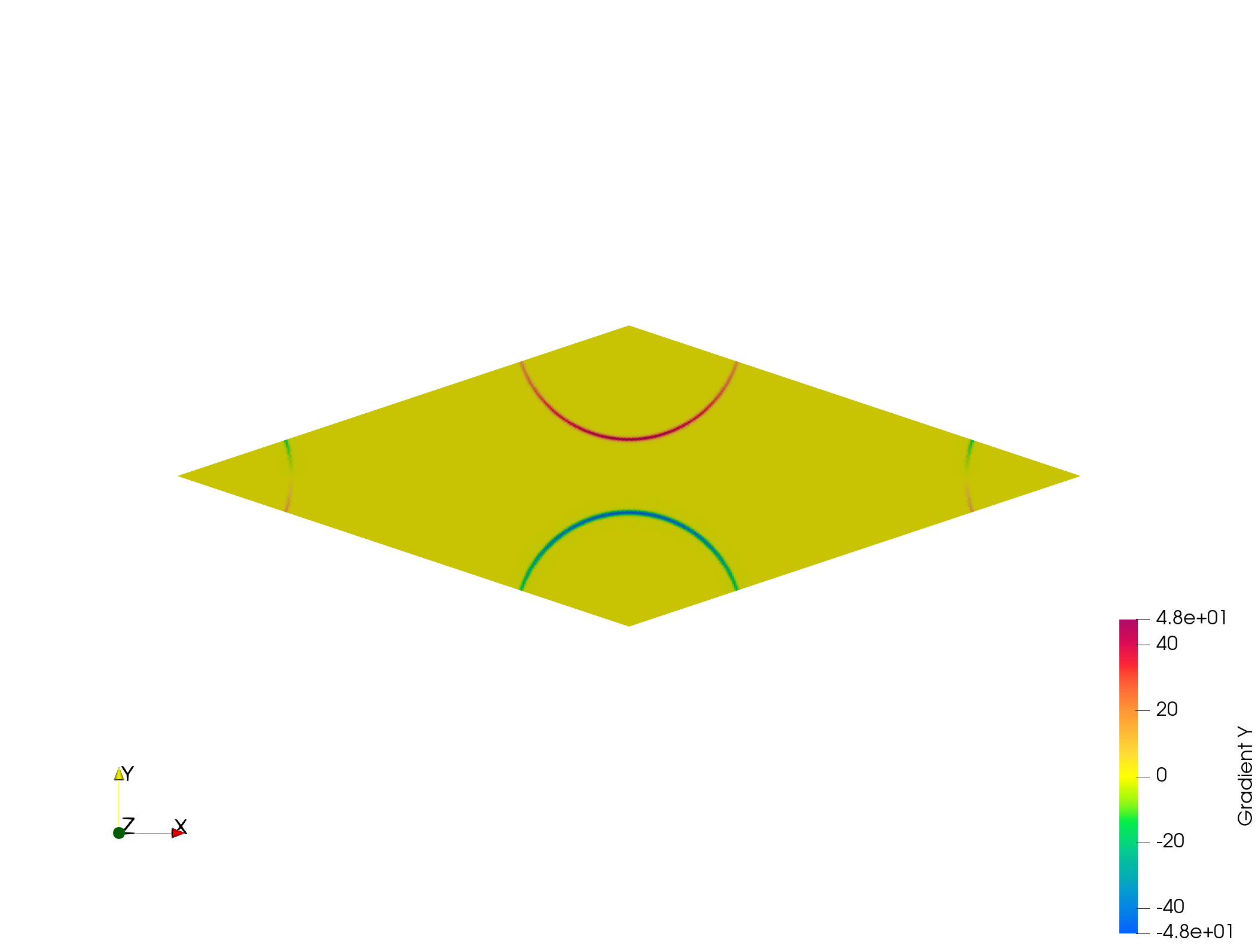}}}\hspace{0.5cm}
{{\includegraphics[height=0.10\textwidth,
trim=300 540 300 540, clip]{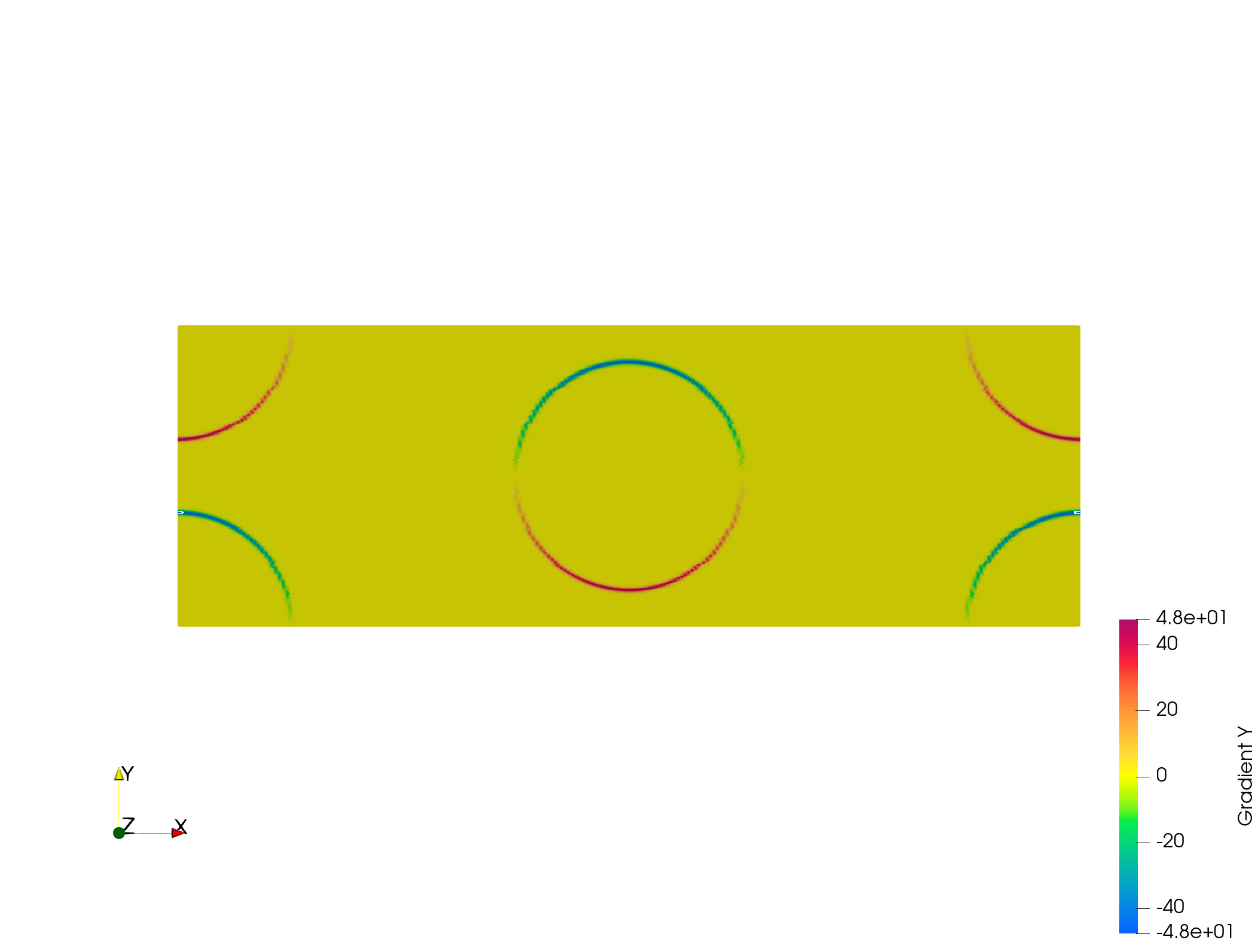}}}\\\vspace{0.1cm}
{{\includegraphics[height=0.10\textwidth,
trim=350 375 350 375, clip]{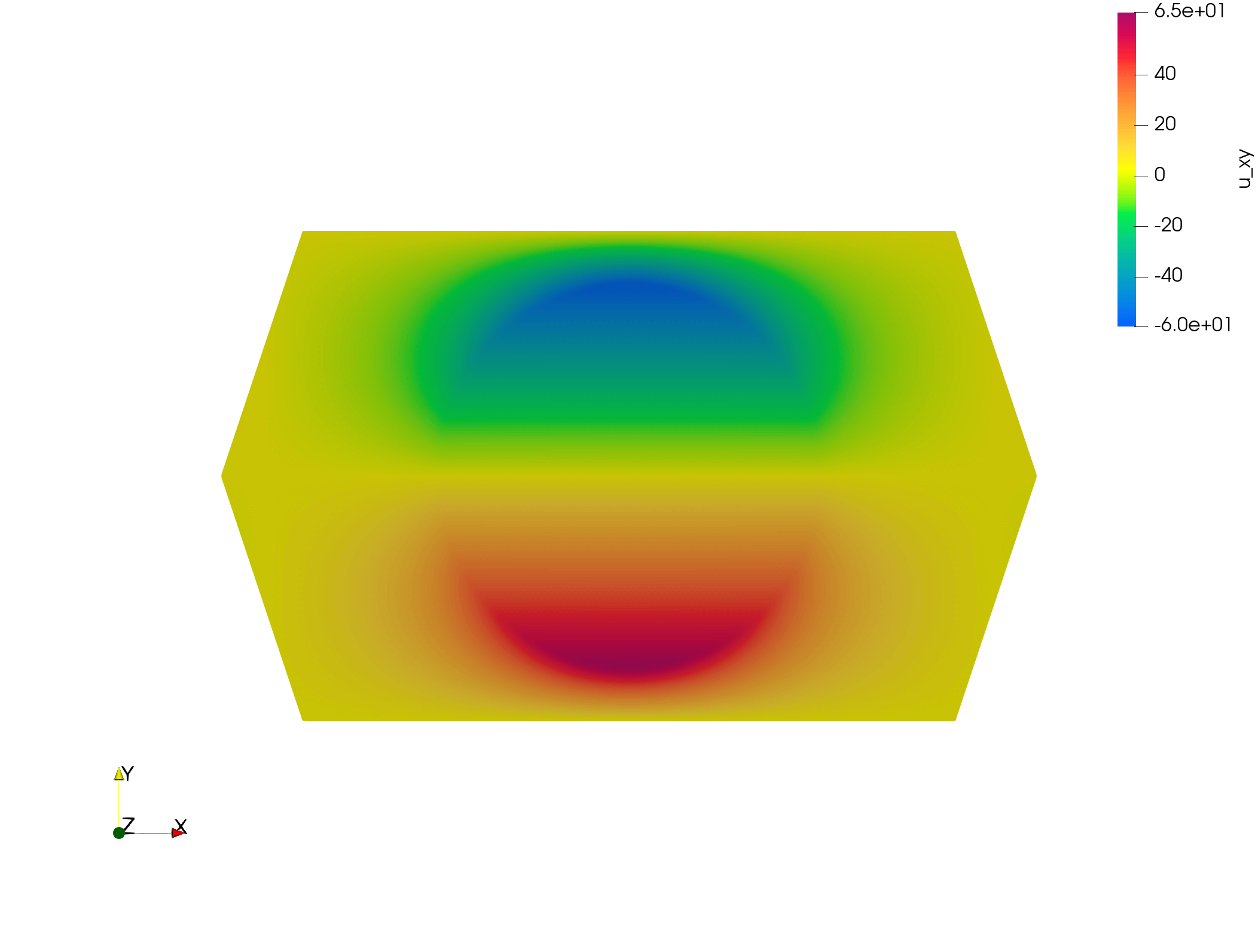}}}\hspace{0.5cm}
{{\includegraphics[height=0.10\textwidth,
trim=300 540 300 540, clip]{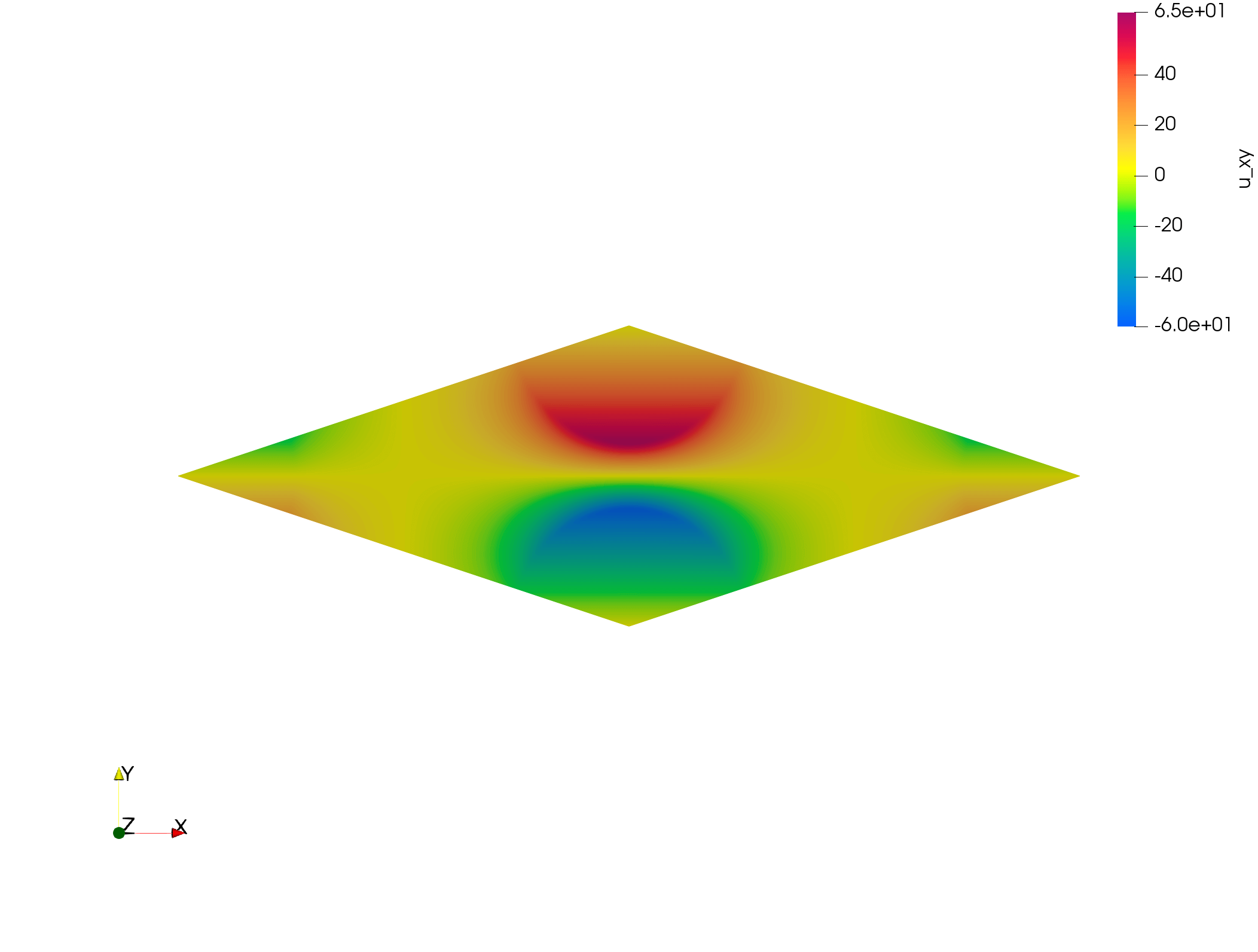}}}\hspace{0.5cm}
{{\includegraphics[height=0.10\textwidth,
trim=300 540 300 540, clip]{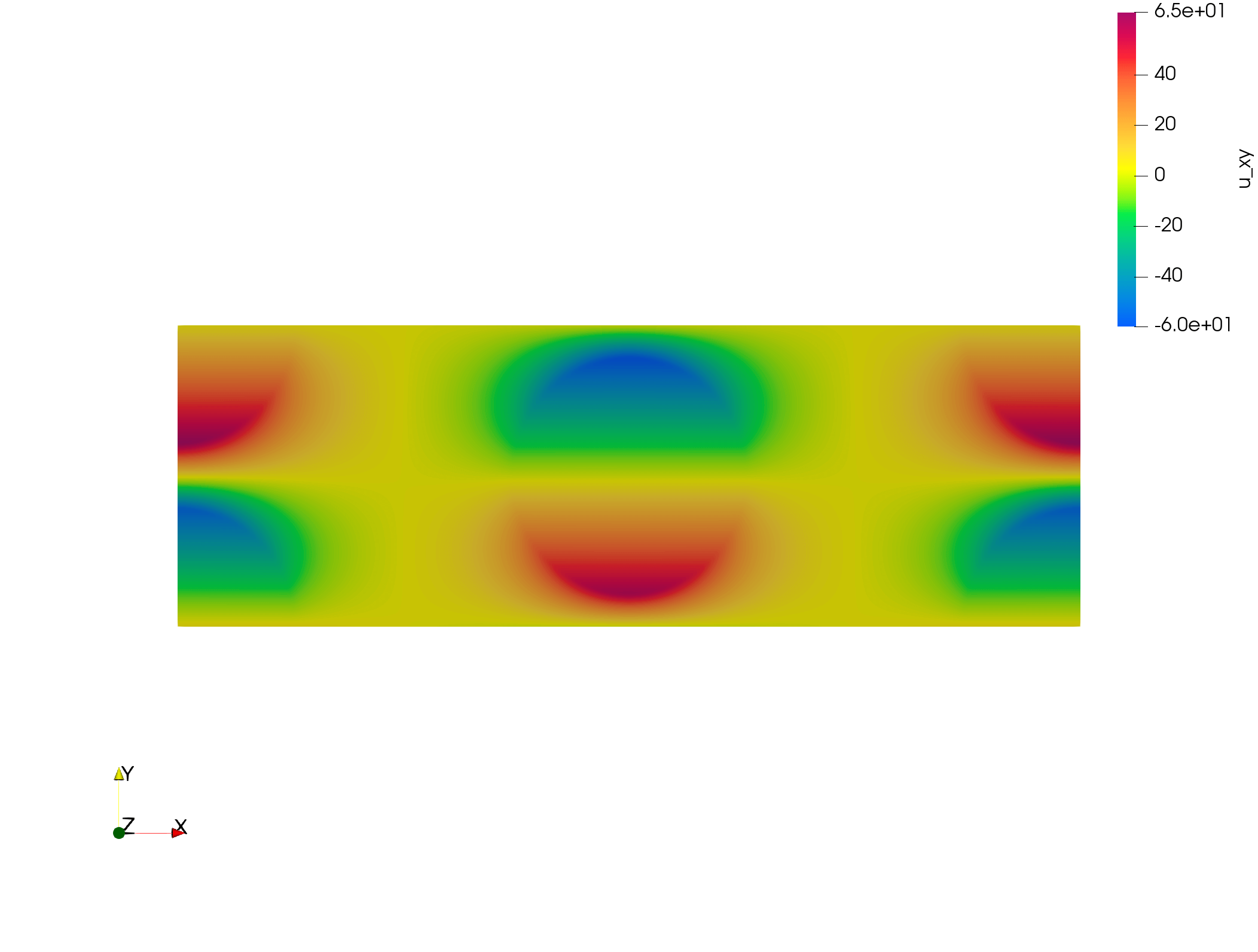}}}\\\vspace{0.1cm}
{{\includegraphics[height=0.10\textwidth,
trim=350 375 350 375, clip]{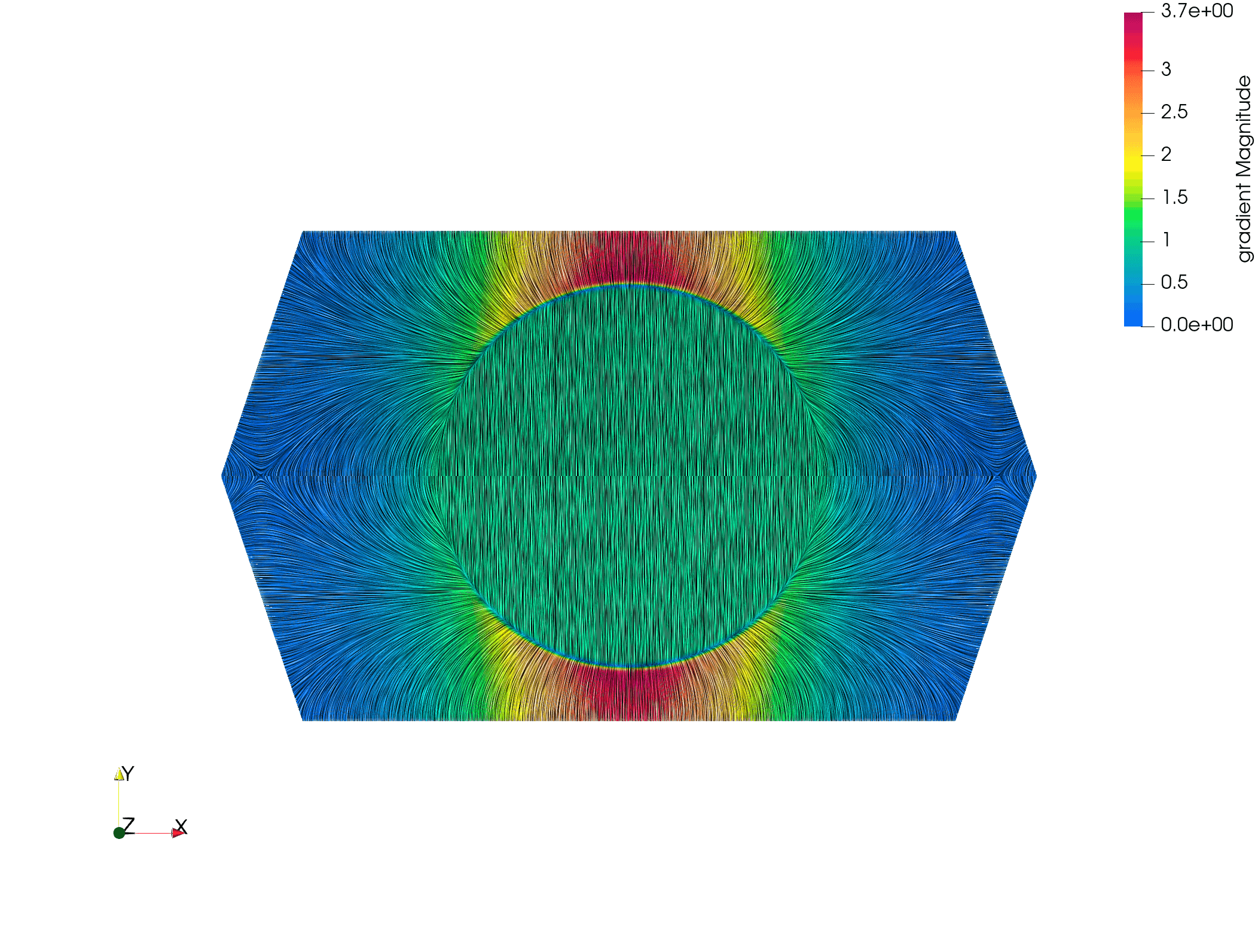}}}\hspace{0.5cm}
{{\includegraphics[height=0.10\textwidth,
trim=300 540 300 540, clip]{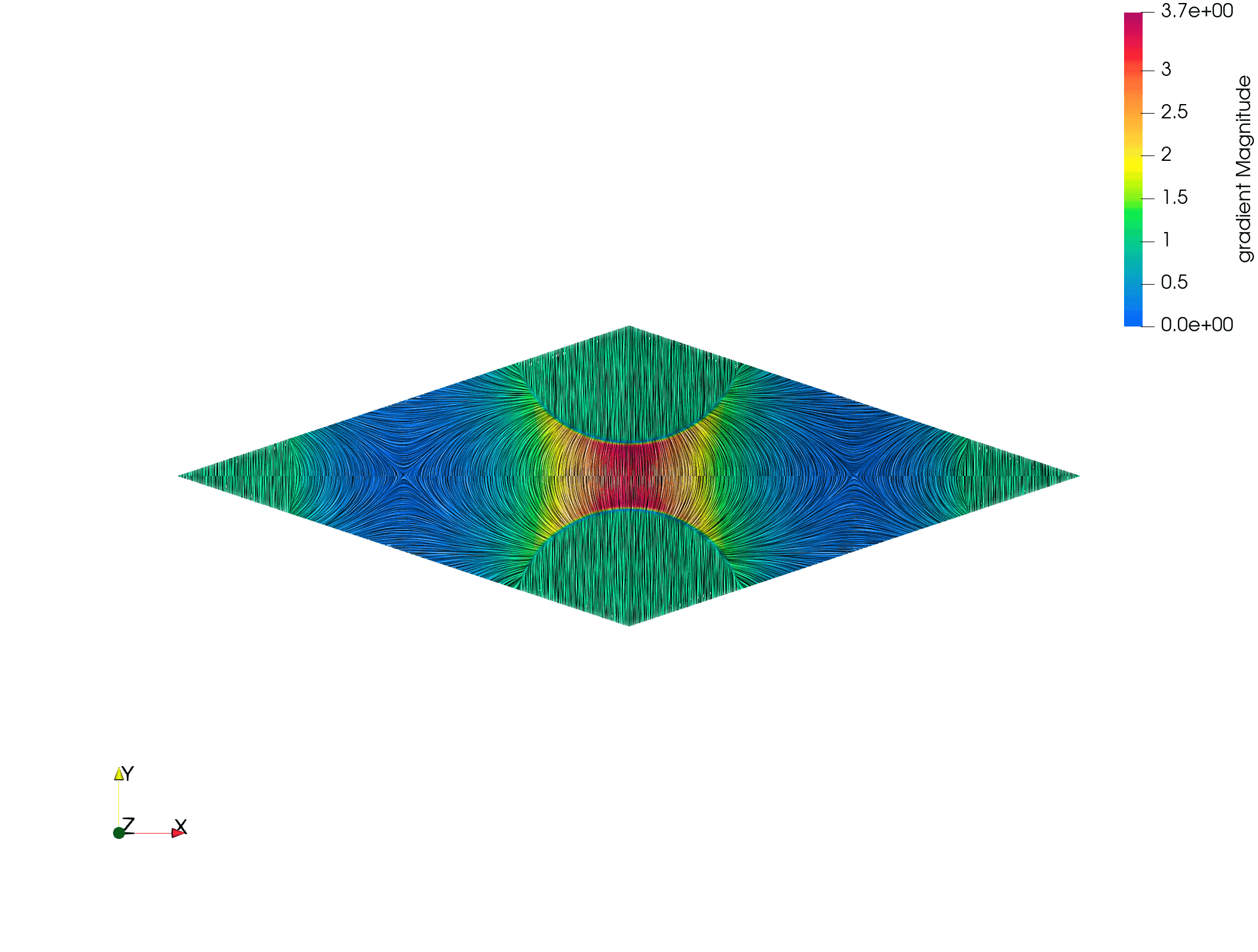}}}\hspace{0.5cm}
{{\includegraphics[height=0.10\textwidth,
trim=300 540 300 540, clip]{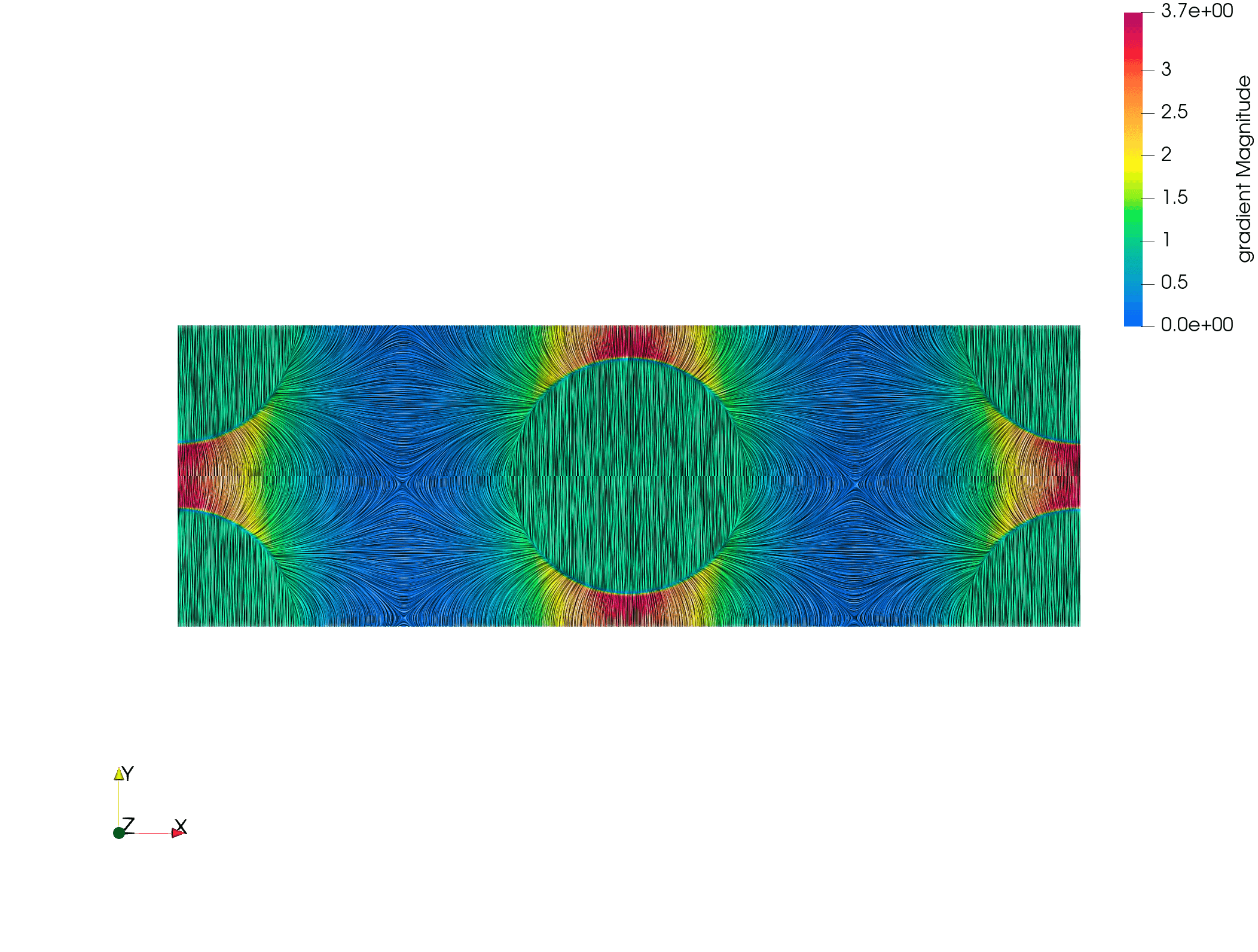}}}\\
min$\,\,$\frame{\includegraphics[width=0.30\textwidth, trim=0 15 0 15, clip]{legend_2}}$\,\,$max
\caption{The contour plots for the input,
the property field $a(\boldsymbol x)$ with $\boldsymbol a(\boldsymbol x)=a(\boldsymbol x)\, \boldsymbol 1$ and  $\xi=1$ (row 1) and the source term $\partial a(\boldsymbol x)/\partial x_1$ (row 2), and the corresponding ANN solution,
$\chi(\boldsymbol x)$ (row 3) and $|\boldsymbol{\nabla}\chi|$ (row 4), and  the source term $\partial a(\boldsymbol x)/\partial x_2$ (row 5), and the corresponding ANN solution,
$\chi(\boldsymbol x)$ (row 6) and $|\boldsymbol{\nabla}\chi|$ (row 7) fields of the cell-problem for periodic unit cells $\mathcal{R}^\mathrm{c}_1$, $\mathcal{R}^\mathrm{c}_2$ and $\mathcal{R}^\mathrm{c}_3$  for the best-performing hyperparameter set for each case.
The intervals [min, max] of the  contour plots are: for the first row $[0,100]$,
for the second and fifth rows $[-49.5,49.5]$,
for the third row $[0,123]$,
for the fourth row $[0,1.5]$,
for the sixth row $[0,125]$,
for the seventh row $[0,3.7]$.}
\label{F:cr_unit_cells_result_fields}
\end{figure*}

\begin{figure*}[htb!]
\centering
{{\includegraphics[width=0.45\textwidth,
trim=0 100 0 80, clip]{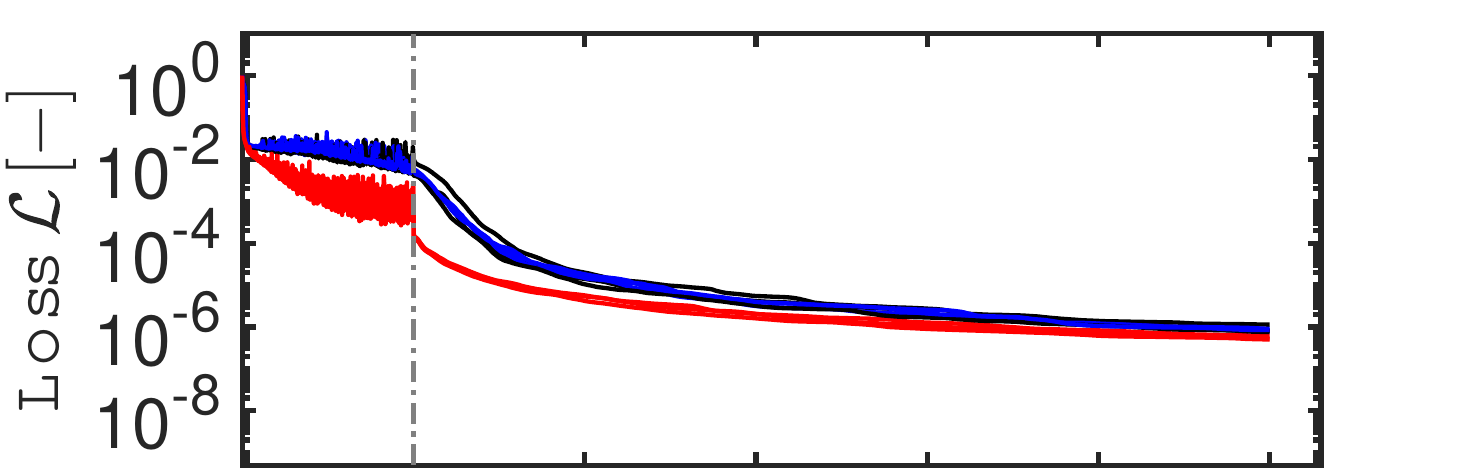}}}
{{\includegraphics[width=0.45\textwidth,
trim=0 100 0 80, clip]{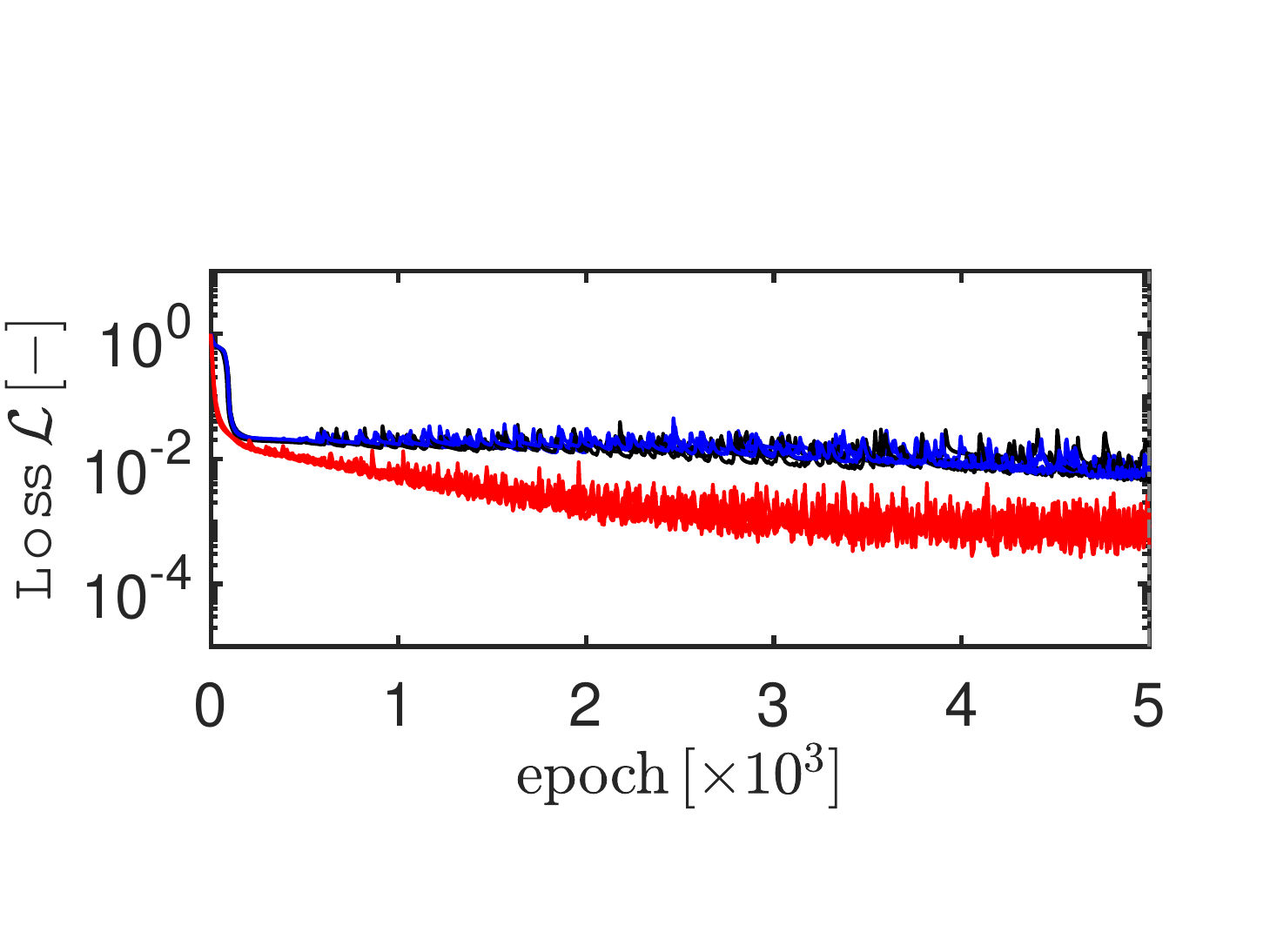}}}\\
{{\includegraphics[width=0.45\textwidth,
trim=0 40 0 80, clip]{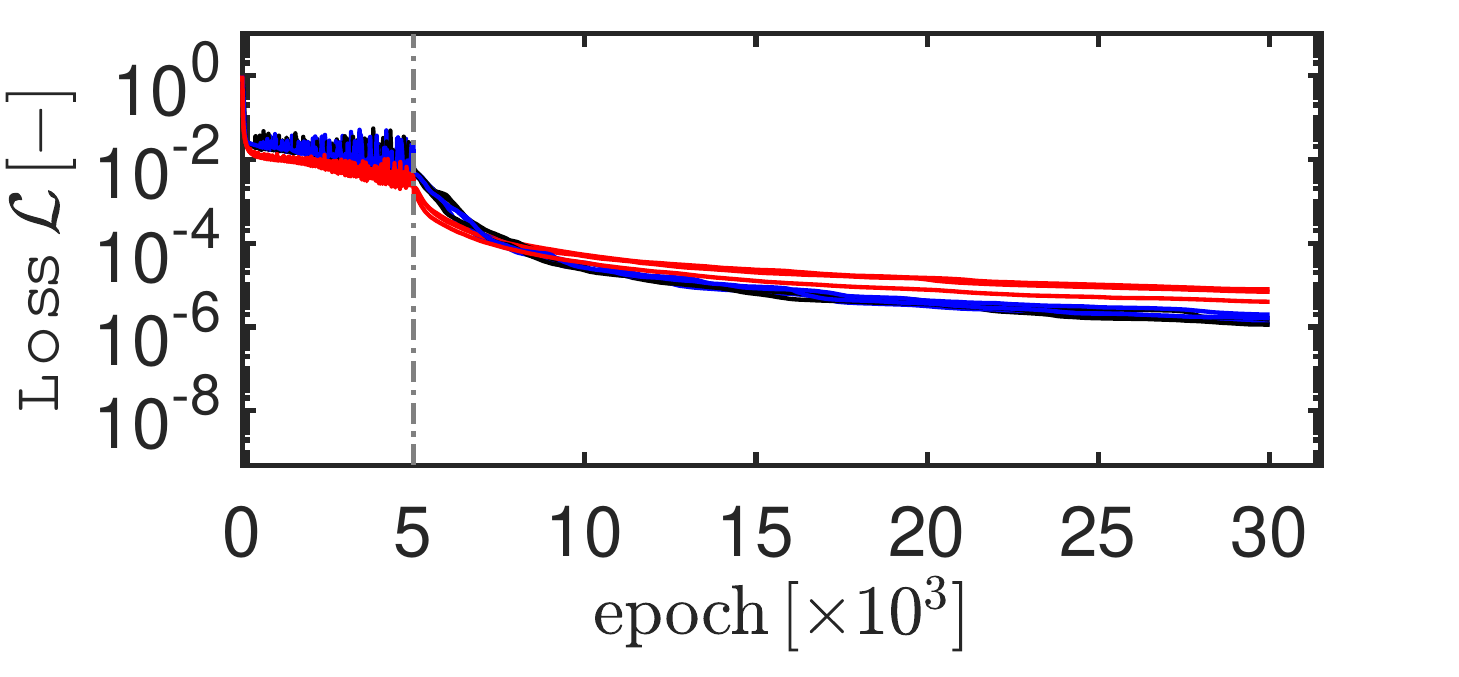}}}
{{\includegraphics[width=0.45\textwidth,
trim=0 40 0 80, clip]{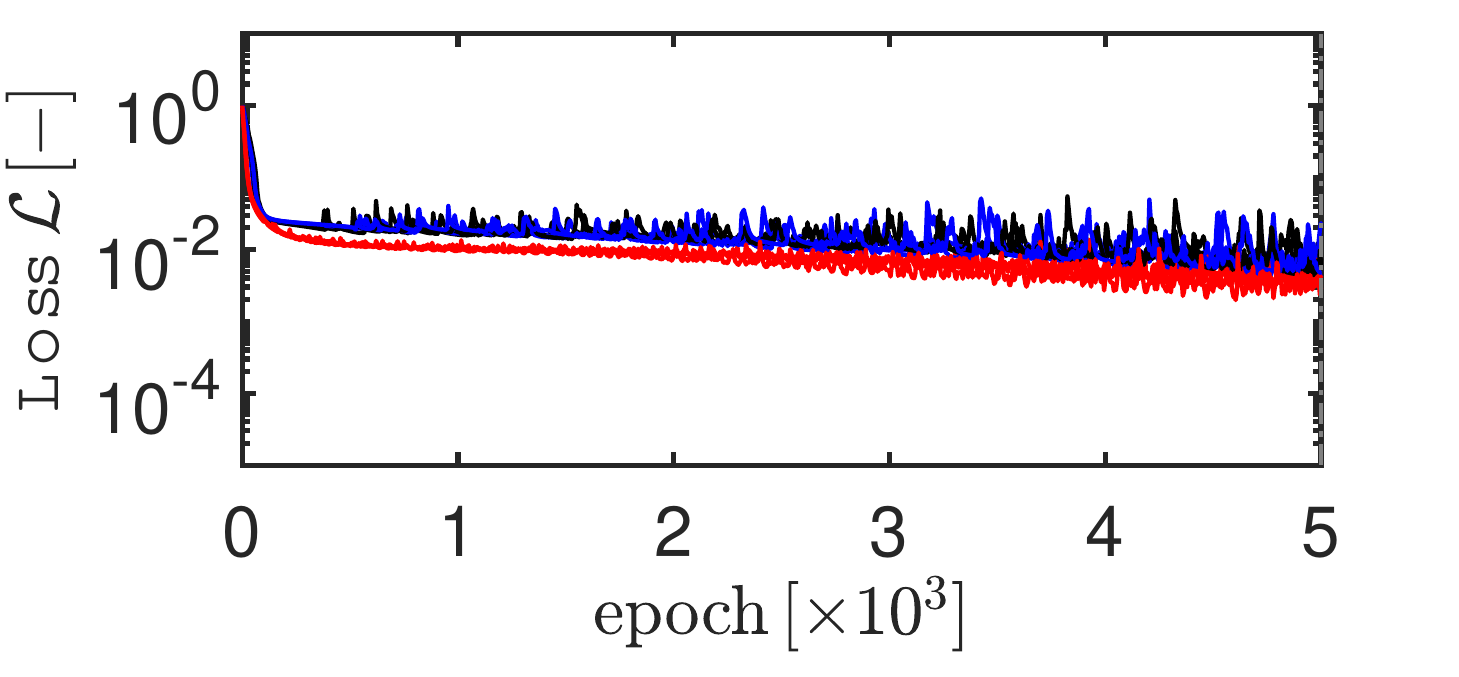}}}\\
\vspace{-10pt}
\caption{Training history of the normalized loss function for the ANN solution considering unit cells  of the cell-problem for periodic unit cells $\mathcal{R}^\mathrm{c}_1$ (black), $\mathcal{R}^\mathrm{c}_2$ (blue) and $\mathcal{R}^\mathrm{c}_3$ (red). For each hyperparameter set, three training sessions are run.  The first 5000 epochs (marked with a vertical dashed line) belong to adam optimizer, whereas the rest to L-BFGS. For $\mathcal{R}^\mathrm{c}_1$ and $\mathcal{R}^\mathrm{c}_2$, a learning rate of 0.001 with 100$\times$6 ANN architecture possessing low-frequency Fourier features is used, whereas, for $\mathcal{R}^\mathrm{c}_3$, a learning rate of 0.010 with 50$\times$3 ANN architecture possessing high-frequency Fourier features with the first 10 integer multiples of the reciprocal base vector.}
\label{F:cr_results_training_histories}
\end{figure*}

\subsubsection{Three-dimensional Simple Cubic Lattice}
Finally, we present an application in 3D for a simple cubic lattice with $L=\rvert\boldsymbol c_1\rvert=\rvert\boldsymbol c_2\rvert=\rvert\boldsymbol c_3\rvert=200\xi$ where $\xi=1$ considering two unit cells $\mathcal C_1$ and $\mathcal C_2$;  see  Fig.\ \ref{fig:BravaisLattice1D2D3D}.

The ANN training batch size corresponds to 512000. A learning rate of 0.010 is used with 50$\times$3 ANN architectures with both low- and high-frequency Fourier features, where the ANN is trained three times for each hyperparameter combination.

\begin{figure*}[htb!]
\centering
\subfigure[$a(\boldsymbol x)$]{
{{\includegraphics[height=0.22\textwidth,
trim=385 35 350 100, clip]{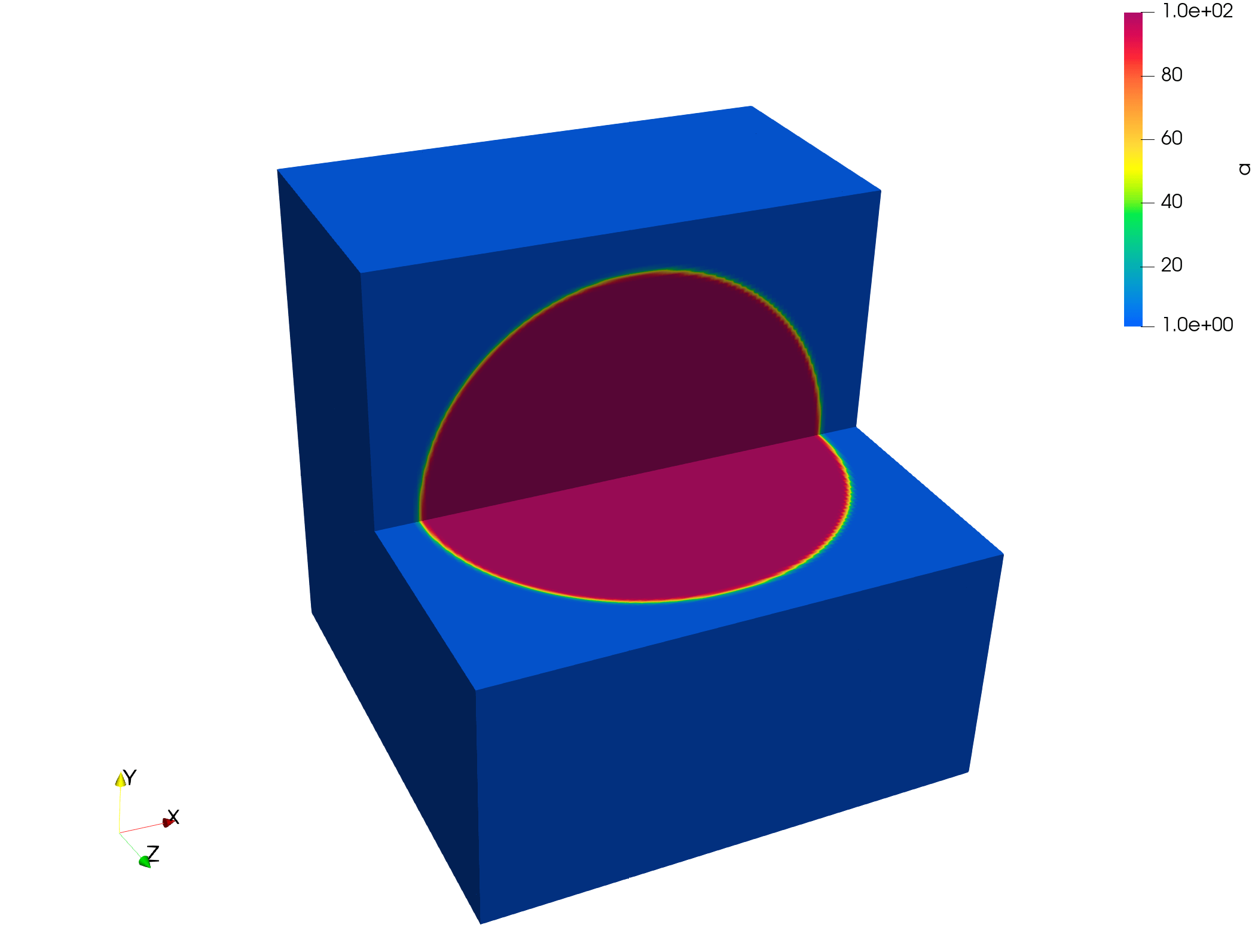}}}
{{\includegraphics[height=0.22\textwidth,
trim=385 35 350 100, clip]{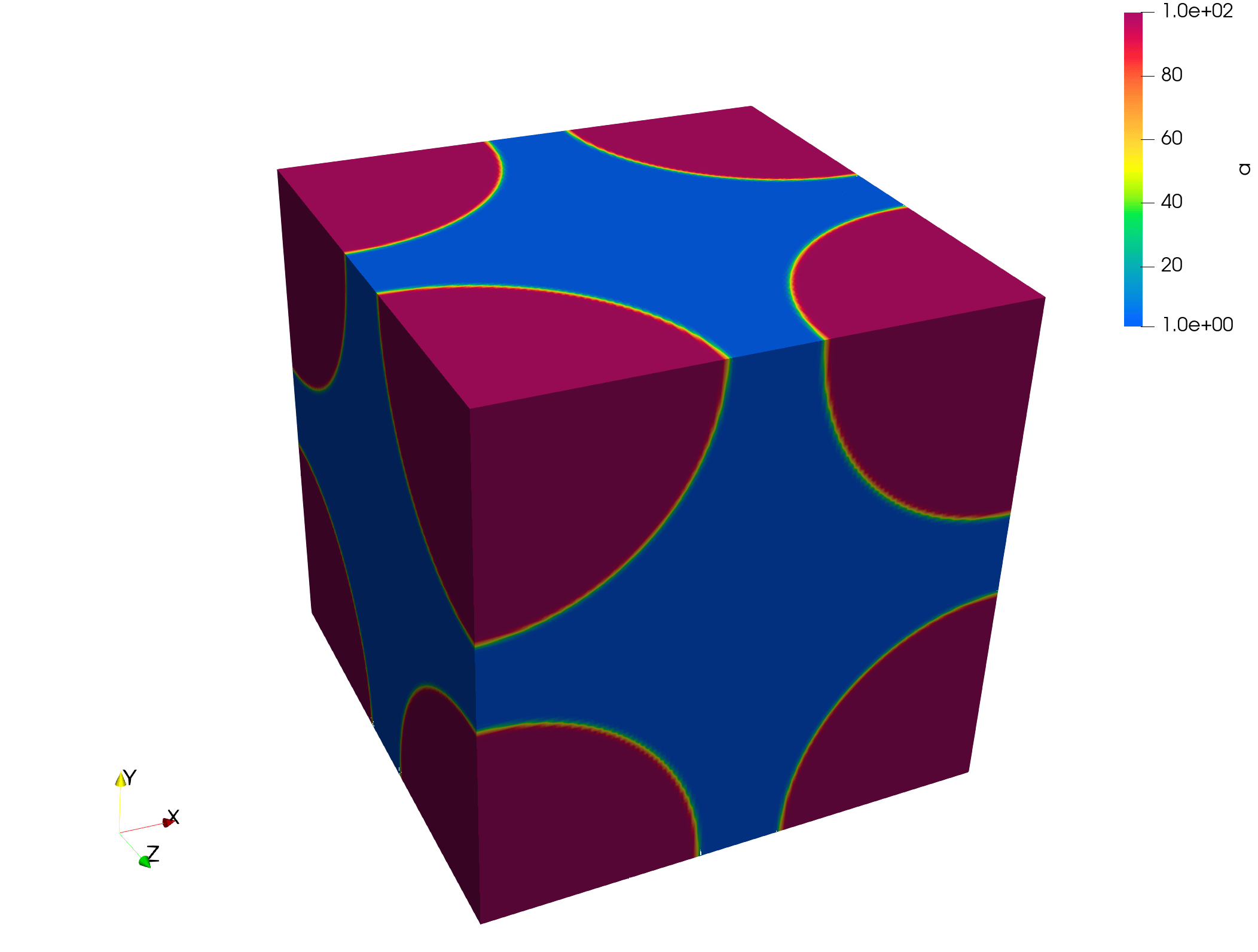}}}
}\hspace{0.6cm}
\subfigure[$f^1=\partial a(\boldsymbol x)/\partial x_1$]{
{{\includegraphics[height=0.22\textwidth,
trim=385 35 350 100, clip]{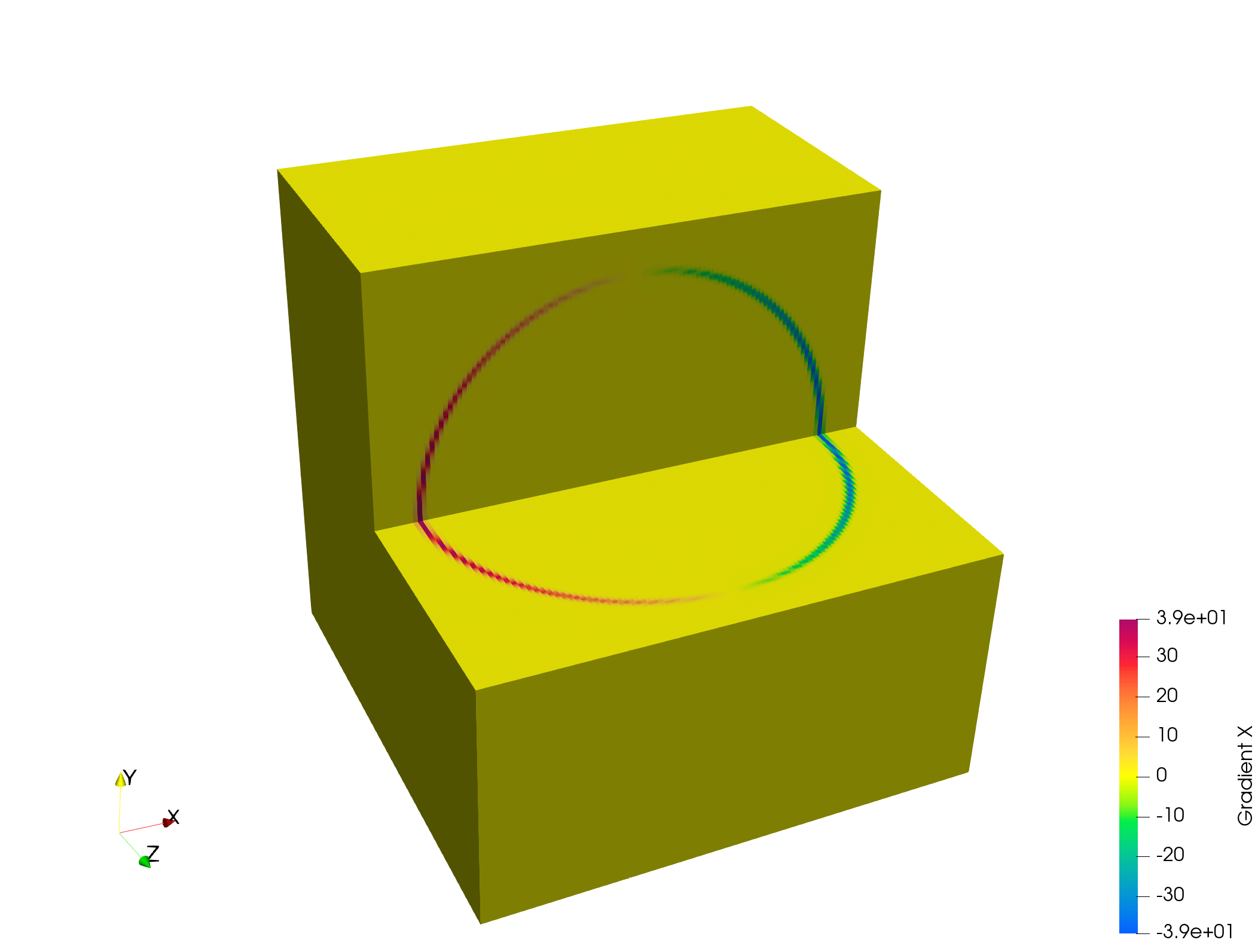}}}
{{\includegraphics[height=0.22\textwidth,
trim=385 35 350 100, clip]{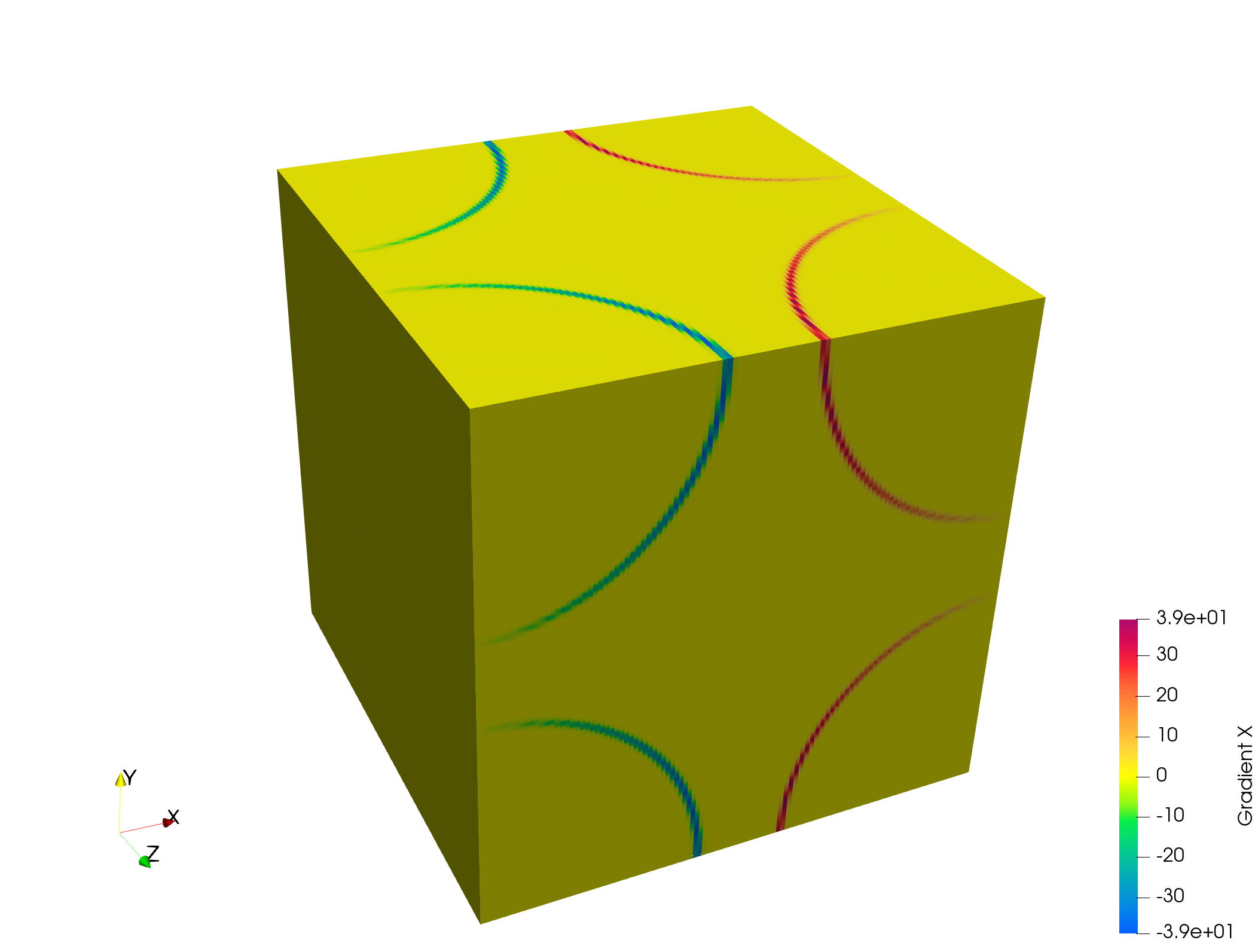}}}
}\\
\subfigure[$\chi^1(\boldsymbol x)$ for $f^1$]{
{{\includegraphics[height=0.22\textwidth,
trim=385 35 350 100, clip]{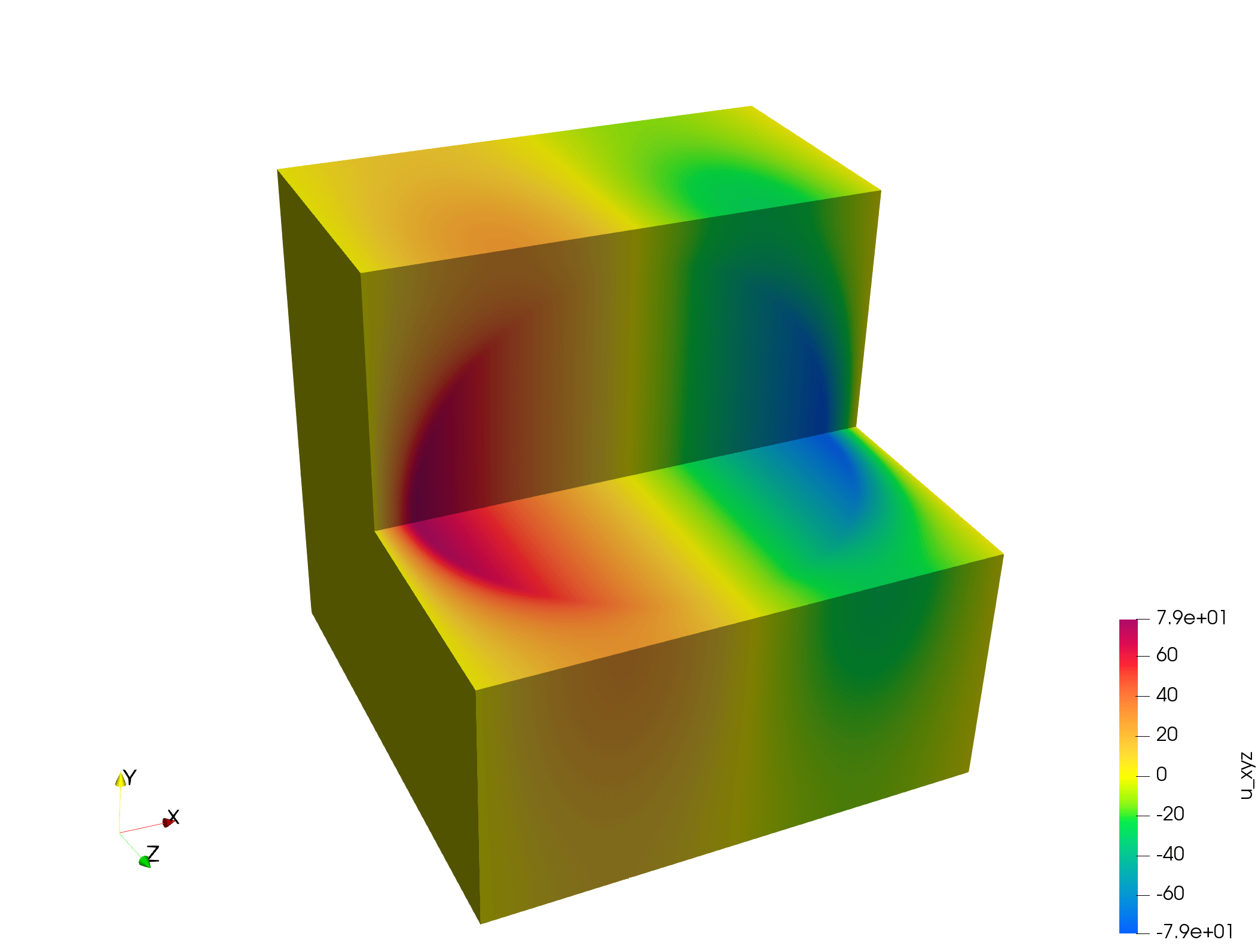}}}
{{\includegraphics[height=0.22\textwidth,
trim=385 35 350 100, clip]{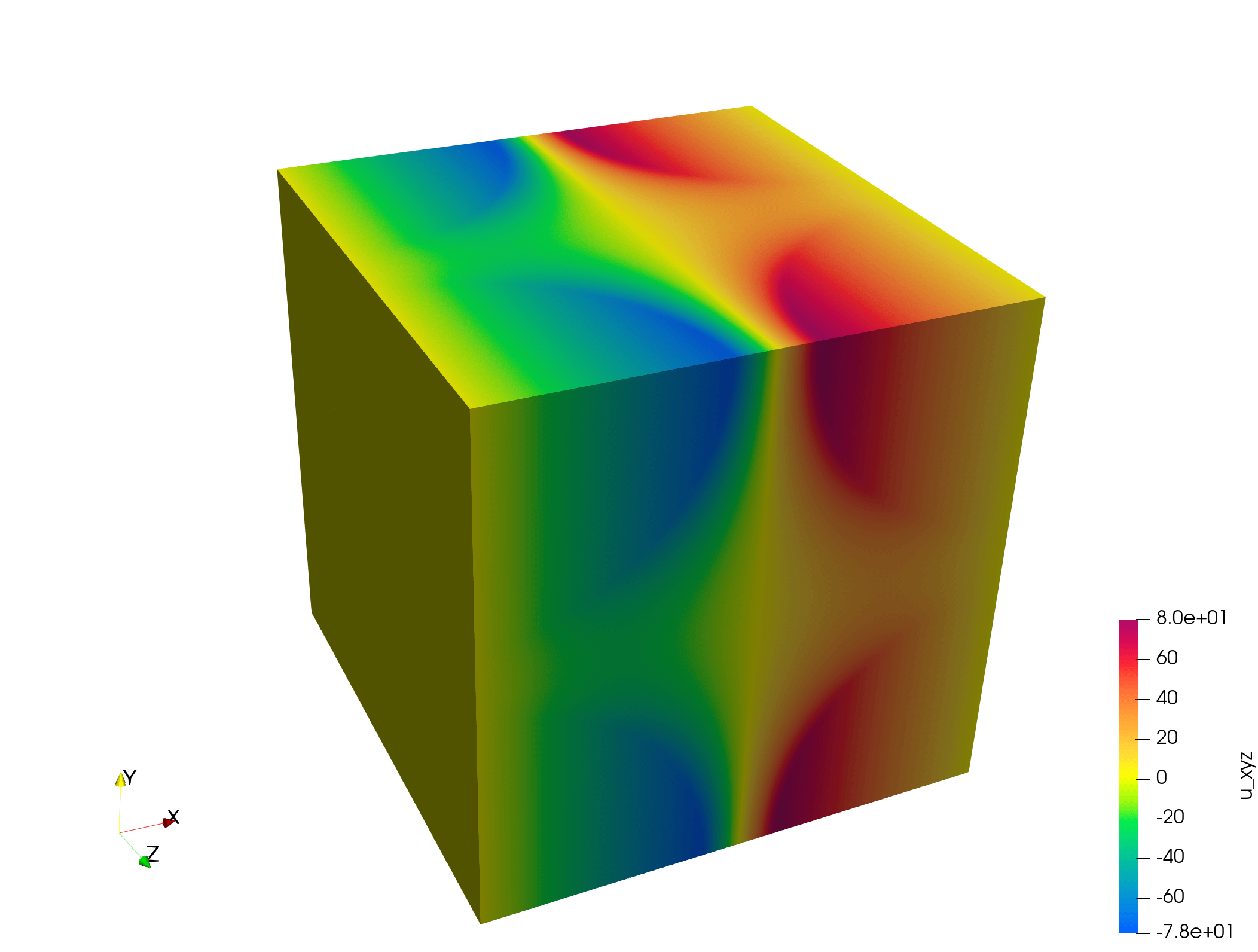}}}
}\hspace{0.6cm}
\subfigure[$|\boldsymbol{\nabla}\chi^1(\boldsymbol x|$ for $f^1$]{
{{\includegraphics[height=0.22\textwidth,
trim=385 35 350 100, clip]{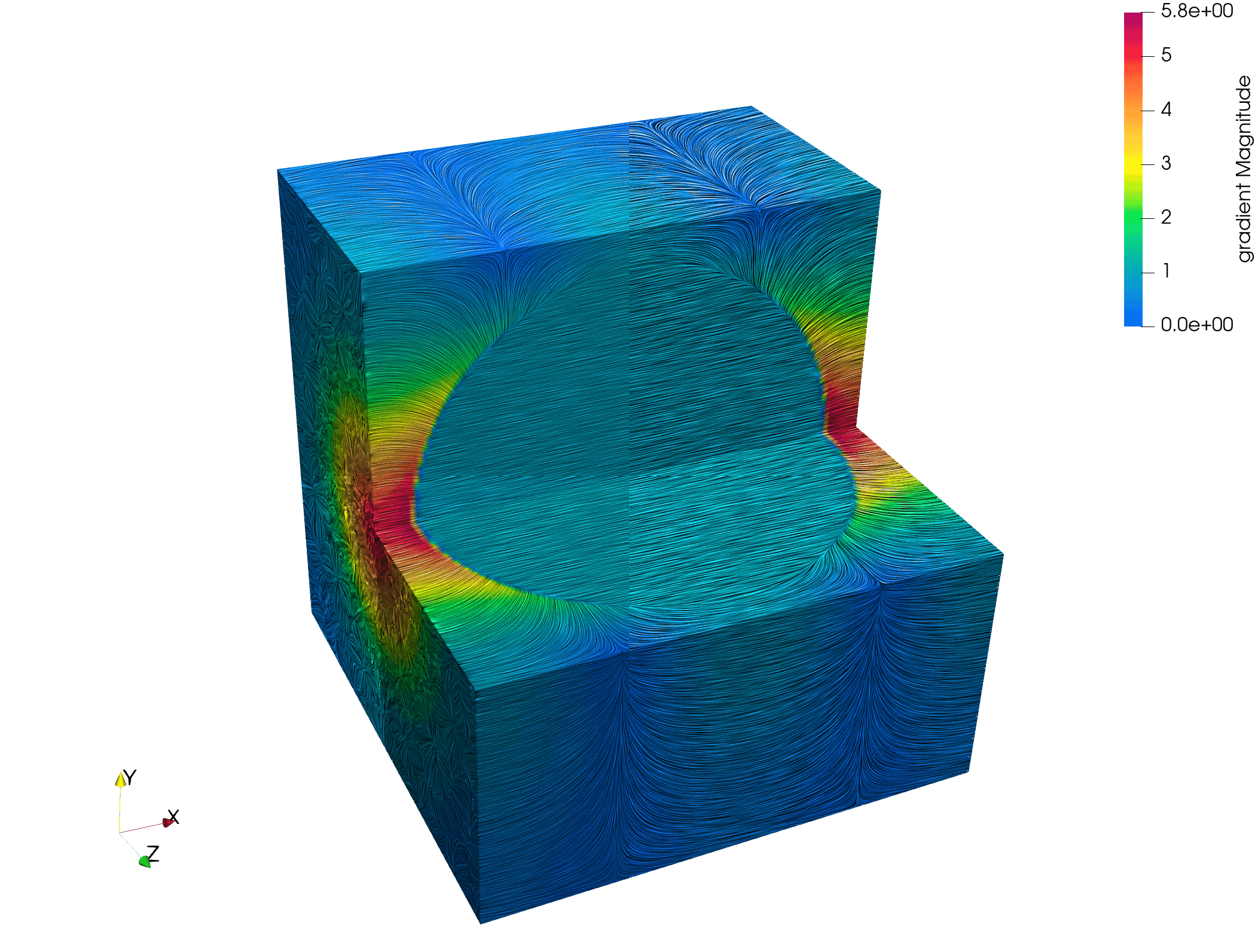}}}
{{\includegraphics[height=0.22\textwidth,
trim=385 35 350 100, clip]{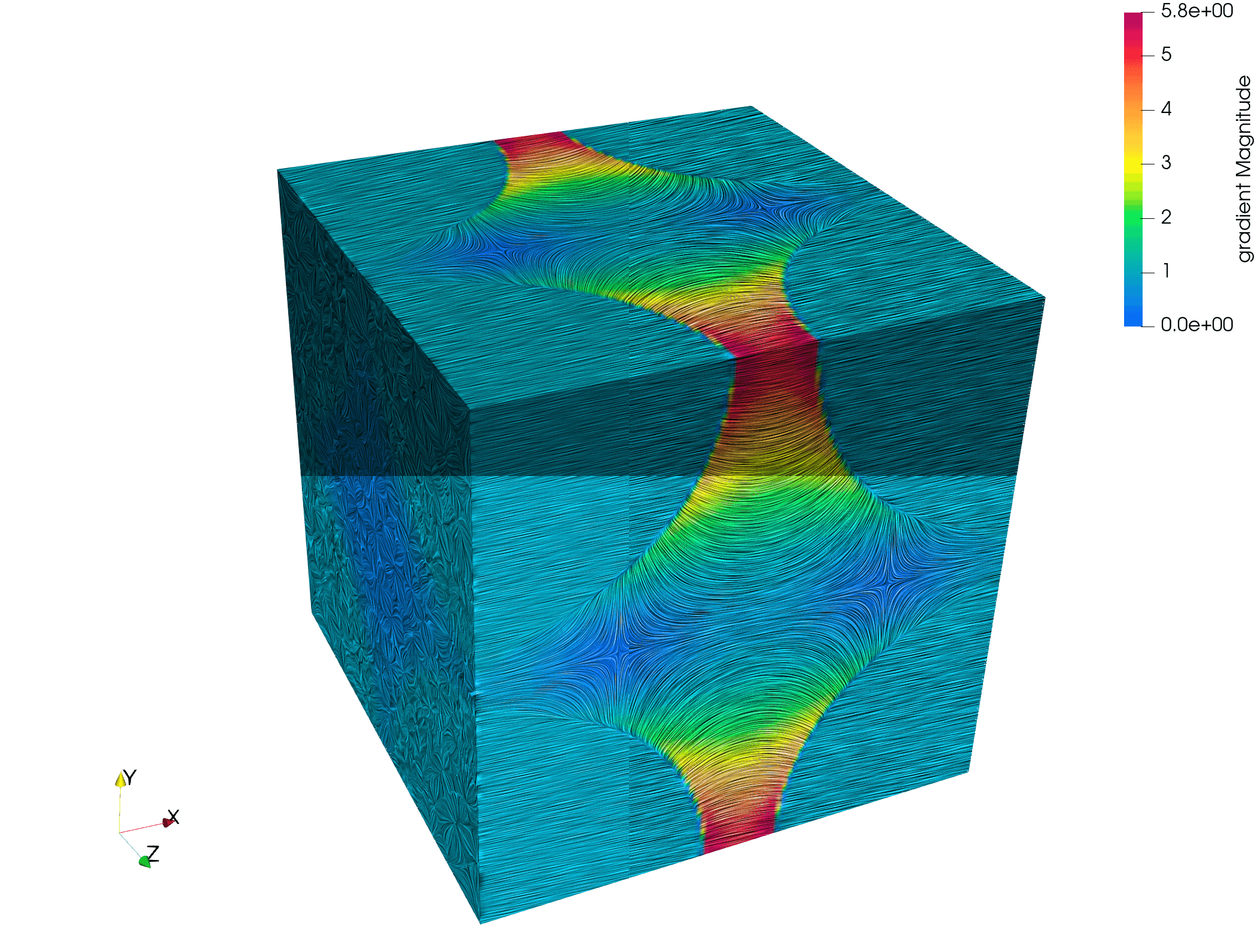}}}}\\
min$\,\,$\frame{\includegraphics[width=0.30\textwidth, trim=0 15 0 15, clip]{legend_2}}$\,\,$max
\caption{The contour plots for the input,
the property field $a(\boldsymbol x)$ with $\boldsymbol a(\boldsymbol x)=a(\boldsymbol x)\, \boldsymbol 1$ and  $\xi=1$ (row 1) and the source term $\partial a(\boldsymbol x)/\partial x_1$, and the corresponding ANN solution for high-frequency Fourier features with the first 10 integer multiples of the reciprocal base vector. The intervals [min, max] of the  contour plots are (a) $[1,100]$, (b) $[-49.5,49.5]$ (c) $[0,158]$  and (d) $[0,5.8]$.}
\label{F:results_3D_C}
\end{figure*}

\begin{figure*}[htb!]
\centering
{{\includegraphics[height=0.22\textwidth,
trim=0 40 0 80, clip]{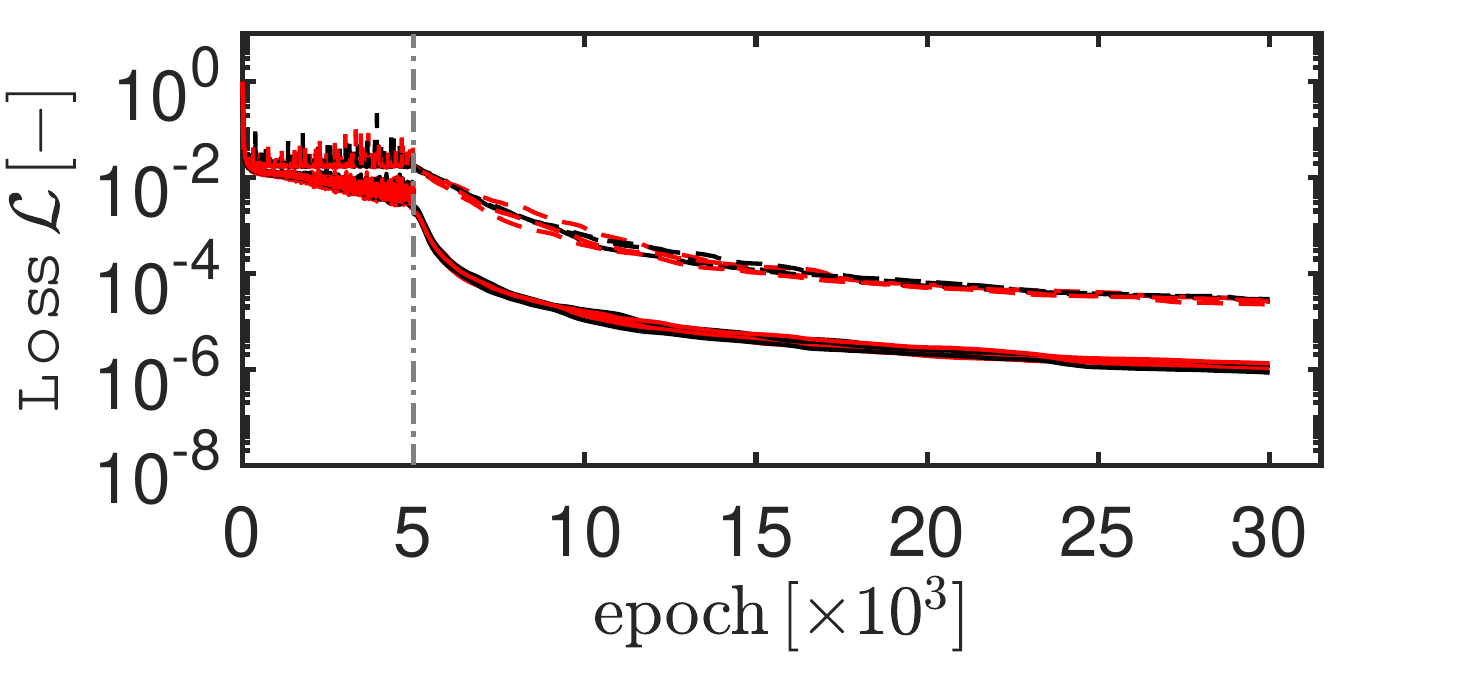}}}
{{\includegraphics[height=0.22\textwidth,
trim=0 40 0 80, clip]{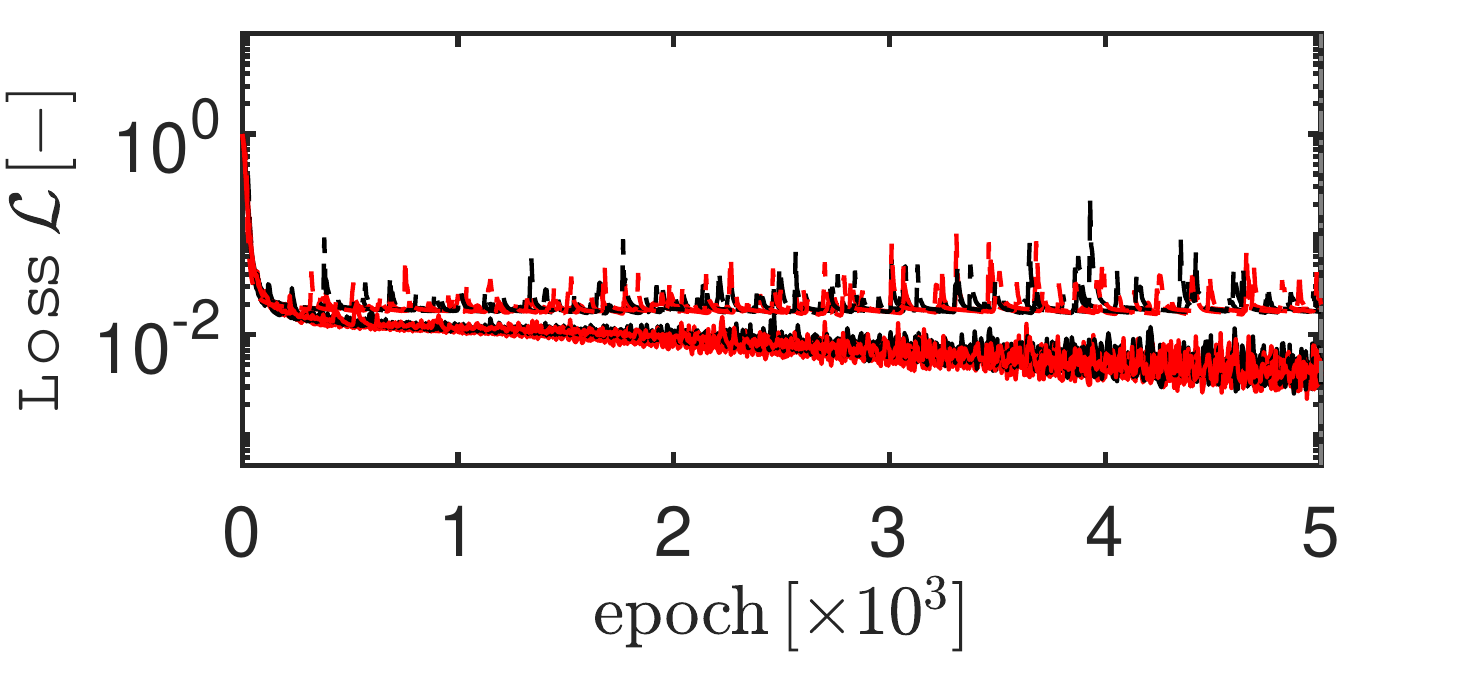}}}
\caption{Training history of the normalized loss function for the 3D simulations with unit cells $\mathcal C_1$ (black) and $\mathcal C_2$ (red). Continuous lines belong to the ANN solution for high-frequency Fourier features, whereas dashed ones are low-frequency. For each hyperparameter set, three training sessions are conducted. TThe First 5000 epochs (marked with a vertical dashed line) belong to adam optimizer, whereas the rest 25000 to L-BFGS.}
\label{F:loss_3D_2}
\end{figure*}

For the symmetry group possessed by this material microstructure, a spherical second-order effective property tensor is anticipated. Using a chosen basis that conforms to cubic dimensions, this gives non-zero and equal diagonal terms and zero off-diagonal terms, resulting in $a^\star_{11} = a^\star_{22}= a^\star_{33}$ and $a^\star_{ij}=0$ for $i\neq j$ and $i,j=1,2,3$. Thus, in our computations, we consider only $x-$gradients of the property field in the source term. The results show that ANNs with high-frequency Fourier features incorporating the first 10 integer multiples of the reciprocal base vector performed best; see Fig.\  \ref{F:loss_3D_2}. The contour plots for the PINN solution $\chi(\boldsymbol x)$
 the flux norm  $|\boldsymbol{\nabla}\chi(\boldsymbol x)|$ fields of the cell-problem for periodic unit cells $\mathcal{C}_1$ and $\mathcal{C}_2$ are given for this hyperparameter set are given in Fig.\ \ref{F:results_3D_C}. As observed from the contour plots, the choice of the selected unit cell does not change the solution.

 The computed effective property tensor components are given in Table\ \ref{T:3D_C_results}3 in Appendix \ref{S:tabulated_results} for $\mathcal{C}_1$ and $\mathcal{C}_2$. For the more accurate solution with high-frequency Fourier features, incorporating the first 10 integer multiples of the reciprocal base the mean and standard deviation for the tabulated effective property matrix components result in $a_{11}=2.4550\pm0.0016$,
$a_{12}=0.0001\pm0.0001$,
$a_{13}=-0.0001\pm0.0002$,
for $\mathcal{C}_1$,
and in
$a_{11}=2.4553\pm0.0020$,
$a_{12}=0.0001\pm0.0000$,
$a_{13}=0.0000\pm0.0002$,
for $\mathcal{C}_2$. These results show a good agreement between the solutions devising different unit cells.
Considering 0.30 inclusion fraction but a sharp interface condition, the truncated series expansion solution of Lam\ \cite{Lam1986} cited in \cite{Ren2018}, given in \ref{section:3D_analytics} gives $a^\star=2.2742$. Similar to our observations for 1D and 2D square lattices with regular circular inclusion distribution for which analytical solutions exist, the ANN solution for the main diagonal term $a^\star_{11}$ results in a slightly larger prediction. This discrepancy is due to the utilized diffuse interface assumption. The following representation in matrix form in two decimal points emerges as a good approximation for the effective property tensor
\begin{equation}
\left[ \bs{a}^{\mathcal{C}}\right]
=\left[
\begin{array}{ccc}
2.46  & 0.00 &  0.00 \\
0.00  & 2.46 &  0.00 \\
0.00  & 0.00 &  2.46 \\
\end{array}%
\right]\,.
\end{equation}

\section{Conclusion}
We applied physics-informed neural networks (PINNs) to determine effective (macroscopic) properties of heterogeneous media using first-order two-scale periodic asymptotic homogenization. A relatively large phase contrast is aimed at. A generic divergence-type elliptic equation is considered to this end which applies to a broad range of physics problems, including electrostatics, magnetostatics, steady-state (time-independent) heat and electrical conduction, and anti-plane elasticity. The unit cell problem for the elliptic equation can be solved up to a constant. The reliance on the standard integral solution for the property tensors on only the gradient of the solution field dismissed the requirement for uniqueness in the solution.
This, together with the absence of Dirichlet boundary conditions, resulted in a lossless boundary condition application. ANN architectures with an input-transfer layer (Fourier feature mapping) materializing reciprocal lattice vectors are considered. This allowed an exact imposition of periodic boundary conditions for periodic domains with arbitrary and nonrectangular shapes pertaining to crystalline arrangements defined with different Bravais lattices and improved convergence of high-frequency functions during training once integer multiples of the reciprocal basis in the Fourier mapping are considered. As illustrative examples, we considered applications in one, two, and three dimensions with various regular and stochastic composites. We summarized the influence of the utilized ANN architecture and hyperparameters on the prediction quality.

\appendix

\section{Implementing Periodic Boundary Conditions in Test Cases}
\label{Apx:PBCs - tests}
In this section, we provide further evaluations of the effectiveness of Fourier features in incorporating periodic boundary conditions. Some of these problems are discussed in Ref.\ \cite{DONG2021110242}. In all the problems training is performed using 5000 adam optimization steps, followed by 25000 L-BFGS optimization steps.
\subsection{Approximation of a One-dimensional Periodic Function}
The first selected problem is the ANN approximation of a 1D periodic function. To this end, we select the following function \cite{DONG2021110242}
\begin{align}
u(x)=\sin(2\pi x+0.25 \pi)+\cos(9\pi x+0.10 \pi)-2\sin(7\pi x+0.33 \pi)\,,
\label{E:1Dperfunc}
\end{align}
which is periodic over the domain $\Omega=\{x:0\leq x \leq 2\}$.

In the solution to this problem, ANN density and depth are taken as 50 and 3, respectively. 512 points are sampled from the domain. The learning rate is taken as 0.001 and 0.010.

\begin{figure*}[htb!]
\centering
{{\includegraphics[height=0.22\textwidth,
trim=0 40 0 80, clip]{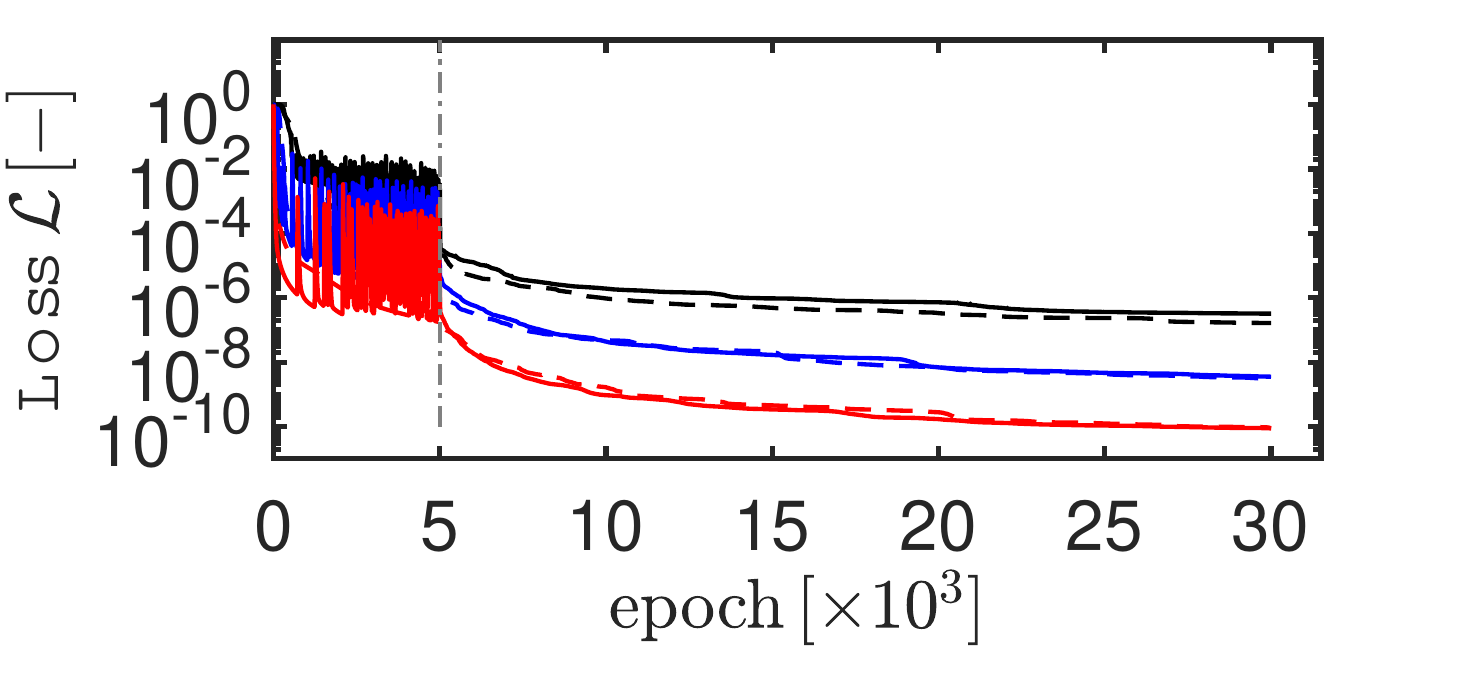}}}
{{\includegraphics[height=0.22\textwidth,
trim=0 40 0 80, clip]{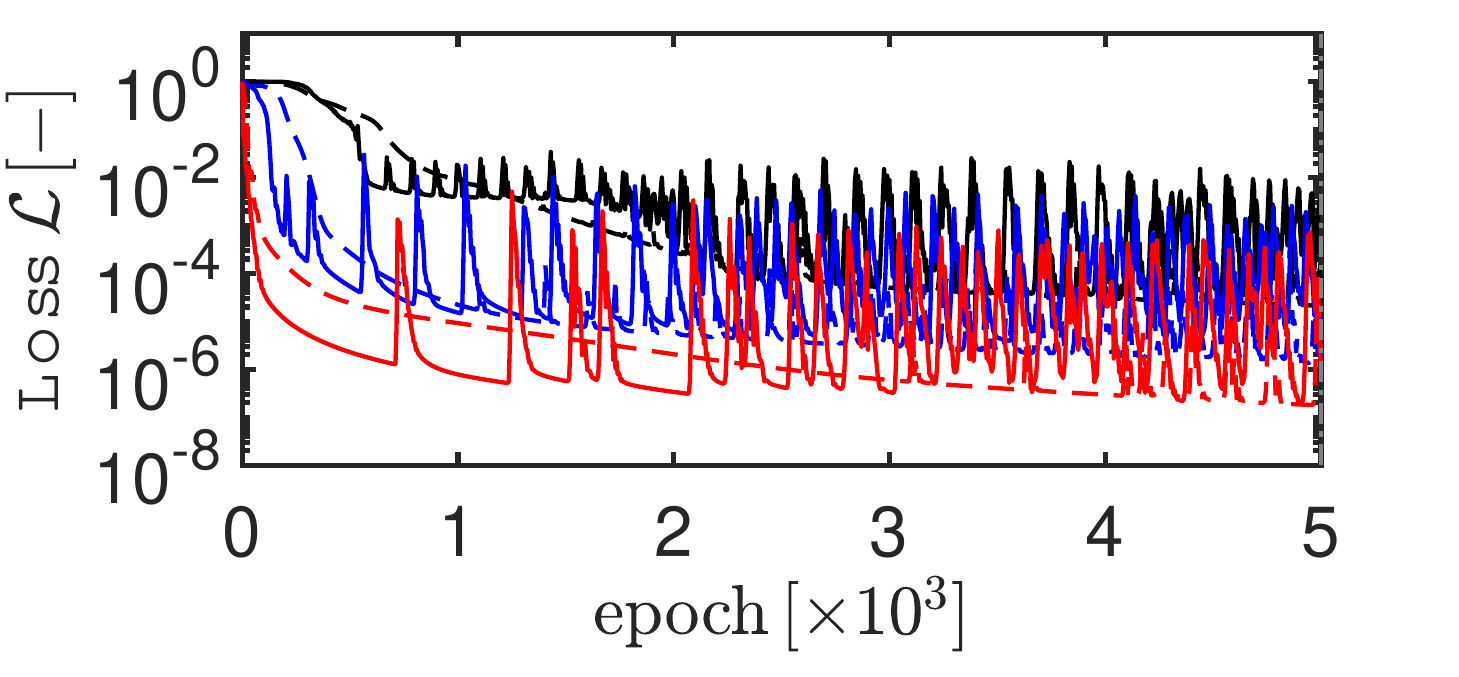}}}
\caption{Training history of the normalized loss function for ANN approximation of the 1D periodic function given in Eq.\ \eqref{E:1Dperfunc}, see Fig.\ \ref{F:function_loss_DONG_NI_1D}. The First 5000 epochs (marked with a vertical dashed line) belong to adam optimizer, whereas the rest to L-BFGS. In the numerical trials learning rates of 0.001 (dashed lines) and 0.010 (continuous lines) are used. Black curves give results for no Fourier feature, whereas dark blue and red for low- and high-frequency Fourier features with a single and the first 10 integer multiples of the reciprocal base vector, respectively.}
\label{F:loss_DONG_NI_1D}
\end{figure*}

\begin{figure*}[htb!]
\centering
\subfigure[DNN approximation to Eq.\ \eqref{E:1Dperfunc}]{
{{\includegraphics[height=0.22\textwidth,
trim=0 40 0 80, clip]{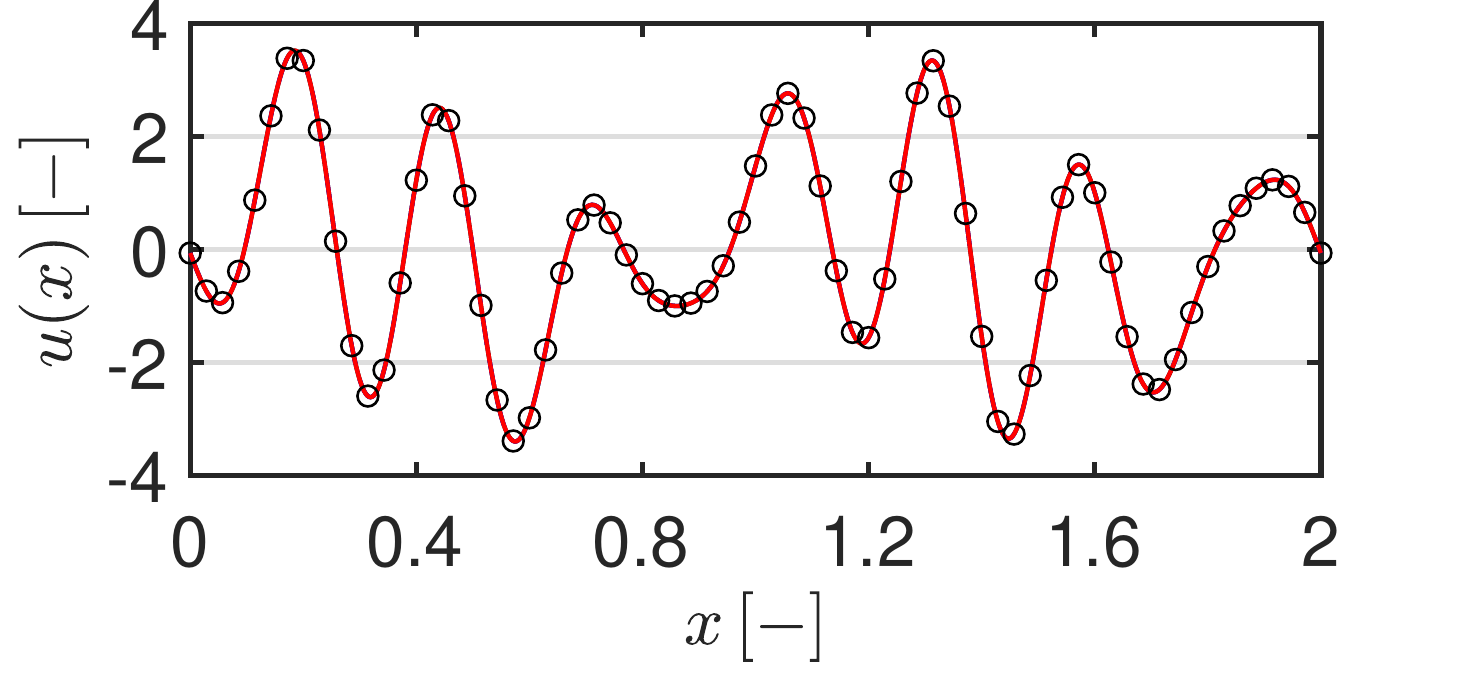}}}
}
\subfigure[absolute error in DNN soln.]{
{{\includegraphics[height=0.22\textwidth,
trim=0 40 0 80, clip]{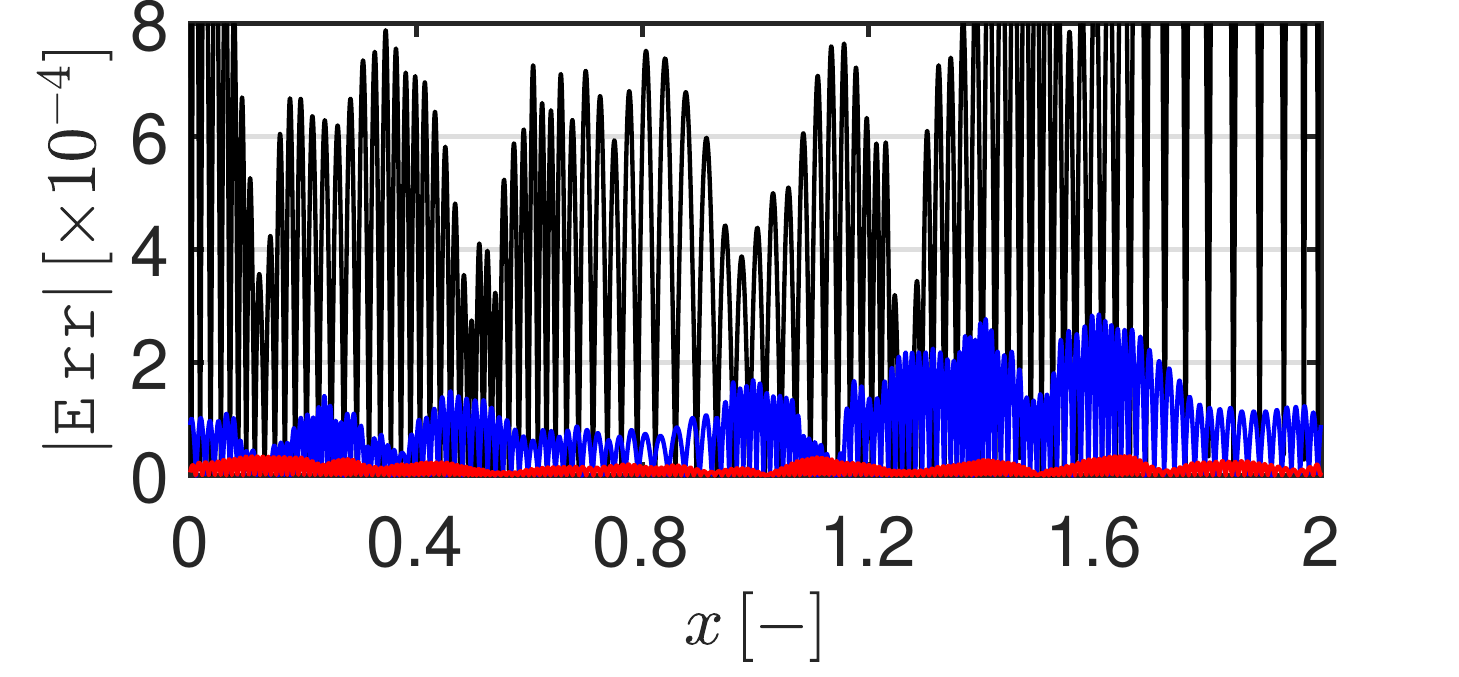}}}
}
\caption{Plots for the 1D periodic function given in Eq.\ \eqref{E:1Dperfunc} and its ANN approximation (left) and absolute error in the ANN approximation  (right) over the problem domain $x\in[0,2]$. Results for a learning rate of 0.010 are demonstrated. Black curves give results for no Fourier feature, whereas dark blue and red for low- and high-frequency Fourier features with a single and the first 10 integer multiples of the reciprocal base vector, respectively.}
\label{F:function_loss_DONG_NI_1D}
\end{figure*}

Figs.\ \ref{F:loss_DONG_NI_1D} and \ref{F:function_loss_DONG_NI_1D} demonstrates that
ANN with high-frequency Fourier features using the first 10 integer multiples of the reciprocal base vector performs the best in approximating the function in both the convergence rate and the final accuracy.
The training loss at the end of 30000 epochs is nearly four orders of magnitude larger for ANNs without Fourier features.

\subsection{Two-dimensional Helmholz Problem with Periodic Boundary Conditions}
In this problem, we solve the following Helmholz equation
\begin{align}
\dfrac{\partial^2 u}{\partial x^2}+\dfrac{\partial^2 u}{\partial y^2}-\lambda u=f(x,y)\,,
\label{E:2D_Helmholz}
\end{align}
which is $x-$ and $y-$periodic over the square domain $\Omega=\{(x,y):0\leq x \leq 4,0\leq y \leq 4\}$. The parameter $\lambda$ is selected as $\lambda=10$. Following Ref.\ \cite{DONG2021110242}, the source term $f(x, y)$ is selected to give the solution $u(x,y)$ of Eq.\ \eqref{E:2D_Helmholz} as
\begin{align}
u(x,y)=-[1.5\cos(\pi x+0.0.4 \pi)+2\cos(2\pi x-0.2 \pi)][1.5\cos(\pi y+0.0.4 \pi)+2\cos(2\pi y-0.2 \pi)]\,.
\label{E:2D_Helmholz_soln}
\end{align}
In the solution to this problem, the ANN density and depth are taken as 50 and 3, respectively.
25600 points are sampled from the domain. The learning rate is taken as 0.010.

\begin{figure*}[htb!]
\centering
{{\includegraphics[height=0.22\textwidth,
trim=0 40 0 80, clip]{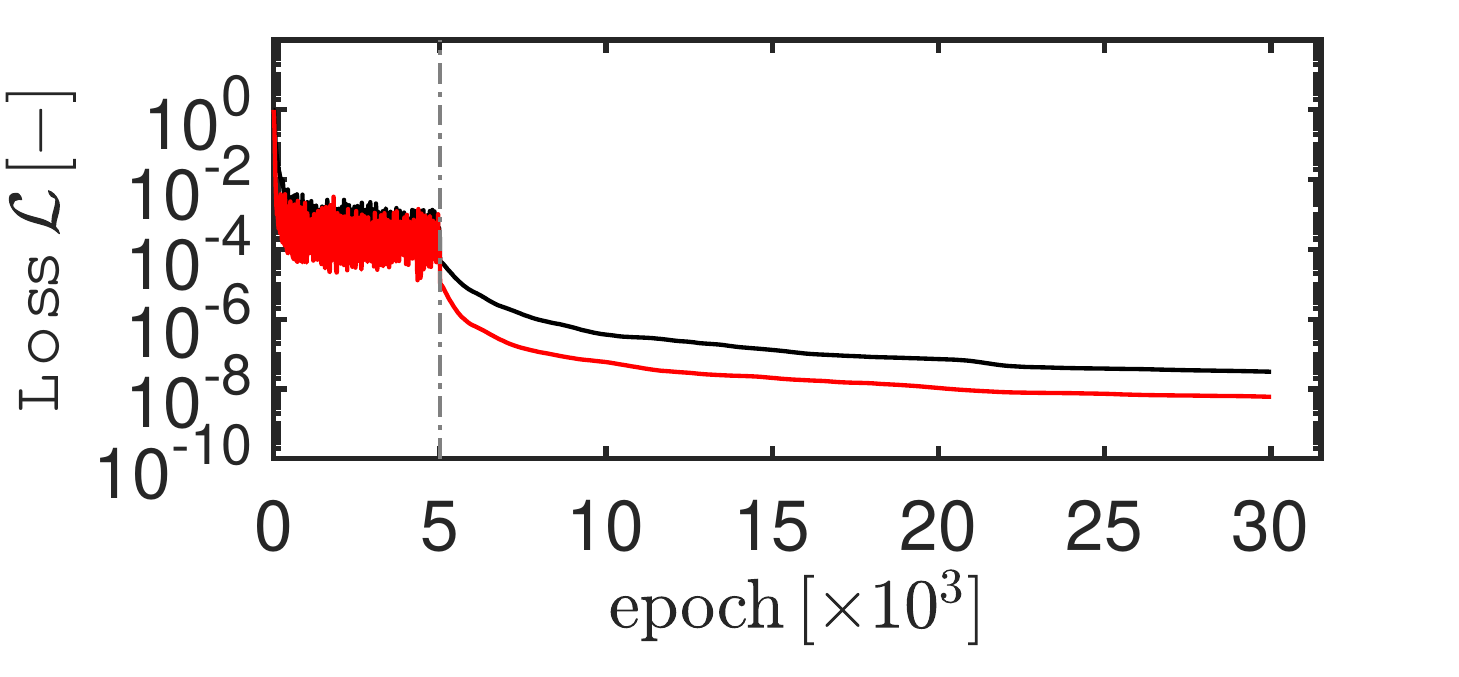}}}
{{\includegraphics[height=0.22\textwidth,
trim=0 40 0 80, clip]{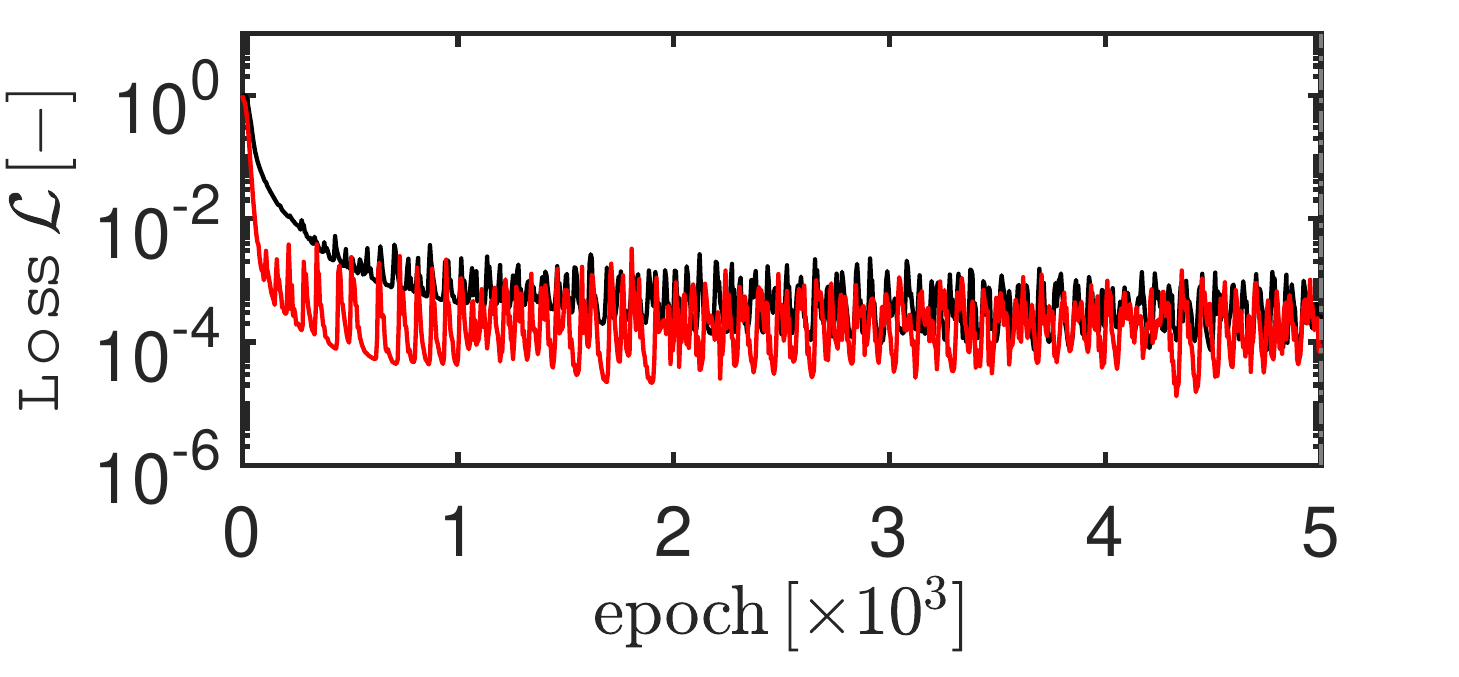}}}
\caption{Training history of the normalized loss function for ANN solution for the 2D Helmholz problem given in Eq.\ \eqref{E:2D_Helmholz_soln}, see Fig.\ \ref{F:2D_Helmholz}. The First 5000 epochs (marked with a vertical dashed line) belong to adam optimizer, whereas the rest to L-BFGS. ANN with low-(black) and high-frequency (red) Fourier features respectively using a single and the first 10 integer multiples of the reciprocal base vector are considered.}
\label{F:loss_2D_Helmholz}
\end{figure*}

\begin{figure*}[htb!]
\centering
\subfigure[low-frequency]{
{\frame{\includegraphics[height=0.22\textwidth,
trim=489 233 489 233, clip]{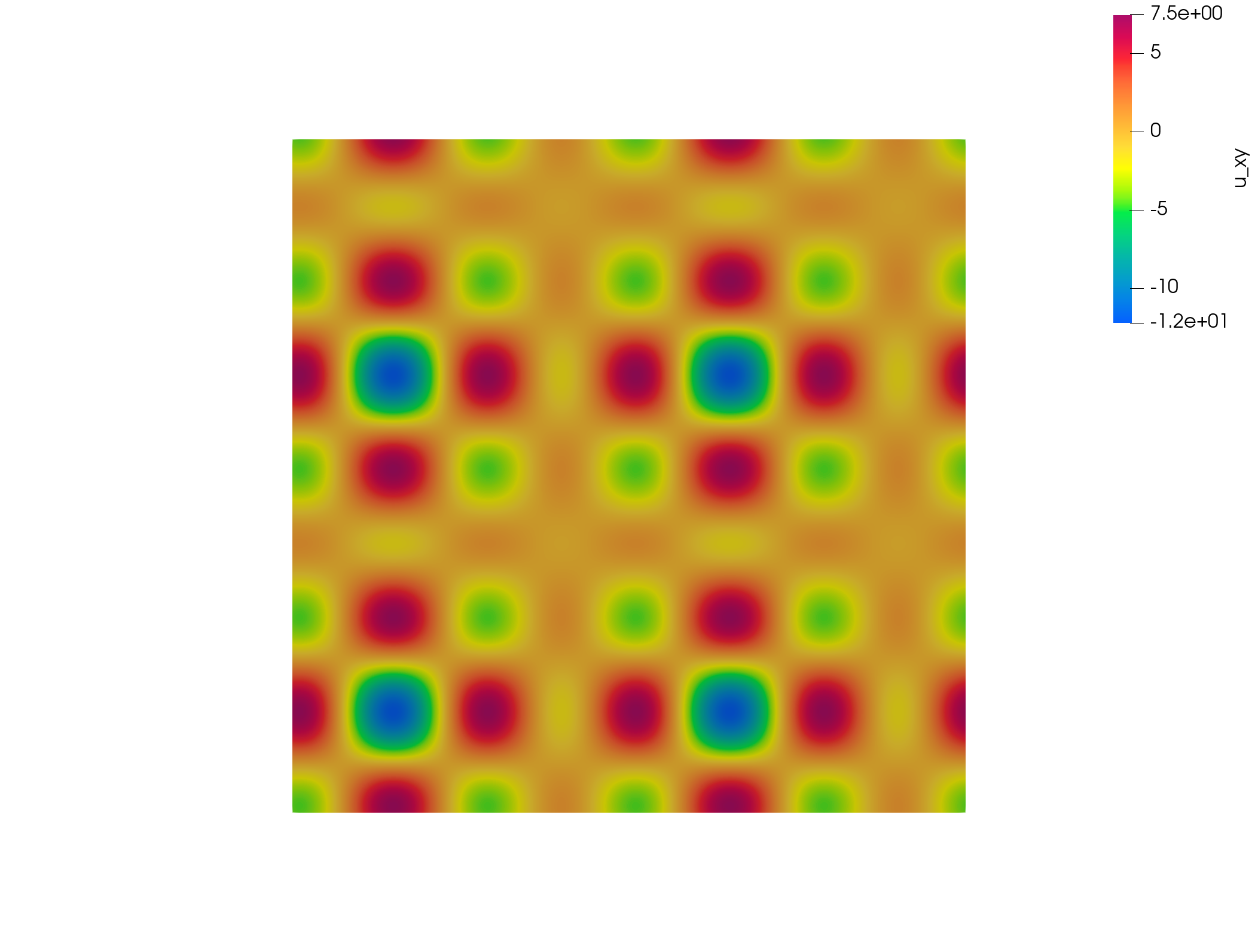}}}
{\frame{\includegraphics[height=0.22\textwidth,
trim=489 233 489 233, clip]{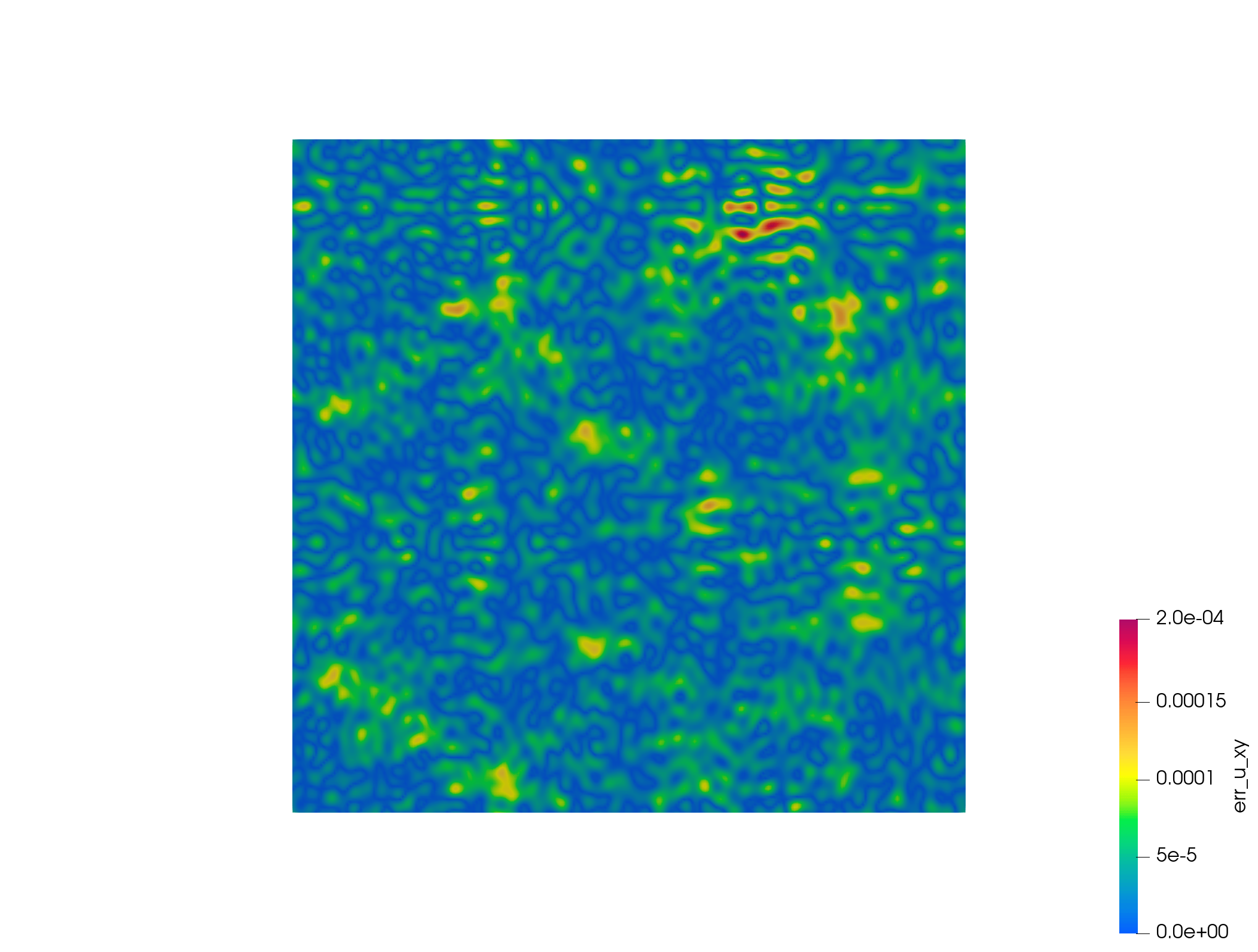}}}
}
\hspace{0.2cm}
\subfigure[high-frequency]{
{\frame{\includegraphics[height=0.22\textwidth,
trim=489 233 489 233, clip]{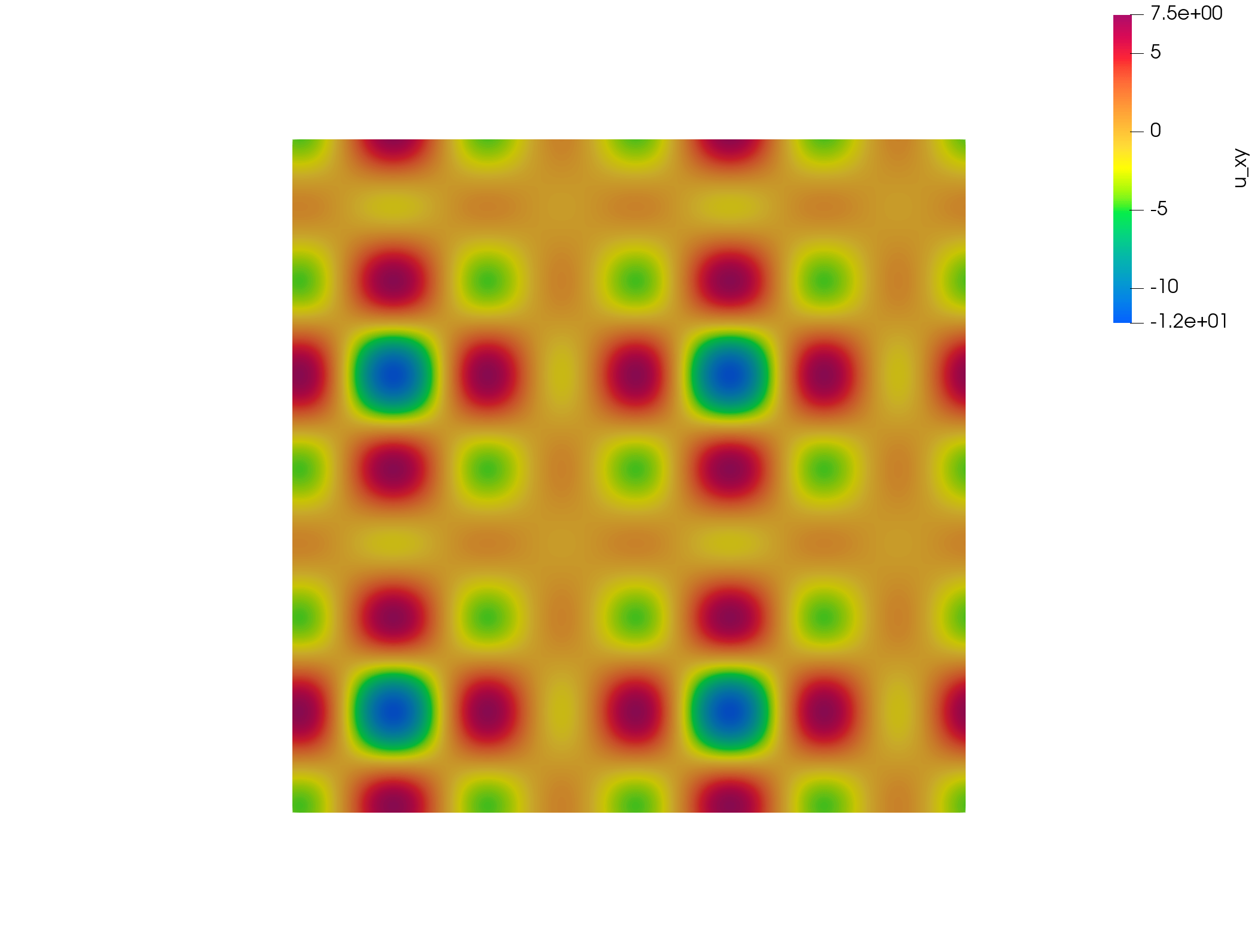}}}
{\frame{\includegraphics[height=0.22\textwidth,
trim=489 233 489 233, clip]{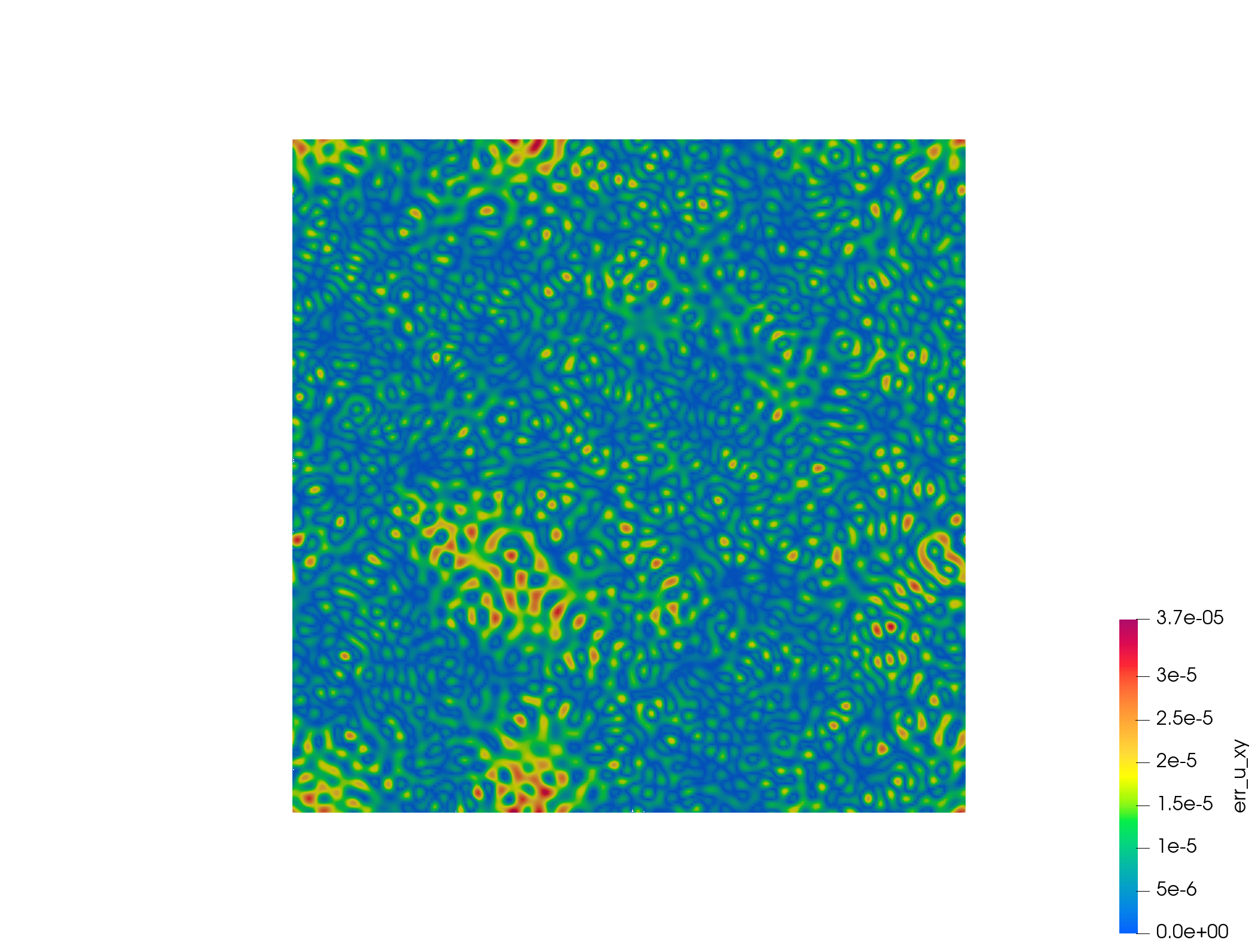}}}
}\\
min$\,\,$\frame{\includegraphics[width=0.30\textwidth, trim=0 15 0 15, clip]{legend_2}}$\,\,$max
\caption{Contours for the ANN solution (left) and its absolute error with respect to the exact solution given in Eq.\ \eqref{E:2D_Helmholz_soln} (right) for the 2D Helmholz problem over $[0,4]\times[0,4]$ problem domain. ANN with low-(a) and high-frequency (b) Fourier features using a single and the first 10 integer multiples of the reciprocal base vector are considered. The intervals [min, max] of the contour plots are $[-12,7.5]$ and $[0,2.0\times10^{-4}]$ for (a) and for $[-12,7.5]$ and $[0,3.7\times10^{-5}]$ (b) for the solution and the absolute error, respectively.}
\label{F:2D_Helmholz}
\end{figure*}

The training loss, the solution, and the error involved with respect to the exact solution are given in Figs.\ \ref{F:loss_2D_Helmholz} and \ref{F:2D_Helmholz}, respectively.
The results demonstrate that
ANN with high-frequency Fourier features using the first 10 integer multiples of the reciprocal base vector performs better in approximating the function in both the convergence rate and the final accuracy.

\subsection{Approximation of a Two-dimensional Periodic Function}
We develop an ANN approximation of a 2D periodic function on a hexagonal lattice in this problem. To this end, we select the following function
\begin{align}
u(x,y)=\tanh(\sin(\rho\,x')+0.1\,\cos(\rho\,x')+0.6\,\cos(\rho\,y'))\,,
\label{E:2Dperfunc}
\end{align}
with the transformed coordinates along hexagonal lattice direct vectors  $x'(x,y)=2\pi/L\,x+2\pi/\sqrt{3}L\,y$ and $y'(x,y)=4\pi/\sqrt{3}L\,y$ which is periodic over the selected hexagonal domain. Here $\rho$ is the frequency. In our solutions, $\rho=1$ and $\rho=10$ are tried. In this problem, $L$ is selected as 200.

In solving this problem, the ANN density and depth are taken as 50 and 3, respectively.
25600 points are sampled from the domain. The learning rate is taken as 0.010.

\begin{figure*}[htb!]
{{\includegraphics[height=0.20\textwidth,
trim=489 150 489 150, clip]{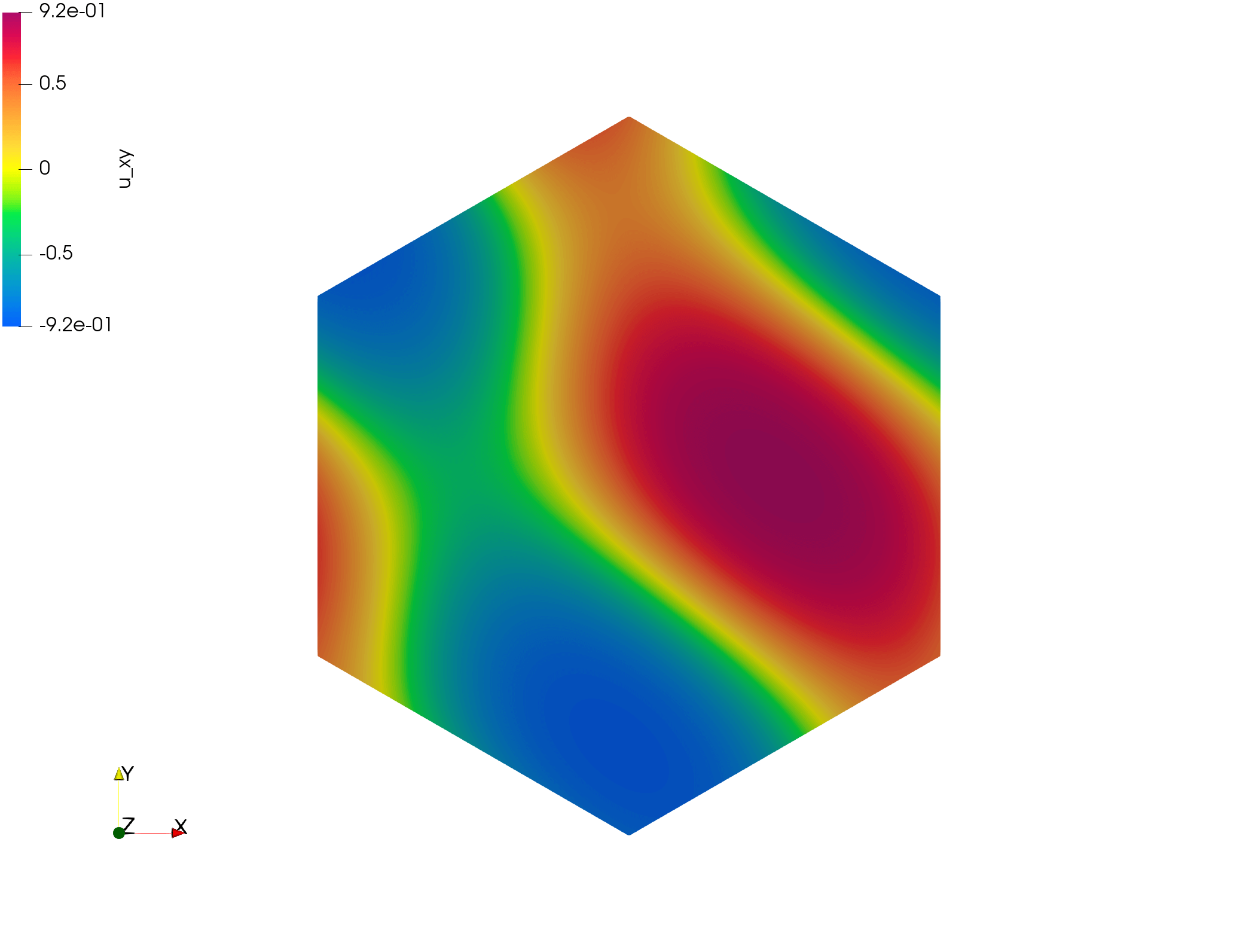}}}
{{\includegraphics[height=0.20\textwidth,
trim=489 150 489 150, clip]{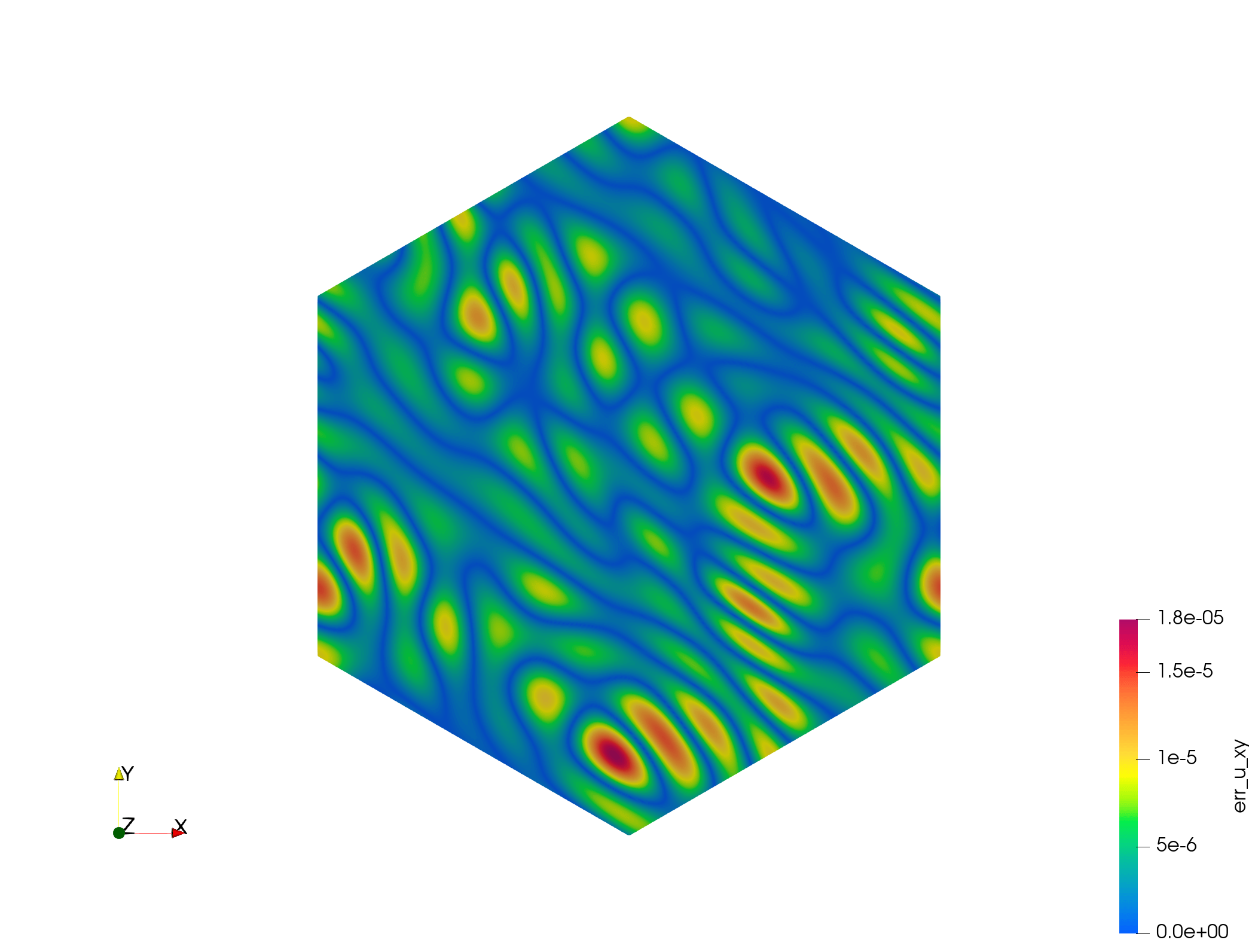}}}
\hspace{0.2cm}
{{\includegraphics[height=0.20\textwidth,
trim=350 300 350 300, clip]{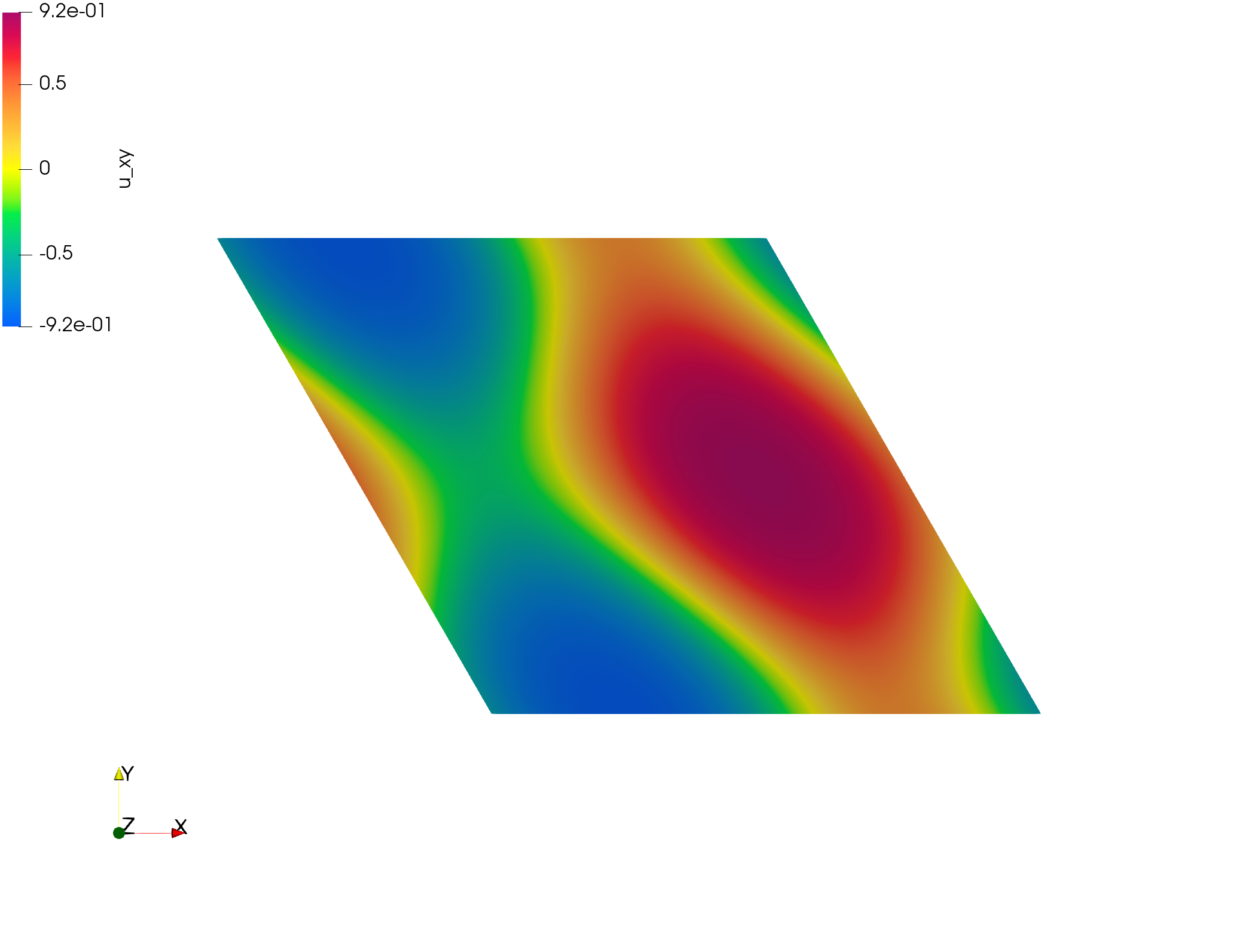}}}
{{\includegraphics[height=0.20\textwidth,
trim=350 300 350 300, clip]{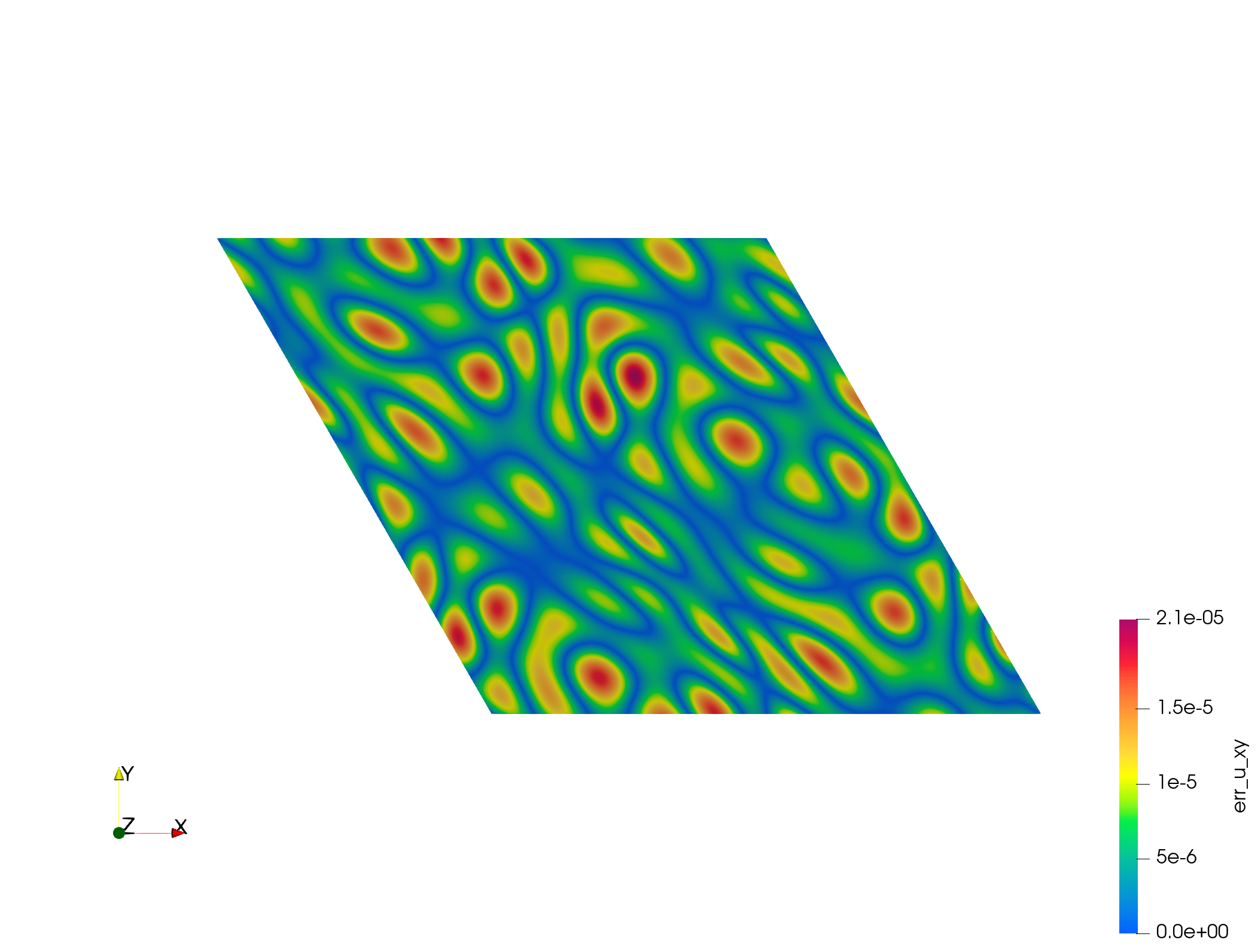}}}
\\
{{\includegraphics[height=0.20\textwidth,
trim=489 150 489 150, clip]{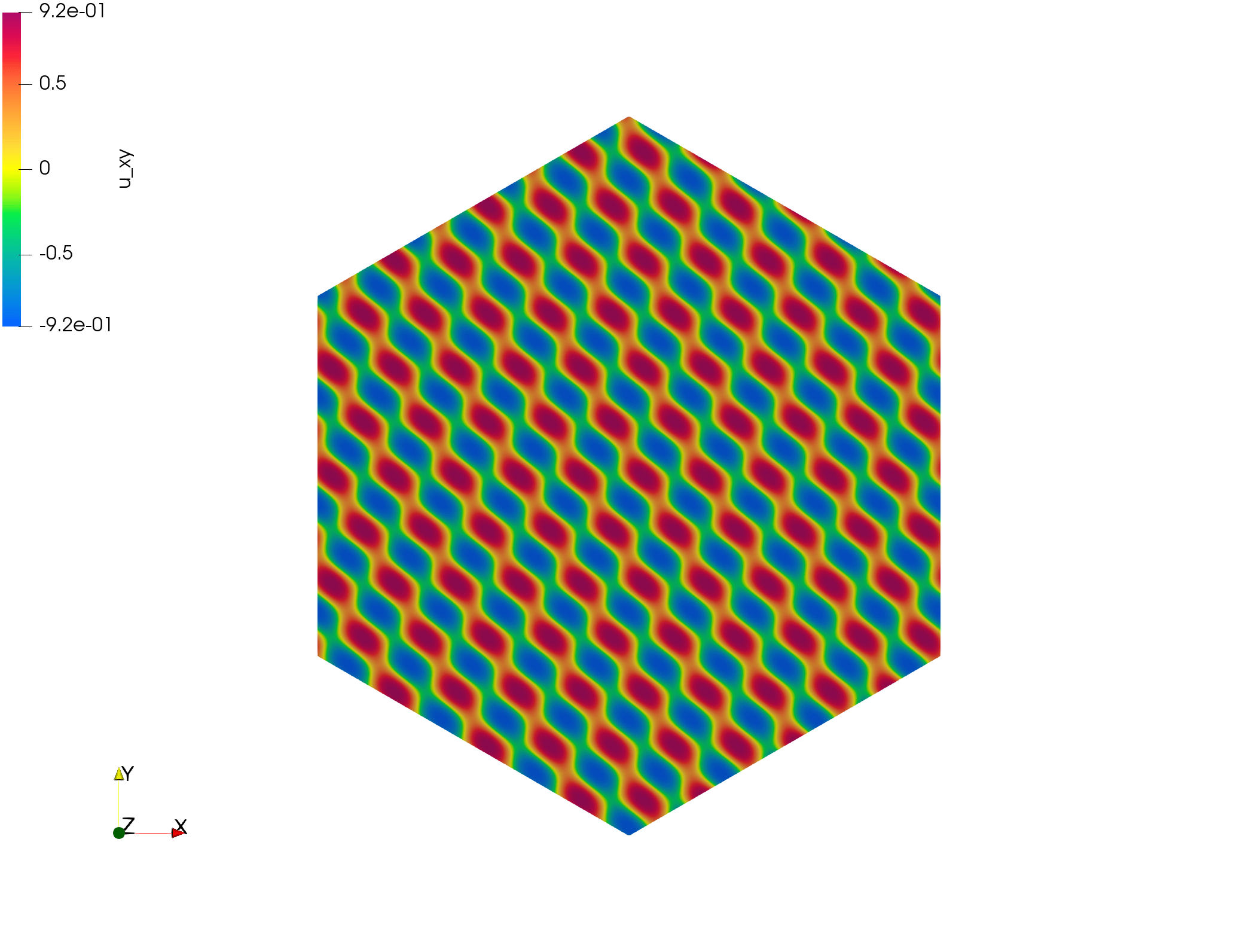}}}
{{\includegraphics[height=0.20\textwidth,
trim=489 150 489 150, clip]{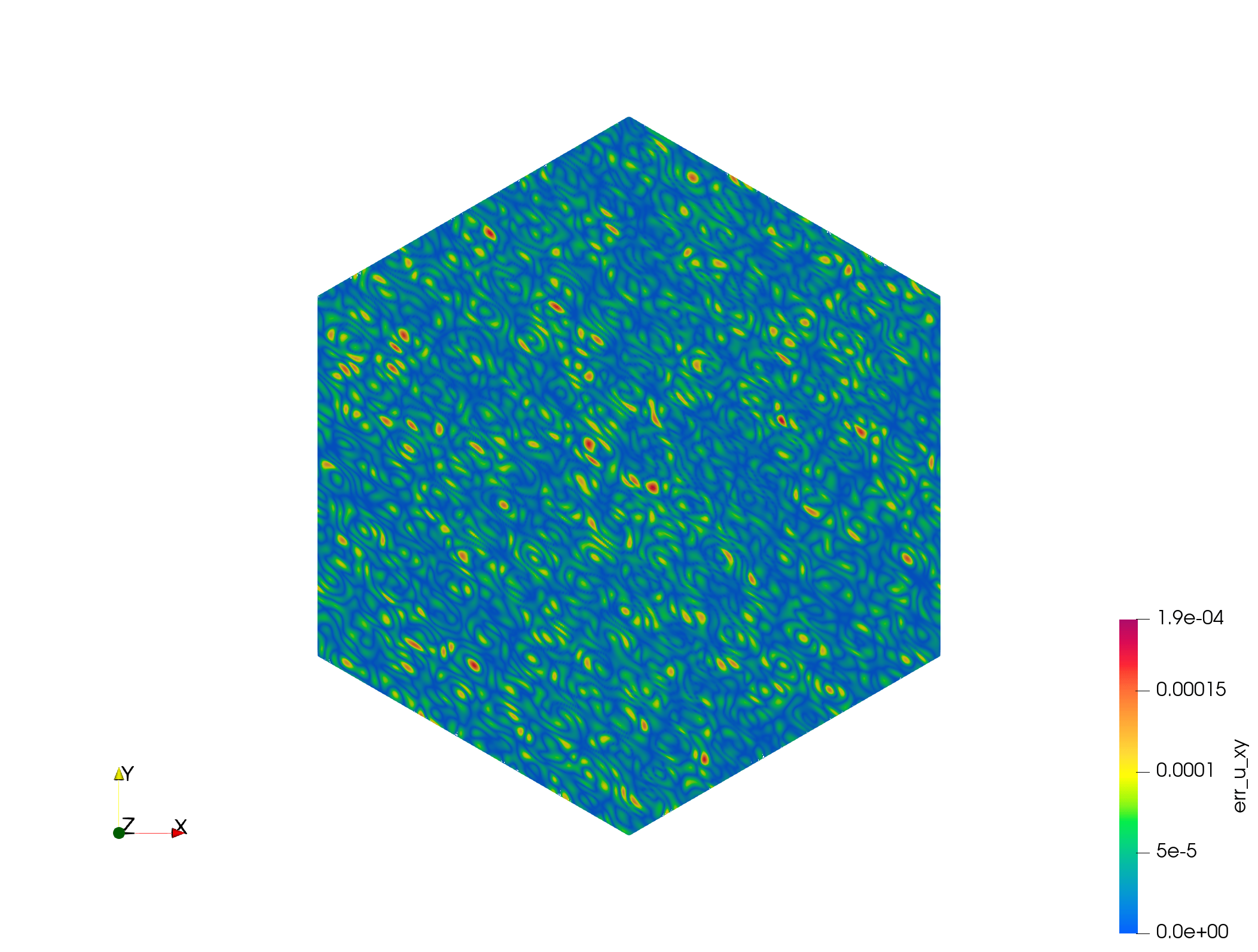}}}
\hspace{0.2cm}
{{\includegraphics[height=0.20\textwidth,
trim=350 300 350 300, clip]{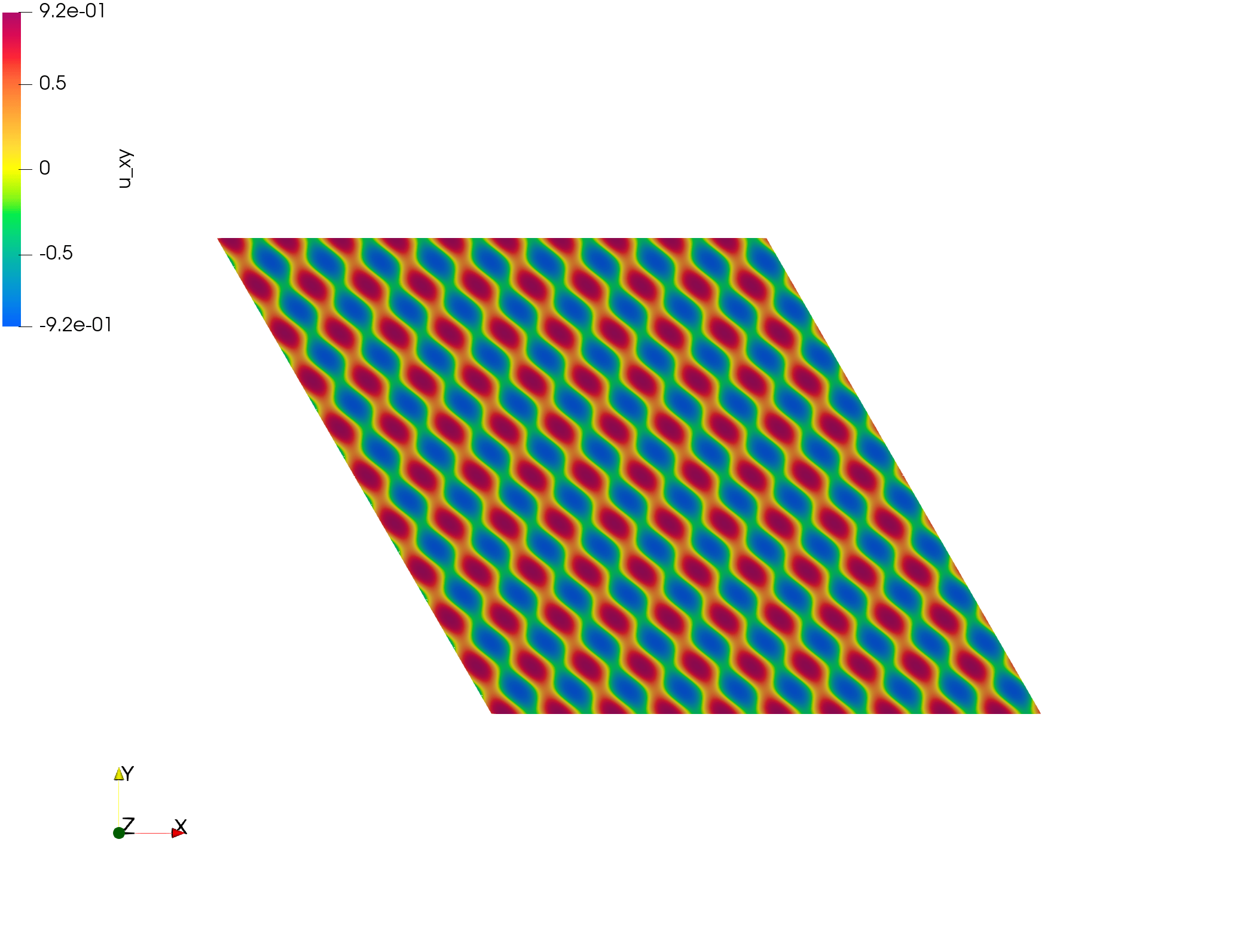}}}
{{\includegraphics[height=0.20\textwidth,
trim=350 300 350 300, clip]{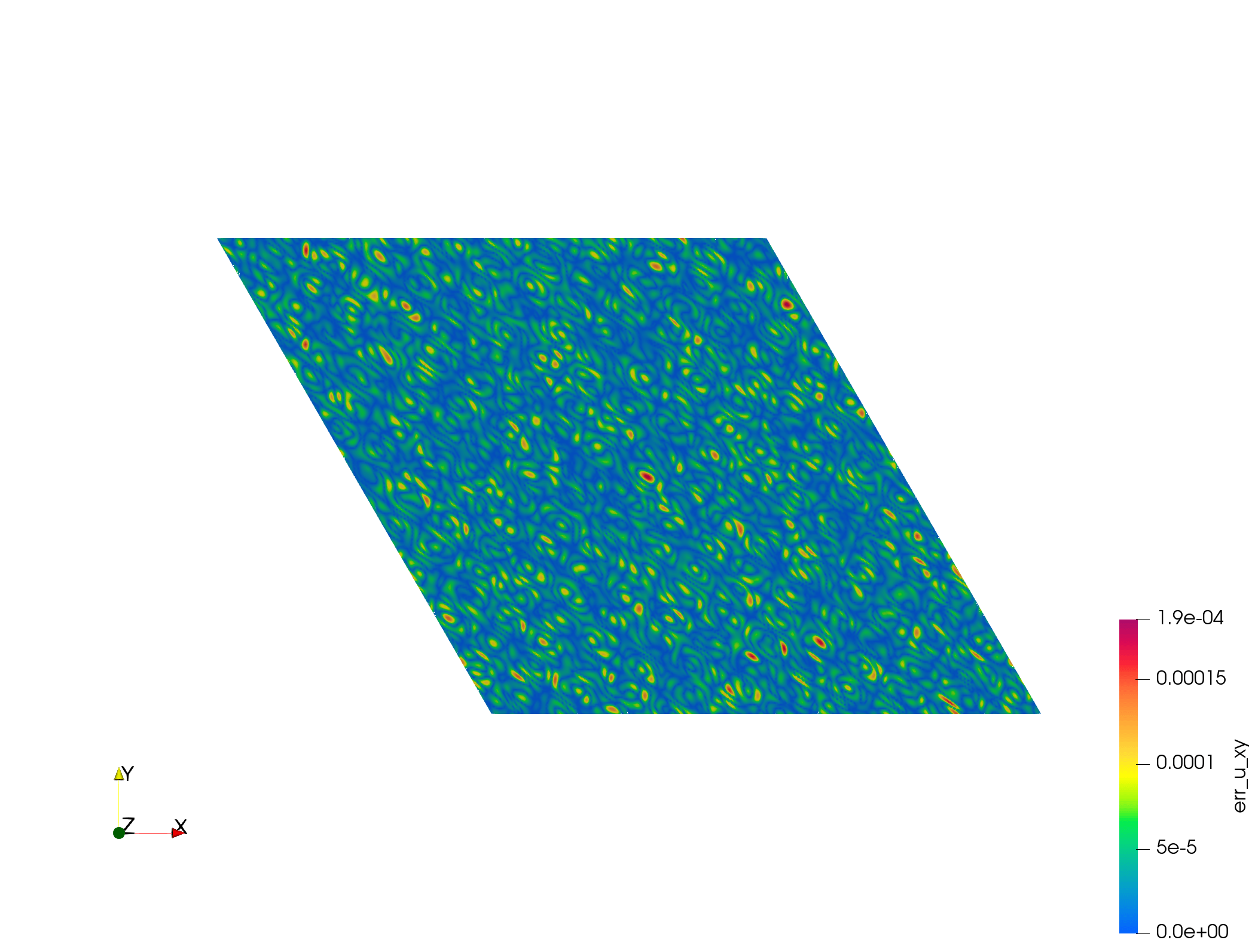}}}\\
\centering
min$\,\,$\frame{\includegraphics[width=0.30\textwidth, trim=0 15 0 15, clip]{legend_2}}$\,\,$max
\caption{Contours for the ANN solution (left) and its absolute error (right) for the approximation of the 2D periodic function given in Eq.\ \eqref{E:2Dperfunc} for $\rho=1$ (top) and $\rho=10$ (bottom).
Two primitive unit cells $\mathcal H_1$ and $\mathcal H_2$ with a hexagonal Bravais lattice are considered. The intervals [min, max] of the contour plots for the top row:
$[-9.2,9.2]$ and $[0,1.8\times10^{-5}]$,  $[-9.2,9.2]$ and $[0,2.1\times10^{-5}]$,
bottom row:
$[-9.2,9.2]$ and $[0,1.9\times10^{-4}]$,  $[-9.2,9.2]$ and $[0,1.9\times10^{-4}]$ for the solution and the absolute error, respectively.}
\label{F:2D_hex_periodic}
\end{figure*}

\begin{figure*}[htb!]
\centering
{{\includegraphics[width=0.45\textwidth,
trim=0 100 0 80, clip]{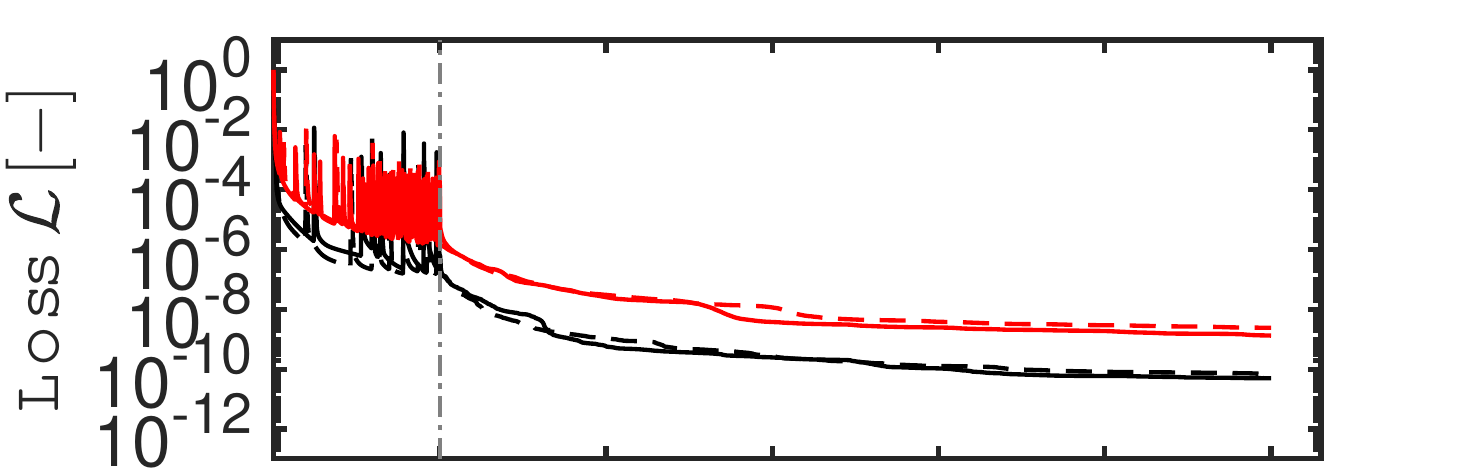}}}
{{\includegraphics[width=0.45\textwidth,
trim=0 100 0 80, clip]{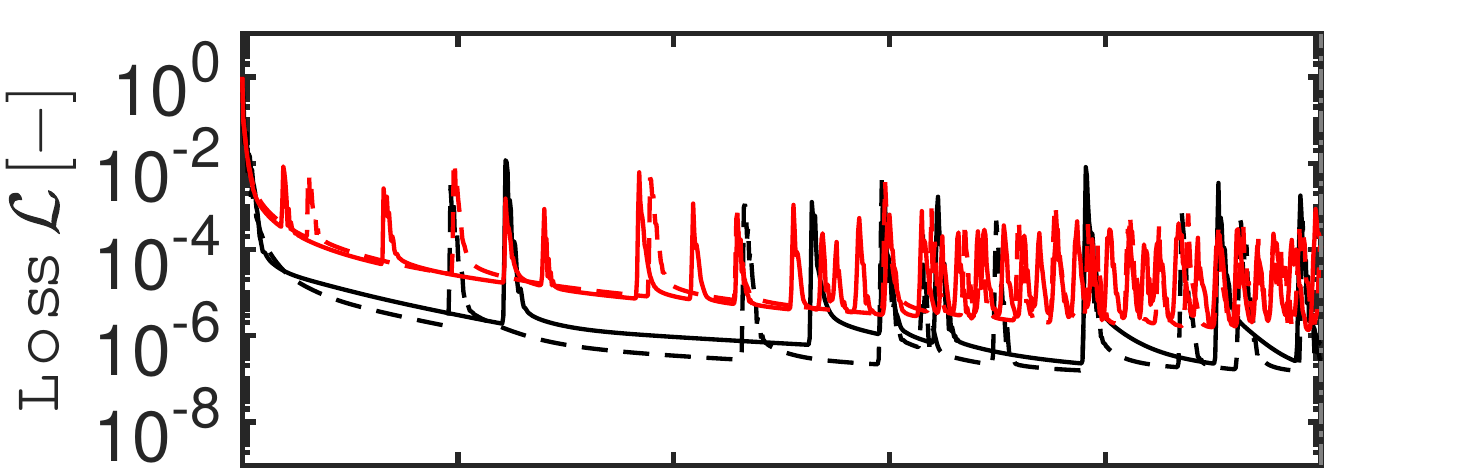}}}\\
{{\includegraphics[width=0.45\textwidth,
trim=0 40 0 80, clip]{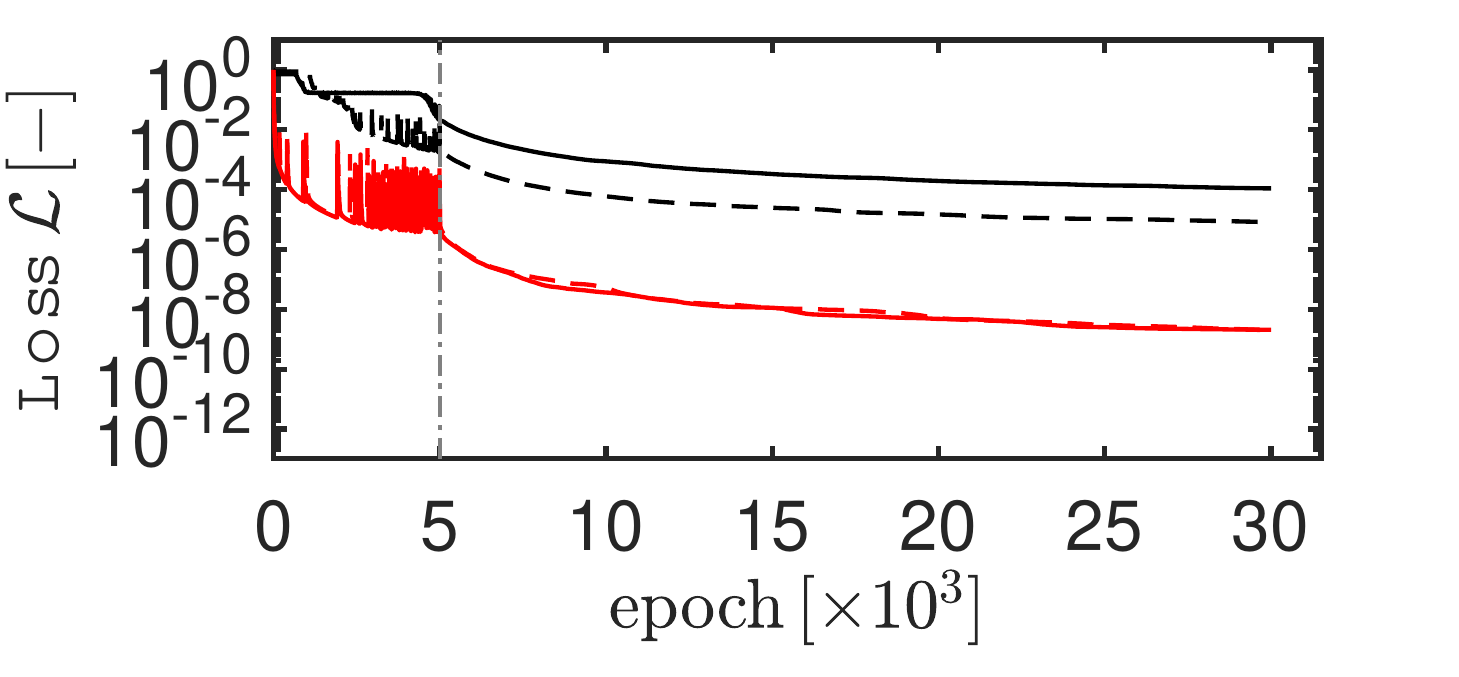}}}
{{\includegraphics[width=0.45\textwidth,
trim=0 40 0 80, clip]{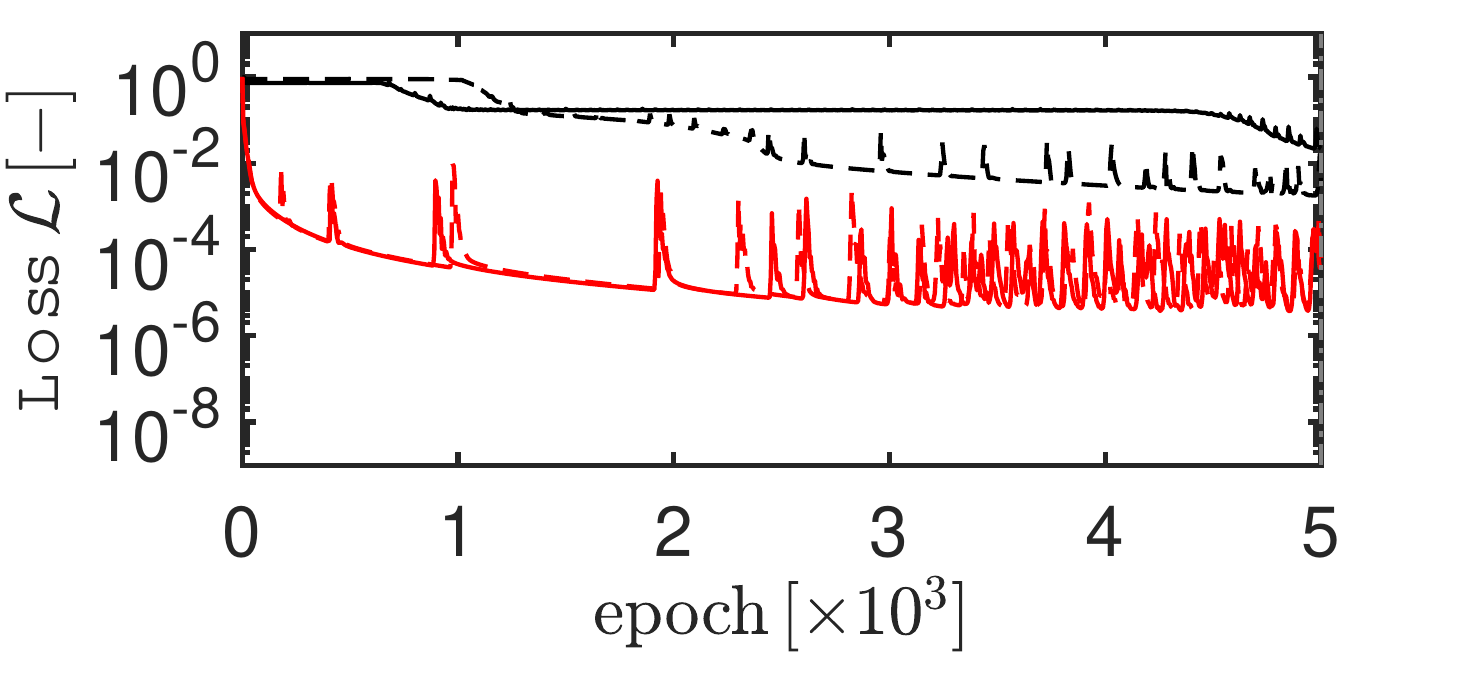}}}\\
\caption{Training history of the normalized loss function for the approximation of the 2D periodic function given in Eq.\ \eqref{E:2Dperfunc} for $\rho=1$ (top) and $\rho=10$ (bottom) and two primitive unit cells $\mathcal H_1$ (continuous) and $\mathcal H_2$ (dashed) considering ANNs with low-(black) and high-frequency (red) Fourier features respectively using a single and the first 10 integer multiples of the reciprocal base vector.}
\label{F:results_hex_periodic_training_histories}
\end{figure*}

The solution and error involved with respect to the exact solution and the training losses are given in Figs.\ \ref{F:2D_hex_periodic} and \ref{F:results_hex_periodic_training_histories}.
The results show that using low-frequency Fourier features improves the performance of the ANN when $\rho=1$ while using high-frequency Fourier features with the first 10 integer multiples of the reciprocal base vector improves performance for $\rho=10$ in both convergence rate and  accuracy.

\setcounter{table}{0}
\renewcommand{\thetable}{B\arabic{table}}

\section{Further Results}
\label{S:tabulated_results}
In this section, additional predictions made using Artificial Neural Networks (ANN) are presented in Tables\ \ref{T:sq_rand_effectiveprops}1, \ref{T:cr_results}2 and \ref{T:3D_C_results}3.

\begin{table}[htb!]
\label{T:sq_rand_effectiveprops}
\caption{ANN predictions with two different  architectures for the source term $f^1$ and $f^2$ for the random and periodic inclusion distribution considering a square lattice. For $f^1$  the subscript  $i=1$, otherwise $i=2$. Results are obtained using high-frequency Fourier features with the first 10 integer multiples of the reciprocal base vector.}
\centering
\begin{tabular}{cccrr}
\hline
source term & architecture & training ID & $a^\star_{1i}$ & $a^\star_{2i}$ \\
\hline
\multirow{6}{*}{$f^1$} &
\multirow{3}{*}{50$\times$3}  & 1  & 2.29216298 & 0.03443467 \\
&                             & 2  & 2.31131626 & 0.02962746 \\
&                             & 3  & 2.30685852 & 0.00065440 \\
&\multirow{3}{*}{100$\times$6}  & 1 & 2.29740728  & $-0.02406874$  \\
&                               & 2 & 2.29496328  & $-0.02575189$  \\
&                               & 3 & 2.29844449  & $-0.02840626$  \\
\hline
\multirow{6}{*}{$f^2$} &

\multirow{3}{*}{50$\times$3} & 1  & $-0.01858249$ & 2.42691608 \\
&                            & 2  & $-0.03353900$ & 2.45424801 \\
&                            & 3  & 0.00227923 & 2.42688176  \\
& \multirow{3}{*}{100$\times$6}  & 1  & $-0.03044871$  & $2.39939146$  \\
&                            & 2 & $-0.02986018$  & $2.40368601$  \\
&                            & 3 & $-0.02852686$  & $2.40053253$  \\
\hline
\end{tabular}
\end{table}

\begin{table}[htb!]
\label{T:cr_results}
\caption{ANN predictions for the source term $f^1$ and $f^2$ considering the three different unit cells and centered rectangular lattice. For $\mathcal{R}^\mathrm{c}_1$ and $\mathcal{R}^\mathrm{c}_2$, a learning rate of 0.001 with 100$\times$6 ANN architecture possessing low-frequency Fourier features is used, whereas, for $\mathcal{R}^\mathrm{c}_3$, a learning rate of 0.010 with 50$\times$3 ANN architecture possessing high-frequency Fourier features with the first 10 integer multiples of the reciprocal base vector. For $f^1$ the subscript $i=1$, otherwise $i=2$.}
\centering
\begin{tabular}{cccrr}
\hline
source term & unit cell & training ID & $a^\star_{1i}$ & $a^\star_{2i}$  \\
\hline
\multirow{9}{*}{$f^1$}
& \multirow{3}{*}{$\mathcal{R}^\mathrm{c}_1$}     & 1 & 1.74271078 &  0.00048956  \\
&                         & 2 & 1.73539798 &  0.00427126  \\
&                         & 3 & 1.74248766 &  0.00074224  \\
& \multirow{3}{*}{$\mathcal{R}^\mathrm{c}_2$}     & 1 & 1.74130306 & $-0.00138061$  \\
&                         & 2 & 1.74509759 &  0.00094600  \\
&                         & 3 & 1.74036560 & $-0.00070005$  \\
& \multirow{3}{*}{$\mathcal{R}^\mathrm{c}_3$}     & 1 & 1.73772220 & $-0.00001145$   \\
&                         & 2 & 1.73706130 & $-0.00000122$   \\
&                         & 3 & 1.73864375 & $-0.00001003$   \\
\hline
\multirow{9}{*}{$f^2$}
& \multirow{3}{*}{$\mathcal{R}^\mathrm{c}_1$}     & 1 & 0.00604348 &  2.27994320   \\
&                         & 2 & 0.00377181 &  2.28445564  \\
&                         & 3 & 0.00283645 &  2.28166015  \\
&\multirow{3}{*}{$\mathcal{R}^\mathrm{c}_2$}     & 1 &  0.00117976 & 2.28162650   \\
&                         & 2 & $-0.00035555$ & 2.28630918  \\
&                         & 3 &  0.00227585 & 2.28578361  \\
&\multirow{3}{*}{$\mathcal{R}^\mathrm{c}_3$}     & 1 & $-0.00120572$ & 2.29766637  \\
&                         & 2 &  0.00162594 & 2.29669590    \\
&                         & 3 &  0.00067097 & 2.28941049   \\
\hline
\end{tabular}
\end{table}

\begin{table}[htb!]
\label{T:3D_C_results}
\caption{ANN predictions for the source term $f^1$ for considering the two different unit cells $\mathcal{C}_1$ and $\mathcal{C}_2$ and simple cubic lattice. A learning rate of 0.010 with 50$\times$3 ANN architecture is utilized with low-frequency and high-frequency Fourier features, respectively possessing a single and first 10 integer multiples of the reciprocal base vector.}
\centering
\begin{tabular}{cccrrr}
\hline
unit cell & Fourier features & training ID & $a^\star_{11}$ & $a^\star_{21}$ & $a^\star_{31}$ \\
\hline
\multirow{6}{*}{$\mathcal{C}_1$}
& \multirow{3}{*}{low-frequency}   & 1 & 2.59757603 &  0.00951802 &  0.00804197 \\
&                         & 2 & 2.63482571 & $-0.00019953$ & $-0.00224011$ \\
&                         & 3 & 2.53020037 &  0.00050227 &  0.00174256 \\
& \multirow{3}{*}{high-frequency}  & 1 & 2.45337680 & 0.00009547 &  0.00006592  \\
&                         & 2 & 2.45661833 & 0.00023308 & $-0.00030972$  \\
&                         & 3 & 2.45504276 & 0.00000776 & $-0.00020233$  \\
\hline
\multirow{6}{*}{$\mathcal{C}_2$}
&\multirow{3}{*}{low-frequency}   & 1 & 2.55916996 &  0.00283198 & 0.00495483 \\
&                         & 2 & 2.65327004 &  0.00095332 & 0.00479847 \\
&                         & 3 & 2.59437703 & $-0.00665691$ & 0.00356463 \\
&\multirow{3}{*}{high-frequency}  & 1 & 2.45392204 & 0.00009346 &  0.00023590  \\
&                         & 2 & 2.45781939 & 0.00010920 & $-0.00023763$  \\
&                         & 3 & 2.45432994 & 0.00011981 &  0.00010465  \\
\hline
\end{tabular}
\end{table}

\section{Analytical Solutions}
\subsection{Two-dimensional Square Circular Disk Distributions}
\label{section:2D_analytics}
Square and periodic arrangement of disks possess isotropic effective properties once the considered property tensors are of second order \cite{Nye1985}. Thus, the effective property tensor $\boldsymbol a^\star$ can be represented by a scalar $a^\star$ where $\boldsymbol a^\star=a^\star\,\boldsymbol 1$. For a matrix with circular inclusions with the square and periodic arrangement and respective properties denoted by $a_{\mathrm{m}}$ and $a_{\mathrm{i}}$, the truncated series expansion solution of Godin\ \cite{Godin2013} cited in \cite{Ren2018}, gives the analytical expression for $a^\star$ as
\begin{equation}
a^\star=\left[\dfrac{1+ \phi_\mathrm{i}\,\lambda(\phi_\mathrm{i})}{1-\phi_\mathrm{i}\,\lambda(\phi_\mathrm{i})}\right]\,a_{\mathrm{m}}
\label{E:analytical_solution}
\end{equation}
with
\begin{equation*}
\lambda(\phi_i)=c_0+c_4\phi^4_\mathrm{i}+c_8\phi_\mathrm{i}^8+c_{12}\phi^{12}_\mathrm{i}+O(\phi^{16}_\mathrm{i})\,.
\end{equation*}
Here, the coefficients $c_i$ are functions of $\alpha=[a_\mathrm{i}-a_\mathrm{m}][a_\mathrm{i}+a_\mathrm{m}]$ with
\begin{align*}
c_0&=\alpha\,,\\
c_4&=0.305827\alpha^3\,,\\
c_8&=\alpha^3[0.0935304\alpha^2+0.0133615]\,,\\
c_{12}&=\alpha^3[0.0286042\alpha^4+0.437236\alpha^2+0.000184643]\,.
\end{align*}

\subsection{Three-dimensional Cubic Sphere Distributions}
\label{section:3D_analytics}
Coming to the sphere distributions in 3D,  the truncated series expansion solution of Lam\ \cite{Lam1986} cited in \cite{Ren2018}, gives the analytical expression for $a^\star$ as
\begin{equation}
a^\star=\left[1+\dfrac{3\phi_\mathrm{i}}{\gamma(\phi_\mathrm{i})}\right]\,a_m
\label{E:analytical_solution_2}
\end{equation}
with
\begin{align*}
\gamma(\phi_\mathrm{i})& =-\dfrac{1}{R_1}
-\phi_\mathrm{i}
+1.3045 R_3 \phi_\mathrm{i}^{10/3}
+0.0723 R_5 \phi_\mathrm{i}^{14/3}\\
& -0.5289 R^2_3 \phi_\mathrm{i}^{17/3}
+0.1526 R_7 \phi_\mathrm{i}^{6}
+O(\phi^{7}_\mathrm{i})\,,
\end{align*}
where
\begin{align*}
R_n=\dfrac{n[a_\mathrm{m}-a_\mathrm{i}]}{
[n+1]a_\mathrm{m}+na_\mathrm{i}}\,.
\end{align*}
\bibliography{bibliography}
\end{document}